%% file: arbeit.tex
\documentclass[11pt,twoside]{report}
\usepackage{epsfig}

\def\lessim{\mathrel {\vcenter {\baselineskip 0pt \kern 0pt
\hbox{$<$} \kern 0pt \hbox{$\sim$} }}}
\def\gessim{\mathrel {\vcenter {\baselineskip 0pt \kern 0pt
\hbox{$>$} \kern 0pt \hbox{$\sim$} }}}

\begin{document}

\setlength{\baselineskip}{20pt}  

\bibliographystyle{unsrt}

\pagenumbering{roman}

\input{title.tex}
\newpage

\vspace{2.0in}					
$\rule{0mm}{8mm}$ \newpage    

\addcontentsline{toc}{chapter}{Abstract}

\setlength{\topmargin}{-0.50in}
\setlength{\oddsidemargin}{0.50in}		
\setlength{\evensidemargin}{0.0in}		
\input{abstract.tex} \newpage			

\addcontentsline{toc}{chapter}{Acknowledgments}
\input{acknow.tex}

\tableofcontents
\listoftables
\listoffigures
\newpage

\pagestyle{headings}
\pagenumbering{arabic}
\chapter{Introduction}
\input{intro.tex}

\chapter{Experimental Apparatus}
\input{apparat.tex}

\chapter{Selection of Candidate Events}
\input{select.tex}

\chapter{Geometric and Kinematic Acceptance}
\input{geom_kin.tex}

\chapter{Efficiency Corrections}
\input{effic.tex}

\chapter{Branching Fraction Measurements}
\input{br_fract.tex}

\chapter{Conclusions}
\input{concl.tex}

\appendix


\chapter{Level 2 Dimuon Triggers}
\input{l2_dimuon_trig.tex}

\chapter{CDF Track Characterization}
\input{trk_char.tex}

\chapter{Pattern Recognition Efficiencies}
\input{patt_rec_eff.tex}

\chapter{The CDF Collaboration}
\label{app:cdf_collab}
\input{cdf_runi_authors.tex}

\newpage
\addcontentsline{toc}{chapter}{Bibliography}
\bibliography{references}

\end{document}

%% file: title.tex
\thispagestyle{empty}
\begin{center}
{\large
{\bf A STUDY OF EXCLUSIVE NONLEPTONIC DECAYS OF $B$ MESONS INTO FINAL STATES OF
STRANGE MESONS AND $1S$ OR $2S$ CHARMONIA}\\
\vspace{1.25in}
\normalsize
{\bf by}\\
\vspace{1.25in}
{\bf Andreas~T.~Warburton}\\
\vspace{1.75in}
A thesis submitted in conformity with the requirements\\
for the degree of Doctor of Philosophy,\\
Graduate Department of Physics,\\
in the University of Toronto\\
\vspace{1.25in}
{\normalsize \copyright\ Copyright by Andreas~T.~Warburton, 1998}
}
\end{center}

%% file: abstract.tex
\begin{center}
{\bf Abstract}
\end{center}
Bound states of heavy quarks can serve as a laboratory for inquiry
into the behaviour of the fundamental strong and electroweak
interactions.  This thesis examines observations of $B^0$,
$\overline{B^0}$, and $B^\pm$ mesons produced in proton-antiproton
collisions at centre-of-mass energies of $\sqrt{s}=1.8$~TeV.  The
$B$-meson decay products are recorded using the Collider Detector at
Fermilab (CDF) located on the Tevatron collider at the Fermi National
Accelerator Laboratory in Batavia, Illinois, USA.

Four $B$-meson decays and their charge conjugates are studied: $B^+\to
J/\psi \, K^+$, $B^0 \to J/\psi \, K^*(892)^0$, $B^+\to\psi(2S)\,K^+$,
and $B^0\to\psi(2S)\,K^*(892)^0$.  Using a data sample corresponding
to $\int\!{\cal L}\,dt = (109 \pm 7)$~pb$^{-1}$, statistically
significant signals are observed in all the channels.  Topological
similarities between the $B$ decays are exploited to measure the six
relative branching fractions (${\cal B}$) of each channel with respect
to the other channels.  The ratios involving the $B^+\to J/\psi\,K^+$
mode are
\begin{eqnarray}
\frac{{\cal B}(B^0\to J/\psi\,K^*(892)^0)}
		{{\cal B}(B^+\to J/\psi\,K^+)} &=&
		1.76\pm 0.14 {\rm [stat]}\pm 0.15 {\rm [syst]}\nonumber
		\\[0.10in]
\frac{{\cal B}(B^+\to\psi(2S)\,K^+)}
		{{\cal B}(B^+\to J/\psi\,K^+)} &=&
		0.558\pm 0.082 {\rm [stat]}\pm 0.056 {\rm [syst]}
		 \nonumber \\[0.10in]
\frac{{\cal B}(B^0\to\psi(2S)\,K^*(892)^0)}
		{{\cal B}(B^+\to J/\psi\,K^+)} &=&
		0.908\pm 0.194 {\rm [stat]}\pm 0.100 {\rm [syst]}. \nonumber
\end{eqnarray}
The indicated uncertainties are statistical and systematic,
respectively.  In ratios involving unlike $B$-meson species, equal
production rates for $B^+$ and $B^0$ mesons have been assumed.

Absolute branching fractions are extracted by normalizing the above
ratio measurements to the world-average value, ${\cal B}(B^+\to
J/\psi\,K^+) = (1.01\pm 0.14)\times 10^{-3}$:
\begin{itemize}
\item[]
${\cal B}(B^0\to J/\psi\,K^*(892)^0) =
	(1.78 \pm 0.14 {\rm [stat]} \pm 0.29 {\rm [syst]})
		\times 10^{-3}$
\item[]
${\cal B}(B^+\to\psi(2S)\,K^+) =
	(0.56 \pm 0.08 {\rm [stat]} \pm 0.10 {\rm [syst]})
		\times 10^{-3}$
\item[]
${\cal B}(B^0\to\psi(2S)\,K^*(892)^0) =
	(0.92 \pm 0.20 {\rm [stat]} \pm 0.16 {\rm [syst]})
		\times 10^{-3}$.
\end{itemize}

The $B^+\to\psi(2S)\,K^+$ and $B^0\to\psi(2S)\,K^*(892)^0$
reconstructions are the first observations of these processes at a
hadron collider.  The branching-fraction ratio measurements are
consistent with phenomenological predictions that employ the
factorization {\it Ansatz}, and the absolute branching-fraction
measurements are consistent with previous world-average limits and
values from $e^+\,e^-$ colliders.  The measured branching fractions
that involve $\psi(2S)$ final states constitute the world's most
precise measurements of these quantities.

%% file: acknow.tex
\begin{center}
{\bf Acknowledgments}
\end{center}

It is impossible to convey the extent of my gratitude to Pekka
Sinervo, my mentor and graduate advisor.  I have been fortunate to
have benefited from Pekka's profound generosity of spirit, superlative
physics expertise, and professional but compassionate approach to all
things.

There are literally hundreds of people who deserve acknowledgment for
their part in making this study possible.
Appendix~\ref{app:cdf_collab} lists the Run 1 CDF collaboration, whose
immeasurable hard work over several years resulted in the successful
construction and operation of the CDF detector.  Not listed in
Appendix~\ref{app:cdf_collab} are the hundreds of technicians,
engineers, and other support personnel who were vital to the outcome
of the physics programme.

I owe Andreas H\"{o}lscher my thanks for introducing me to the topic
of exclusive $B$ decays into charmonium.  George Sganos, who pioneered
many of the techniques used in this dissertation, provided me with an
abundance of computer code and energetic advice about the {\it modus
operandi} required to advance through a personal physics analysis
effort while completing service responsibilities for the
Collaboration.  My other immediate colleagues at Toronto, Robert
Cropp, Szymon Gadomski, Bj\"{o}rn ``$\underline{\scriptstyle {\rm
q}}$uantum $\underline{\scriptstyle {\rm
c}}$hromo$\underline{\scriptstyle {\rm d}}$ynamics'' Hinrichsen,
Andrew Robinson, and Wendy Taylor, were always helpful and supportive,
even when I was too preoccupied to reciprocate.

I would like to thank William Trischuk for his much appreciated good
humour, support, and valuable advice.  I also am grateful to Michael
Luke and Bob Orr for their encouragement and helpful contributions to
this work.

Thank you to my companion in the trenches, Hyunsoo Kim, for his
friendship, help, and willingness to endure my ramblings at any hour
of the day.  I also thank my CDF colleagues Bill Ashmanskas, Kostas
Kordas, Petar Maksimovi\'{c}, and Benn Tannenbaum for friendship and
good times in Chicago.

I gratefully acknowledge the financial support of the Natural Sciences
and Engineering Research Council of Canada, the W.~C. Sumner Memorial
Foundation, and the University of Toronto.  I thank the officers of
Massey College for electing me as a Junior Fellow; my membership in
the Massey community has yielded several social and intellectual
opportunities beyond the world of physics.

I am grateful to Nicole Schulman for her patience and camaraderie,
especially in waging the War of the Squirrels.  Thank you to Marc Ozon
for his unconditional friendship, perennial generosity, and help of
all kinds.  Words cannot describe my gratitude to Colleen Flood for
her love and forbearance and to my dear family, Rennie, Ruth, Mark,
and Rolf, who have always tolerated and supported me in my endeavours.
{\it Ein besonderes Dankesch\"{o}n meinem Grossmueti, Hedi
Eichenberger-Wiederkehr, f\"{u}r sein mutiges Korrekturlese-Angebot!}

\begin{flushright}
Andreas
\end{flushright}

%% file: intro.tex

At the high-energy frontier of experimental particle physics, the last
quarter of this century has seen the discovery of a new generation of
heavy and (apparently) fundamental constituents of matter, the bottom
and top quarks.  This dissertation describes a study of some of the
properties of the bottom-quark member of this new generation,
properties that can aid in the understanding of matter and the
fundamental forces that act upon it.  The present chapter opens with a
chronology of experimental results that puts into context and
motivates the study of exclusive nonleptonic decays of $B$ mesons,
composite particles that consist of a bottom quark and a
light antiquark.

An understanding of the properties of all types of matter, be it
particulate or astronomical in scale, appears to require the existence
of agents, or forces, to mediate its interactions.  Unlike the
infinite-range forces of quotidian experience, namely electromagnetism
and gravity, the `weak' and `strong' nuclear interactions did not
receive recognition until the decades near the turn of this century.
Fifteen years before Rutherford reported evidence for the atomic
nucleus in 1911~\cite{rutherford:nucleus}, Becquerel demonstrated the
existence of spontaneous radioactivity emitted by phosphorescent
substances~\cite{becquerel:radioactivity}, a harbinger of a hitherto
unknown `weak' force.  The `strong' force, the first evidence of which
was reported by Chadwick and Bieler in studies of $\alpha$-particle
collisions with hydrogen nuclei~\cite{chadwick_bieler:strong}, was
postulated to bind together the known baryons
(protons~\cite{rutherford:proton} and
neutrons~\cite{chadwick:neutron}) into atomic nuclei.  The concept of
the `meson' was introduced by Yukawa in 1935 as a hypothetical
mediator of the strong nuclear force experienced between proximate
nucleons~\cite{yukawa:meson}.  Yukawa's meson, called the pion
($\pi$), was eventually discovered in cosmic rays in
1947~\cite{perkins_powell:pion} after having been mistaken for the
muon, the first evidence of which was reported a decade
earlier~\cite{neddermeyer_anderson:muon}.  By the 1950s, the evolution
of cosmic-ray experimental techniques and the advent of the modern
particle accelerator made possible the study of a more massive type of
matter in the form of `strange' mesons and baryons.  Whereas the
production of the new strange matter was thought to involve the strong
interaction, the long lifetime of the strange particles
($\sim$10$^{-10}$~s) indicated that the weak interaction played a
primary r\^{o}le in their decay~\cite{pais:Vparticles}.  The
subsequent ``explosion'' of new particle discoveries during the 1960s
and 1970s hinted that two large classes of particles, the mesons and
baryons, were composite, just as the multitude of different atomic
elements foretold the existence of subatomic structure in the early
years of this century.

\section{The Fifth Quark}
\label{sect:fifth_quark}

The notion of quarks found its origins in the early 1960s in the
course of searches for an organizing principle to describe the
proliferation of hadronic particles and resonances observed by the
experiments of that time~\cite{cahn_goldhaber}.  To this end,
Gell-Mann~\cite{gell-mann:su3} and, independently,
Ne'eman~\cite{neeman:su3} refined an application of the SU(3)
representation\footnote{ SU(3) denotes the `special unitary' group
in three dimensions, where the matrix operator that effects
transitions between members of this group is unitary and has
determinant $+1$ (special).}, which was originally formulated in terms
of ``fundamental'' $p$, $n$, and $\Lambda$ baryons~\cite{sakata:su3},
to introduce an organizational framework of the known baryons and
mesons.  The SU(3) approximate-symmetry interpretation was further
extended by Gell-Mann~\cite{gell-mann:quarks} and, independently,
Zweig~\cite{zweig:quarks} with a hypothesis that hadrons consisted of
`quarks'.  Gell-Mann and Zweig hypothesized three `flavours' of
quarks: up ($u$), down ($d$), and strange ($s$).

Although the quark idea met with immediate success by explaining the
taxonomy of the experimentally observed particles and resonances,
evidence for quarks as dynamical objects was to come from future
experiments.  Studies of the deep inelastic scattering of electrons by
protons, where the incoming electron scatters off the target proton to
produce a massive hadronic recoil system, were able to probe the
structure of nucleons with measurements of the differential scattering
cross section as a function of the recoiling hadronic invariant mass
for different values of four-momentum transfer between the electron
and proton.  In the recoil invariant mass region beyond the
resonances, the cross section exhibited only a weak dependence on the
momentum transfer between the electron and the hadronic
state~\cite{bloom_breidenbach_miller:dis}.  This observation, which
was called `scale invariance' because the cross section appeared to be
independent of the momentum transfer and to depend on a dimensionless
quantity that related the momentum of the recoiling hadronic system to
the incident proton momentum, gave credence to quark parton models of
nucleon structure advanced by Bjorken~\cite{bjorken:parton} and
Feynman~\cite{feynman:parton}, which predicted this scaling behaviour.
Further confirmation of this parton picture was supplied six years
later by observations of azimuthal asymmetries in the production of
hadrons using electron-positron annihilation~\cite{schwitters:parton}.

In spite of the parton model's successful description of deep
inelastic scattering results, the quark idea suffered from a
theoretical deficiency when used to classify, for example, the
$\Delta^{++}$ baryon\footnote{In this dissertation, references to
specific charge states imply the additional charge-conjugate state,
unless obviated by context or noted otherwise.}
resonance~\cite{anderson:deltapp}.  The $\Delta^{++}$ baryon, thought
in the quark model to be made up of three $u$ quarks, each in the same
state, appeared to be symmetric under the interchange of any of the
constituent quarks, thereby violating the Pauli
principle~\cite{pauli:pep,fermi:fermi-dirac}.  An exact SU(3)
symmetry, that of a new quantum degree of freedom for quarks, termed
`colour', was hypothesized~\cite{greenberg-han-nambu-bogolyubov} to
resolve the conflict with Fermi statistics by rendering the
$\Delta^{++}$ wave function antisymmetric.  With this new colour
symmetry, there also emerged a theory of the strong force called
quantum chromodynamics, or QCD, which described the strong
interactions as being between spin-1/2 quarks and mediated by spin-1
gluons~\cite{weinberg:qcd,fritzsch:qcd,gross:qcd}.  It was postulated
that all flavours of quarks and antiquarks were each endowed with one
of three colours and anticolours, respectively.  Unlike photons, which
do not carry electric charge, the gluon mediators carried colour and
were thought to form a colour SU(3) octet.

QCD also postulated that all naturally occurring particles are colour
SU(3) singlets.  This was motivated largely by the inability of the
experiments to produce isolated quarks.  In QCD, since gluons are
themselves colour sources, they are self-interacting; this property
incites the QCD coupling to grow in strength as the separation between
two colour sources increases.  If two quarks within a hadron or a pair
of hadrons are made to recede from each other in an energetic process
({\it e.g.}, in a particle accelerator), the potential energy accrued
by the increased interquark separation will make it energetically
favourable for pairs of quarks to be produced from the vacuum and to
interact with the receding particles and with each other.  This
process continues until all the quarks are once again confined inside
hadrons producing `jets' of particles.  Evidence of such jets arising
from energetic quarks was first reported in 1975 in $e^+\,e^-$
annihilation studies~\cite{hanson:jets}.  As predicted by QCD, jets
due to gluon {\it Bremsstrahlung} in processes like $e^+\,e^- \to
\gamma^* \to q\,\bar{q}\,g$ were subsequently discovered at higher
centre-of-mass
energies~\cite{barber-brandelik-berger-bartel:gluon-jets}.

An explanation of the revolutionary and largely unexpected 1974
discovery of the $J/\psi$ meson in $e^+\,e^-$ annihilation and
$p$-$Be$ fixed-target experiments~\cite{jpsi_discovery} proved to be
one of the quark parton model's greatest achievements.  The unusually
high mass and long lifetime~\cite{pdg96} of the $J/\psi$ meson
indicated the presence of fundamentally new physics; in turn, the
quark parton model established the observation as a manifestation of a
fourth quark, charm ($c$), in a bound state with its antiquark to form
the $J/\psi$ meson.

The interpretation of the $J/\psi$ meson as a $c\bar{c}$ bound state
was buttressed by the discovery of the $\psi^\prime$, or $\psi(2S)$,
meson in its $e^+\,e^-$ decay channel~\cite{abrams:psi2s_discovery}, a
resonance that was identified immediately as a radial excitation of
the $J/\psi$ state.  Quite analogously to electrodynamic explanations
of the states of positronium observed two decades previous, QCD was
able to predict the charmonium $c\bar{c}$ bound states and their
narrow widths.  The subsequent experimental observation of the decay
$\psi(2S)\to J/\psi\,\pi^+\,\pi^-$~\cite{abrams:pipi} served to
complement the dilepton channels in clarifying the spectroscopy of the
charmonium system.

The discovery of charm brought the count of known fundamental fermions
to four quarks and four leptons (the electron, $e^-$, and muon,
$\mu^-$, and their associated neutrinos) in two generations, thereby
vindicating theoretical prejudice toward a lepton-quark
symmetry~\cite{GIM}.  This symmetry was temporarily broken by the
discovery of the $\tau^-$ lepton in $e^+\,e^-$
collisions~\cite{perl:tau}, setting the stage for a third generation
of fundamental fermions.

\subsection{Discovery}

Evidence for the $b$ quark, often referred to as the `bottom' or
`beauty' quark, was initially obtained with techniques analogous to
those used in the discovery of charmonium.  In 1977, a significant
excess in the rate of dimuon production was observed in 400-GeV
proton-nucleus collisions by a fixed-target experiment at
Fermilab~\cite{lederman:upsilon}.  The original enhancement, observed
near 9.5 GeV/$c^2$, was interpreted as arising due to decays of
bottomonium, a $b\bar{b}$ bound state ($\Upsilon$), and was rapidly
confirmed and resolved into two resonances, namely the $\Upsilon(1S)$
and $\Upsilon(2S)$ mesons~\cite{upsilon:confirm}.

Extensions of nonrelativistic potential models tuned on the $c\bar{c}$
system were used to calculate the dielectronic partial width of the
$\Upsilon$ state, $\Gamma(\Upsilon\to e^+\,e^-)$, which depended on
the electric charge carried by the $b$ and $\bar{b}$ quarks.
Comparisons of the calculated $\Gamma(\Upsilon\to e^+\,e^-)$ with the
area under the observed $\Upsilon$ line shape suggested that the $b$
quark would join its $d$ and $s$-quark counterparts in possessing an
electric charge of $-1/3$.  The emerging pattern of generations
implied by the properties of the observed quarks\footnote{Analogous
to the need for the charm quark in the second quark generation to
cancel out unobserved strangeness-changing neutral currents in the
electroweak theory (GIM mechanism~\cite{GIM}), the lack of observed
flavour-changing neutral currents ({\it e.g.}, $B^0\to \mu^+\,\mu^-$)
in the $b$-quark system inferred the existence of a weak isospin
partner to the third-generation $b$ quark.  Moreover, the weak isospin
of the $b$ quark was measured, via angular asymmetries in $e^+\,e^-
\to b\,\bar{b}$ production, to be $T_3 = -1/2$~\cite{bartel:a_fb},
suggesting a doublet structure similar to that for the lighter
generations.} helped presage the 1995 discovery of a sixth quark, top
($t$), observed in 1.8-TeV $p\bar{p}$
collisions~\cite{abe:topprdl,abe:topprl_1ab,abachi:topprl_1ab}.

\subsection{Electroweak Interactions}

Whereas the weak interactions of leptons have been observed to be
strictly intragenerational~\cite{lept_num}, quarks in their
mass-eigenstate generations,
\begin{equation}
\left(\begin{array}{c} u \\ d \end{array}\right)
\left(\begin{array}{c} c \\ s \end{array}\right)
\left(\begin{array}{c} t \\ b \end{array}\right),
\end{equation}
may interact weakly with quarks in generations other than their own.
The typical notation used to describe the degree of this `mixing' is
the Cabibbo-Kobayashi-Maskawa (CKM)
matrix~\cite{cabibbo,kobayashi-maskawa},
\begin{equation}
\label{eqn:ckm}
\left(\begin{array}{c} d^\prime \\ s^\prime \\ b^\prime \end{array}\right)
=
\left(\begin{array}{ccc}
	V_{ud} & V_{us} & V_{ub} \\
	V_{cd} & V_{cs} & V_{cb} \\
	V_{td} & V_{ts} & V_{tb}
\end{array}\right)
\left(\begin{array}{c} d \\ s \\ b \end{array}\right),
\end{equation}
which, by convention, leaves the $+2/3$-charged quarks unmixed; the
states $\left(d^\prime, s^\prime, b^\prime\right)$ are the weak
eigenstates.  Under the constraints that there be three quark
generations and that the CKM matrix be unitary, the mixing can be
parameterized with three angles and one complex phase.  The pursuit of
measurements~\cite{pdg96} to constrain the CKM matrix and observe the
charge-parity\footnote{`Parity' refers to the quantum mechanical
operator that inverts a spatial displacement vector, {\it i.e.}, ${\bf
r} \to -{\bf r}$.  Parity violation was first observed in studies of
$\beta$ decay in polarized Co$^{60}$ atoms~\cite{cswu:parity}.} ($CP$)
violation in the $b$-quark system (a result of a non-zero complex
phase) constitutes a major component of many experimental programmes
at modern particle accelerators~\cite{pp_experiments}.

Contemporary understanding of the weak interactions has its
foundations in Fermi's field theory of $\beta$ decay, introduced in
1934~\cite{fermi:beta_decay}.  The idea of the four-fermion
interaction was retained for several years before Sakurai introduced
the universal $V-A$ (vector and axial-vector current)
modification~\cite{sakurai:V-A} to accommodate the experimentally
observed parity violation.  Unfortunately, the $V-A$ Fermi theory had
the shortcomings of unitarity violation (the nonconservation of
probability in predicted cross sections that grew quadratically as a
function of centre-of-mass momentum) and
nonrenormalizability\footnote{A theory is renormalizable if the
predicted amplitudes of physical processes remain finite at all
energies and for all powers of the coupling constant, often at the
expense of the introduction of a finite number of arbitrary
experimentally-determined parameters~\cite{perkins:intro}.} in its
predictions of cross sections at high energies.  The `standard model'
of electroweak interactions, developed primarily by Glashow, Weinberg,
and Salam in the 1960s~\cite{standard_model} and based on the gauge
group $SU(2) \times U(1)$, hypothesized four intermediate gauge fields
to avoid these difficulties: the $W^+$, $W^-$, $Z^0$, and $\gamma$
bosons.  The first three of these were thought to be endowed with mass
via the Higgs mechanism~\cite{higgs}; the fourth ($\gamma$) boson was
the massless photon of the electromagnetic interaction.  The weak and
electromagnetic forces were thence consolidated into a single theory.
Soon after the Glashow-Weinberg-Salam standard model had been proven
to be renormalizable~\cite{thooft:renorm}, neutral weak-current
interactions, predicted by the existence of the $Z^0$ boson in the
theory, were discovered in a neutrino-antineutrino experiment using
the Gargamelle bubble chamber at CERN in
1973~\cite{hasert:gargamelle}.  Complete confirmation of the
electroweak theory, however, came a decade later with the discovery of
the $W^\pm$ and $Z^0$ intermediate vector bosons~\cite{WZ_discovery}.

The term in the standard-model electroweak Lagrangian that plays a
large r\^{o}le in the $b$-quark decays investigated in this
dissertation represents the charged-current weak interaction between
the fermion fields~\cite{pdg96}:
\begin{equation}
\label{eqn:sm_lagrangian}
{\cal L}_{\rm CC} =
-\frac{g}{2\sqrt{2}}\sum_{i}\overline{\xi}_i\,\gamma^\mu (1-\gamma^5)
(T^+ W_\mu^+ + T^- W_\mu^-)\,\xi_i,
\end{equation}
where $g$ is the $SU(2)$ gauge coupling constant, $\gamma^\mu
(1-\gamma^5)$ are the Dirac $\gamma$ matrices representing the $V-A$
current, $T^+$ and $T^-$ are the weak-isospin raising and lowering
operators, respectively, $W^\pm_\mu$ are the massive weak charged
boson fields, and the index $i$ represents the fermion families.  In
charged-current weak interactions of the $b$ quark, the fermion
fields, $\xi_i$, are either left-handed $SU(2)$ doublets,
${\displaystyle \xi_3 =
\left(\begin{array}{c} t \\ b^\prime \end{array}\right)_L}$, or
right-handed $SU(2)$ singlets, ${\displaystyle \xi_3 =
\left(b^\prime\right)_R}$, where $b^\prime$ is the weak eigenstate
defined in Equation~\ref{eqn:ckm}.

\subsection{Hadroproduction}
\label{sect:hadroproduction}

In the present study, $b$ and $\bar{b}$ quarks were produced in
1.8-TeV collisions of protons ($p$) and antiprotons ($\bar{p}$) by way
of the inclusive process
\begin{equation}
\label{eqn:b_bbar_hadroproduction}
p(k_p) + \bar{p}(k_{\bar{p}}) \to b(k_b) + \bar{b}(k_{\bar{b}}) + X,
\end{equation}
where $X$ denotes the `underlying event' and $k_p$ and $k_{\bar{p}}$
($k_b$ and $k_{\bar{b}}$) are the momenta of the baryons ($b$ quarks).
Note that the $p$ ($\bar{p}$) baryons each comprise several partons:
the $uud$ ($\bar{u}\bar{u}\bar{d}$) `valence' quarks, gluons, and many
`sea' quark-antiquark pairs\footnote{The `underlying event' refers to
the aggregate product of lower-energy interactions between those
partons not directly involved in the ``hard'' scattering part of
Equation~\ref{eqn:b_bbar_hadroproduction}.}.

A perturbative QCD formula for the invariant differential
hadroproduction cross section of a $b$ quark with energy $E_b$ and
mass $m_b$ can be expressed by convolving the partonic cross section
($\hat{\sigma}$) with the parton distribution functions of the two
hadron reactants in the form~\cite{nde:bxsec}
\begin{equation}
\label{eqn:diff_bxsec}
\frac{E_b\,d^3\sigma}{d^3 k_b} =
\sum_{i,j} \int_0^1\! dx_1 \int_0^1\! dx_2
\left[\frac{E_b\,d^3\hat{\sigma}_{ij}}{d^3k_b}
(x_1k_p, x_2k_{\bar{p}}, k_b; m_b, \mu, \Lambda)\right]
F_i^p(x_1, Q^2) \, F_j^{\bar{p}}(x_2, Q^2),
\end{equation}
where $x_1k_p$ and $x_2k_{\bar{p}}$ are the momenta of the incoming
partons, $F^{p,\bar{p}}_{i,j}$ are the parton distribution functions
for the $i^{\rm th}$ and $j^{\rm th}$ parton in the $p$ and $\bar{p}$
baryons, respectively, and $Q^2$ is the square of the four-momentum
transfer.  The parameter $\mu$ represents the energy scale of the
process and, by assumption, $|Q| \equiv \mu$.  The quantity $\Lambda$
is an experimentally-determined parameter used in the description of
the dependence of the strong coupling constant, $\alpha_s$, on the
energy scale, $\mu$.

Integrating Equation~\ref{eqn:diff_bxsec} over the momentum $k_b$
yields the total cross section for the production of a $b$ quark,
\begin{equation}
\label{eqn:tot_bxsec}
\sigma(s) = \sum_{i,j} \int_0^1\! dx_1 \int_0^1\! dx_2\,\hat{\sigma}_{ij}
(x_1x_2s; m_b, \mu, \Lambda)\,F_i^p(x_1,\mu^2)\,F_j^{\bar{p}}(x_2,\mu^2),
\end{equation}
where $s$ is the square of the centre-of-mass energy of the colliding
proton and antiproton.  The threshold condition for $b\bar{b}$
production is met when the square of the parton-parton centre-of-mass
energy, $\hat{s} \equiv x_1x_2s$, satisfies the condition $\hat{s} =
4m_b^2$.

\begin{figure}[p]
\begin{center}
\leavevmode
\hbox{%
\epsfxsize=6.0in
\epsffile{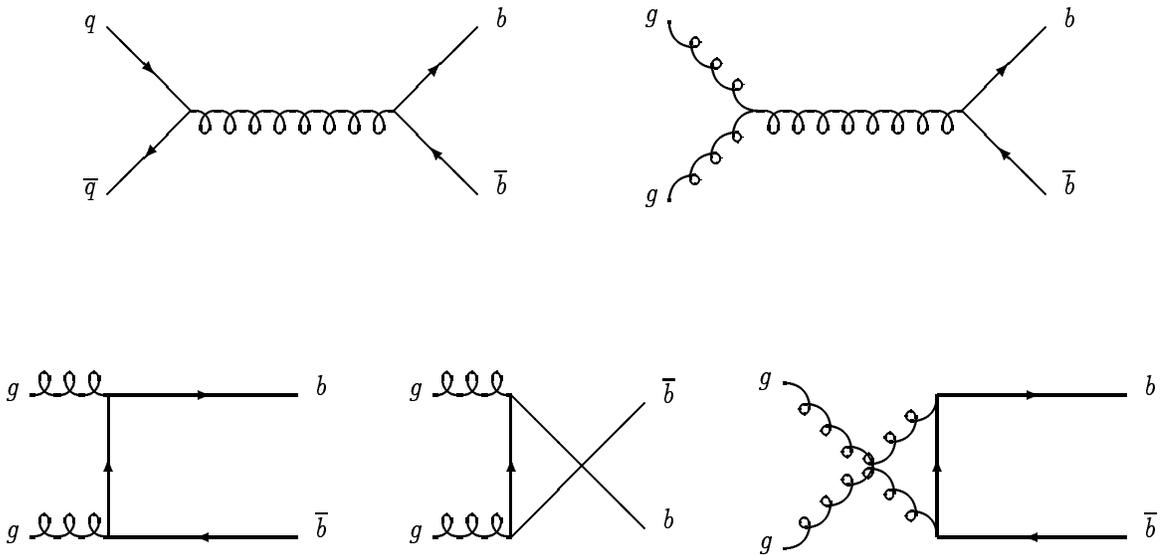}}
\end{center}
\caption
[Feynman diagrams for ${\cal O}(\alpha_s^2)$ $b\bar{b}$ hadroproduction.]
{Feynman diagrams for the lowest-order (${\cal O}(\alpha_s^2)$)
mechanisms of $b\bar{b}$ hadroproduction.}
\label{fig:alpha_s_2_feyn}
\end{figure}

\begin{figure}[p]
\begin{center}
\leavevmode
\hbox{%
\epsfxsize=6.0in
\epsffile{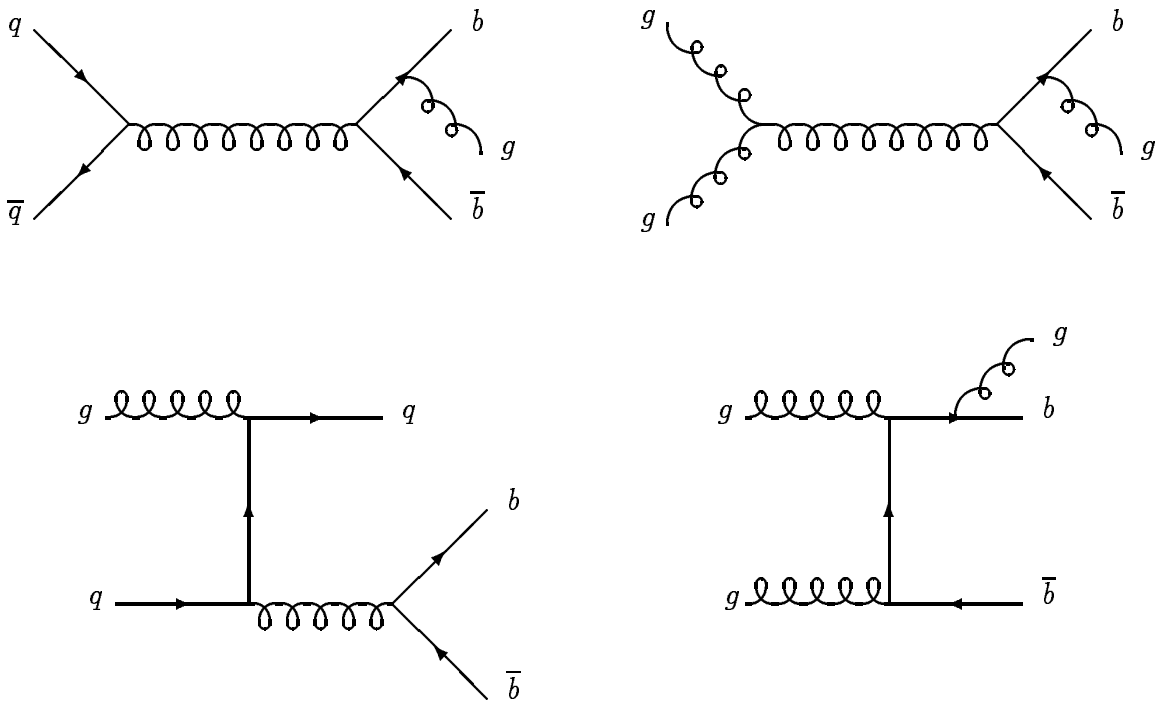}}
\end{center}
\caption
[Feynman diagrams for ${\cal O}(\alpha_s^3)$ $b\bar{b}$ hadroproduction.]
{Examples of some of the Feynman diagrams for the next-to-leading
order (${\cal O}(\alpha_s^3)$) mechanisms of $b\bar{b}$
hadroproduction.}
\label{fig:alpha_s_3_feyn}
\end{figure}

The heavy mass of the $b$ quark makes possible QCD calculations of
$\hat{\sigma}_{ij}$ as a perturbation series in powers of the running
strong coupling constant, $\alpha_s$.  The first terms in the series
that contribute to the cross section are ${\cal O}(\alpha_s^2)$
quark-antiquark annihilation or gluon-gluon fusion processes:
\begin{eqnarray}
\label{eqn:alpha_s_2}
q + \bar{q} & \to & b + \bar{b} \\
g + g & \to & b + \bar{b}. \nonumber
\end{eqnarray}
Figure~\ref{fig:alpha_s_2_feyn} depicts Feynman diagrams of these lowest-order
$b\bar{b}$ production mechanisms~\cite{ellis:ppbar}.
The next-to-leading ${\cal O}(\alpha_s^3)$ terms in the perturbative
series arise from processes~\cite{ellis:ppbar} like
\begin{eqnarray}
q + \bar{q} & \to & b + \bar{b} + g     \nonumber \\
g + g & \to & b + \bar{b} + g		\\
g + q & \to & b + \bar{b} + q		\nonumber \\
g + \bar{q} & \to & b + \bar{b} + \bar{q},  \nonumber
\end{eqnarray}
some examples of which are illustrated with Feynman diagrams in
Figure~\ref{fig:alpha_s_3_feyn}.
Due to interference with diagrams containing virtual gluons, the two
processes in Equation~\ref{eqn:alpha_s_2} can also contribute at
${\cal O}(\alpha_s^3)$.  For high energies, {\it i.e.}, when $k_{\rm T}(b)
\gg m_b$, where $k_{\rm T}(b)$ is the momentum
of the $b$ quark projected onto a plane perpendicular to the axis of
the two incoming partons, some of the next-to-leading order ${\cal
O}(\alpha_s^3)$ mechanisms can contribute to the cross section by
amounts comparable to the ${\cal O}(\alpha_s^2)$
contributions~\cite{nde:bxsec}.

In Equations~\ref{eqn:diff_bxsec} and \ref{eqn:tot_bxsec}, there is a
degree of arbitrariness in the value of the renormalization scale,
$\mu$, that contributes a relatively large uncertainty to QCD
predictions of $b$-quark production because they are not calculated to
all orders in $\alpha_s$.  The value of $\mu$ is typically assigned to
be near a physical scale, such as $m_b$ or ${\displaystyle
\sqrt{m_b^2 + k_{\rm T}^2(b)}}$; however, these choices of $\mu$ are
``bootstrapped'' because the fact that $b$ quarks are confined inside
hadrons requires that extractions of $m_b$ depend on the
renormalization scheme used, model-specific definitions of $m_b$, and
the value of $\mu$ itself~\cite{pdg96,luke:mb}.

Nason, Dawson, and Ellis (NDE) have calculated ${\displaystyle
\frac{d^2\sigma}{dy_b\,dk_{\rm T}^2(b)}}$, the inclusive differential
$b$-quark production cross section, as a function of rapidity, $y_b$,
and $k_{\rm T}(b)$~\cite{nde:bxsec}.  Rapidity is a measure of the
polar angle of a particle's trajectory, usually with respect to the
collision axis, and is defined for a $b$ quark as
\begin{equation}
y_b \equiv \frac{1}{2} \ln\left[\frac{E_b + k_z(b)}{E_b - k_z(b)}\right],
\end{equation}
where $k_z(b)$ is the projection of the $b$-quark momentum onto the
beam axis.  The rapidity variable is useful to descriptions of
high-energy particle production because the shape of the
particle-multiplicity distribution, $dN/dy_b$, is Lorentz-invariant;
reference frame transformations amount to linear shifts in the origin
of $y_b$~\cite{pdg96}.

Table~\ref{tab:accel_bbbar_xsec} compares the maximum $b$-quark
production cross section ($\sigma^{\rm max}_{b\bar{b}}$) applicable
to the present study (Tevatron) with that in other collision types at
different energies.  Although the rates of $b\bar{b}$ hadroproduction
exceed those of leptoproduction by three orders of magnitude, they
constitute a significantly smaller fraction ($\sigma_{b\bar{b}} /
\sigma_{\rm tot}$) of the total cross section ($\sigma_{\rm tot}$)
than do the analogous $b\bar{b}$ leptoproduction fractions.  The
consequentially low signal-to-background ratios for $b$-quark
production at hadron colliders pose experimental challenges that must
be overcome.

\begin{table}
\begin{center}
\begin{tabular}{|c||c|c|c|c|}        \hline
\rule[-3mm]{0mm}{8mm}
Collision Type & $p\bar{p}$ & $pp$ & \multicolumn{2}{c|}{$e^+\,e^-$} \\ \hline
\rule[-3mm]{0mm}{8mm}
Accelerator & Tevatron & LHC & CESR & LEP, SLC \\ \hline\hline
\rule{0mm}{5mm}
$\sqrt{s}$\ \ [GeV] & 1\,800 & 14\,000 & 10.58 & 92 \\
\rule{0mm}{5mm}
$\sigma^{\rm max}_{b\bar{b}}$\ \ [$\mu$b] & 20$-$40 & $\sim$200 &
	$\sim 1.1 \times 10^{-3}$ & $\sim 9 \times 10^{-3}$ \\
\rule[-3mm]{0mm}{8mm}
$\sigma_{b\bar{b}} / \sigma_{\rm tot}$ & $4 \times 10^{-4}$ &
	$2 \times 10^{-3}$ & 0.25$-$0.33 & $\sim 0.2$ \\ \hline
\end{tabular}
\end{center}
\caption
[Cross sections for $b\bar{b}$ production at various colliders.]
{Comparison of hadroproduction and leptoproduction cross sections for
$b\bar{b}$ production at various centre-of-mass
energies~\cite{schwarz:heavy_flav}.  The parameters of the five
accelerators are summarized in Reference~\cite{pdg96}.}
\label{tab:accel_bbbar_xsec}
\end{table}

\section{The $B$ Meson}
\label{sect:bmeson}

Analogously to studies of the charm sector, the discovery of the
bottomonium states lead naturally to the expectation that, at masses
somewhat greater than those of the lowest-lying $\Upsilon$ resonances,
``open'' bottom meson production would occur in the form of bare $B^+$
($\bar{b}u$) and $B^0$ ($\bar{b}d$) mesons.  First evidence for
$B$-meson production, reported in 1981 by the CLEO
collaboration~\cite{cleo:bmeson}, was obtained through observations of
increases in the single electron and muon inclusive cross sections in
$e^+\,e^-$ collisions.  The enhancements were attributed to inclusive
semileptonic decays of $B$ mesons via the processes $B \to
X\,\ell\,\nu_\ell$, where $\ell$ denotes either the $e$ or $\mu$
lepton flavour and $X$ represents the remaining hadronic system.

Prior to the first evidence for the existence of $B$ mesons, however,
Fritzsch argued that ``\ldots the only realistic method to discover
the $B$ mesons'' was through the reconstruction of their decays to
charmonium states~\cite{fritzsch:bmesons}.  Fritzsch's assertion,
which launched a considerable amount of theoretical work on the
subject (see, for example, References~\cite{wise_degrand:btopsi}), was
motivated by the following points: $c\bar{c}$ mesons are readily
produced in weak $B$-meson decays; the $\mu^+\,\mu^-$ and $e^+\,e^-$
decay modes of the $c\bar{c}$ states can be identified easily, most
notably in hadronic collision environments with their sizeable
backgrounds; and the relative heaviness of the charmonium states
forces the remaining hadronic system in each $B$ decay to be
relatively simple because of the attendant restricted multiparticle
phase space~\cite{fritzsch:bmesons}.  In spite of theoretical
expectations that $B$ mesons would first be reconstructed in their
charmonium final states, it is interesting to note that the first full
reconstruction of $B$ decays was achieved using final states
containing $D^0$ and $D^*(2010)^+$ mesons, such as $B^- \to
D^0\,\pi^-$ and $B^- \to
D^*(2010)^+\,\pi^-\,\pi^-$~\cite{behrends:bmesons}.  More mention of
the mechanism of charmonium production in $B$-meson decays, which is
the subject of this investigation, is given in
Section~\ref{sect:nleptonic_decays} and thereafter.

\subsection{Hadronization}
\label{sect:hadronization}

The `fragmentation' of a $b$ quark into a colour-singlet hadron, in
this case a $B$ meson, is a long-distance, nonperturbative QCD
process.  Models of fragmentation typically employ a parameter $z_f$,
where ${\displaystyle z_f \equiv \frac{E_B + p_{\parallel}(B)}{E_b +
k(b)}}$~\cite{chrin:epsilon}, which represents the fraction of
available energy-momentum carried by the $B$ meson.  The symbols $E_B$
and $p_{\parallel}(B)$, respectively, represent the hadronized
$B$-meson energy and the momentum component parallel to the direction
of fragmentation.  Simple kinematical arguments support the claim that
functions describing $b \to B\,q$ processes, where $q$ is the
light-quark counterpart to the $\bar{q}$ quark in the $B$ meson, peak
at high values of $z_f$ due to the expectation that the majority of
the $b$-quark momentum is transferred to the $B$
meson~\cite{suzuki:fragment,bjorken:fragment}.  Peterson {\it et al.}
noted that the quantum mechanical amplitude describing fragmentation
would be inversely proportional to the energy transfer of the process,
$\Delta E \equiv E_B + E_q - E_b$~\cite{peterson:frag}.  Assuming $m_b
\approx m_B$, then
\begin{equation}
\label{eqn:deltaE}
\Delta E = \sqrt{m_b^2 + z_f^2k(b)^2} + \sqrt{m_q^2 + (1-z_f)^2k(b)^2}
- \sqrt{m_b^2 + k(b)^2}.
\end{equation}
The fragmentation function, $D^B_b(z_f)$, was estimated from
Equation~\ref{eqn:deltaE} with the expression
\begin{equation}
D^B_b(z_f)\  \propto \ \frac{1}{z_f\,(\Delta E)^2}\  =
\ \frac{1}{z_f\,\left[1-\frac{1}{z_f}-\frac{\epsilon_b}{1-z_f}\right]^2},
\end{equation}
where the factor of $z_f$ in the denominator arose from longitudinal phase
space and the quantity $\epsilon_b \sim m_q^2/m_b^2$ was the ``Peterson
parameter''~\cite{peterson:frag}.

The Peterson parameterization benefits from its dependence on a single
experimentally determined parameter ($\epsilon_b$).  Chrin has
estimated a value of $\epsilon_b = 0.006 \pm 0.002$, which is based on
a survey of several experimental $e^+\,e^-$
observations~\cite{chrin:epsilon}.  The search continues for a better
understanding of $b$-quark fragmentation and its sensitivity to the
type of collision environment in which the $b$ quark was produced and
the flavour(s) of the non-$b$ quark(s) constituting the final-state
hadrons.

\subsection{Nonleptonic $B$-Meson Decays}
\label{sect:nleptonic_decays}

Nonleptonic decays of $B$ mesons feature a rich interplay of the weak
and strong interactions.  Knowledge of $B$-meson decay rates may be
used to obtain angle and phase information in the CKM matrix
(Equation~\ref{eqn:ckm}) by way of the charged-current electroweak
interaction described in Equation~\ref{eqn:sm_lagrangian}.  The
practical extraction of information about the weak processes, however,
is confounded by the fact that quarks are necessarily confined inside
bound colour-singlet states, requiring the invocation of quantum
chromodynamics to complete the description.  The relatively heavy mass
of the $b$ quark, nominally 4.1 to 4.5~GeV/$c^2$~\cite{pdg96} or about
5 times the mass of a proton, is such that QCD descriptions can be
decoupled into `short' and `long-distance' dynamical
regimes~\cite{wirbel:dbmesons} (see Section~\ref{sect:factorization}).
Short-distance effects exploit the asymptotic-freedom property of QCD,
which enables the perturbative calculation of corrections to the
electroweak Hamiltonian due to exchanges of hard
gluons~\cite{cteq:qcd_handbook}.  In the context of nonleptonic decays
of $B$ mesons, long-distance QCD effects generally involve the
hadronization of the decay products and include the exchange of soft
gluons, the creation of $q\bar{q}$ pairs from the vacuum, and
interactions of the hadrons in the final state.  That long-distance
QCD processes are nonperturbative has so far thwarted calculations of
these effects from first principles; nevertheless, the separation of
the decay mechanism into short and long-distance components is
currently the most successful approach to understanding many $B$-meson
decay processes.  Experimental measurements of $B$-meson decay
properties are clearly essential to the pursuit of an understanding of
the dynamical complexities of the underlying phenomena.

\begin{figure}
\begin{center}
\leavevmode
\hbox{%
\epsfysize=1.5in
\epsffile{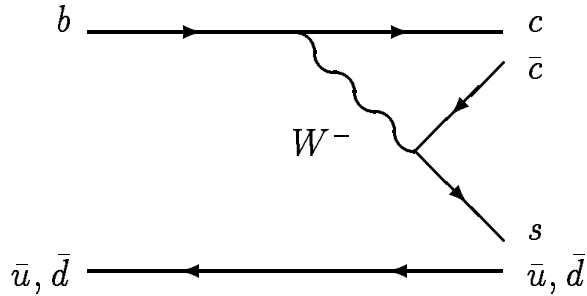}}
\end{center}
\caption
[Feynman diagram of an exclusive colour-suppressed $B$-meson decay.]
{Feynman diagram of the `spectator-internal' weak decay mechanism
for a $B$ meson (here either a $b\bar{u}$ ($B^-$) or $b\bar{d}$
($\overline{B^0}$) state) decaying to charmonium ($J/\psi$ or
$\psi(2S)$) and a strange meson ($K^-$ or $\overline{K}^*(892)^0$).
The decay is colour-suppressed because it only occurs when the
$\bar{c}s$ pair, itself a colour singlet, conspires with the $c$ quark
and the $\bar{u}$ or $\bar{d}$ quark to form colour-singlet $c\bar{c}$
and $s\bar{u}$ or $s\bar{d}$ mesons.  In this mechanism, the $\bar{u}$
or $\bar{d}$ quark is assumed to be a `spectator' of the weak
process.}
\label{fig:bdecay_feyn}
\end{figure}

This dissertation concerns the experimental study of exclusive
nonleptonic decays of $B$ mesons into $J/\psi$ or $\psi(2S)$
vector-meson and $K^+$ pseudoscalar-meson or $K^*(892)^0$
vector-meson\footnote{The $K^*(892)^0$ vector meson was the first
meson resonance to be observed~\cite{alston:k*892,chinowsky:k*_spin}.}
final states.  Figure~\ref{fig:bdecay_feyn} exhibits the
weak-interaction aspect of these decays, which are categorized as
`spectator-internal' processes by virtue of two assumptions: (i) the
light-quark (spectator) constituent of the $B$ meson fails to
participate in any process before hadronization and final-state
interactions occur and (ii) the quark with which the spectator forms a
colour singlet in the final state derives from an ``internally
produced'' $W$ boson.  These decays are considered to be
colour-suppressed because they can only occur when the $W$ bosons'
hadronic decay products, themselves constituting colour singlets,
conspire with the charm quarks from the flavour-changing decays and
the spectator quarks to form colour-singlet charmonium and strange
mesons, respectively.  Finally, these processes are CKM-favoured, as
their rates depend upon $\left|V_{cb}\right|^2$ (refer to
Equation~\ref{eqn:ckm}), where $\left|V_{cb}\right| = 0.041 \pm
0.003$~\cite{pdg96}.

Gourdin, Kamal, and Pham state that ``it is generally believed that
the best place to study the importance of colour-suppressed processes
in $B$-meson decay is to look at final states involving a charmonium
and a strange meson''~\cite{gourdin:difficulties}.  Precise
measurements of properties of this class of decays can improve
knowledge of wave functions in the $B$ system~\cite{deshpande:bpsik},
the same wave functions that are involved in, {\it e.g.}, the rare
processes $B^0
\to K^*(892)^0\,\gamma$ and $B \to
K\,e^+\,e^-$~\cite{deshpande:bpsik,ahmady:nonleptonic}.  Experimental
observations of $B \to \psi\,K$ decays also provide input to and tests
of theoretical and phenomenological methodologies and hypotheses.
Some of these issues are discussed in
Section~\ref{sect:factorization}.


\subsection{The Factorization {\it Ansatz}}
\label{sect:factorization}


The factorization {\it Ansatz}\, model was first used by Bauer, Stech,
and Wirbel (BSW) to describe exclusive semileptonic decays of heavy
mesons~\cite{bsw:semi}.  The techniques were subsequently extended to
encompass nonleptonic heavy-flavoured meson
decays~\cite{bsw:nonleptonic}.  A fundamental assumption in the BSW
model is the distinction, noted in
Section~\ref{sect:nleptonic_decays}, between two disparate time (or
distance) scales in the decay dynamics.  In the nonleptonic weak decay
of a $B$ meson, the $b$ quark decays with a time scale of $\tau \sim
1/M_W \sim 10^{-26}$~s.  It is assumed that all other partons in the
system (including the spectator quark, sea quarks, and soft gluons)
suspend any interaction until the longer QCD time scale, $\tau \sim
1/\Lambda_{\rm QCD} \sim 10^{-23}$~s, indicating the time when
confinement forces become important, takes effect.

Using the mode $B^+ \to J/\psi\,K^+$ as an example, the amplitude of
the decay (${\cal M}$) can be expressed as a superposition of local
operators, $O_i$, with scale-dependent short-distance Wilson
coefficients, $C_i(\mu)$, in the form
\begin{eqnarray}
{\cal M}(B^+ \to J/\psi\,K^+) &=& \langle J/\psi, K^+ |{\cal H}_{\rm eff}|
	B^+\rangle \\ \nonumber
	&=& -\frac{G_F}{\sqrt{2}} \sum_{i} C_i(\mu)
		\langle J/\psi, K^+ | O_i | B^+\rangle,
\end{eqnarray}
where $G_F$ is the Fermi coupling constant and ${\cal H}_{\rm eff}$ is
the effective Hamiltonian embodying the ${\cal O}(\alpha_s)$
hard-gluon corrections to the weak decay of the $b$
quark~\cite{wirbel:ddstar}:
\begin{equation}
\label{eqn:heff}
{\cal H}_{\rm eff} \equiv
-\frac{G_F}{\sqrt{2}} V_{cb}
        \left\{C_1(\mu)\left[(\bar{c}b)^\alpha\,
		(\bar{s}^\prime c)_\alpha\right]+
              C_2(\mu)\left[(\bar{s}^\prime b)^\alpha\,(\bar{c}c)_\alpha\right]
              \right\}.
\end{equation}
The primed fields represent weak eigenstates, which were defined in
Equation~\ref{eqn:ckm}.  Similar to the form of
Equation~\ref{eqn:sm_lagrangian}, Equation~\ref{eqn:heff} uses the
notation
\begin{equation}
(\bar{q}_2 q_1)^\alpha \equiv \sum_{\lambda} \bar{q}_{2\lambda}\,\gamma^\alpha
                       \,(1-\gamma^5)\,q_{1\lambda}
\end{equation}
for the current terms, where $\lambda$ is the colour index.  Currents
proportional to the Wilson coefficient $C_1(\mu)$ are charged, as
exemplified in Figure~\ref{fig:bdecay_feyn}, and currents proportional
to $C_2(\mu)$ are effectively neutral.  In principle, gluonic
penguin\footnote{Penguin decays have the characteristic that the $W$
boson is reabsorbed by the quark line from which it was emitted,
yielding an effective net flavour-changing neutral current process.  A
photon, gluon, or $Z^0$ boson is emitted from the resulting loop.}
interactions can also contribute to Equation~\ref{eqn:heff}; however,
these are highly suppressed due to the fact that the production of a
$J/\psi$ meson from gluons necessarily involves no less than three
gluons~\cite{barger_phillips}.

The factorization {\it Ansatz}\, is an approximation whereby one of
the currents in the (current) $\times$ (current) form expressed in
Equation~\ref{eqn:heff} is assumed to be asymptotic, thereby enabling
the factorization of the amplitude in terms of hadronic ($H$) currents
instead of quark currents~\cite{bsw:nonleptonic,browder:review}:
\begin{equation}
\label{eqn:amplitude}
{\cal M}(B^+\to J/\psi\,K^+) = -\frac{G_F}{\sqrt{2}}\,a_2\,V_{cb}
	\left\{\langle J/\psi | (\bar{c}c)_\alpha^H | 0\rangle \cdot
               \langle K^+ | (\bar{s}^\prime b)^\alpha_H | B^+ \rangle\right\},
\end{equation}
where the quantity $a_2$ is a new scale-independent coefficient native
to spectator-internal mechanisms (see Figure~\ref{fig:bdecay_feyn})
and constructed via a linear combination of the Wilson coefficients
evaluated at a scale defined by $\mu \simeq m_b$,
\begin{equation}
\label{eqn:a_2}
a_2 \simeq C_2(m_b) + \xi C_1(m_b).
\end{equation}
The new parameter $\xi$ is an {\it ad hoc} colour factor, often taken
na\"{\i}vely to be $\sim$1/3 due to the three degrees of colour
freedom\footnote{As the choice of the notation $a_2$ suggests, there
is another scale-independent coefficient, $a_1$, defined by $a_1
\simeq C_1(m_b) + \xi C_2(m_b)$, which represents a separate
`spectator-external' class of $B$-meson decays.}.  Since the colour
structure in Equation~\ref{eqn:a_2} can easily be destroyed by
long-distance soft-gluon dynamics, $a_2$ is typically treated as a
free parameter~\cite{bsw:nonleptonic,wirbel:ddstar}.
Equation~\ref{eqn:amplitude} suggests that a knowledge of the $B\to
\psi\,K$ decay rates can furnish estimates of the magnitude of
the $a_2$ parameter~\cite{gourdin:a1_a2,neubert:new_look}.

The asymptotic $\langle J/\psi | (\bar{c}c)_\mu^H | 0\rangle$
single-particle matrix element in Equation~\ref{eqn:amplitude} is
proportional to the $J/\psi$ decay constant, thus reducing the
amplitude calculation to a determination of the hadronic form factors
constituting the $\langle K^+ | (\bar{s}^\prime b)^\mu_H | B^+
\rangle$ matrix element.  Several different approaches to modeling
the hadronic form factors, which typically employ measurements from
experiments involving semileptonic decays, exist in the literature and
will not be discussed here in detail.  One elegant and successful
method has been to exploit the heavy-quark symmetries that arise in
the $m_b \to \infty$ limit~\cite{isgur_wise_luke:hqet}.  The
application of heavy-quark symmetries to the determination of $B
\to K$ form factors, however, is hampered by the relative lightness of
the $s$ quark.  Moreover, the observed absence of tree-level
flavour-changing neutral-current decays complicates the use of
experimental inputs in estimates of $B \to K$ form factors.  Highly
model-dependent {\it Ans\"{a}tze}\, have been used to estimate $B \to
K$ form factors from, {\it e.g.}, $D \to K^{(*)}\,\ell\,\nu$
semileptonic decay
measurements~\cite{deandrea:predictions,aleksan:critical}.  Two
principal sources of significant theoretical uncertainty are present
in all these estimations: the uncertainties in the numerical values of
the form factors at zero momentum transfer between the $B$ parent and
the $K$ daughter ($q^2 = 0$) and the uncertainties in the assumed
parameterizations of their $q^2$
extrapolations~\cite{kamal_pham:cabibbo}.

It is important to point out that, unlike for semileptonic decays
where the amplitude can be decomposed into leptonic and hadronic
currents, there is no theoretical justification for the application of
the factorization {\it Ansatz}\, to colour-suppressed $B\to \psi\, K$
decay modes~\cite{bigi:nonleptonic,aleksan:critical}.  Tests of the
validity of factorization for $B\to \psi\, K$ decays are challenged by
the difficulty in isolating factorization effects from form-factor
effects.  Part of the ambiguity can be removed by confronting the data
with calculated observables that involve {\it ratios} of form factors,
thereby purging some reliance on the absolute values of the form
factors, but nevertheless retaining a dependence on assumptions about
their $q^2$ extrapolations~\cite{kamal:probing}.  Tests of
factorization have consisted of requiring that the models reconcile
ratio-of-branching-fraction measurements ({\it e.g.}, ${\cal B}(B\to
J/\psi\,K) / {\cal B}(B\to\psi(2S)\,K)$) with measurements of
longitudinal polarization fractions ({\it e.g.}, ${\displaystyle
\frac{\Gamma_L}{\Gamma} (B^0\to J/\psi\,K^*(892)^0)}$).  Factorization
models have been shown to be unable to account simultaneously for
measurements of these two
observables~\cite{gourdin:difficulties,aleksan:critical}.  While
inadequacies in the form-factor approximations are clearly possible,
it has been suggested that the discrepancies may be due to
nonfactorizable contributions to the decay
amplitudes~\cite{cheng:nonfactorization,soares:nonfactorization,kamal:nonfactorization,al-shamali:nonfactorization,neubert_stech:heavy_flavours_2}.

\subsection{Theoretical Predictions}
\label{sect:predictions}


The foregoing synopsis of factorization techniques points out some of
the difficulties and uncertainties in constructing a reliable
theoretical and phenomenological picture of exclusive $B$-meson decays
to charmonium and strange-meson final states.  Although the present
tendency is for the experimental data to drive phenomenological
investigations of form-factor assumptions and nonfactorizable
contributions to the $a_1$ and $a_2$ parameters, there nevertheless
exist some recent branching-fraction predictions in the literature.

The three sets of factorization predictions considered in
Table~\ref{tab:predictions} were selected because they all included
treatments of the $\psi(2S)$ final states.  The Deshpande {\it et al.}
results~\cite{deshpande:bpsik} made use of form factors from BSW, who
calculated solutions to a relativistic harmonic oscillator potential
model and assumed a single-pole\footnote{The generic monopole
form-factor formula is ${\displaystyle F(q^2) = F(0)/(1-q^2/m^2)}$,
where $m$ represents the pole
mass~\cite{bsw:semi,deandrea:predictions}.}  $q^2$
dependence~\cite{bsw:semi,bsw:nonleptonic}.  The Deandrea {\it et al.}
calculations employed form factors that were derived from experimental
exclusive semileptonic $D$-meson decay measurements and that were also
taken to possess a monopole $q^2$
dependence~\cite{deandrea:predictions}.  The Cheng {\it et al.}
predictions\footnote{Note that the Cheng {\it et al.} predictions were
compared in Reference~\cite{cheng:nonfactorization} with preliminary
versions~\cite{warburton:warsaw} of the measurements described in this
dissertation.} for the $\psi(2S)$ final states were based both on an
$a_2$ parameter that was computed from experimental measurements of
the analogous $J/\psi$ modes and on form factors that were calculated
explicitly over the entire physical $q^2$
range~\cite{cheng:nonfactorization}.

\begin{table}
\begin{center}
\begin{tabular}{|l||c|c|c|}        \hline
\multicolumn{1}{|c||}{Decay}   &
  \multicolumn{3}{c|}{Branching-Fraction Predictions
					(${\cal B}$)~$[\times 10^{-3}]$} \\
\multicolumn{1}{|c||}{Channel} &
	Deshpande~\cite{deshpande:bpsik} &
	Deandrea~\cite{deandrea:predictions} &
	Cheng~\cite{cheng:nonfactorization} \\ \hline\hline
\rule{0mm}{5mm}
$B^+\to J/\psi\,K^+$        & 0.84 & $1.1 \pm 0.6$   & n/a \\
\rule{0mm}{5mm}
$B^0\to J/\psi\,K^*(892)^0$ & 1.63 & $1.6 \pm 0.5$   & n/a \\
\rule{0mm}{5mm}
$B^+\to\psi(2S)\,K^+$       & 0.33 &$0.37 \pm 0.19$ & $0.52$ \\
\rule[-3mm]{0mm}{8mm}
$B^0\to\psi(2S)\,K^*(892)^0$& 1.27 &$0.74 \pm 0.23$ & $0.76$ \\ \hline\hline
\multicolumn{1}{|c||}{Year Published} & 1990 & 1993 & 1997 \\ \hline
\end{tabular}
\end{center}
\caption
[Theoretical branching-fraction predictions.]
{Theoretical branching-fraction predictions, based on the
factorization {\it Ansatz}, of the decay modes under study.  The
Deandrea {\it et al.} predictions assumed $B$-meson lifetimes of
$\tau_{B^+} = \tau_{B^0} = 1.4 \times 10^{-12}$~s and the Cheng {\it
et al.} predictions used the world-average~\cite{pdg96} measured
lifetimes: $\tau_{B^+} = 1.62 \times 10^{-12}$~s and $\tau_{B^0} =
1.56 \times 10^{-12}$~s.}
\label{tab:predictions}
\end{table}

\section{Dissertation Overview}
\label{sect:overview}

\begin{figure}
\unitlength=1mm
\begin{center}
\begin{picture}(105,45)(0,0)
\put(0,40){\makebox(0,0){\bf $B^{+}$ }}
\put(5,40){\vector(1,0){8}}
\put(21,40){\makebox(0,0){\bf $J/\psi$ }}
        \put(28,40){\makebox(0,0){\bf $K^{+}$ }}
\put(19,36){\line(0,-1){6}}
\put(19,30){\vector(1,0){8}}
  \put(33,30){\makebox(0,0){\bf $\,\,\,\mu^+\,\mu^-$ }}

\put(60,40){\makebox(0,0){\bf $B^{+}$ }}
\put(65,40){\vector(1,0){8}}
\put(81,40){\makebox(0,0){\bf $\psi(2S)$ }}
        \put(91,40){\makebox(0,0){\bf $K^{+}$ }}
\put(79,36){\line(0,-1){6}}
\put(79,30){\vector(1,0){8}}
  \put(93,30){\makebox(0,0){\bf $\,\,\,\mu^+\,\mu^-$ }}
\put(79,26){\line(0,-1){6}}
\put(79,20){\vector(1,0){8}}
  \put(97,20){\makebox(0,0){\bf $\,\,\,J/\psi\,\pi^+\,\pi^-$ }}
\put(90,16){\line(0,-1){6}}
\put(90,10){\vector(1,0){8}}
  \put(104,10){\makebox(0,0){\bf $\,\,\,\mu^+\,\mu^-$ }}
\end{picture}
\caption
[Schematic diagrams of the $B^+$ decay modes.]
{Schematic diagrams of the $B^+$ decay modes reconstructed in this study.}
\label{fig:topo1}
\end{center}
\end{figure}
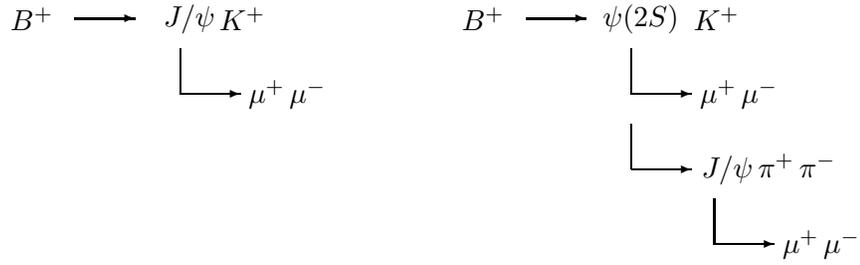

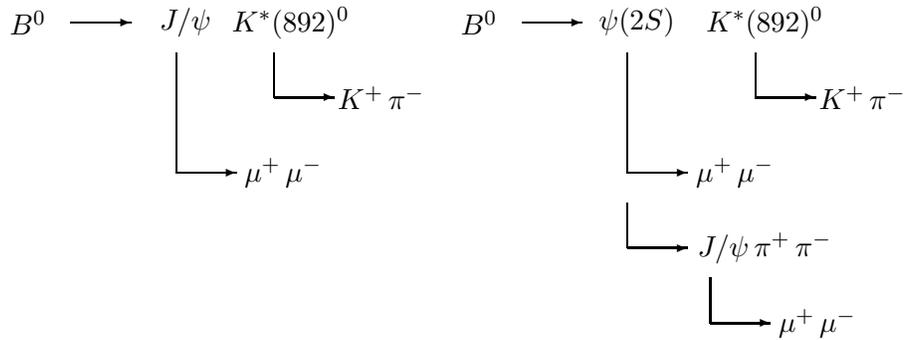
\begin{figure}
\unitlength=1mm
\begin{center}
\begin{picture}(105,55)(0,0)
\put(0,50){\makebox(0,0){\bf $B^{0}$ }}
\put(5,50){\vector(1,0){8}}
\put(21,50){\makebox(0,0){\bf $J/\psi$ }}
        \put(35,50){\makebox(0,0){\bf $K^*(892)^0$ }}
\put(19,46){\line(0,-1){16}}
\put(19,30){\vector(1,0){8}}
  \put(33,30){\makebox(0,0){\bf $\,\,\,\mu^+\,\mu^-$ }}
\put(32,46){\line(0,-1){6}}
\put(32,40){\vector(1,0){8}}
  \put(46,40){\makebox(0,0){\bf $\,\,\,K^+\,\pi^-$ }}

\put(60,50){\makebox(0,0){\bf $B^{0}$ }}
\put(65,50){\vector(1,0){8}}
\put(81,50){\makebox(0,0){\bf $\psi(2S)$ }}
        \put(98,50){\makebox(0,0){\bf $K^*(892)^0$ }}
\put(96,46){\line(0,-1){6}}
\put(96,40){\vector(1,0){8}}
  \put(110,40){\makebox(0,0){\bf $\,\,\,K^+\,\pi^-$ }}
\put(79,46){\line(0,-1){16}}
\put(79,30){\vector(1,0){8}}
  \put(93,30){\makebox(0,0){\bf $\,\,\,\mu^+\,\mu^-$ }}
\put(79,26){\line(0,-1){6}}
\put(79,20){\vector(1,0){8}}
  \put(97,20){\makebox(0,0){\bf $\,\,\,J/\psi\,\pi^+\,\pi^-$ }}
\put(90,16){\line(0,-1){6}}
\put(90,10){\vector(1,0){8}}
  \put(104,10){\makebox(0,0){\bf $\,\,\,\mu^+\,\mu^-$ }}
\end{picture}
\caption
[Schematic diagrams of the $B^0$ decay modes.]
{Schematic diagrams of the $B^0$ decay modes reconstructed in this study.}
\label{fig:topo2}
\end{center}
\end{figure}

This dissertation describes searches for the decay channels listed in
Table~\ref{tab:predictions} and relates details of their full
reconstruction and branching-fraction (${\cal B}$) measurements, as
observed in 1.8-TeV proton-antiproton collisions.
Figures~\ref{fig:topo1} and
\ref{fig:topo2} schematically illustrate the topologies considered in
the analysis of $B^+$ and $B^0$ mesons, respectively.  Both species of
charmonium meson are sought in their dimuon modes, and the $\psi(2S)$
meson is supplementally reconstructed in its hadronic-cascade
$\psi(2S)\to J/\psi\,\pi^+\,\pi^-$ mode (refer to
Section~\ref{sect:fifth_quark}).

Just as the examination of relative, as opposed to absolute, branching
fractions can simplify theoretical approaches to $B$-meson decay
(\cite{kamal:probing} and Section~\ref{sect:factorization}), the use
of candidate event yields to compute ratios of branching fractions can
have several benefits on the experimental side.  Consider an
expression describing the requisite ingredients to measure the
experimental absolute branching fraction of, for example, the decay
$B^\pm\to\psi(2S)\,K^\pm$, where the $\psi(2S)$ meson decays in its
$\psi(2S)\to J/\psi\,\pi^+\,\pi^-$ channel:
\begin{eqnarray}
\label{eqn:br_fract}
\lefteqn{{\cal B}(B^\pm\to\psi(2S)\,K^\pm) =} \\
 & & \frac{N(B^\pm\to\psi(2S)\,K^\pm)} {2\varepsilon {\cal A} f_u
\cdot \sigma\left(p\,\bar{p}\to b(\bar{b})\,X\right) \cdot \int\!{\cal L}\,dt
\cdot {\cal B}(\psi(2S)\to J/\psi\,\pi^+\,\pi^-) \cdot {\cal
B}(J/\psi\to\mu^+\mu^-)}, \nonumber
\end{eqnarray}
where $N(B^\pm\to\psi(2S)\,K^\pm)$ denotes the yield of candidate
decays reconstructed, $\varepsilon$ represents the product of several
reconstruction efficiencies (Chapter~\ref{chapt:effic}), ${\cal A}$ is
the geometric and kinematic acceptance correction factor
(Chapter~\ref{chapt:geom_kin}), $f_u$ is the fragmentation fraction,
or the probability that a $b$ quark will hadronize with a $u$ quark
into a $B^\pm$ meson, $\sigma\left(p\,\bar{p}\to b(\bar{b})\,X\right)$
is the $b$-quark hadroproduction cross section ($X$ denotes the
underlying event), $\int\!{\cal L}\,dt$ signifies the time-integrated
luminosity of the data sample, and ${\cal B}(\psi(2S)\to
J/\psi\,\pi^+\,\pi^-)$ and ${\cal B}(J/\psi\to\mu^+\mu^-)$ are
charmonium branching fractions.  The factor of two in the denominator
of Equation~\ref{eqn:br_fract} accounts for the fact that both $B^+$
and $B^-$ meson candidates are reconstructed.

As the schematic decay chains in Figures~\ref{fig:topo1} and
\ref{fig:topo2} connote, several topological similarities exist
amongst the decays under scrutiny.  The construction of ratios of
branching fractions like the one expressed in
Equation~\ref{eqn:br_fract} can therefore exploit these congruities by
accommodating the cancelation of several common reconstruction
efficiencies and systematic uncertainties.  Moreover, other quantities
with sizeable measurement uncertainties concomitantly divide to unity
in these ratios.  Since the data sample
(Section~\ref{sect:data_sample}) is common to modes in both the
numerator and denominator of a ratio of branching fractions, the
$\int\!{\cal L}\,dt$ factors are no longer important.  Depending on
which modes make up a given ratio, some of the charmonium and
$K^*(892)^0$ branching-fraction factors can divide out of the
expression, thereby reducing the overall systematic uncertainty.  The
cancelation of the $\sigma\left(p\,\bar{p}\to b(\bar{b})\,X\right)$
factors is expected to be especially beneficial, as theoretical models
(Section~\ref{sect:mc_gen}) and experimental measurements of the
$b$-quark differential hadroproduction cross section versus $p_{\rm T}$,
both of which rely on some assumptions, have been observed to have
consistent shapes but only marginally consistent
normalizations~\cite{abe:bxsec}.

An example of a branching-fraction ratio to be measured in this study
is given by extending Equation~\ref{eqn:br_fract} to the expression
\begin{equation}
\label{eqn:ratio_br_fract}
\frac{{\cal B}(B^+\to\psi(2S)\,K^+)} {{\cal
B}(B^0\to J/\psi\,K^*(892)^0)} =
\frac{N(B^+\to\psi(2S)\,K^+)}{N(B^0\to J/\psi\,K^*(892)^0)} \cdot
\frac{\varepsilon^\prime_R {\cal A}^\prime}{\varepsilon_R {\cal A}} \cdot
\frac{f_d}{f_u} \cdot
\frac{{\cal B}(K^*(892)^0 \to K^+\,\pi^-)}
{{\cal B}(\psi(2S)\to J/\psi\,\pi^+\,\pi^-)},
\end{equation}
where the primed quantities refer to the $B^0\to J/\psi\,K^*(892)^0$
reconstruction, the subscripts $R$ indicate that the efficiency
products have been reduced from their absolute values due to the
cancelation of common factors in the ratio, and $f_d$ is the
fragmentation fraction for the $b\to B^0$ hadronization process.
Direct measurements of branching-fraction ratios such as that in
Equation~\ref{eqn:ratio_br_fract} may be made with the assumption that
$f_u = f_d$.  Measurements of $f_u/f_d$ that assumed isospin symmetry
have confirmed this hypothesis in $p\bar{p}$ collisions, up to an
uncertainty of 21\%~\cite{abe:george_prd}.  Reference~\cite{pdg96}
also assumes $f_u = f_d$ on the grounds that the $B^+$ and $B^0$ meson
masses are nearly equal and that the CLEO collaboration has measured
$f_u/f_d = 1.13\pm 0.20$~\cite{cleo:fufd}.  Finally, the
world-average~\cite{pdg96} $B^+\to J/\psi\,K^+$ branching fraction
will be used to extract measurements of those absolute branching
fractions listed in Table~\ref{tab:predictions} from the appropriate
measured ratios of branching fractions.

Table~\ref{tab:ratios_to_measure} lists the branching-fraction ratios
investigated.  Numerators and denominators containing $\psi(2S)$
mesons are composed of contributions from two separate $B$-meson
reconstructions, namely those involving the $\psi(2S)\to\mu^+\,\mu^-$
and $\psi(2S)\to J/\psi\,\pi^+\,\pi^-$ decay modes.

\begin{table}
\begin{center}
\begin{tabular}{|l||ccc|}        \hline
\multicolumn{1}{|c||}{$R^i_j \equiv {\cal B}(i)/{\cal B}(j)$}
		      &$B^+\to J/\psi\,K^+$ & $B^0\to J/\psi\,K^*(892)^0$&
			$B^+\to \psi(2S)\,K^+$	 \\ \hline\hline
\rule{0mm}{7mm}
			$B^0\to J/\psi\,K^*(892)^0$ &
			$R^{J/\psi\,K^*(892)^0}_{J/\psi\,K^+}$ & & \\
\rule{0mm}{7mm}
$B^+\to\psi(2S)\,K^+$ & $R^{\psi(2S)\,K^+}_{J/\psi\,K^+}$ &
			$R^{\psi(2S)\,K^+}_{J/\psi\,K^*(892)^0}$ & \\
\rule[-5mm]{0mm}{12mm}
$B^0\to\psi(2S)\,K^*(892)^0$ & $R^{\psi(2S)\,K^*(892)^0}_{J/\psi\,K^+}$&
			$R^{\psi(2S)\,K^*(892)^0}_{J/\psi\,K^*(892)^0}$&
			$R^{\psi(2S)\,K^*(892)^0}_{\psi(2S)\,K^+}$ \\ \hline
\end{tabular}
\end{center}
\caption
[Branching-fraction ratios measured in the analysis.]
{The branching-fraction ratios measured for the various $B$-meson final
states.  The ratio $R^i_j$ is located in the $i^{\rm th}$ row and the
$j^{\rm th}$ column, and the $i$ and $j$ indices refer to the numerators
and denominators of the ratios, respectively.  Note that the ratios
containing $\psi(2S)$ mesons are composed of contributions from two
separate $B$-meson reconstructions.}
\label{tab:ratios_to_measure}
\end{table}

Chapter~\ref{chapt:apparat} describes the acceleration and detection
apparatus used to produce $B$ mesons and record their decays,
respectively.  Techniques invoked to reconstruct $B$-meson candidates
and reject background processes are discussed in
Chapter~\ref{chapt:select}, whereas Monte Carlo methods employed to
determine the geometric and kinematic acceptance corrections receive
treatment in Chapter~\ref{chapt:geom_kin}.  Chapter~\ref{chapt:effic}
traces the reckoning of several efficiencies and their associated
systematic uncertainties, both of which are used in
Chapter~\ref{chapt:br_fract} to calculate ratios of branching
fractions from the observed yields of the decays under study.  The
implications of the measurements are discussed further in
Chapter~\ref{chapt:br_fract} and conclusions are offered in
Chapter~\ref{chapt:concl}.

%% file: apparat.tex
\label{chapt:apparat}

The apparatus employed in this study resides at the Fermi National
Accelerator Laboratory (FNAL) in Batavia, Illinois, USA.  The
laboratory, which is commonly known as Fermilab, is owned by the
United States Department of Energy and is operated under a contract
with the Universities Research Association, Incorporated.
Section~\ref{sect:tevatron} in this chapter briefly describes the
sequence of accelerators, culminating in the Tevatron synchrotron,
that ultimately accelerate and collide beams of protons against those
of antiprotons at centre-of-mass energies of 1.8~TeV.

In this study, observation of the ensuing collision products was
achieved through the use of one of two general-purpose particle
detectors situated at different interaction regions on the Tevatron
collider ring.  The Collider Detector at Fermilab (CDF detector) is an
azimuthally and forward-backward symmetric device that consists of
several tracking, calorimeter, and muon subsystems.
Section~\ref{sect:cdfoverview} provides an overview of the CDF
detector and Sections~\ref{sect:tracking} through
\ref{sect:daq} describe those subsystems apposite to the present
analysis: the tracking, muon, trigger, and data acquisition systems.

\section{The Fermilab Tevatron $p\bar{p}$ Collider}
\label{sect:tevatron}
\begin{figure}
\begin{center}
\leavevmode
\hbox{%
\epsfxsize=6.0in
\epsffile{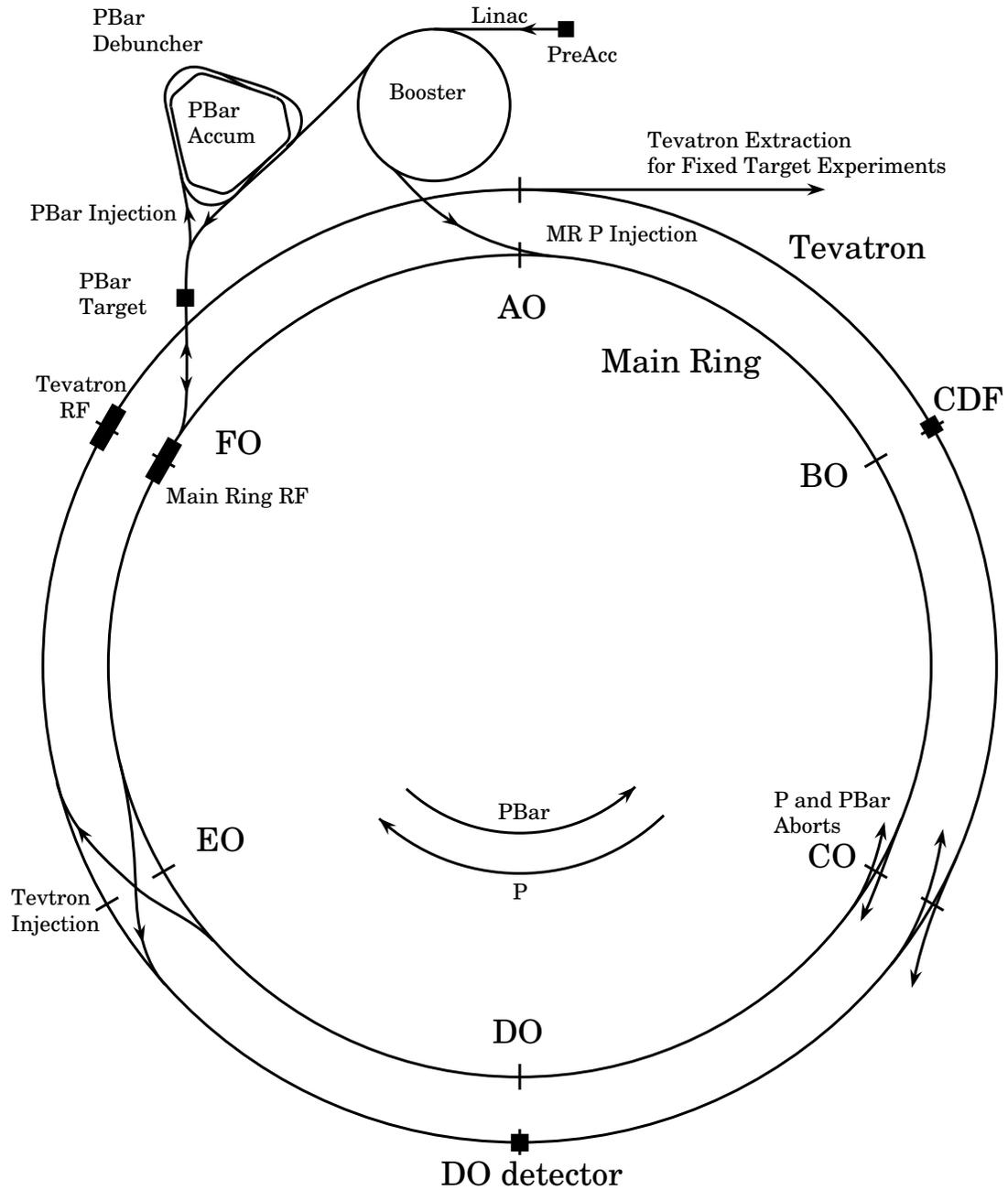}}
\end{center}
\caption
[Schematic diagram of the Tevatron collider and its affiliated
accelerators.]
{Schematic diagram of the $\sim$6.28-km circumference Tevatron
$p\bar{p}$ collider and its affiliated accelerators.  For simplicity,
the Main Ring and Tevatron are diagrammed as coplanar.  This figure is
not drawn to scale.}
\label{fig:tevatron}
\end{figure}

The acceleration of protons and antiprotons to energies of 900 GeV is
accomplished at Fermilab by a synergism of six particle accelerators.
The Cockroft-Walton~\cite{cockroft-walton} pulsed ion source begins
the sequence by converting gaseous $H_2$ molecules to $H^-$ ions,
which are subsequently subjected to a 750-keV electric potential.  The
$H^-$ ions then enter a 150-m linear accelerator, or Linac, where they
are accelerated to energies of 400~MeV by a sequence of drift-tube
induced oscillating electric
fields~\cite{sanford:fermilab,lederman:tevatron}.  Refer to
Figure~\ref{fig:tevatron} for a schematic diagram that depicts the
Linac and the other accelerators.

At the end of the Linac stage, the $H^-$ ions are guided into the
Booster, an alternating gradient synchroton~\cite{mcmillan:synch} with
a circumference of $\sim$470~m, in which they make $\sim$16\,000
revolutions and acquire an energy of 8~GeV.  During injection into the
Booster, both electrons are stripped from the $H^-$ ions by passing
the ions through a carbon foil.  The 8-GeV protons are injected from
the Booster into the $\sim$6.28-km circumference Main Ring proton
synchroton where they are accelerated to typical energies of
$\sim$150~GeV under the guidance of $1\,014$ conventional
copper-coiled dipole and quadrupole magnets~\cite{sanford:fermilab}.
Please refer to Figure~\ref{fig:tevatron}.

The Tevatron, located directly below and in the same tunnel as the
Main Ring, is another synchrotron, but one that is distinct from the
others in that it has magnet coils consisting of superconducting
niobium-titanium (Nb-Ti) alloy filaments embedded in copper instead of
the conventional copper coils used in the Booster and Main Ring
magnets~\cite{palmer:sc_magnets,wilson:tevatron}.  The increased
magnetic fields produced by the Nb-Ti magnets enable the Tevatron to
accelerate protons to energies of nearly 1~TeV.

Antiprotons are produced using 120-GeV protons that are extracted from
the Main Ring and are made to strike a 7-cm thick nickel or copper
target.  A liquid lithium lens focuses the antiprotons and directs
them to the Debuncher, which is a ring 520-m in circumference where
the antiproton beam aperture and energy distribution are reduced by
means of stochastic cooling~\cite{vdmeer:cooling} and
debunching~\cite{ruggiero:debunch} techniques, respectively.  The
antiprotons are then transferred to the Accumulator ring, which is
concentric with the Debuncher, for storage and further cooling.  Once
enough antiprotons have been accumulated and the Tevatron has already
been filled with 150-GeV proton bunches, antiprotons in the
Accumulator are `shot' into the Main Ring, boosted to 150~GeV,
injected into the Tevatron in counter-rotation to the proton bunches,
and then accelerated along with the protons to 900~GeV.  The
counter-rotating beams are made to collide at interaction regions such
as B\O\ (shown in Figure~\ref{fig:tevatron}) where 1.8-TeV collisions
occur near the geometric centre of the CDF detector.

\section{An Overview of the CDF Detector}
\label{sect:cdfoverview}

The Collider Detector at Fermilab (CDF detector) is a general-purpose
device designed to study the physics of $p\bar{p}$ collisions at
centre-of-mass energies near 2.0~TeV.  A comprehensive description of
the CDF detector and its subsystems is given in
References~\cite{abe:cdfnim_overview,tkaczyk:svxnim,amidei:svxnim,azzi:svxpnim,abe:topprdl,abe:wmass92}
and citations therein.

\begin{figure}
\begin{center}
\leavevmode
\hbox{%
\epsfxsize=6.0in
\epsffile{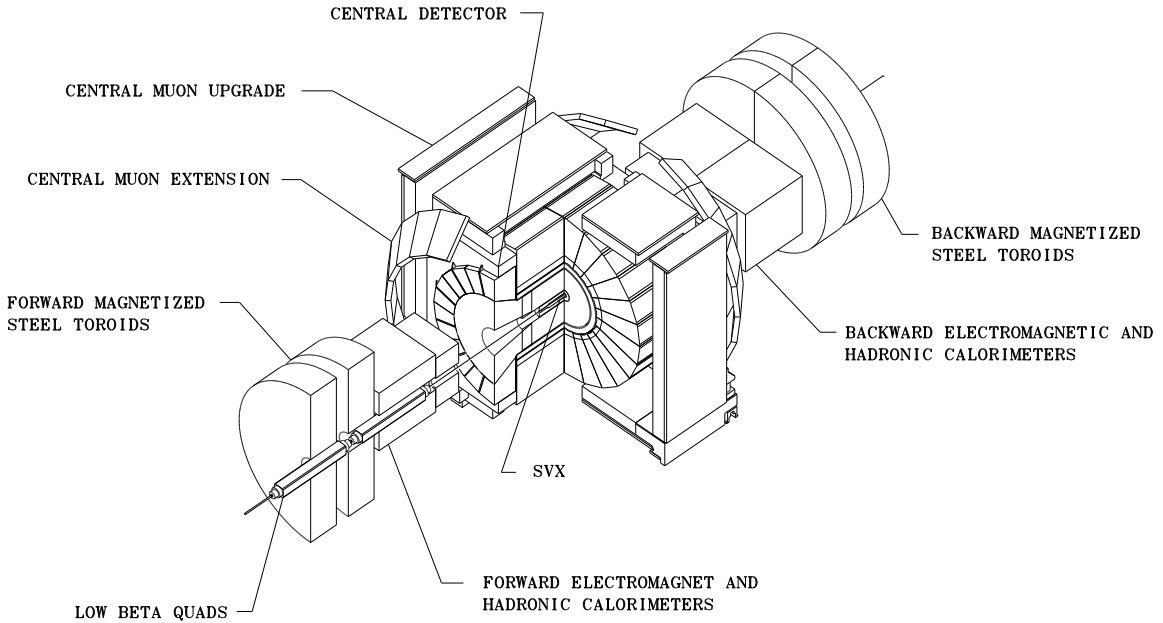}}
\end{center}
\caption
[Isometric view of the CDF detector.]
{An isometric view of three quarters of the CDF detector.}
\label{fig:cdf_iso}
\end{figure}

The basic design goals of the CDF detector, pictured in
Figure~\ref{fig:cdf_iso}, were to identify leptons and measure the
momenta and energies of particles originating from the B\O\
interaction region.  Since the phase space for high energy hadronic
collisions is typically described by rapidity (refer to
Section~\ref{sect:hadroproduction}), transverse momentum, and
azimuthal angle, it is natural that the CDF detector components have
an approximately cylindrical symmetry and uniform segmentation in
pseudorapidity and azimuthal angle\footnote{The CDF coordinate system
is right-handed with $x$ pointing out of the Tevatron ring, $y$
vertical, and $z$ in the proton beam direction.  The polar angle,
$\theta$, is measured with respect to the proton direction; the
pseudorapidity, $\eta$, is defined by
$\eta\equiv-\ln(\tan(\theta/2))$, with $\theta$ measured assuming a
$z$-vertex position of zero; the azimuthal angle is represented by
$\varphi$ and defined with respect to the plane of the Tevatron; and
the transverse displacement coordinate is denoted by $r$.}.  Tracking
detectors, which detect charged particles and measure their momenta,
reside nearest the interaction region and inside a $\sim$1.4-T
magnetic field.  The field is generated by a large electromagnet that
consists of $1\,164$ turns of Nb-Ti/Cu superconductor that constitute
a solenoid 4.8-m in length, 1.5-m in radius, and 0.85 radiation
lengths in radial thickness.  The tracking systems surround an
evacuated beryllium beam pipe that is 3.8-cm in diameter, has walls
0.5-mm thick, and forms part of the Tevatron.
Section~\ref{sect:tracking} describes the CDF tracking systems in some
detail.

\begin{figure}[t]
\begin{center}
\leavevmode
\hbox{%
\epsfxsize=6.0in
\epsffile{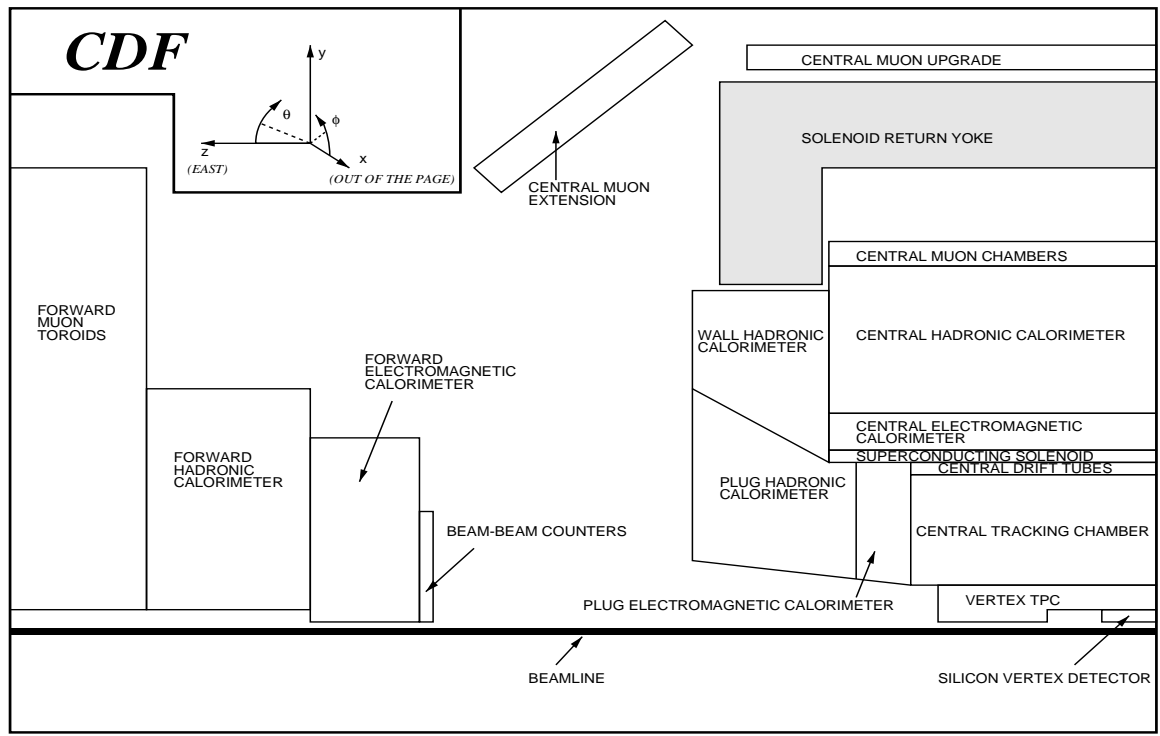}}
\end{center}
\caption
[Schematic elevation view of the CDF detector.]
{A schematic side-elevation view of one quadrant of the CDF detector.
The CDF detector, with the exception of the central muon upgrade and
extension subsystems, is largely cylindrically symmetric about the
interaction region, which is located on the beamline at the right side
of the figure.}
\label{fig:cdf_xsec}
\end{figure}

The detector is divided into a central region ($|\eta| < 1.1$), two
end plug regions ($1.1 < |\eta| < 2.4$), and two forward-backward
regions ($2.2 < |\eta| < 4.2$).  Refer to Figure~\ref{fig:cdf_xsec}
for a schematic elevation-view of one quadrant of the CDF detector.
The tracking volume and solenoid are surrounded by sampling
calorimeters that measure electromagnetic and hadronic energy flow
from the collision point for particles with $|\eta| < 4.2$.  The
calorimeter systems are segmented into projective $\eta$-$\varphi$
`towers', each of which points back towards the nominal interaction
region and has an electromagnetic shower counter in front of a
corresponding hadronic calorimeter cell.  In high energy $p\bar{p}$
collisions, such a projective tower geometry is motivated by the
importance of reconstructing jets, which are defined as collections of
particles that have similar trajectories in $\eta$-$\varphi$ space and
are typically assumed to originate from a single high energy quark or
gluon.

Each of the three main CDF detector regions has an electromagnetic
calorimeter; these are the central electromagnetic (CEM), the plug
electromagnetic (PEM), and the forward electromagnetic (FEM)
calorimeters.  Behind the CEM there are two hadronic calorimeters: the
central (CHA) and wall hadronic (WHA) calorimeters.  The PEM and FEM
have a corresponding plug (PHA) and forward hadronic (FHA) calorimeter
behind each of them, respectively.  Table~\ref{tab:calorimeters}
summarizes some selected properties of the CDF calorimeter systems.

Proportional chambers located between the solenoid and the CEM
constitute the central preradiator detector, which samples the early
$r$-$\varphi$ development of electromagnetic showers induced in the
material of the superconducting solenoid coils.  Other proportional
chambers with strip and wire readout are located inside the CEM
calorimeter at a depth of $\sim$6X$_0$, the approximate point where an
electromagnetic shower is most fully developed.  These central
electromagnetic strip detectors measure the positions of showers in
both the $z$ and $r$-$\varphi$ views.

\begin{table}
\begin{center}
\begin{tabular}{|l||c|c|c|c|c|c|c|}	\hline
\rule[-3mm]{0mm}{8mm}
Property & CEM & CHA & WHA & PEM & PHA & FEM & FHA \\ \hline\hline
\rule[-3mm]{0mm}{8mm}
$|\eta|$ & $0 - 1.1$ & $0 - 0.9$ & $0.7 - 1.3$ & $1.1 - 2.4$ & $1.3 - 2.4$ &
	$2.2 - 4.2$ & $2.3 - 4.2$ \\ \hline
\rule[-3mm]{0mm}{8mm}
$\Delta\eta$ & $\sim0.1$ & $\sim0.1$ & $\sim0.1$ & $0.09$ & $0.09$ &
	$0.1$	& $0.1$ \\ \hline
\rule[-3mm]{0mm}{8mm}
$\Delta\varphi$ & $15^\circ$ & $15^\circ$ & $15^\circ$ & $5^\circ$ &
	$5^\circ$ &$5^\circ$ &$5^\circ$ \\ \hline
\rule[-3mm]{0mm}{8mm}
Active & \multicolumn{3}{c|}{plastic scintillator} &
	\multicolumn{4}{c|}{gas chambers with cathode pad readout}\\\hline
\rule[-3mm]{0mm}{8mm}
Absorber & Pb & Fe & Fe & Pb & Fe & Pb/Sb & Fe \\ \hline
\rule[-3mm]{0mm}{8mm}
Thickness & 18X$_0$ & 4.5$\lambda_0$ & 4.5$\lambda_0$ & $18-21$X$_0$ &
	5.7$\lambda_0$ & 25X$_0$ & 7.7$\lambda_0$ \\ \hline
\end{tabular}
\end{center}
\caption
[Properties of the CDF calorimeters.]
{Selected properties of the CDF calorimeter systems.  Shown are the
pseudorapidity coverage and segmentation, the azimuthal segmentation,
the active medium, the type of absorber, and the thickness in
radiation lengths (X$_0$) and interaction lengths ($\lambda_0$) of the
electromagnetic and hadronic calorimeters, respectively.}
\label{tab:calorimeters}
\end{table}

\section{The Tracking Systems}
\label{sect:tracking}

The reconstruction of exclusive $B$-meson decays relies heavily on
precise measurements of the daughter particle decay vertices, momenta,
and charges.  The CDF detector's main tracking capabilities consist of
four distinct but complementary tracking subsystems.  These systems,
listed in order of increasing distance from the interaction region,
are the silicon vertex detector, the vertex time projection chamber,
the central tracking chamber, and the central drift tube array.
Figure~\ref{fig:cdf_xsec} illustrates the positions of the tracking
subsystems, both relative to each other and to the rest of the CDF
detector.

\subsection{The Silicon Microstrip Vertex Detector}
\label{sect:svx}

\begin{figure}
\begin{center}
\leavevmode
\hbox{%
\epsfxsize=4.5in
\epsffile{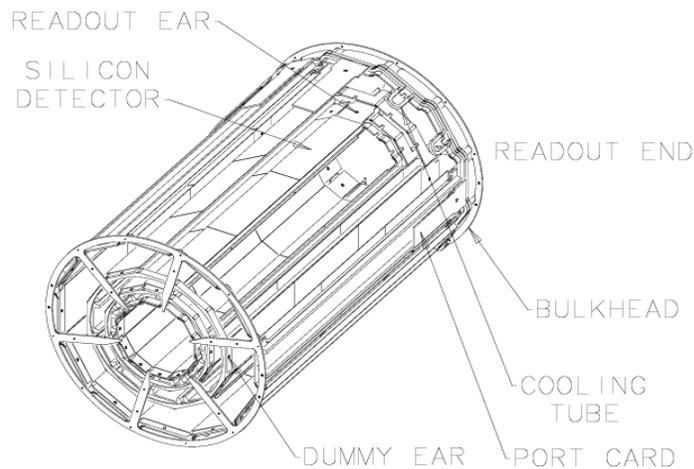}}
\end{center}
\caption
[Isometric view of one of the two SVX barrels.]
{Isometric view of one of the two silicon microstrip vertex detector
(SVX) barrels.  The dummy-ear sides of both barrels are conjoined
(with an effective gap of 2.15~cm) at the $z = 0$ position inside the
CDF detector.}
\label{fig:svx_iso}
\end{figure}

The silicon microstrip vertex detector
(SVX)~\cite{tkaczyk:svxnim,amidei:svxnim} enables the identification
in the $r$-$\varphi$ plane of secondary vertices displaced from the
$p\bar{p}$ collision point resulting from the weak decays of $b$
quarks.  Installed in the CDF detector in 1992, the SVX was the first
detector of its kind to be operated successfully in a hadron collider
environment.  In 1993, a more radiation-hard and low-noise version of
the SVX, the SVX$^\prime$~\cite{azzi:svxpnim}, was commissioned for
the 1994-1995 Tevatron collider run\footnote{Unless noted otherwise,
references to the SVX apply to the SVX$^\prime$ as well.} (refer to
Section~\ref{sect:data_sample}).  The SVX consists of two identical
cylindrical modules, one of which is pictured in
Figure~\ref{fig:svx_iso}, each comprising four concentric cylindrical
layers with radii of 3.0, 4.3, 5.7, and 7.9~cm.  The SVX$^\prime$ has
the same overall configuration as the SVX, except that the innermost
layer has a slightly smaller radius of 2.9~cm.  Since the luminous
$p\bar{p}$ interaction region is rather elongated in the $z$ direction
(with a Gaussian distribution having a standard deviation of
$\sigma\sim30$~cm), approximately 40\% of $p\bar{p}$ collision
vertices lie outside the acceptance of the SVX, which has an active
length of 51.1~cm.

\begin{figure}
\begin{center}
\leavevmode
\hbox{%
\epsfysize=3.1in
\epsffile{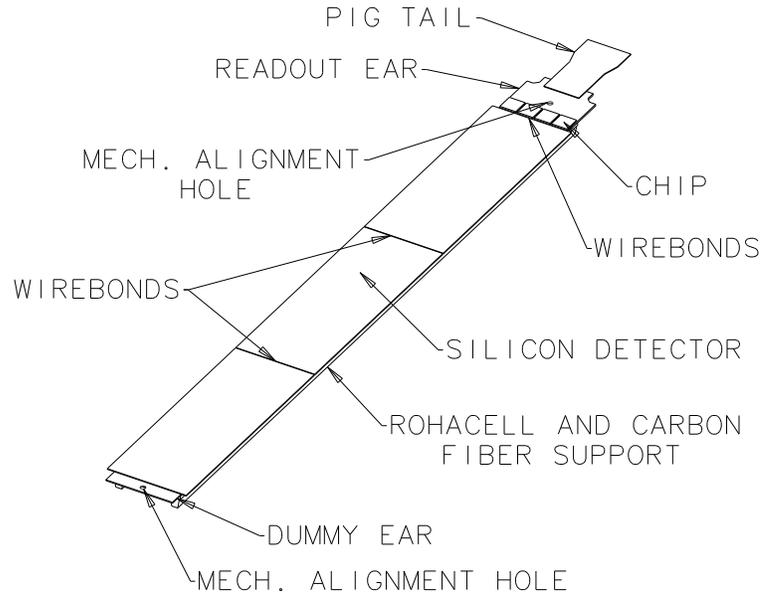}}
\end{center}
\caption
[Layout of an SVX ladder module.]
{Layout of a typical ladder module in the third layer of the SVX.
Three single-sided silicon microstrip detectors are wire bonded
together to constitute each ladder module.}
\label{fig:svx_ladder}
\end{figure}

In both of the SVX barrels, the four layers are each segmented into
twelve `ladders' that subtend approximately 30$^\circ$ in azimuth
and are oriented parallel to the beam axis.
Figure~\ref{fig:svx_ladder} depicts a typical ladder situated in the
third layer of the SVX.  The ladder substrates are fabricated from a
light-weight foam (Rohacell) reinforced with strips of carbon fibre.
Three single-sided 8.5-cm long silicon microstrip detectors are
electrically bonded together with aluminum wire along the $z$
direction to form a 25.5-cm active silicon region on each ladder
module.

Silicon microstrip detectors are a kind of solid ionization chamber in
which incident charged particles dislodge electrons via ionization.
In their most basic form, the detectors consist of an $n$-doped
silicon wafer, typically 300-$\mu$m thick, with strips of $p$-type
material on one side~\cite{kleinknecht:part_detectors}.  If a
reverse-biased potential is applied to a strip, a $p$-$n$ junction
diode is set up, and electron-hole pairs created by the passage of a
charged particle migrate to their respective electrodes, thereby
manifesting an electronic signal that resolves the location of the
particle's trajectory as a function of the strip separation, or
`pitch'.

The silicon strip pitch of the inner three SVX layers is 60~$\mu$m and
that for the outermost layer is 55~$\mu$m.  The average position
resolutions for the SVX and the SVX$^\prime$ were measured to be
13~$\mu$m and 11.6~$\mu$m, respectively, and the high transverse
momentum (asymptotic) impact parameter resolution was determined to be
17~$\mu$m for the SVX and 13~$\mu$m for the SVX$^\prime$.  Adjacent
ladders in a given layer slightly overlap each other to provide full
azimuthal coverage; this is achieved with a $3^\circ$ rotation of the
ladders about their major axes.  The SVX$^\prime$ has all four of its
layers overlapped; however, the innermost layer of the SVX suffers
from a $1.26^\circ$ gap in $\varphi$ between adjacent ladder modules.

As shown in Figure~\ref{fig:svx_ladder}, the outside end of each
ladder has a small circuit board that contains the front end readout
chips, which each serve 128 channels.  Because the ladder widths
increase with increasing $r$, the number of readout chips on a given
ladder module depends upon the layer in question.  The innermost
layer, for example, has two readout chips per ladder module whereas
the outermost layer has six chips per ladder module.  The total number
of instrumented strips in the SVX is $46\,080$.

The silicon strips used in the SVX$^\prime$ are AC coupled whereas
those for the SVX are DC coupled.  The SVX$^\prime$ consequently
benefits from a marked reduction in noise compared to the SVX, for
which strip-to-strip leakage current variations have to be subtracted
by cycling the front end readout chips through two successive charge
integrations.  An additional advantage of the AC-coupled SVX$^\prime$
design is that the readout chip preamplifiers, which have 40\% more
gain than their SVX counterparts, will not saturate, even after
significant radiation damage has increased the silicon microstrip
leakage currents appreciably.  The SVX$^\prime$ readout chips were
fabricated using radiation hard 1.2-$\mu$m CMOS technology.  They
therefore have a design absorbed-dose limit of 10~kGy compared with
200~Gy for the SVX, which uses more conventional 3-$\mu$m CMOS
electronics~\cite{kleinfelder:svx_cmos}.

The readout electronics typically generate $\sim$53.1~W of heat in
each of the two SVX barrels.  Cooling pipes transport chilled
de-ionized water at a temperature of 13~$^\circ$C and a flow rate of
10~g/s to the beryllium bulkhead (see Figure~\ref{fig:svx_iso}) and
the readout circuit boards (see Figure~\ref{fig:svx_ladder}) to
maintain an operating temperature near 20~$^\circ$C.  The cooling
circuit runs at a subatmospheric pressure to minimize the potential
damage due to a breach in the cooling pipes.  Controlling the
temperature not only minimizes leakage currents in the silicon
microstrips and prevents damage to the front end electronics, but it
also discourages thermal gradients in the mechanical support structure
that can distort the internal alignment of the SVX.

\subsection{The Vertex Time Projection Chamber}
\label{sect:vtx}

A vertex time projection drift chamber (VTX) surrounds the SVX (refer
to Figure~\ref{fig:cdf_xsec}).  It was designed to measure the
trajectories of charged particles in the $r$-$z$ plane in a
pseudorapidity range $|\eta| \lessim 3.0$.  The VTX is important for
the determination of the $z$ position of the primary vertex and the
identification of multiple interactions in the same beam crossing.
The VTX resolution of a primary vertex location along the beamline,
nominally 2~mm~\cite{huffman:pc}, depends on the number of detected
tracks originating from that location and the number of primary
$p\bar{p}$ interactions in the event.

\begin{figure}
\begin{center}
\leavevmode
\hspace*{0.45in}
\hbox{%
\epsfxsize=6.0in
\epsffile{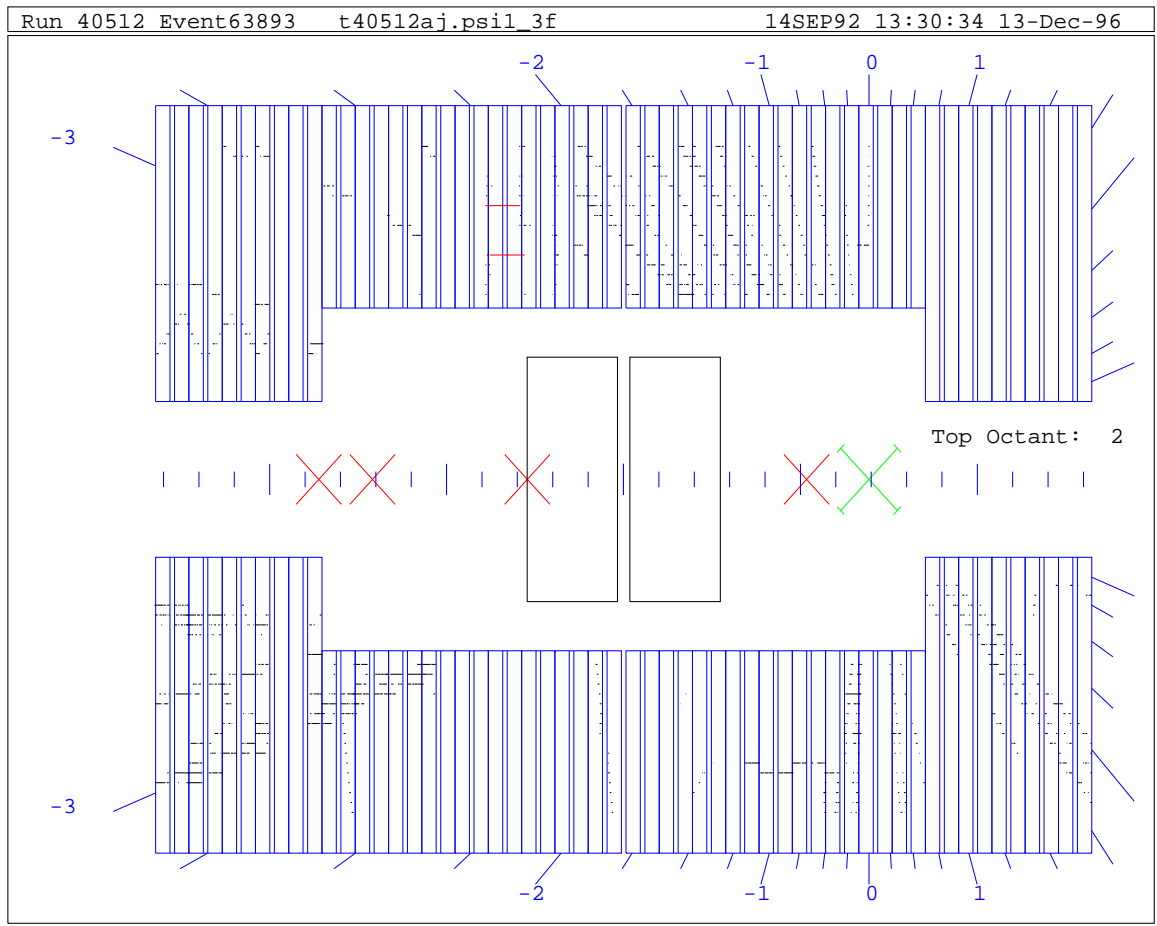}}
\end{center}
\caption
[Event display of the vertex time projection chamber (VTX).]
{Event display of two 45$^\circ$ octants in the vertex time projection
chamber (VTX).  The beamline runs through the middle of the diagram,
across the page.  Shown are the 28 drift modules containing hits that
form tracks due to the traversal of charged particles from the
interaction region.  The hits represent the times that the drifting
ionization electrons arrive near the sense wires.  The crosses along
the beamline represent $z$ vertex candidates found by the VTX
software; the single seriffed cross represents an especially high
quality vertex that consequently defines the zero location of the
event pseudorapidity coordinates, as indicated by the $\eta$ scale
denoted around the outside of the figure.  The two rectangles drawn
near the geometric centre of the VTX represent active regions of the
SVX.  This event is a dramatic example of a case where the primary
vertex is so distant from $z = 0$ that the SVX is of little
consequence.  Also interesting is the helical trajectory of a particle
in the upper VTX octant at $\eta \sim -3.5$, a result of the large
axial magnetic field.}
\label{fig:df_vtx}
\end{figure}

The VTX, which extends 132~cm in each $z$ direction and has a radius
of 22~cm, consists of 28 drift modules, each containing two drift
regions separated by an aluminum central high voltage grid.  Endcaps
on each side of the drift modules are azimuthally segmented into
octants and are rotated in $\varphi$ by 15$^\circ$ with respect to
adjacent modules (in $z$) to eliminate inefficiencies near module
boundaries.  Within each octant, 16 or 24 sense wires, depending on
the $z$ position of the module, are oriented tangentially, thereby
providing tracking information in the $r$-$z$ view.  The $z$ location
of a track with respect to a given wire in a given module is
determined by the drift time, and the $r$ information is determined
from the radial location of the wire.  The electric field is
maintained near 1.6~kV/cm and the gas used is a 50:50 admixture of
argon and ethane.  Figure~\ref{fig:df_vtx} is an event display diagram
showing hits from charged particles in two VTX octants.  The VTX
vertex-finding algorithm has calculated a primary vertex $z$ position
in this event based on a fit to extrapolations of tracks reconstructed
from sense wire hits.  The $z$ vertex position is represented by the
seriffed cross in Figure~\ref{fig:df_vtx}.

\subsection{The Central Tracking Chamber}
\label{sect:ctc}

The most prominent subsystem in the CDF detector is the central
tracking chamber, or CTC.  It is the only CDF tracking device that can
perform three dimensional momentum and position measurements, both of
which are essential to the reconstruction of exclusive $B$-meson
decays.  The CTC, as indicated in Figure~\ref{fig:cdf_xsec}, surrounds
the VTX and SVX subsystems and has a coaxial bicylindrical geometry
with a 3\,201.3-mm length (including the endplates), a 2\,760.0-mm
outer diameter, and a 554.0-mm inner diameter.  Aluminum is used in
the construction of the outer cylinder; carbon fibre reinforced
plastic constitutes the inner cylinder wall.

\begin{figure}
\begin{center}
\leavevmode
\hbox{%
\epsfxsize=5.0in
\epsffile{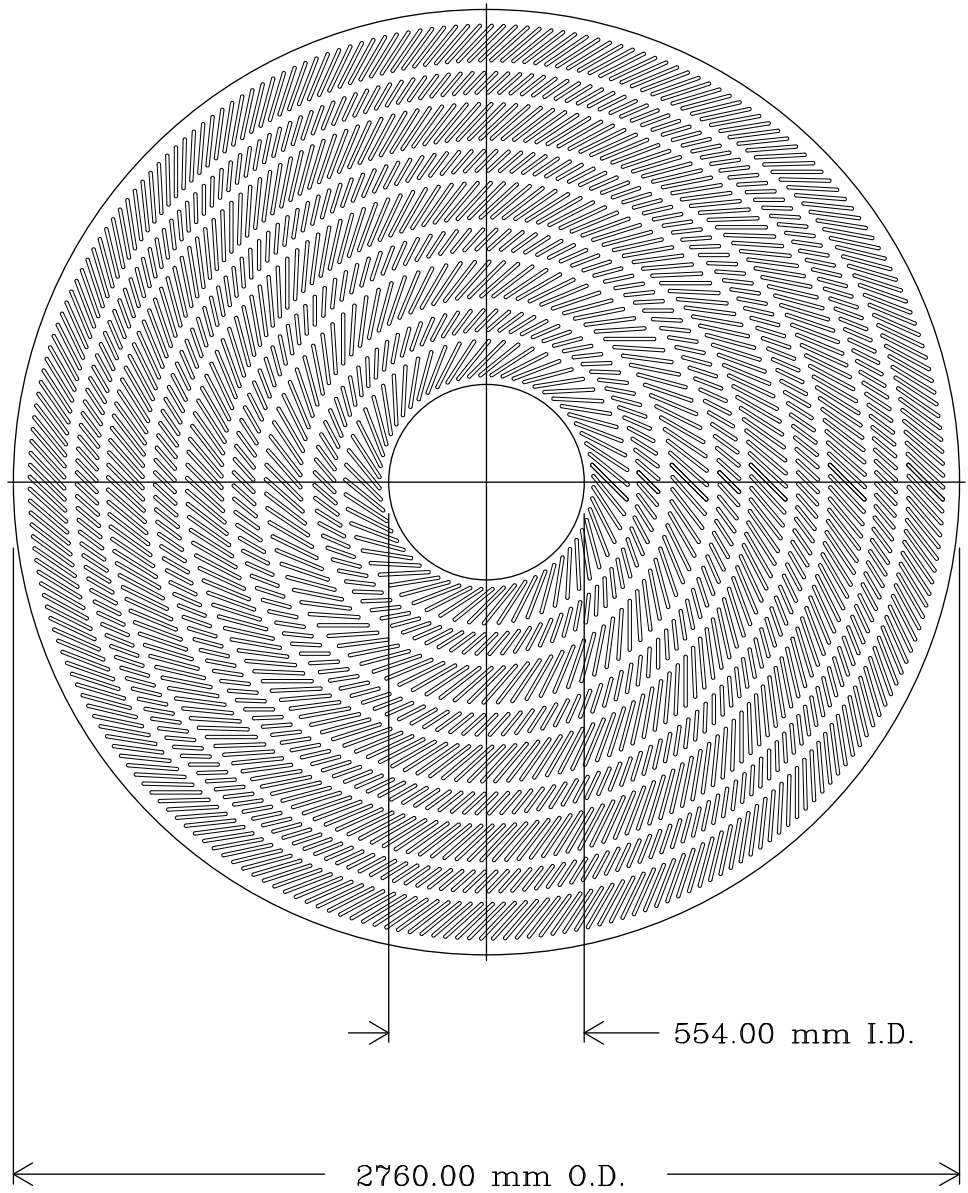}}
\end{center}
\caption
[A central tracking chamber (CTC) endplate.]
{One aluminum endplate of the central tracking chamber (CTC), viewed
from along the beam axis.  The wire slot locations for the alternating
axial and stereo superlayers are apparent.}
\label{fig:ctc_endplate}
\end{figure}

The CTC is a drift chamber that contains 84 layers of 40-$\mu$m
diameter gold-plated tungsten sense wires arranged into nine
`superlayers', five of which have their constituent sense wires
oriented parallel to the beam axis (axial superlayers), and four of
which have their wires canted at angles of either $+3^\circ$ or
$-3^\circ$ with respect to the beamline (stereo superlayers).  The
innermost and outermost sense wires have radii of 309~mm and
1\,320~mm, respectively.  The axial and stereo superlayers alternate
with increasing radius and each consists of twelve and six sense wire
layers, respectively.  The configuration is illustrated in
Figure~\ref{fig:ctc_endplate}, which shows the wire slot locations in
the aluminum endplates.  The majority of the CTC pattern recognition
is done using data from the axial layers, which provide tracking
information in the $r$-$\varphi$ view.  The stereo layers furnish
tracking information in the $r$-$z$ view.

The superlayers are functionally segmented into open drift cells.  A
drift cell contains a superlayer of (either 12 or 6) sense wires
alternating with (either 13 or 7) stainless steel potential wires,
which serve to control the gas gain on the sense wires.  Two planes of
stainless steel field wires on either side of the sense wire
superlayers define the fiducial boundaries of each drift cell and
control the strength of the electric field in the $\lessim$40-mm drift
regions.  The number of cells in each superlayer increases with $r$
such that the drift distance, which translates into a maximum drift
time of $\sim$800~ns, is approximately constant for all cells in the
CTC.  To keep the electric field uniform throughout the fiducial
volume of every drift cell, extra stainless steel shaper and guard
wires are located near the cell perimeters, bringing the number of
wires in the CTC to a total of $36\,504$.  This translates to a total
wire tension of 245~kN and a combined wire length that is in excess of
110~km.

As is evident in Figure~\ref{fig:ctc_endplate}, the CTC drift cells
are tilted such that the angle between the radial direction and the
electric field direction is approximately 45$^\circ$.  Such a large
cant angle is necessary to offset the 45$^\circ$ Lorentz angle, which
results from the combined effects of the $\sim$1.4-T magnetic field,
the argon-ethane-alcohol gas mixture used (in the proportions
$49.6\%:49.6\%:0.8\%$), and the relatively low $\sim$1.35-kV/cm
electric field.  The drift trajectories in the CTC are therefore
approximately parallel to the azimuthal direction.  Every high
transverse-momentum track passes close to at least one sense wire in
each superlayer.


Preamplifiers mounted on the endplates of the CTC are connected to the
sense wires, whose analog signals are amplified further, shaped, and
discriminated by circuitry mounted on the solenoid return yoke (see
Figure~\ref{fig:cdf_xsec}). The discriminator signals undergo
time-to-digital (TDC) conversion in a counting room located at the end
of 70~m of flat cable.  The 1-ns resolution TDCs can record $>7$ hits
per wire per event.  The CTC double track resolution is $<5$~mm due to
the approximately 100-ns minimum separation between two resolved hits.
The CTC has a single hit resolution of $<200$~$\mu$m, and the overall
momentum resolution of the combined SVX-CTC system is $\delta p_{\rm
T} / p_{\rm T} = \left[\left(0.0009\,p_{\rm T}\right)^2 +
\left(0.0066\right)^2\right]^{1/2}$, where $p_{\rm T}$ is the transverse
momentum measured in units of GeV/$c$.  Section~\ref{sect:track_effic}
and Appendix~\ref{app:patt_rec_eff} provide further discussion of the
performance of the CTC, and Figure~\ref{fig:event_display_text}
contains an event-display diagram of a sample event showing
reconstructed CTC track candidates.

\subsection{The Central Drift Tube Array}

The central drift tube array, or CDT, is situated at a radius of
1.4~m, between the outer cylinder of the CTC and the inner wall of the
solenoid cryostat, as indicated in Figure~\ref{fig:cdf_xsec}.  The CDT
system consists of stainless steel circular tubes; these are 1.27-cm
in diameter, 3-m in length, and $2\,016$ in number.  Closely packed
into three layers, the tubes are each strung with 50-$\mu$m diameter
stainless steel anode wires.  By virtue of its charge division
capability on the anode wires, the CDT can provide tracking
information in both the $r$-$\varphi$ and $r$-$z$ views.  For the
analysis described in this thesis, CDT tracking information was not
used explicitly in the reconstruction of particle tracks; however, the
CDT was used to identify cosmic ray muons as coincident hits with
$\Delta\varphi\sim 180^\circ$.  Cosmic ray muons were used to perform
the initial relative alignments of the SVX, VTX, and CTC subsystems
within the CDF detector.

\section{The Muon Chambers}
\label{sect:muonchambers}

\begin{figure}
\begin{center}
\leavevmode
\hbox{%
\epsfxsize=5.5in
\epsffile{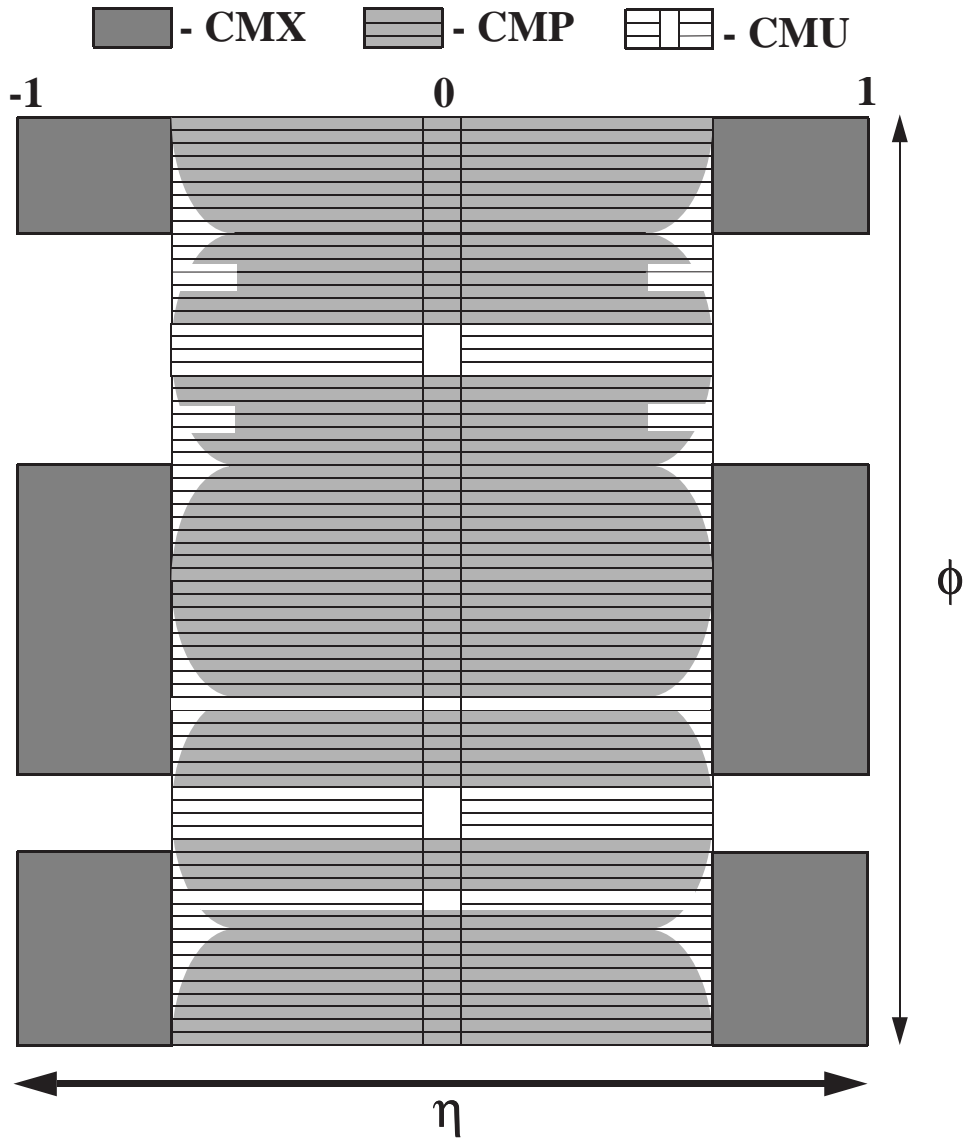}}
\end{center}
\caption
[Central muon $\eta$-$\varphi$ coverage map.]
{Coverage of the central muon subsystems in pseudorapidity ($\eta$)
and azimuth ($\varphi$)~\cite{lewis:cmp_cmx}.  The lack of CMX
coverage at $\varphi \sim 90^\circ$ and $\varphi \sim 270^\circ$
results from the interference due to the Main Ring bypass beampipe and
the concrete collision hall floor, respectively.}
\label{fig:muon_coverage}
\end{figure}

The ability to identify muons and their trajectories is essential to
the reconstruction of $J/\psi$ and $\psi(2S)$ mesons in the dimuon
channels.  Muon identification can be achieved by exploiting the
relatively high muon critical energy\footnote{The muon critical energy
is the energy at which losses due to radiation and ionization are
equal~\cite{pdg96}.}, which is several hundred GeV in materials such
as iron~\cite{pdg96}, significantly higher than the critical energy
for other ionizing particles.  This ability of the muon to penetrate
matter thus motivates the location of the muon subsystems in the outer
regions of the CDF detector that can only be reached by those charged
particles that originate from the interaction region and that
penetrate the intervening material.  This material, consisting
primarily of the calorimeters, serves to filter out the majority of
hadrons and electrons before they reach the muon subsystems.  Refer to
Figure~\ref{fig:cdf_xsec} for the locations of the three central muon
subsystems used in this analysis: the central muon detector (CMU), the
central muon upgrade detector (CMP), and the central muon extension
(CMX).  A map of the $\eta$-$\varphi$ muon detection coverage in the
central region is shown in Figure~\ref{fig:muon_coverage}.  The
forward muon toroid subsystem, shown in Figures~\ref{fig:cdf_iso} and
\ref{fig:cdf_xsec}, is not used in this study due to its poor
intrinsic momentum resolution and the lack of overlap in acceptance
between it and the CTC and SVX tracking systems.

\begin{figure}
\begin{picture}(200,250)(0,0)
\put(0,0){\includegraphics{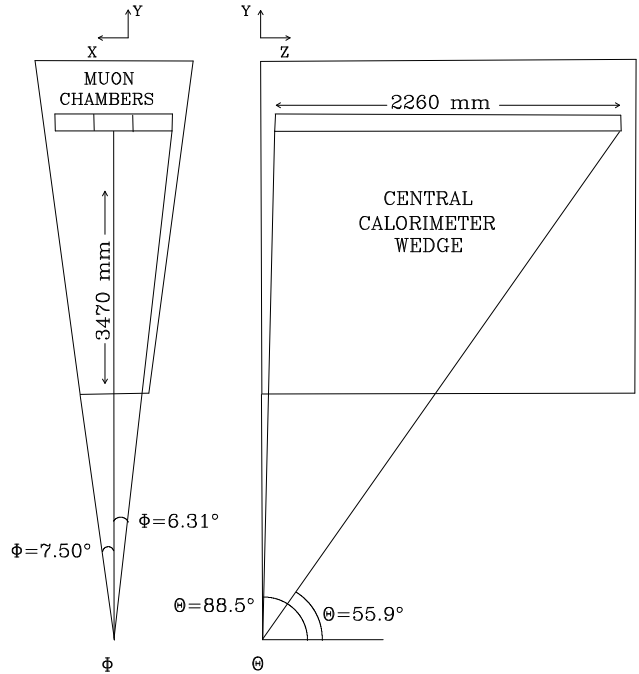}}
\end{picture}
\caption
[Layout of a central muon (CMU) wedge on a central calorimeter wedge.]
{The layout of a central muon (CMU) wedge with respect to a central
calorimeter wedge in both the azimuthal (left figure) and polar (right
figure) views.}
\label{fig:cmu_tower}
\end{figure}

\subsection{The Central Muon Detector}
\label{sect:cmu}

\begin{figure}
\begin{center}
\leavevmode
\hbox{%
\epsfxsize=6.0in
\epsffile{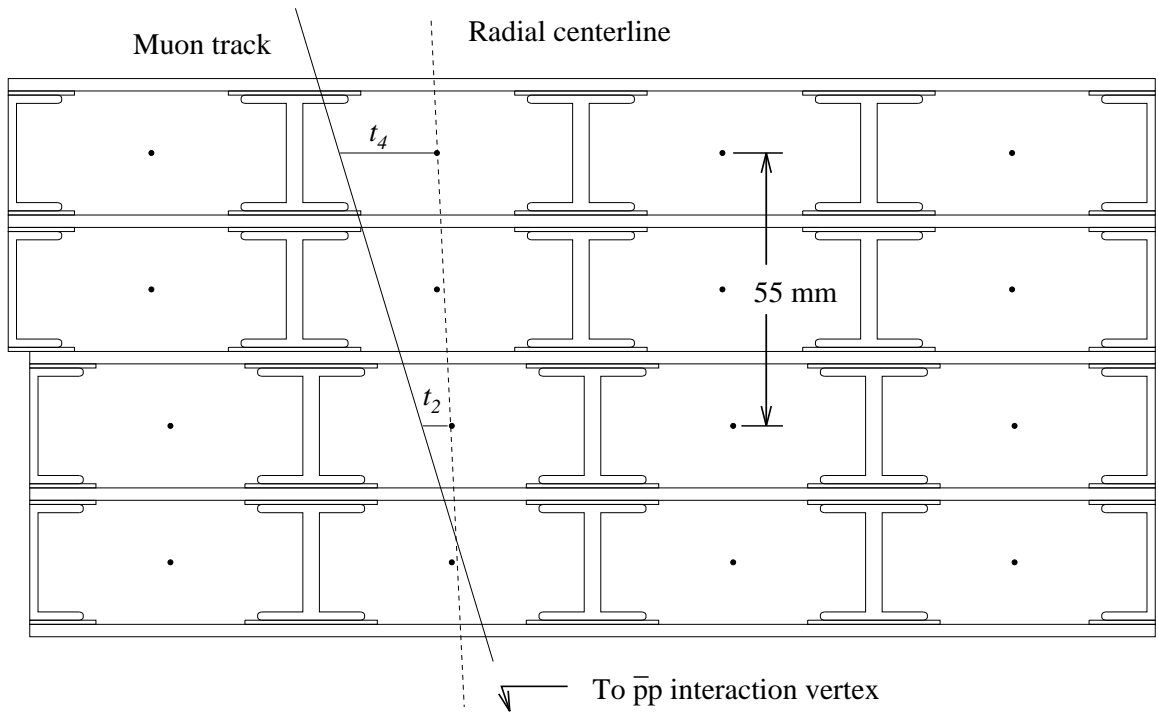}}
\end{center}
\caption
[Layout of a central muon detector (CMU) module.]
{Layout of a central muon detector (CMU) module showing the four
towers, each with four layers of rectangular drift cells.  The
outermost and second innermost cells in each tower are oriented such
that their sense wires lie on a radial that originates from the
geometric centre of the CDF detector.  The other two drift cells are
offset to determine which side of the radial the track passed.  The
quantities $t_2$ and $t_4$ represent drift electron arrival times;
their difference, $\left|t_4 - t_2\right|$, determines the angle
between the candidate muon track and the radial, thus providing a
crude but fast measurement of transverse momentum that can be used in
a low level trigger.  Analogous information from $t_1$ and $t_3$
yields a second independent measurement.}
\label{fig:cmu_layout}
\end{figure}

The CMU covers the region $55^\circ\leq\left|\theta\right|\leq
90^\circ$ and resides on the outer edge of the central hadronic
calorimeter, 3\,470~mm from the beam axis, as indicated in
Figure~\ref{fig:cmu_tower}.  Each 12.6$^\circ$ azimuthal wedge
comprises three modules, each subtending 4.2$^\circ$ in $\varphi$.  A
CMU module, shown in Figure~\ref{fig:cmu_layout}, consists of four
towers, each with four layers of rectangular drift cells.  The
outermost and second innermost cells in each tower are oriented such
that their sense wires lie on a radial that originates from the centre
of the CDF detector.  The innermost and second outermost drift cells
lie on another radial that is offset from the first by 2~mm at the
midpoint (in $r$) of the CMU.  The offset cells in each tower resolve
the side of the radial, in azimuth, on which the track passed.  As
indicated in Figure~\ref{fig:cmu_layout}, the absolute difference in
drift electron arrival times for a pair of cells having sense wires on
the same radial determines the angle between the candidate muon track
and that radial.  This angle can be related to the transverse momentum
of a muon candidate and is therefore exploited by the trigger system
(refer to Section~\ref{sect:trigger_l1}).

\begin{figure}
\begin{center}
\leavevmode
\hbox{%
\epsfxsize=6.0in
\epsffile{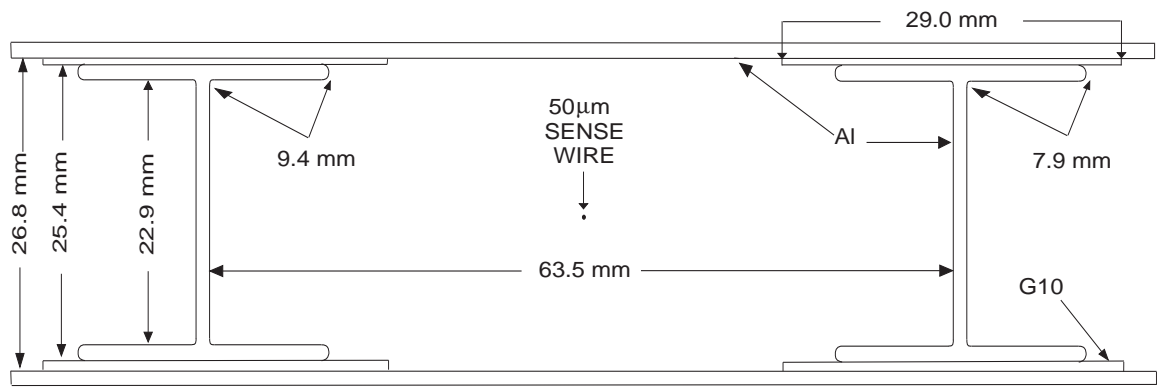}}
\end{center}
\caption
[Layout of a central muon detector (CMU) drift cell.]
{Layout of a central muon detector (CMU) drift cell, showing the
0.79-mm aluminum top and bottom plates and the aluminum I-beams that
separate adjacent towers.}
\label{fig:cmu_cell}
\end{figure}

A drift cell in the CMU, shown in Figure~\ref{fig:cmu_cell}, is
rectangular with dimensions 63.5~mm $\times$ 26.8~mm
$\times$~2\,261~mm and has a single 50-$\mu$m stainless steel sense
wire strung through its centre.  The drift cells are operated in
limited streamer mode using a 50:50 admixture of argon and ethane gas,
and potentials of $+3\,150$~V on the sense wires and $-2\,500$~V on
the I-beams, which are electrically isolated from the top and bottom
aluminum plates by 0.62 mm of insulation.  The position of a muon
candidate track along the sense wire ($z$) direction can be determined
with a resolution of 1.2~mm using charge division electronics.  The
position resolution in the drift ($\varphi$) direction is 250~$\mu$m.

\subsection{The Central Muon Upgrade Detector}
\label{sect:cmp}

An average of 5.4 pion interaction lengths lies between the CMU and
the $p\bar{p}$ collision region, resulting in approximately 1 in 220
high energy hadrons traversing the calorimeters unchecked.  This
`noninteracting punch-through' results in an irreducible false muon
background rate.  The central muon upgrade detector (CMP), shown in
Figures~\ref{fig:cdf_iso} and \ref{fig:cdf_xsec}, was commissioned to
contend with this punch-through hadron rate~\cite{lewis:cmp_cmx}.  The
CMP surrounds the central region of the CDF detector with 630~tons of
additional steel.  The geometry is box-like, with the return yoke of
the solenoid providing the absorption steel on the top and bottom, and
two retractable 60-cm thick slabs arranged as vertical walls on each
side.  The additional absorption material brings the number of pion
interaction lengths to 7.8 at $\theta = 90^\circ$.
Figure~\ref{fig:muon_coverage} illustrates the variation in
pseudorapidity coverage with azimuth caused by the geometry of the
CMP.

\begin{figure}
\begin{center}
\leavevmode
\hbox{%
\epsfxsize=6.0in
\epsffile{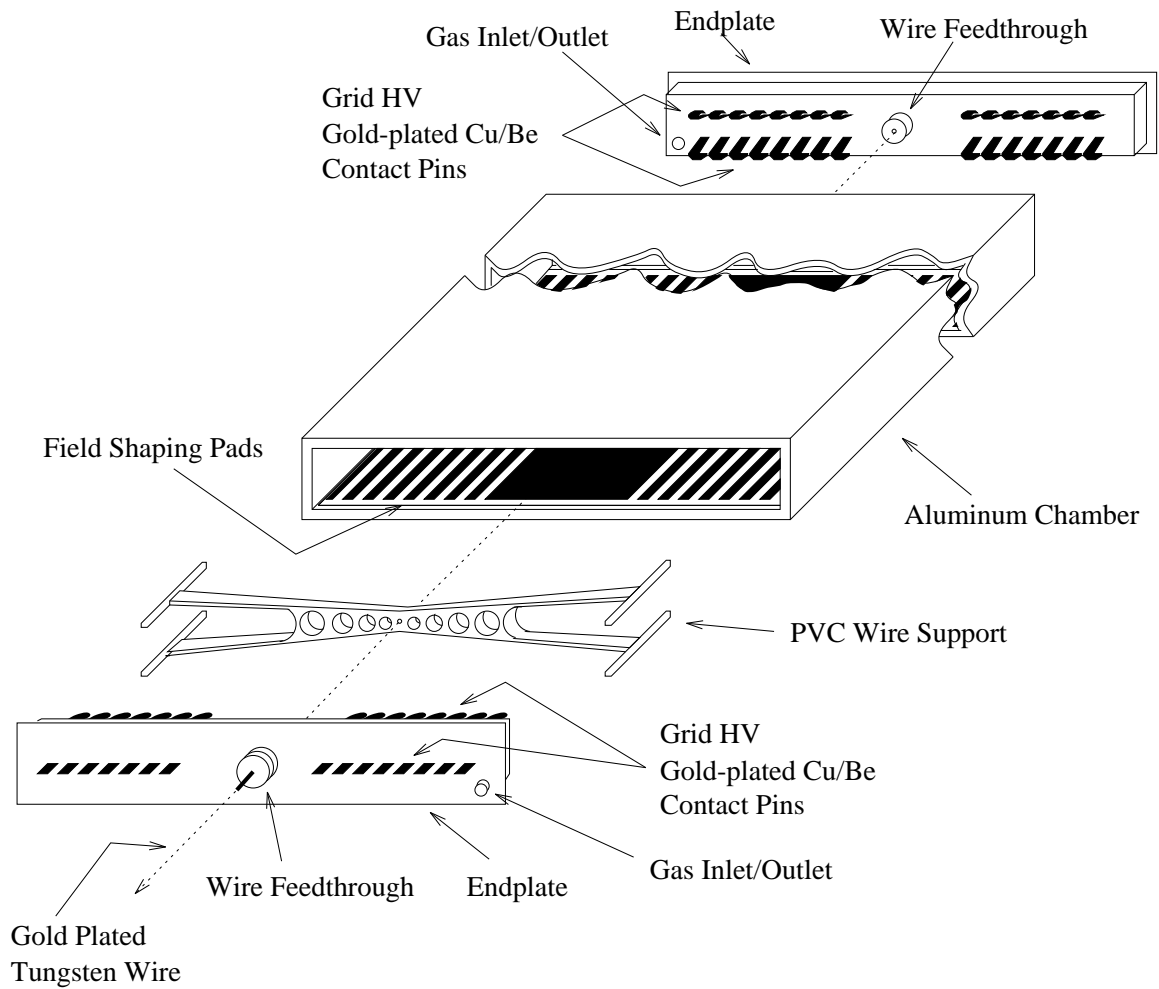}}
\end{center}
\caption
[Schematic of a CMP or CMX proportional drift cell.]
{Schematic drawing of a proportional drift cell used in both the
central muon upgrade detector (CMP) and the central muon extension
(CMX)~\cite{lewis:cmp_cmx}.}
\label{fig:cmp_cmx_tube}
\end{figure}

The active planes of the CMP consist of four layers of half-cell
staggered single-wire drift tubes operating in proportional mode.
Each drift cell, of which there are 864 in the CMP, consists of a
rectangular extruded aluminum tube 25.4-mm high, 152.4-mm wide, and
with a length that depends upon where the tube is mounted.
Figure~\ref{fig:cmp_cmx_tube} is a schematic drawing of the components
of a CMP proportional drift cell.  The anode, a 50-$\mu$m gold-plated
tungsten wire, is biased to a potential of $+5\,600$~V, the wide
central field-shaping cathode pad is biased to $+3\,000$~V, and the
eight narrow field-shaping strips have decrementally decreasing
voltages from the centre of the cell out to the edges.  The maximum
drift time is 1.4~$\mu$s.

\subsection{The Central Muon Extension}
\label{sect:cmx}

The central muon extension, or CMX, provides additional pseudorapidity
acceptance in the region $0.65\leq\left|\eta\right|\leq 1.0$.  Shown
in Figures~\ref{fig:cdf_iso} and \ref{fig:cdf_xsec}, the CMX modules
possess geometries that correspond to sides of the frusta of two
cones, each with a base at $z=0$ and an axis along either the proton
or antiproton direction.  The azimuthal coverage of the CMX is not
continuous; due to the floor of the collision hall, there is a
90$^\circ$ gap in $\varphi$ at the bottom of the CDF detector, and,
due to the Main Ring bypass beampipe, there is a 30$^\circ$ gap at the
top of the detector.  The 1\,536 proportional drift cells that
constitute the CMX modules are shorter than, but otherwise identical
to, those used in the CMP (Figure~\ref{fig:cmp_cmx_tube}).  No
additional absorber was added between the CMX and the interaction
region; however, the smaller polar angle through the hadronic
calorimeter and magnet return yoke yields a shielding thickness of 6.2
pion interaction lengths at $\theta = 55^\circ$.

\begin{figure}
\begin{center}
\leavevmode
\hbox{%
\epsfxsize=6.0in
\epsffile{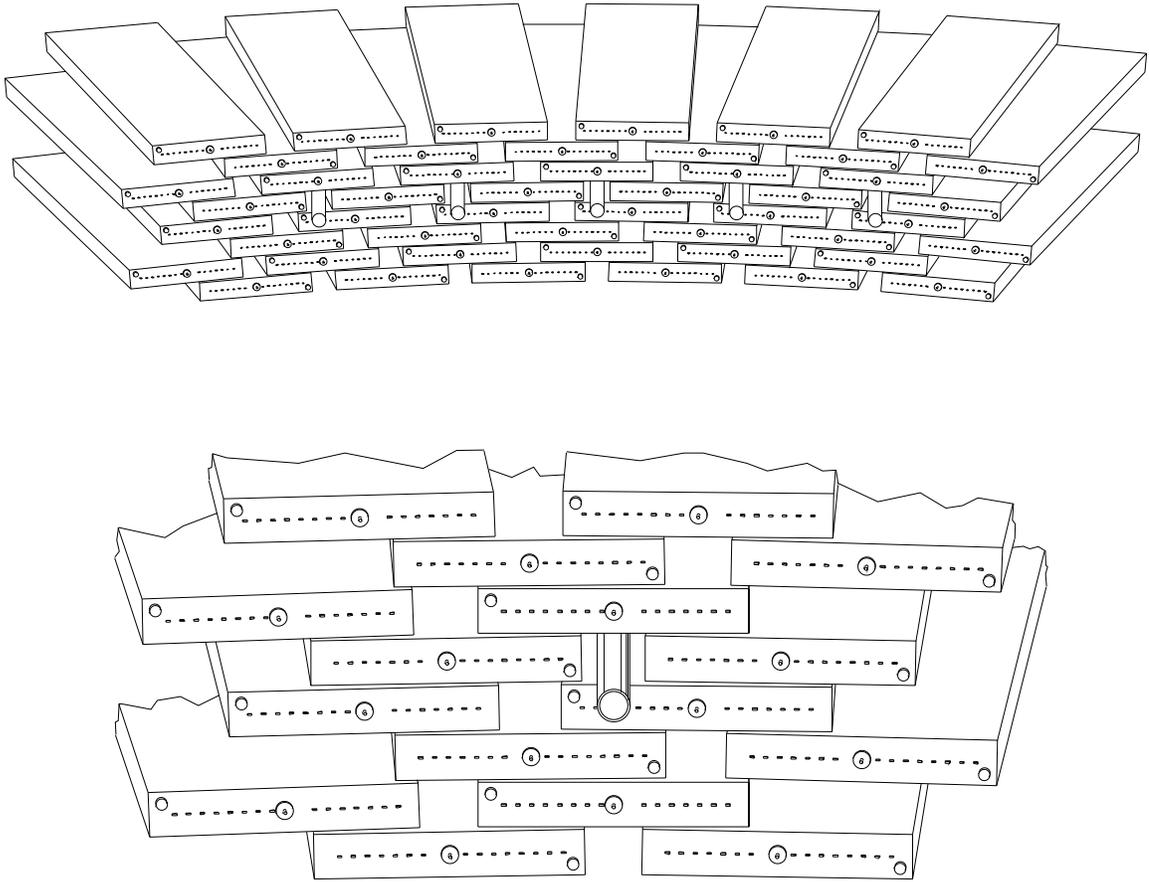}}
\end{center}
\caption
[Cell layout in a central muon extension (CMX) module.]
{Proportional drift cell layout in a 15$^\circ$ module of the central
muon extension (CMX) (top figure).  Also shown is a close-up view of
the staggered cell configuration about one of the threaded rods used
to fix the CMX to its support arch (bottom
figure)~\cite{lewis:cmp_cmx}.}
\label{fig:cmx_tube_layout}
\end{figure}

The CMX is organized into four stacks, two on the proton side and two
on the antiproton side of the CDF detector.  Each stack consists of
eight modules, which each subtend 15$^\circ$ in azimuth.  A module
comprises 48 proportional drift cells that are grouped in eight
half-cell staggered layers of six tubes each.  Refer to
Figure~\ref{fig:cmx_tube_layout} for an illustration of the
interleaved geometry necessary to arrange the rectangular cells along
a conical surface.  The invisibility of intermodule boundaries is an
advantage of this interleaved configuration.

The maximum CMX drift time, 1.4~$\mu$s, is such that the spread of
arrival times due to background particles is short by comparison.
Background rejection and a high speed trigger are provided by an array
of scintillation counters mounted on the inner and outer sides of each
CMX module.  Four such scintillators, each with its own
photomultiplier tube, are located on both sides of every 15$^\circ$
CMX module, for a total of 256 scintillation counters.  Background
effects are vetoed in the trigger by requiring that both the inner and
outer scintillators adjacent to a CMX hit produce pulses that are
coincident with the $p\bar{p}$ beam crossing to within a few
nanoseconds.

\section{The Trigger Systems}
\label{sect:trigger}

In the course of data acquisition for this study, proton and
antiproton bunch crossings in the Tevatron collider occurred every
3.5~$\mu$s, corresponding to a crossing frequency of 286~kHz.  With
typical instantaneous luminosities of ${\cal L} =
10^{31}$~cm$^{-2}$s$^{-1}$ and a measured $p\bar{p}$ total cross
section of $\sigma_{\rm tot} = 80.03 \pm
2.24$~mb~\cite{abe:ppbar_tot_xs} at $\sqrt{s} = 1.8$~TeV, at least one
$p\bar{p}$ interaction per beam crossing was expected.  Moreover, the
low signal-to-noise ratio of the $b\bar{b}$ hadroproduction processes
(refer to Section~\ref{sect:hadroproduction}) made the implementation
of a trigger system essential.

A CDF event, which amounted to the digitized information from a single
beam crossing that could be read out from the CDF detector at a given
time, had a data length of $\sim$165~kB.  Such an event size could
only be reliably written out to several 8-mm magnetic tapes at a rate
of approximately 10~Hz.  This constituted the principal limitation to
the CDF data acquisition rate and necessitated a trigger system that
could both accommodate the $p\bar{p}$ interaction rate and select
interesting physics events with a $\sim$30\,000:1 rejection factor.
The CDF trigger
\cite{abe:cdfnim_overview,amidei:trignim} consisted of three
successive levels, each of which imposed a logical ``{\sc or}'' of a
limited number of programmable selection criteria that collectively
reduced the data rate exposed to the next higher trigger level.  The
reduction in the rate presented to the higher trigger levels provided
time for more sophisticated analyses of potential events with the
accrual of less dead time\footnote{In this context, `dead time' refers
to the amount of time that the CDF detector was unable to consider
subsequent $p\bar{p}$ collisions.}.

\subsection{Level 1}
\label{sect:trigger_l1}

The Level 1 trigger required less time than the 3.5-$\mu$s beam
crossing period to reach a decision on whether or not a given event
was suitable for consideration by the higher trigger levels; it
therefore incurred no dead time.  Such speed of operation was achieved
by the analog read-out and processing of data from selected detector
components with dedicated {\sc fastbus}-based electronics.  At an
instantaneous luminosity of ${\cal L} = 5 \times
10^{30}$~cm$^{-2}$s$^{-1}$, the Level 1 trigger acceptance rate was
approximately 1~kHz~\cite{abe:topprdl}.  Although it could have been
configured to base its decisions on information from several different
CDF subsystems, the Level 1 trigger primarily used signals from the
calorimeters and the muon systems.

The calorimetry component of the Level 1 trigger considered analog
signals on dedicated cables from the scintillator photomultiplier
tubes in the central calorimeter subsystems and from cathode pads in
the plug and forward calorimeter subsystems (refer to
Table~\ref{tab:calorimeters}).  For the purposes of the trigger, the
calorimeters were logically segmented into `trigger towers' with
$\Delta\varphi = 15^{\circ}$ and $\Delta\eta = 0.2$.  For each
subsystem listed in Table~\ref{tab:calorimeters}, the individual tower
minimum-energy thresholds could be specified to the trigger, which
summed the deposited energies, weighted by the polar angle $\theta$ to
determine the transverse energy $E_{\rm T} \equiv E\,\sin\theta$, for all
those trigger towers that were above these thresholds.  If the total
$E_{\rm T}$, measured in this manner, exceeded a given global threshold,
then the Level 1 trigger accepted the event.  There was also a similar
Level 1 calorimetry trigger that had significantly reduced tower
energy thresholds, but was prescaled by a factor of 20 or 40,
depending on the data-taking period (Run 1A or Run 1B, respectively;
refer to Section~\ref{sect:data_sample}).

The muon component of the Level 1 trigger exploited the relative drift
electron arrival times ($\Delta t$) between pairs of drift cell layers
in a given CMU module, as described in Section~\ref{sect:cmu}.  The
two cells constituting each of these pairs were separated in $r$ by
one drift cell, as shown in Figure~\ref{fig:cmu_layout}.  The trigger
logic operated on objects, called `muon stubs', that were defined by
the existence of any wire pair in a 4-tower 4.2$^\circ$ muon detector
module (see Figure~\ref{fig:cmu_layout}) with a $\Delta t$ less than a
value corresponding to a given minimum transverse momentum ($p_{\rm
T}$) requirement.

Out of a total of seven Level 1 triggers that involved muon
candidates, two were directly relevant to this analysis because they
specifically identified dimuon candidates in the central region of the
CDF detector.  One of these dimuon triggers ({\sc two\-\_cmu\-\_3pt3})
required that two CMU muon stubs exist, whereas the other ({\sc
two\-\_cmu\-\_cmx\-\_3pt3}) required that the two stubs each be in
either CMU or CMX modules.  In the case of the CMX stubs, the trigger
required coincident hits in the CMX scintillators (see
Section~\ref{sect:cmx}).  The Level 1 system made no further
requirements on the positions of the two muon stubs, except for a
criterion that they be located in noncontiguous modules.  That is, at
least one muon module lacking a muon stub must have resided between
the two modules where muon stubs were observed.  If this was not the
case, then the two adjacent stubs were merged into a single muon stub
for the purposes of the Level 1 trigger.  The minimum $p_{\rm T}$
requirement on each stub in these two triggers was nominally
3.3~GeV/$c$.  Section~\ref{sect:trigsim_eff} discusses the
efficiencies of the Level 1 low-$p_{\rm T}$ central muon triggers.

\subsection{Level 2}
\label{sect:level2}

In a $p\bar{p}$ beam crossing for which the Level 1 trigger did not
fire, a timing signal from the Tevatron announcing the occurrence of
the next beam crossing would cause the stored signals in the CDF
detector to be cleared in preparation for the next crossing.  If the
Level~1 trigger did fire, then subsequent timing signals were
inhibited from clearing information stored in the CDF detector for a
period of up to 20~$\mu$s, during which the Level 2 trigger made its
decision and five disregarded beam crossings could occur.  At an
instantaneous luminosity of ${\cal L} = 5 \times
10^{30}$~cm$^{-2}$s$^{-1}$, the typical Level 2 trigger output rate
was approximately 12~Hz~\cite{abe:topprdl}.

With the increased processing time, the Level 2 trigger system could
perform simple tracking calculations and determine basic topological
features of the event by considering, with greater sophistication, the
same dedicated calorimetry and muon signals used in Level~1.
Specifically, expeditious electromagnetic and hadronic
transverse-energy clustering was performed at Level 2 by applying
`seed' and `shoulder' thresholds to all the calorimeter trigger
towers.  If a given tower energy exceeded the seed threshold, which is
higher than the shoulder threshold, then a cluster was formed.
Adjacent trigger towers were iteratively included in this cluster if
they both exceeded the shoulder threshold and were not part of another
cluster.

High speed track pattern recognition was achieved in Level 2 with the
central fast tracker (CFT), a hardware track finder that detected
high-$p_{\rm T}$ charged particles in the CTC
(Section~\ref{sect:ctc}).  The CFT measured transverse momentum and
azimuth, since it only examined hits in the five axial CTC
superlayers.  For a given traversal of an axial superlayer by a
charged particle, the CFT considered two types of timing information:
prompt and delayed hits.  Prompt hits, gated $\leq$80~ns after the
beam crossing time, were due to the short drift times caused by
charged particles traversing the plane of sense wires in a superlayer.
Pairs of delayed hits, one on each side (in $\varphi$) of a given
superlayer, were recorded by a gate that occurred 500-650~ns after the
beam crossing.  The absolute prompt and delayed drift times provided
information on a track's trajectory, whereas the relative drift times
furnished measurements of curvature, and hence $p_{\rm T}$.  After all
the drift hits were recorded, the CFT sought to construct tracks by
first examining hits in the outermost superlayer.  For each sense wire
in the outer layer with a prompt hit, the CFT looked to the inner
layers for `roads', or hit patterns, that matched patterns in a
look-up table that had eight $p_{\rm T}$ bins and two $\varphi$ bins,
one for each sign of curvature.  The $p_{\rm T}$ bins
ranged\footnote{The CFT $p_{\rm T}$-bin thresholds were changed
between Runs 1A and 1B.} from $\sim$3 to $\sim$30~GeV/$c$, and the
transverse momentum resolution was $\delta p_{\rm T} / p_{\rm T} \sim
0.035 \times p_{\rm T}$, where $p_{\rm T}$ is in units of GeV/$c$.

The Level 2 trigger system organized the energy clusters, CFT tracks,
and muon stubs into clusters called `physics objects'.  These included
jets, $\Sigma E_{\rm T}$ (total transverse energy), electrons,
photons, taus, muons, and neutrinos (whose signature is missing
transverse energy, $E\!\!\!\!/_{\rm T}$).  For the majority of the
data-taking period, a custom-built `Jupiter' module accessed the
clusters and made a Level 2 decision to accept or reject events.  The
Jupiter module had two separate processor boards, one that loaded
clusters into memory, and another that checked the triggers by
imposing several requirements on the physics objects.  In the last
year of data acquisition, these processors were replaced with
commercially available AXP (Alpha) processors manufactured by Digital
Equipment Corporation.

All of the dimuon selection triggers at Level 2 imposed a matching
criterion between at least one of the two Level 1 muon stubs and a CFT
track.  Early in the data-taking period, the requirement was that the
stub and the extrapolated track have an azimuthal separation in the
transverse plane that was $\leq$15$^\circ$.  This criterion was later
tightened to $\Delta\varphi\leq 5^\circ$ to reduce further the trigger
rate due to accidental coincidences.  The various dimuon Level 2
triggers used in this analysis are listed, along with their
prerequisite Level 1 triggers, in Appendix~\ref{app:l2_dimuon_trig}.
Section~\ref{sect:trigsim_eff} discusses the efficiencies of the Level
2 low-$p_{\rm T}$ central muon triggers.

\subsection{Level 3}
\label{sect:level3}

The Level 3 trigger~\cite{carroll:l3_trig} was a flexible, high-level,
software-based computer processor `farm' that could reconstruct
several events in parallel.  When the Level 2 trigger accepted an
event, the channels in the CDF detector with valid data were digitized
and read out by the data acquisition (DAQ) system.  The DAQ
electronics subsequently transported the event data to the Level 3
processor farm.  Over the course of the data-taking period, both the
Level 3 trigger system and the DAQ system (see Section~\ref{sect:daq})
underwent several significant changes.  Although most of these changes
were effected in the interval between the Run 1A and Run 1B
collider running periods, not all of them were implemented for physics
data taking from the beginning of Run 1B.  Unless specifically noted
otherwise, the following descriptions of the Level 3 trigger and the
DAQ system pertain to the upgraded configurations used for data taking
later in Run 1B.

The Level 3 computing farm consisted of 64 commercial processors that
were manufactured by Silicon Graphics, Inc.\ and that ran under the
IRIX operating system, a flavour of {\sc unix}.  Half of these processors
were R3000 Power Servers and half were R4400 Challenge machines.  The
farm processors received data fragments read out by the DAQ system for
a given beam crossing and `built' these fragments into a contiguous
event.  Prior to the logging of the events to disk or 8-mm magnetic
tape, the processors reconstructed and characterized these events for
later selection using a configurable trigger table and optimized
executables of compiled {\sc fortran} computer codes.  A given event
data buffer did not necessarily reside on the same farm computer as
the reconstruction executable and processor that were operating on it.

In Run 1A, only 48 R3000 Power Servers were used, and the event
fragments were already built into events by the Run 1A DAQ system
prior to their reception by the Level 3 farm (refer to
Section~\ref{sect:daq}).  Every processor had two dedicated, but
separate, buffers, which each had the capacity to contain an entire
event.  A separate `farm steward' computer communicated with the rest
of the DAQ system, controlled the initiation and cessation of event
processing on the farm CPUs\footnote{CPU is an acronym for `central
processing unit'.}, maintained performance statistics of farm
activities, and provided Level 3 status information.  In the Run 1B
trigger system, however, the duties of the farm steward were absorbed
into other programmes executing on the farm computers.

For the purposes of analyses involving central muons, the Level 3
executables reconstructed muon stubs and CTC tracks using algorithms
that were largely identical to those employed in the off-line event
reconstruction (refer to Section~\ref{sect:data_reduct}); however,
because three-dimensional track reconstruction constituted most of the
Level 3 execution time, only the faster of two tracking algorithms
used in the off-line code was engaged in the trigger.  Two Run 1A
dimuon triggers were used to form the data samples for the present
analysis.  One of these, that which contained dimuon candidates from
the decay $J/\psi\to\mu^+\,\mu^-$, required two oppositely charged
muon candidates with a combined invariant mass in the range $2.8 -
3.4$~GeV/$c^2$.  The other trigger, which accepted dimuon candidates
from the decay $\psi(2S)\to\mu^+\,\mu^-$, had no opposite-charge
requirement and selected dimuon candidates with an invariant mass in
the range $2.8 - 4.0$~GeV/$c^2$.  In Run 1B, both the $J/\psi$ and
$\psi(2S)$ dimuon modes were accepted by the same trigger, one that
imposed no opposite-sign charge requirement and had an invariant mass
window of $2.7 - 4.1$~GeV/$c^2$.

In addition, the Level 3 dimuon triggers used in this study placed
position matching requirements between the muon stubs and their
associated CTC tracks.  The algorithm extrapolated the CTC track to
the appropriate muon subsystem and determined the difference in
position between the projected track and the muon stub in both the
$r$-$\varphi$ plane and the $z$ direction, correcting for energy loss
and multiple scattering as a function of $p_{\rm T}$.  The Run 1A
$J/\psi$ tracks were required to match the muon stubs within 4
standard deviations ($\sigma$) of the combined multiple scattering and
measurement uncertainties, whereas the $\psi(2S)$ tracks had a
matching requirement of $\sim$6$\sigma$.  The Run 1B Level 3 muon
matching requirement was $\sim$3$\sigma$.

\section{The Data Acquisition System}
\label{sect:daq}

The data acquisition (DAQ) pipeline, at its lowest, or `front end',
level, began with the readout of analog signals from channels in the
various subsystems of the CDF detector.  In general, these analog
pulses were preamplified and transported to an electronics crate on
the detector where they underwent further amplification and, depending
on the subsystem, pulse shaping and discrimination.  Digitization of
these analog signals was accomplished with either analog-to-digital or
time-to-digital conversion electronics.

\begin{figure}
\input{diagrams/biery_daq_fig.tex}
\caption
[Schematic diagram of the data acquisition pipeline.]
{A schematic drawing of the principal elements of the CDF data
acquisition pipeline~\cite{biery:daq_figure}.  The individual
components are described in the text.  Scramnet is the fibre optic
reflective memory network.}
\label{fig:biery_daq}
\end{figure}
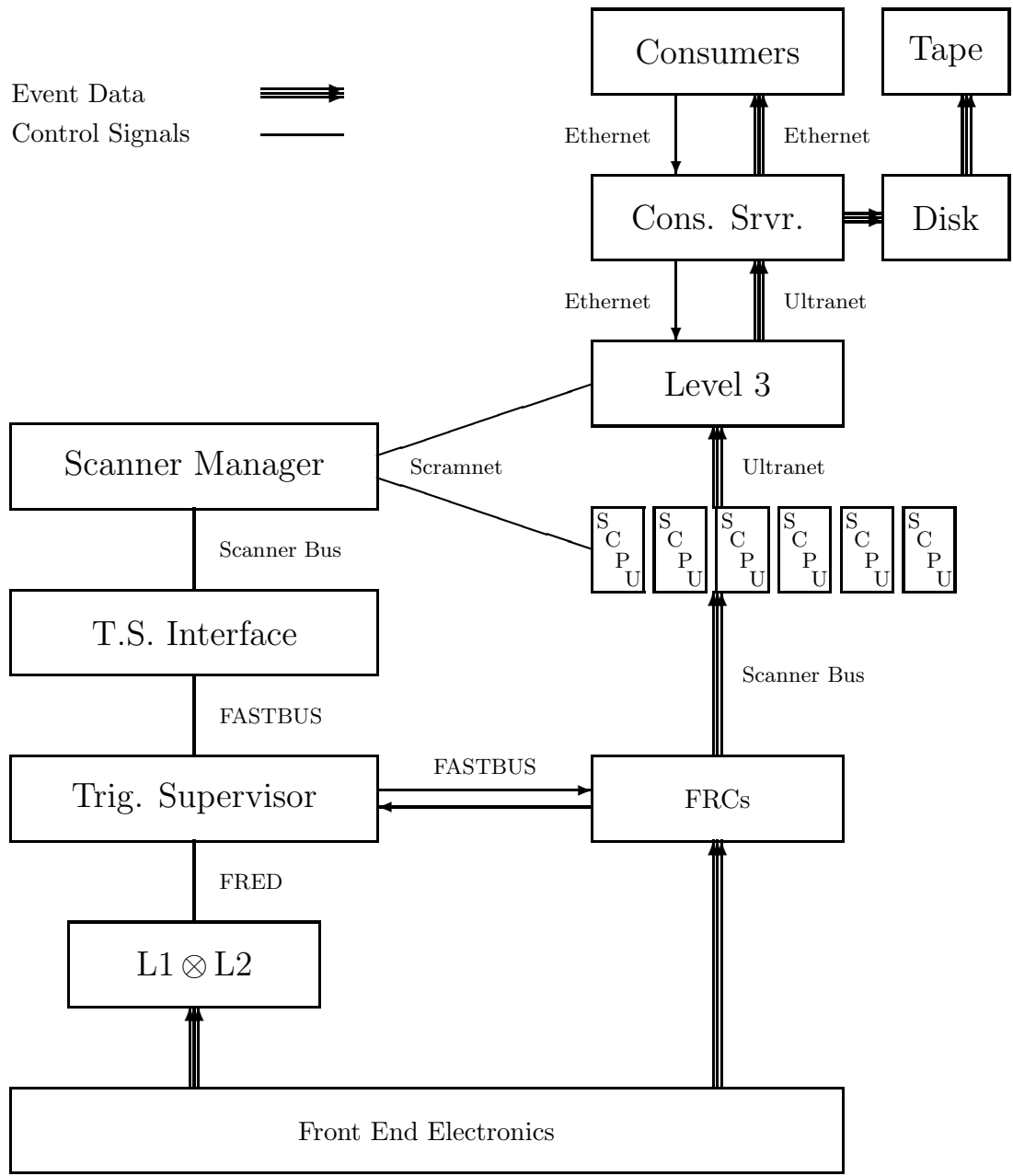

Dedicated signal cables communicated synoptical event information from
the front end electronics to the Level 1 (L1) and Level 2 (L2)
triggers, as illustrated schematically in Figure~\ref{fig:biery_daq}.
The Level 1 and Level 2 event acceptance decisions were coordinated by
a programmable {\sc fastbus} device known as FRED\footnote{The
acronymic or abbreviational origins of FRED are
unknown~\cite{amidei:trignim}.}, which acted as the interface between
the trigger and the trigger supervisor board.  The trigger supervisor,
also a {\sc fastbus} module, initiated and monitored the readout of
data from the front end electronics using {\sc fastbus} readout
controllers, or FRCs, which were single-width modules that contained
MIPS R3000 processors.  The FRCs sent their data over a 16-bit
parallel scanner bus to six scanner CPUs, or SCPUs.  The SCPUs, which
were VME\footnote{VME stands for `Versa Module Eurocard' and is a
crate-based electronics package scheme.}-based Motorola 68030
processors running the VxWorks operating system, transported the event
fragments to the Level 3 processor nodes via a commercial 256~Mbit/s
serial Ultranet hub.  The scanner manager, also a VME-based 68030 CPU,
used a fibre optic reflective memory network to control the flow of
data between the FRCs and the Level 3 trigger system, ensuring that
all fragments of a given event were destined for the same Level 3
node.  The trigger supervisor interface, also shown in
Figure~\ref{fig:biery_daq}, facilitated communication between the
scanner manager and the trigger supervisor.  It consisted of two FRCs
that resided in the same crate as the trigger supervisor and were
connected to the scanner manager via a scanner bus.

Events accepted by the Level 3 trigger were passed via Ultranet to the
consumer server (see Figure~\ref{fig:biery_daq}), a process running on
a dedicated two-processor Silicon Graphics, Inc.\ Challenge~L machine.
The consumers, which received events from the consumer server over
ethernet, were diagnostic applications that monitored trigger rates,
luminosity conditions, detector performance, and a selected number of
known high-rate physics processes.  Another consumer, an on-line
interactive event display system, provided CDF control-room personnel
with graphical physics and detector performance diagnostics (refer to
Figures~\ref{fig:df_vtx} and \ref{fig:event_display_text} for examples
of two of the system's displays).  Data logger programmes, one for
each output data stream, ran on the consumer server machine and wrote
accepted events to local disk.  The disk-resident data logger events
were subsequently written to 8-mm tape by a tape-staging programme
that also ran on the consumer server computer.

In contrast, the Run 1A DAQ system lacked FRCs, using custom-built
scanner modules instead.  A hardware event builder~\cite{sinervo:evb},
rather than SCPUs, constructed complete events and communicated them
to the Level 3 trigger.  {\sc fastbus}, in lieu of the fibre optic
reflective memories, delivered all system control messages, and the
r\^{o}le of the Run 1B scanner manager was served in Run 1A by a
buffer manager running on a MicroVAX~II computer.

%% file: diagrams/biery_daq_fig.tex
 \thicklines
 \setlength{\unitlength}{1.2cm}
 \begin{center}
 \begin{picture}(12.7,13.5)
  \put(00.0,00.0){\framebox(10.0,1.0){Front End Electronics}}
  \put(07.0,04.0){\framebox(3.0,1.0){FRCs}}
  \footnotesize
  \multiput(07.00,07.0)(0.75,0.0){6}{\framebox(0.6,1.0){}}
  \multiput(07.12,07.85)(0.75,0.0){6}{\makebox(0,0)[c]{S}}
  \multiput(07.24,07.625)(0.75,0.0){6}{\makebox(0,0)[c]{C}}
  \multiput(07.36,07.375)(0.75,0.0){6}{\makebox(0,0)[c]{P}}
  \multiput(07.48,07.15)(0.75,0.0){6}{\makebox(0,0)[c]{U}}
  \Large
  \put(07.0,09.0){\framebox(3.0,1.0){Level 3}}
  \put(07.0,11.0){\framebox(3.0,1.0){Cons. Srvr.}}
  \put(10.5,11.0){\framebox(1.5,1.0){Disk}}
  \put(10.5,13.0){\framebox(1.5,1.0){Tape}}
  \put(07.0,13.0){\framebox(3.0,1.0){Consumers}}
  \put(00.7,02.0){\framebox(3.0,1.0){L1$\,\otimes\,$L2}}
  \put(00.0,04.0){\framebox(4.4,1.0){Trig. Supervisor}}
  \put(00.0,06.0){\framebox(4.4,1.0){T.S. Interface}}
  \put(00.0,08.0){\framebox(4.4,1.0){Scanner Manager}}

  \put(02.15,01.0){\vector(0,1){1.0}}
  \put(02.20,01.0){\vector(0,1){1.0}}
  \put(02.25,01.0){\vector(0,1){1.0}}
  \put(02.2,03.0){\line(0,1){1.0}}
  \put(02.2,05.0){\line(0,1){1.0}}
  \put(02.2,07.0){\line(0,1){1.0}}
  \put(04.4,08.63){\line(3,1){2.6}}
  \put(04.4,08.37){\line(3,-1){2.6}}

  \put(08.45,01.0){\vector(0,1){3.0}}
  \put(08.50,01.0){\vector(0,1){3.0}}
  \put(08.55,01.0){\vector(0,1){3.0}}
  \put(08.45,05.0){\vector(0,1){2.0}}
  \put(08.50,05.0){\vector(0,1){2.0}}
  \put(08.55,05.0){\vector(0,1){2.0}}
  \put(08.45,08.0){\vector(0,1){1.0}}
  \put(08.50,08.0){\vector(0,1){1.0}}
  \put(08.55,08.0){\vector(0,1){1.0}}
  \put(08.95,10.0){\vector(0,1){1.0}}
  \put(09.00,10.0){\vector(0,1){1.0}}
  \put(09.05,10.0){\vector(0,1){1.0}}
  \put(08.95,12.0){\vector(0,1){1.0}}
  \put(09.00,12.0){\vector(0,1){1.0}}
  \put(09.05,12.0){\vector(0,1){1.0}}
  \put(08.0,11.0){\vector(0,-1){1.0}}
  \put(08.0,13.0){\vector(0,-1){1.0}}
  \put(10.0,11.45){\vector(1,0){0.5}}
  \put(10.0,11.50){\vector(1,0){0.5}}
  \put(10.0,11.55){\vector(1,0){0.5}}
  \put(11.45,12.0){\vector(0,1){1.0}}
  \put(11.50,12.0){\vector(0,1){1.0}}
  \put(11.55,12.0){\vector(0,1){1.0}}
  \put(04.4,04.6){\vector(1,0){2.6}}
  \put(07.0,04.4){\vector(-1,0){2.6}}

  \footnotesize
  \put(02.5,03.5){\makebox(0,0)[l]{FRED}}
  \put(02.5,05.5){\makebox(0,0)[l]{FASTBUS}}
  \put(02.5,07.5){\makebox(0,0)[l]{Scanner Bus}}
  \put(08.8,06.0){\makebox(0,0)[l]{Scanner Bus}}
  \put(08.8,08.5){\makebox(0,0)[l]{Ultranet}}
  \put(09.3,10.5){\makebox(0,0)[l]{Ultranet}}
  \put(09.3,12.5){\makebox(0,0)[l]{Ethernet}}
  \put(07.7,10.5){\makebox(0,0)[r]{Ethernet}}
  \put(07.7,12.5){\makebox(0,0)[r]{Ethernet}}
  \put(04.8,08.5){\makebox(0,0)[l]{Scramnet}}
  \put(05.7,04.8){\makebox(0,0)[b]{FASTBUS}}

  \normalsize
  \put(0.0,13.0){\makebox(0,0)[l]{Event Data}}
  \put(0.0,12.5){\makebox(0,0)[l]{Control Signals}}
  \put(3.0,12.95){\vector(1,0){1.0}}
  \put(3.0,13.00){\vector(1,0){1.0}}
  \put(3.0,13.05){\vector(1,0){1.0}}
  \put(3.0,12.50){\line(1,0){1.0}}
 \end{picture}
 \end{center}

%% file: select.tex
\label{chapt:select}

This chapter discusses the means by which charmonia and $B$-meson
candidates are extracted from the backgrounds that are characteristic
of hadron-hadron collision environments.  First, the data set is
defined.  This is followed by a general description of the software
machinery used to reduce the data sample.  Finally, specific details
about the criteria used to identify candidate signal events and reject
background sources are presented.  The charmonia and $B$-meson
candidate invariant mass distributions will be presented in
Chapters~\ref{chapt:effic} and
\ref{chapt:br_fract}, respectively.

\section{The Data Sample}
\label{sect:data_sample}

The experimental data used in this investigation were recorded, using
the CDF detector, in the years 1992 through 1995.  During this period,
the Fermilab Tevatron collided protons with antiprotons at a
centre-of-mass energy of $\sqrt{s}\sim$1.8~TeV.

The 1992-1995 Tevatron running period, which is generally known as Run
1, comprised two separate data-taking intervals: Run 1A and Run 1B.
Run 1A consisted of approximately nine months of physics running, which
commenced on 26~August~1992 and ended on 30~May~1993.  The data
collected during Run 1A correspond to a time-integrated luminosity of
$\int\!{\cal L}\,dt = (19.5 \pm 1.0)$~pb$^{-1}$.

In the intervening period between Run 1A and Run 1B, the CDF detector
underwent several improvements.  As described in
Chapter~\ref{chapt:apparat}, the SVX was replaced with the
SVX$^\prime$, new trigger requirements and hardware were implemented,
and a significantly improved data acquisition system was brought
on-line.  Most components of the upgraded trigger and data acquisition
hardware were piggybacked onto the existing system to facilitate, if
the need arose, the immediate reversion to their analogous Run 1A
counterparts.  In this way, the AXP Level 2 trigger processors and the
upgraded data acquisition system underwent testing early in Run 1B,
but were only used for physics data collection several months into the
Run 1B Tevatron running period.

Physics-quality data collection in Run 1B began on 18~January~1994 and
ended on 24~July~1995.  The recorded time-integrated luminosity was
$\int\!{\cal L}\,dt = (89 \pm 7)$~pb$^{-1}$, yielding a consolidated
Run 1 total of $\int\!{\cal L}\,dt = (109 \pm 7)$~pb$^{-1}$.  It
should be noted that some operating conditions were not constant over
the course of the Run 1 data-taking period.  The temperature and
humidity in the collision hall, for example, exhibited a strong
seasonal dependence.  The CDF solenoid, described in
Chapter~\ref{chapt:apparat}, was operated at slightly different
magnetic field strengths (refer to Section~\ref{sect:mag}).  The most
significant operating variations during Run 1, however, were those
associated with the performance of the Tevatron accelerator.  While
the centre-of-mass energy remained constant at 1.8~TeV, the
instantaneous luminosity delivered to the B\O\ $p\bar{p}$ interaction
region varied from a mean value of ${\cal L} = 3.5 \times
10^{30}$~cm$^{-2}$s$^{-1}$ in Run 1A to a mean value of ${\cal L} =
8.0 \times 10^{30}$~cm$^{-2}$s$^{-1}$ in Run 1B.  Peak instantaneous
luminosities in Run 1B reached figures of ${\cal L} \sim 2.6
\times 10^{31}$~cm$^{-2}$s$^{-1}$.  Throughout the Run 1 period, the
Tevatron experienced several service interruptions due to causes that
included power outages, quenching magnets, beam optics adjustments, a
liquid $N_2$ procurement delay, and scheduled maintenance and
cost-saving shutdowns.

The present study treated the entire Run 1 data sample.  Where
necessary, known run-dependent effects were accounted for in the
analysis.  With these corrections, the data set could be regarded as a
single uniform sample.

\section{The Data Reduction Procedure}
\label{sect:data_reduct}

From the Level 3 trigger system to the final selection at the physics
analysis stage, several iterations of data processing and reduction
occurred.  The same event-driven software paradigm was applied at each
of these stages.  This uniform software control framework provided a
flexible means of combining a virtually unlimited number of
independent subprogrammes, or analysis `modules'\footnote{ Analysis
modules were typically coded in {\sc fortran}.}, into a single
executable computer programme.  For a given programme, the user could
invoke, in any order, any subset of the constituent modules.  Events
were handled sequentially by this subset of modules, with the output
of one subprogramme used as input for the next one in the analysis
path.  A standard interface was used to modify the adjustable
parameters for a given module.  Modules also could be configured to
function as filters, which would abort the analysis of an event in the
middle of processing if that event failed any specified filter
criteria.  The possible input sources for this analysis framework
included disk and magnetic-tape files, on-line event buffers, Monte
Carlo generators, and user-written input modules.  A single programme
could support multiple output streams, and any module could function
in several analysis paths with different parameters used in each
implementation.

A typical event was first processed by this standard software
framework on-line in the Level 3 trigger system (refer to
Section~\ref{sect:level3}).  A single optimized computer programme
running on a trigger farm node could implement all of the Level 3
triggers and pipe selected events to any of three output streams.

Off-line, events written to 8-mm tape by the data acquisition system
(refer to Section~\ref{sect:daq}) entered the `production and
splitting' farms, which consisted of 64 Silicon Graphics, Inc.\ and 37
IBM RS6000 computers running the {\sc unix} operating system.  The
production programmes contained modules that reconstructed
higher-level physics objects from low-level detector output data
structures, much in the same way as the Level 3 trigger did.  Since
run-dependent, as opposed to event-dependent, information was known at
the off-line production stage, and since execution speed was less
critical, the production processes could make use of better
calibration and alignment constants, more comprehensive run-condition
information, and more sophisticated tracking algorithms than were
possible in the Level 3 trigger system.  The production farm nodes
also `split' events into several data sets defined using physics
analysis criteria and typically based on one or more Level 3 triggers.
The rate capacity of the production and splitting farm computer system
was $\sim 1.3 \times 10^6$~events/week.

The third principal stage in the data reduction procedure was the
application of analysis-specific algorithms to the reconstruction of
the candidate physics processes under investigation.  In the present
study, the existence of similarities in the reconstruction techniques
of different meson candidates lent itself to a software design
philosophy that employed a small number of highly-configurable and
reusable analysis modules~\cite{sganos:pc}.  For example, it is clear
that the reconstruction of the decays $J/\psi\to\mu^+\,\mu^-$ and
$\psi(2S)\to\mu^+\,\mu^-$ should use identical coding algorithms
governed by different kinematic parameters.  To this end, a generic
`dimuon finder' module was used to reconstruct $J/\psi$ and dimuon
$\psi(2S)$ meson candidates.  A separate `two-track finder' module was
used to seek decays of the form $K^*(892)^0\to K^+\,\pi^-$ by
iterating through charged-track non-muon pair combinations in a given
event.  Finally, a higher-level `charmonium-parent finder' module
could be configured to seek combinations of charmonium candidates,
often reconstructed using the `dimuon finder', and either charged
tracks in the event or meson objects identified using the `two-track
finder'.

Particle types, kinematic selection criteria, and constraints used for
geometric and kinematic fitting could all be specified to the analysis
modules at run time using the standard user interface.  Communication
between a given module and another module downstream in the analysis
path was achieved through data structures identical to those used to
represent the event information itself.  This enabled the same module
to be used more than once, but in different capacities, in a single
analysis path.

This modular philosophy was demonstrated by, for example, the
reconstruction of the decay $B^+\to\psi(2S)\,K^+$ via the $\psi(2S)\to
J/\psi\,\pi^+\,\pi^-$ channel.  First, the `dimuon finder' module was
used to find $J/\psi$ candidates in the event.  Next, the `two-track
finder' module, configured to seek candidate dipions with the expected
kinematic criteria, was invoked.  The `charmonium-parent finder'
module then combined the $J/\psi$ and dipion candidates into
$\psi(2S)$ candidates, applying the appropriate selection criteria and
fit constraints.  Further down the analysis chain, the
`charmonium-parent finder' module was subsequently reused, with very
different operating parameters, to form $B^+$ meson candidates by
constructing combinations of the $\psi(2S)$ candidates with
hitherto-unused tracks in the event.  This recursive recycling of
modular coding logic enhanced the internal consistency of the
analysis.

\section{Magnetic Field Considerations}
\label{sect:mag}

Since the present branching-fraction study relied heavily on
charged-particle trajectory reconstruction, an understanding of the
magnetic field conditions for a given event in the CDF detector was
essential, although somewhat less so than for mass measurement
analyses.  In Run 1A investigations of the $B^0_s$
meson~\cite{abe:bsmass} and $W^+$ boson~\cite{abe:wmass92} masses, the
absolute momentum scale was calibrated by using the decays
$J/\psi\to\mu^+\,\mu^-$, $\Upsilon\to\mu^+\,\mu^-$, and
$Z^0\to\mu^+\,\mu^-$.  The ensuing nominal magnetic field value used
for Run 1A was 1.4127~T.  For the Run 1B portion of the data sample, a
nominal magnetic field value of 1.4116~T was used; however, this value
was corrected on a run-dependent\footnote{A `run', in this context,
refers to a period of uninterrupted data collection, typically lasting
several hours.} basis using a database of $\sim$1\,200 magnetic field
measurements performed over the course of Run 1B.  These corrections
were typically $\lessim$0.17\%~\cite{abe:lambdab_mass}, and
measurements of the magnetic field in the central detector had
uncertainties of $2 \times 10^{-4}$~T~\cite{abe:wmass92}.  Local
residual magnetic field nonuniformities, both in magnitude and
direction, were surveyed in the CDF central tracking volume before the
solenoid magnet was installed~\cite{cnh:magnet_nim}; these were
corrected for in the reconstruction of track helices in both Runs 1A
and 1B.

\section{Primary Vertex Considerations}
\label{sect:pv}

A knowledge of the primary vertex, or point of origin, of the decay
process under examination was important to the analysis.  Primary
vertex information was used in the calculation of primary-secondary
decay lengths, momentum-pointing fit constraints, and isolation
selection criteria.  Two distinct aspects of the primary vertex
determination will be discussed here: how the positions of vertices in
a given event were measured, and how a single vertex was selected in
events where more than one vertex was observed.

The transverse and longitudinal components of the primary vertices
were measured in two very different ways.  The longitudinal ($z$)
coordinate was established on an event-by-event basis using data from
the VTX detector (refer to Section~\ref{sect:vtx}), where vertex
quality was determined on the basis of the number of VTX hits used to
identify the vertex.  As the measurement uncertainty calculated by the
vertex-finding software was deemed to be unreliable due to the
resolution of the VTX subsystem, a fixed uncertainty of $\sigma_z =
0.3$~cm was assumed for all events.  The transverse ($x$-$y$)
coordinates of the primary vertices in an event were typically
calculated using the measured run-averaged beam position.  The
rationale for this was that the transverse beam position typically
varied less than 10~$\mu$m in either the $x$ or $y$ directions over
the course of a single data-taking run and that any event-by-event
transverse coordinate measurements would be biased by fluctuating
track multiplicities and event topologies in individual events.  The
slopes and intercepts of the run-averaged beam position were therefore
combined with the event-by-event $z$ locations of the vertices in an
event to determine the transverse positions of those vertices.  The
transverse coordinate measurement uncertainties were fixed to
$(\sigma_x , \sigma_y) = (25 , 25)$~$\mu$m, corresponding to the
observed circular beam spot size in the transverse plane.  It should
be noted that, in the unusual cases where no run-dependent beam
position information or no VTX $z$-vertex information was available
for a given event, the lacking vertex coordinates were computed using
the available SVX and CTC track information for that event.

During the latter stages of Run 1B, when instantaneous luminosities
often exceeded those in Run 1A by an order of magnitude, the primary
vertex multiplicities also increased dramatically.  Whereas in Run 1A
the average number of high-quality vertices in a given event was
$\sim$1.6 with $\sim$3\% of events having at least four such vertices,
Run 1B events averaged $\sim$2.9 high-quality vertices with $\sim$5\%
of events having at least eight such vertices.  Once a pair of muon
candidates had been identified using the `dimuon finder' machinery
described in Section~\ref{sect:data_reduct} (which was a procedure
that did not directly employ primary vertex information), the $z$
coordinates of the two candidates were used to select a single vertex.
Specifically, of those vertices possessing the highest quality
classification for the given event, the vertex that had the shortest
longitudinal displacement from either of the two muon candidates was
chosen as the primary vertex corresponding to the muon objects.  The
position coordinates of muon candidates, as opposed to those of other
tracks used in this analysis, were used to select primary vertices
because the two muon candidate tracks constituted an unambiguous part
of the final state under study.

\section{Track Quality Criteria}
\label{sect:trk_quality}

The imposition of quality requirements on the tracks used in the
analysis was intended to reduce those backgrounds arising from poor
track measurements in the CDF detector.  Track candidate fits
reconstructed for this study were required to have used at least four
hits in each of at least two axial CTC superlayers (refer to
Section~\ref{sect:ctc} for a description of the CTC).  These track
fits also had to use at least two hits in each of at least two stereo
CTC superlayers.  No requirement was made on which two of the five
axial and four stereo superlayers were to be used in the fit.

In this study, information from the VTX and CTC subsystems was
employed in the reconstruction of track paths.  For all such tracks,
the helical trajectories were extrapolated back into the SVX where
associated hits were sought using a road algorithm.  If a sufficient
number of good SVX hits was found, then the track was refit using all
of the relevant VTX, CTC, and SVX information and the resulting track
helix was used.  Performance disparities in the two silicon microstrip
vertex detectors used in Runs 1A and 1B (refer to
Section~\ref{sect:svx}) motivated two different associated SVX hit
requirements: $\geq$3 hits in the SVX, and $\geq$2 hits in the
SVX$^\prime$.  Moreover, track fits that made use of SVX information
were only considered by this analysis if the SVX $\chi^2$/hit, defined
as the increase in the track fit $\chi^2$ per SVX hit due to the
inclusion of SVX hits in the CTC track fit, satisfied the condition
$\chi^2$/hit~$\leq$~6.0.  This requirement was similar to one used in
CDF studies of the top quark~\cite{abe:topprdl}.

Tracks possessing transverse momenta $p_{\rm T} \lessim 250$~MeV/$c$
were not reconstructed in the production stage of the data reduction
process (refer to Section~\ref{sect:data_reduct}).  Useful
measurements of tracks with $p_{\rm T} < 250$~MeV/$c$ would have been
difficult due to the number of track helices that subtended
$\geq$360$^\circ$ in azimuth while inside the CTC (`loopers') and due
to the dearth of available hits populating the outer superlayers of
the CTC.  A study of the low-$p_{\rm T}$ pattern recognition
efficiency (see Appendix~\ref{app:patt_rec_eff}) indicated that a
requirement of $p_{\rm T} > 400$~MeV/$c$ would ensure that candidate
tracks would be in a $p_{\rm T}$ region with a relatively constant and
measurable tracking efficiency.  This $p_{\rm T}$ requirement was
imposed for all tracks prior to any corrections for
multiple-scattering and energy-loss effects.

Similarly, tracks with trajectories at high absolute pseudorapidities
tended to deposit fewer hits in the CTC, and therefore were
reconstructed less efficiently.  Requirements on the radii at which
extrapolations of track helices intersected one of the endplate planes
of the CTC, $r^{\rm exit}_{\rm CTC}$, were used to remedy this problem
by diminishing the pseudorapidity acceptance.  For example, a cut of
$r^{\rm exit}_{\rm CTC} > 132.0$~cm on this CTC exit radius would have
required that the track in question traverse all nine superlayers of
the CTC.  The tracking efficiency study outlined in
Appendix~\ref{app:patt_rec_eff}, however, showed that a requirement of
$r^{\rm exit}_{\rm CTC} > 110.0$~cm, corresponding to the radial
position of the outer edge of the second-outermost axial superlayer,
defined a set of tracks with high efficiency while minimizing the
reduction in geometric acceptance.  This requirement was imposed for
all non-muon track candidates after multiple-scattering and energy
corrections had been implemented.  Muon candidates were not subjected
to an $r^{\rm exit}_{\rm CTC}$ cut in order to accept candidates
identified by the CMX subsystems, which represented $\sim$22\% of the
total central muon fiducial acceptance.  Since the muon candidates in
all the channels had comparable $\eta$ distributions, any muon
tracking inefficiencies resulting from this lack of a CTC exit radius
criterion divided to unity in the final calculations of the ratios of
branching fractions.


\section{Optimizing the Kinematic Selection Criteria}
\label{sect:optimize}

The kinematic selection criteria were chosen to achieve accurate and
unbiased measurements of the $B$-candidate event yields while
maximizing both the rejection of the background and the statistical
significance of the signal.  Wherever possible, uniform criteria were
used to reconstruct the different decay modes.  This served both to
maximize internal consistencies in the study and to reduce the
magnitude and number of systematic uncertainties.

In order to minimize any bias caused by fluctuations in the background
levels in either the signal or sideband regions of the invariant mass
distributions, all of the $B$-meson decay modes were used in the
determination of the kinematic selection criteria.  The figure of
merit that was maximized was the signal significance, ${\displaystyle
S \equiv \frac{N_s}{\sqrt{N_s + N_b}}}$.  The $N_s$ symbol denotes the
number of signal events calculated with a Monte Carlo procedure (refer
to Chapter~\ref{chapt:geom_kin}) for a given decay mode and a given
time-integrated luminosity.  The expected number of background events,
$N_b$, was estimated from the data by extrapolating the background
rate in the observed $B$-meson invariant mass sidebands to the
background yields under the signal regions, defined to lie within
three standard deviations of the observed means of the $B$-meson
resonances.  The resultant optimal kinematic selection requirements,
which are detailed in the remaining sections of this chapter, varied
little within each of the charged and neutral $B$-decay mode
categories.

The optimum kinematic cuts determined by maximizing $S$ for each of
the decay channels were confirmed by maximizing a different measure of
statistical significance derived entirely from the data,
${\displaystyle
\frac{N_s}{\sigma(N_s)}}$, where ${\displaystyle \sigma(N_s)}$
was taken to be the event-yield uncertainty returned by the fit of the
given invariant mass distribution to a single fixed-width Gaussian
signal and a linear background parameterization.  Although this latter
method was more sensitive to statistical fluctuations, its conclusions
were consistent with those of the $S$-maximization technique.

\section{Muon Candidate Selection}
\label{sect:muon_select}

The minimum transverse momentum required of a muon to traverse the
central calorimeters and the solenoid magnet at $\eta \sim 0$ and
reach the CMU subsystem was $\sim$1.4~GeV/$c$.  A raw kinematic
requirement of $p_{\rm T} > 1.4$~GeV/$c$ was therefore placed on all muon
candidates in the analysis.  Because the numerators and denominators
in the ratio-of-branching-fraction calculations involved muons with
identical selection criteria, there was no need to demand that all
candidate muons populate the plateau region of the trigger efficiency
parameterizations (refer to Section~\ref{sect:trig_effects}).  As a
result, the fact that the lower edge of this $p_{\rm T}$ requirement fell in
inefficient regions of the measured Level 1 and Level 2
$p_{\rm T}$-dependent trigger efficiency distributions was of little
consequence.  A higher $p_{\rm T}$ requirement would have unnecessarily
weakened the statistical significance of those $B$-meson decays with a
$\psi(2S)$ candidate in the final state.

A major source of muon background was that due to charged kaons and
pions decaying to yield muons within the CDF tracking volume.  In some
cases, the charged kaon or pion tracks were reconstructed in the CTC,
and the daughter muons registered stubs in the muon systems.  A second
major source of muon background was that due to hadronic
`punch-through' particles (see Section~\ref{sect:cmp}), namely hadrons
that passed through the calorimeters and entered the muon systems.

In order to reduce these backgrounds, a track candidate in the CTC,
when extrapolated out to the muon chambers, was required to match the
position of a muon stub.  This condition was only met if the muon stub
and the extrapolated CTC track in question matched within three
standard deviations of the multiple-scattering and measurement
uncertainties in both the transverse ($r$-$\varphi$) and longitudinal
($z$) planes.  In cases where a CTC track could be extrapolated to
stubs in more than one muon subsystem, Boolean ``{\sc or}'' operations
were implemented to combine the appropriate matching requirements.  In
Run 1A, the muon matching criteria for $J/\psi$ dimuons had an
efficiency of $(98.66 \pm 0.04)$\% and increased the $J/\psi$
signal-to-background ratio from 3.08 to 3.61 in the $3.0 -
3.2$~GeV/$c^2$ invariant mass interval~\cite{sganos:thesis};
comparable effects were expected for Run 1B.

\section{Charmonium Reconstruction}
\label{sect:charm_reconstruct}

The three charmonium decay modes used in this study were
$J/\psi\to\mu^+\,\mu^-$, $\psi(2S)\to\mu^+\,\mu^-$, and $\psi(2S)\to
J/\psi\,\pi^+\,\pi^-$, where in the latter case the $J/\psi$ meson was
reconstructed in its dimuon mode.  These channels were used over other
charmonium decay modes because the $\mu^+\,\mu^-$ final state could
easily be identified using the CDF trigger system while also making
possible the rejection of several background processes.  This section
describes the selection of charmonium candidates; the resulting
inclusive charmonium invariant mass distributions are presented in
Section~\ref{sect:cl_efficiencies}, where they are used in the
determination of certain efficiency corrections.

\subsection{Dimuon Charmonium Decays}
\label{sect:dimuon_select}

Dimuon charmonium candidates were formed using the `dimuon finder'
module (see Section~\ref{sect:data_reduct}) by considering all the
muon candidates in a given event that met the criteria outlined in
Section~\ref{sect:muon_select}.  The two candidates constituting a
muon pair were required to possess charges of opposite sign.

Background contributions were reduced significantly by performing a
least-squares fit of the two muon track candidates and applying a
`vertex constraint', which forced the two tracks to originate from a
common point in space~\cite{marriner:cdf1996}.  In this fit, an
initial approximation to the track parameters was found, corrected for
multiple scattering and $dE/dx$ (energy-loss) effects, and then
adjusted under the vertex constraint so as to minimize the $\chi^2$.
The confidence level, $CL(\chi^2)$, of the fit was required to exceed
0.01.

When $J/\psi$ and $\psi(2S)$ candidates were used in the
reconstruction of an exclusive $B$ final state, an additional fit was
performed on the dimuon system to improve the invariant-mass
resolution of the $B$ candidate.  In this case, the fit was done with
the simultaneous application of a vertex constraint and a `mass
constraint'~\cite{marriner:cdf1996}, which required that the dimuon
mass equal the appropriate world average $J/\psi$ or $\psi(2S)$ mass
of 3.09688~GeV/$c^2$ or 3.68600~GeV/$c^2$, respectively~\cite{pdg96}.
The confidence level of each of these vertex-plus-mass constrained
fits was also required to satisfy the condition $CL(\chi^2) > 0.01$.

\subsection{Hadronic Cascade $\psi(2S)$ Decays}
\label{sect:pipi_select}

Candidates for the decay $\psi(2S)\to J/\psi\,\pi^+\,\pi^-$ were
chosen by combining $J/\psi$ dimuons, which were selected as described
in Section~\ref{sect:dimuon_select}, with dipions identified using the
`two-track finder' module described in Section~\ref{sect:data_reduct}.
The two pion candidates were required to meet the track quality
criteria detailed in Section~\ref{sect:trk_quality}, to have charges
of opposite sign, and to have an invariant mass, prior to any vertex
or vertex-plus-mass constrained fits, in the range $0.35 <
M(\pi^+\,\pi^-) < 0.61$~GeV/$c^2$.  The lower limit of this mass range
was motivated by the expected dipion invariant mass distribution for
$\psi(2S)$ decays, discussed in Section~\ref{sect:mpipi}.  The upper
limit corresponded to the maximum kinematically-allowed dipion
invariant mass, defined by the $\psi(2S) - J/\psi$ mass difference
with an allowance for measurement uncertainty.

In order to reduce background effects, a vertex-constrained fit was
performed on the four-track dimuon-dipion system, with the two
candidate muon tracks simultaneously constrained to form an invariant
mass equal to the world average $J/\psi$ mass~\cite{pdg96}.  The
confidence level of this fit was required to exceed 0.01.  As in the
case of the dimuon charmonia described in
Section~\ref{sect:dimuon_select}, when an hadronic cascade decay of
the $\psi(2S)$ meson was used as a daughter process in the
reconstruction of another exclusive decay, the four-track system was
simultaneously subjected to an additional mass constraint, namely the
world average $\psi(2S)$ mass~\cite{pdg96}.  Again, the probability of
this fit was required to satisfy the criterion $CL(\chi^2) > 0.01$.

\section{$B$-Meson Candidate Reconstruction}

The reconstruction of the decay modes $B^+\to\psi(2S)\,K^+$,
$B^0\to\psi(2S)\,K^*(892)^0$, $B^+\to J/\psi\,K^+$, and $B^0\to
J/\psi\,K^*(892)^0$ was achieved by combining one of the three
sought-for charmonium meson candidates in a given event with either
$K^+$ or $K^*(892)^0$ candidate mesons, as appropriate.  In spite of
the facts that six different decay chains were reconstructed and that
the final-state charged-particle track multiplicities forming these
chains ranged from three to six tracks, the methods used to identify
candidate $B$-meson decays were kept as uniform as possible.  This was
due in large part to the modularity provided by the data reduction
procedure described in Section~\ref{sect:data_reduct}.  The following
sections describe the selection of kaon candidates, the selection of
$B$-meson candidates, and the treatment of multiple invariant mass
hypotheses.

\subsection{Kaon Candidate Selection Criteria}

The CDF detector lacked the ability to differentiate kaon candidates
from pion candidates, a fact that necessitated the consideration of
all eligible tracks and the assignment of an appropriate kaon or pion
mass hypothesis to these tracks.  A consequence of these `blind'
hypothetical mass assignments was the amassment of significant
combinatorial backgrounds.  Fortunately, the charged particle
inclusive cross section in $p\bar{p}$ collisions is a rapidly falling
function of transverse momentum~\cite{abe:charged_particles}, thus
enabling the use of reasonably efficient minimum-$p_{\rm T}$ criteria
to remove a significant fraction of the combinatorial background.

The $p_{\rm T} > 400$~MeV/$c$ cut (see
Section~\ref{sect:trk_quality}), imposed to avoid low-$p_{\rm T}$
tracking inefficiencies (see Appendix~\ref{app:patt_rec_eff}), also
served to reduce combinatorial backgrounds caused by the lack of
particle identification.  The optimization scheme described in
Section~\ref{sect:optimize}, however, supported the application of
higher $p_{\rm T}$ cuts to combat these combinatorial backgrounds.
The criteria applied were therefore $p_{\rm T}(K^+) > 1.5$~GeV/$c$ for
kaon candidates used in the reconstruction of $B^+$ candidates and
$p_{\rm T}(K^*(892)^0) > 2.0$~GeV/$c$ for the two-track $K$-$\pi$
candidates used in the reconstruction of $B^0$ candidates.  Note that
the two tracks constituting each $K^*(892)^0$ candidate did not have
individual $p_{\rm T}$ cuts imposed on them beyond the universal
400~MeV/$c$ track quality criterion.

In a given event, $K^*(892)^0$ candidates were reconstructed by
considering all track pairs with charges of opposite sign.  For each
pair of tracks considered, both mass assignment hypotheses were
initially retained.  To reduce combinatorial backgrounds further, the
invariant mass of each $K$-$\pi$ candidate, $M(K^+\pi^-)$, was
required to fall within a window centred at the world average
$K^*(892)^0$ mass, 896.10~MeV/$c^2$~\cite{pdg96}.  The size of this
invariant mass window was 160~MeV/$c^2$, a requirement that was
estimated to be $\sim$80.5\% efficient under the assumptions that the
$K^*(892)^0$ resonance could be described by a
Breit-Wigner~\cite{breit-wigner} line shape with an intrinsic width of
$\Gamma = 50.5$~MeV~\cite{pdg96} and that the experimental resolution
was significantly less than $\Gamma$.  Finally, no constrained fits
were used to select and reconstruct $K^*(892)^0$ candidates
explicitly, although the two constituent tracks did undergo some
adjustments in subsequent global constrained fits of $B$-meson
candidates (see Section~\ref{sect:bmeson_select}).  The invariant mass
window criterion described above was imposed on the $K$-$\pi$
candidates after they had undergone global constrained fits.

\subsection{$B$-Meson Candidate Selection Criteria}
\label{sect:bmeson_select}

$B$ meson candidates were reconstructed by forming combinations of
charmonium candidates ($J/\psi$, $\psi(2S)$ dimuon, or $\psi(2S)$
hadronic cascade modes) with either $K^+$ or $K^*(892)^0$ candidates.
In a manner similar to that used in the reconstruction of charmonia, a
least-squares kinematic fit was performed on the $B$ candidate
daughter tracks under the constraints that all the tracks originate
from a single common secondary vertex and that the charmonium
candidates possess an invariant mass equal to the applicable world
average mass~\cite{pdg96}.  It was possible to constrain all the $B$
decay daughters to a single decay point because the distances traveled
by charmonium and $K^*(892)^0$ mesons before decaying were negligible
compared to the decay vertex resolution of the CDF
detector~\cite{marriner:cdf1996}.

In the case of $B$-meson reconstruction, a `momentum-pointing'
constraint was used in the kinematic fits.  This additional constraint
required the flight path direction of the $B$ candidate to be parallel
to its momentum in the transverse ($r$-$\varphi$) plane.  The pointing
constraint was not performed in three dimensions due to the large
uncertainty on the $z$ component of the primary vertex position (refer
to Section~\ref{sect:pv}).  The confidence level of the global
least-squares fit of each $B$ meson candidate, with vertex-plus-mass
and momentum-pointing constraints applied, was required to exceed
0.01.

The kinematic selection optimization, discussed in
Section~\ref{sect:optimize}, resulted in a $p_{\rm T} > 6.0$~GeV/$c$
transverse momentum requirement for $B^+$ candidates and a $p_{\rm T} >
9.0$~GeV/$c$ requirement for $B^0$ candidates.

\begin{figure}
\begin{center}
\leavevmode
\hbox{%
\epsfysize=4.1in
\epsffile{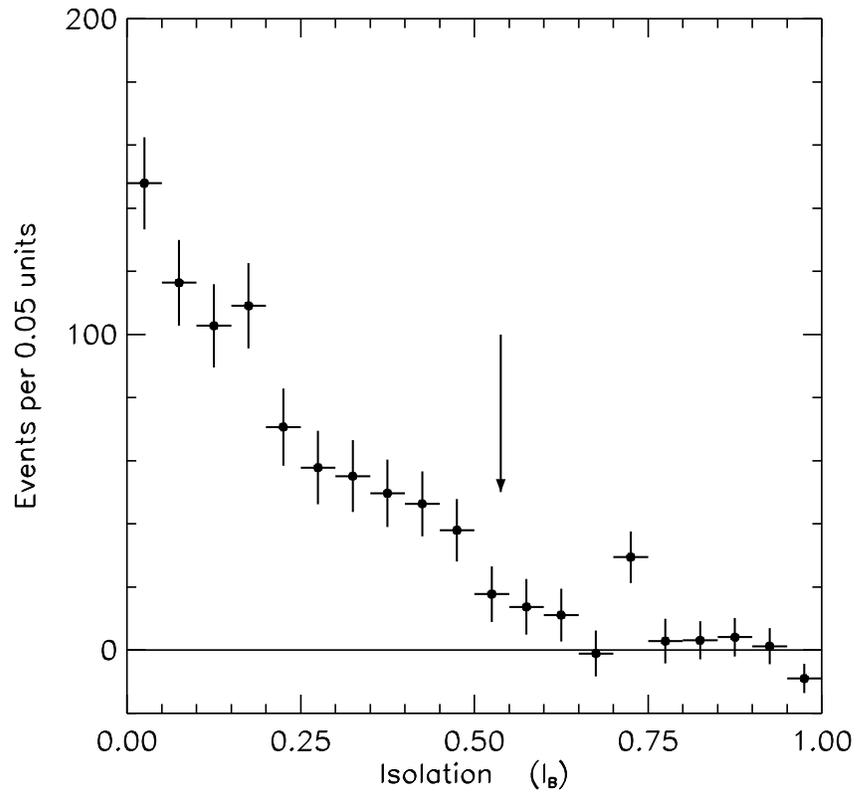}}
\end{center}
\caption
[Distribution of the $B$-meson isolation variable, $I_B$.]
{The distribution of the isolation variable $I_B$ for candidate
$B^+\to J/\psi\,K^+$ decays.  A background subtraction was performed
using the sidebands in the $J/\psi\,K^+$ invariant mass distribution.
The arrow marks the point below which candidates were accepted.}
\label{fig:isolation}
\end{figure}

Motivated by the understanding that $b$ quarks were expected to
`fragment' in a way that imparted most of their momentum to the
ensuing $B$ meson (References~\cite{suzuki:fragment,bjorken:fragment}
and Section~\ref{sect:hadronization}), an isolation criterion was
imposed on the $B$ meson candidates.  This requirement was expressed
using the variable
\begin{eqnarray}
\label{eqn:isolation}
I_B & \equiv & \frac{\sum\limits_{i \in\!\!\!/ B}^{R} \vec{p}_i \cdot
\hat{p}_B} {\left| \vec{p}_B \right|},
\end{eqnarray}
where $\vec{p}_B$ was the 3-momentum of the $B$-meson candidate and
the $\vec{p}_i$ were the momenta of additional particles, other than those
constituting the $B$ candidate, contained within a cone of radius $R
\equiv
\sqrt{\left(\Delta\varphi\right)^2 + \left(\Delta\eta\right)^2} \leq
1.0$ and with its axis collinear with the $B$-candidate 3-momentum
direction.  In addition, non-$B$ tracks were only included in the sum
in Equation~\ref{eqn:isolation} if their longitudinal displacement
parameter, $z_0$, lay within 5~cm of the primary vertex location
corresponding to the $B$ candidate in question.  This improved the
efficiency of the isolation criterion by not discarding $B$ candidates
that were unisolated due to tracks from other primary vertices in the
event.  Figure~\ref{fig:isolation} depicts the distribution of $I_B$
for candidate $B^+\to J/\psi\,K^+$ decays, after the combinatorial
background under the $B^+$ decay signal was statistically subtracted
using events in the $J/\psi\,K^+$ invariant mass sideband regions.
The criterion imposed was $I_B < 7/13$, which resulted from the
optimization procedure described in Section~\ref{sect:optimize}.
Figure~\ref{fig:isolation} illustrates that the $I_B$ requirement was
relatively efficient.

The long lifetimes of $B^+$ and $B^0$ mesons, ($1.64 \pm 0.05$)~ps and
($1.55 \pm 0.05$)~ps~\cite{pdg96}, respectively, made possible the
rejection of background with a cut on the proper decay length, $c\tau_B$,
defined by
\begin{eqnarray}
c\tau_B & \equiv & \frac{\vec{p}_{\rm T} \cdot \vec{x}_{\rm T}}{p_{\rm T}^2}\ m_B,
\end{eqnarray}
where $\vec{x}_{\rm T}$ was the distance between the primary and secondary
($B$-candidate) vertex, projected onto the transverse plane.  The
quantity $m_B$ represented the invariant mass of the candidate $B$
meson.  In the present analysis, a proper decay length requirement of
$c\tau_B > 100$~$\mu$m was imposed on $B$-meson candidates.

\begin{figure}
\begin{center}
\leavevmode
\hbox{%
\epsfxsize=6.0in
\epsffile{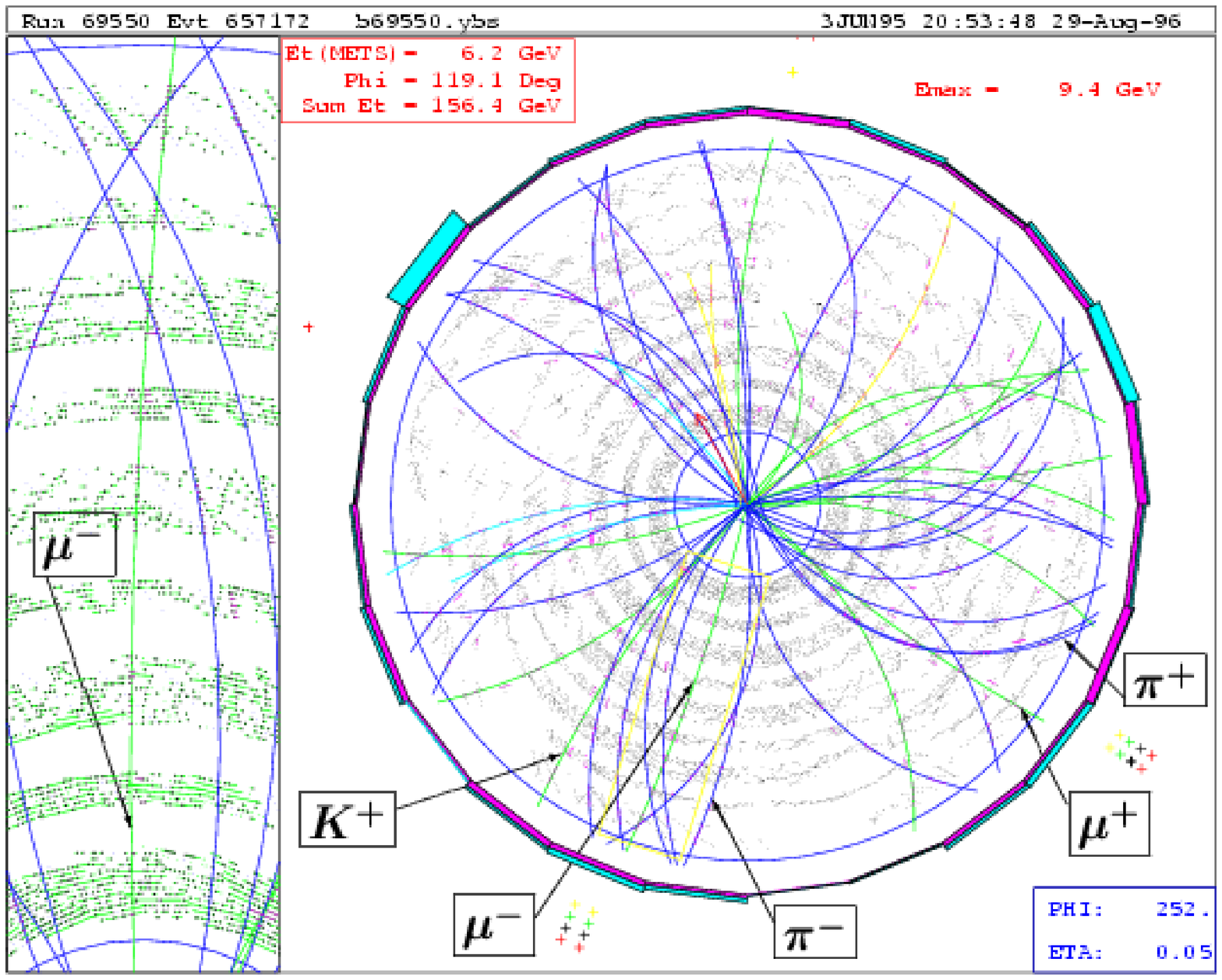}}
\end{center}
\caption
[Event-display diagram of an observed candidate $B^+$ decay.]
{A sample event-display diagram of an observed candidate $B^+$ decay.
The view in this figure is similar to that in
Figure~\ref{fig:ctc_endplate}.  CTC sense wire hits are visible for
all nine superlayers, and the transverse projections of the
reconstructed candidate track trajectories are indicated by smooth
curves.  The inner circle denotes the boundary between the CTC and
VTX, and the bars outside the CTC represent energy deposition in the
15$^\circ$ central calorimeter wedges.  Near the top of the figure are
listed the missing transverse energy ($E\!\!\!\!/_{\rm T}$) and its
azimuthal direction, the total transverse energy for the event, and
the maximum energy recorded in a single calorimeter wedge.  The
``$\scriptscriptstyle +$'' symbols represent stubs in the central muon
chambers.  The data reduction algorithms and event selection criteria
identified this event as a candidate for the decay
$B^+\to\psi(2S)\,K^+$, where the $\psi(2S)$ meson was reconstructed in
its $\psi(2S)\to J/\psi\,\pi^+\,\pi^-$ mode and the $J/\psi$ meson was
reconstructed in its $J/\psi\to\mu^+\,\mu^-$ mode.  The identified
track candidates are indicated in the figure.  The azimuth and
pseudorapidity coordinates in the bottom right corner are those of the
track with the highest $p_{\rm T}$, the $\mu^-$ candidate, a close-up
view of which is illustrated in the left panel of the figure.}
\label{fig:event_display_text}
\end{figure}

Figure~\ref{fig:event_display_text} depicts a sample event display of
a candidate $B$-meson decay that was identified as such by the
data reduction algorithms and selection criteria described in this
chapter.

\subsection{Multiple Invariant Mass Combinations}
\label{sect:kst_rc_wc}

Multiple $B$-meson candidates that passed all the selection criteria
for a given event were a significant source of background.  Such
additional combinations usually resulted from other tracks that formed
a vertex with the two muons in the event and were therefore more
prevalent in reconstructions with higher daughter-track
multiplicities, such as those involving the $\psi(2S)\to
J/\psi\,\pi^+\,\pi^-$ mode.  Another obvious source of multiple
combinations was in the case of $K^*(892)^0$ candidates, where the
$K$-$\pi$ mass assignment ambiguity and the size of the $K^*(892)^0$
mass window cut required by the natural width of the $K^*(892)^0$
meson increased the probability of additional invariant mass
combinations in the event.

In the case of the $K^*(892)^0$ candidates, three treatments of the
$K$-$\pi$ mass hypothesis ambiguity were considered: (a) random
selection of one of the mass hypotheses, (b) selection of the
hypothesis with $M(K^+\pi^-)$ nearest the accepted world average
$K^*(892)^0$ pole value~\cite{pdg96}, and (c) parameterization of the
relative contributions and resolutions of the right and wrong mass
assignment hypotheses in the signal region.  All of these techniques
would have required Monte Carlo calculations.

In this analysis, the last of the above three treatments was adopted
in conjunction with a general procedure that made use of the $\chi^2$
probability of the global constrained least-squares fit, $CL(\chi^2)$,
to remove multiple combinations.  The prescription used was the
following:
\begin{enumerate}

\item
All selection criteria were applied, including the $M(K^+\pi^-)$
requirement.

\item
For a given event, only the candidate possessing the highest global
$CL(\chi^2)$ value returned by the constrained fit was retained.

\item
For events where both $K^*(892)^0$ $K$-$\pi$ mass hypotheses survived
all of the selection criteria, the two $CL(\chi^2)$ values were often
similar.  If one of the $M(K^+\pi^-)$ combinations satisfied the
highest-$CL(\chi^2)$ criterion, and a second combination that differed
only in the $K$-$\pi$ mass assignment existed, then the second entry
was also retained.

\item
The signal observed in the $B$ candidate invariant mass distribution
of those modes containing a $K^*(892)^0$ candidate was fit using two
Gaussian distributions with their relative widths and amplitudes
constrained by ratios determined using the Monte Carlo calculations
described in Chapter~\ref{chapt:geom_kin}.  The constraints are listed
in Table~\ref{tab:kst_rc_wc}.  Refer to Section~\ref{sect:yields} for
a synopsis of the $B$-meson candidate event yields.

\end{enumerate}

\begin{table}
\begin{center}
\begin{tabular}{|l|l|c|c|}    \hline
$B$ Mode & $c\bar{c}$ Mode  & $\frac{\xi_{WC}}{\xi_{RC}}$ &
\rule{0mm}{0.19in} $\frac{\sigma_{WC}}{\sigma_{RC}}$ \\[0.05in]  \hline\hline
\rule{0mm}{5mm}
$B^0\to J/\psi\,K^*(892)^0$ & $J/\psi \to \mu^+\,\mu^-$ & 0.0683 & 3.623 \\
\rule{0mm}{5mm}
$B^0\to\psi(2S)\,K^*(892)^0$ & $\psi(2S)\to\mu^+\,\mu^-$ & 0.0456 & 5.750 \\
\rule[-3mm]{0mm}{8mm}
$B^0\to\psi(2S)\,K^*(892)^0$ & $\psi(2S)\to J/\psi\,\pi^+\,\pi^-$ & 0.0429 &
		6.327 \\ \hline
\end{tabular}
\end{center}
\caption
[Monte Carlo constraints on $K$-$\pi$ mass assignment combinations.]
{The relative amplitudes ($\xi$) and widths ($\sigma$) of the right
($RC$) and wrong ($WC$) $K$-$\pi$ combinations, as determined by fits
to Monte Carlo invariant mass distributions using double-Gaussian
signal parameterizations.}
\label{tab:kst_rc_wc}
\end{table}

%% file: geom_kin.tex
\label{chapt:geom_kin}

In branching fraction analyses, a knowledge of the proportion of
candidates that go unobserved due to both the fiducial detector
geometry and the kinematic selection criteria is essential.  Since the
data themselves could not be used to determine these acceptances and
efficiencies, it was necessary to make use of Monte
Carlo~\cite{james:mc} calculations.

This chapter describes the Monte Carlo machinery used to generate
bottom quarks, hadronize these quarks into $B$ mesons, decay the $B$
mesons into final-state particles, simulate the signatures left by the
particles in the CDF detector, and simulate the first two levels of
the triggers relevant to the study.  Neither the underlying event
(refer to Equation~\ref{eqn:b_bbar_hadroproduction}) nor multiple
$p\bar{p}$ interactions were modeled in these Monte Carlo studies.

\section{The Monte Carlo Generation of $B$ Mesons}
\label{sect:mc_gen}

Single $b$ quarks were generated according to an inclusive transverse
momentum ($k_{\rm T}(b)$) spectrum based on a next-to-leading order QCD
calculation~\cite{nde:bxsec} that used the Martin-Roberts-Stirling
MRSD\O\ parton distribution functions~\cite{mrsd0:pdf}, a
renormalization scale of $\mu = \mu_0 \equiv \sqrt{m_b^2 + k^2_{\rm T}(b)}$,
and a $b$-quark mass of $m_b = 4.75$~GeV/$c^2$.  Refer to
Section~\ref{sect:hadroproduction} for an account of $b$-quark
hadroproduction.  The $b$ quarks were produced in the rapidity range
$-1.1 < y_b < 1.1$ with $k_{\rm T}(b) > 5.0$~GeV/$c$ and fragmented into $B$
mesons according to a model that used the Peterson fragmentation
function~\cite{peterson:frag} with the Peterson $\epsilon_b$ parameter
defined to be 0.006~\cite{chrin:epsilon} (refer to
Section~\ref{sect:hadronization}).  Flavour was conserved in both the
production and fragmentation of $b$ quarks.

\section{The Monte Carlo Decay of $B$ Mesons}
\label{sect:mc_decay}

Decays of Monte Carlo generated $B$ mesons into charmonium and kaon
final states were performed using a modified version of the CLEO {\sc
qq} Monte Carlo programme~\cite{avery:cleoqq}.  Properties of the
relevant particles, including mass, lifetime, and intrinsic width,
were updated in the programme to reflect the current world-average
values~\cite{pdg96}.  For the pseudoscalar~$\to$~vector~$+$~vector
decays, the decay helicities were nominally set to the central value
of the world-average longitudinal polarization fraction measured for
the decay $B^0\to J/\psi\,K^*(892)^0$, $\Gamma_L/\Gamma = 0.78 \pm
0.07$~\cite{browder:review,kamal:nonfactorization,abe:bjpkst_helicity}.
Finally, a customized {\sc qq} Monte Carlo matrix element was
constructed to model correctly the observed kinematics of the pions in
the charmonium decay $\psi(2S)\to J/\psi\,\pi^+\,\pi^-$.  The
following section describes this customization.

\subsection{The $\psi(2S)\to J/\psi\,\pi^+\,\pi^-$ Monte Carlo Matrix Element}
\label{sect:mpipi}

The dipion kinematics in the decay $\psi(2S)\to J/\psi\,\pi^+\,\pi^-$
were studied to search for techniques to remove background efficiently
and to ensure that the pion tracks were simulated correctly.  Whereas
the generated {\sc qq} transverse momentum distribution of the dipion
system was consistent with that in the data, the simulated {\sc qq}
dipion invariant mass was not in agreement with the data.

The default {\sc qq} Monte Carlo programme determined the kinematics
of the $J/\psi\,\pi^+\,\pi^-$ decay products using pure three-body
phase space; however, at the time of the discovery of the $\psi(2S)\to
J/\psi\,\pi^+\,\pi^-$ decay channel, it was observed that the angular
distributions of the pions were isotropic~\cite{abrams:pipi}.  In
subsequent phenomenological investigations of the decay amplitude, the
absence of any observed angular correlations made apparent the fact
that the amplitude had a strong dependence on the invariant mass of
the $\pi^+\,\pi^-$ pair~\cite{schwinger:pipi,brown:pipi,pham:pipi}.

A set of customized routines was therefore written to model the
$\psi(2S)\to J/\psi\,\pi^+\,\pi^-$ matrix element in the {\sc qq}
Monte Carlo programme using the phenomenological prescription.  For
the purposes of the algorithm, the decay was considered to have the
logical form $\psi(2S)\to J/\psi\,{\cal P}$, where ${\cal P}$ was
taken to be a dipion pseudostate (${\cal P} \to \pi^+\,\pi^-$).  In
the first step, the invariant mass of ${\cal P}$ was calculated using
the ${\cal P}$ mass distribution~\cite{pham:pipi},
\begin{equation}
\frac{dN}{dm_{\cal P}} \propto \left(m_{\cal P}^2 -
                                     4m_\pi^2\right)^{\frac{5}{2}}
\sqrt{\left(m_{\psi(2S)}^2 - m_{J/\psi}^2 - m_{\cal P}^2\right)^2
      - 4m_{J/\psi}^2m_{\cal P}^2},
\end{equation}
which is in a form similar to that used in
Reference~\cite{coffman:jpsi_fraction}.  The mass of the ${\cal P}$
pseudostate was extracted using the Von Neumann acceptance-rejection
Monte Carlo method~\cite{pdg96}.  The second and third steps in the
algorithm consisted of decaying, using phase space alone, the modes
$\psi(2S) \to J/\psi\,{\cal P}$ and ${\cal P}\to\pi^+\,\pi^-$,
respectively.  The $J/\psi\to\mu^+\,\mu^-$ decay was performed using
the default {\sc qq} matrix element for that decay.

\begin{figure}
\begin{center}
\leavevmode
\hbox{%
\epsfysize=3.3in
\epsffile{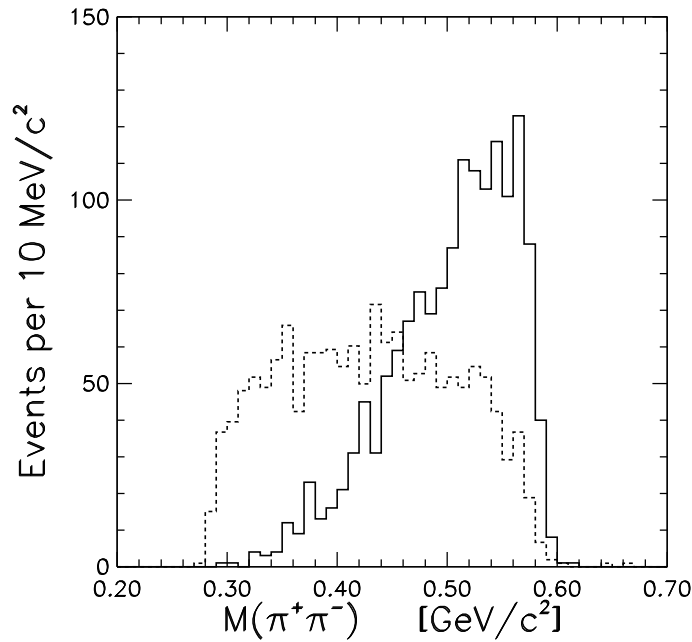}}
\end{center}
\caption
[Dipion invariant mass distributions using two Monte Carlo models.]
{The dipion invariant mass distribution calculated using a
phenomenologically-prescribed {\sc qq} Monte Carlo matrix element
(solid histogram).  The dashed histogram illustrates the results of a
pure phase-space calculation.}
\label{fig:mc_cust_def}
\end{figure}

\begin{figure}
\begin{center}
\leavevmode
\hbox{%
\epsfxsize=5.0in
\epsffile{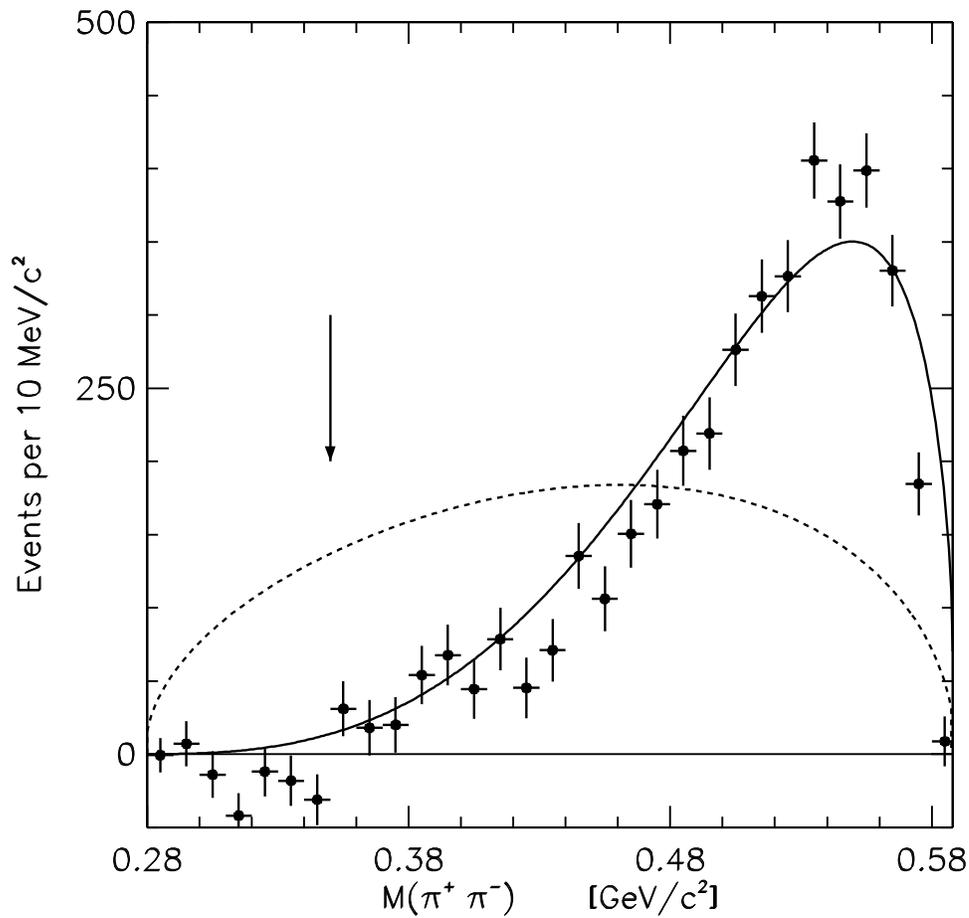}}
\end{center}
\caption
[Dipion invariant mass distribution observed in $\psi(2S)\to
J/\psi\,\pi^+\,\pi^-$ candidate decays.]
{The observed background-subtracted distribution of the dipion
invariant mass (points) in decays of the form $\psi(2S)\to
J/\psi\,\pi^+\,\pi^-$.  The arrow indicates where a cut at
0.35~GeV/$c^2$ was subsequently imposed.  The solid curve represents a
phenomenological prediction due to Pham {\it et al.}~\cite{pham:pipi};
the broken curve describes a pure phase-space parameterization.}
\label{fig:mpipi}
\end{figure}

Figure~\ref{fig:mc_cust_def} illustrates the
phenomenologically-inspired and phase-space {\sc qq} dipion invariant
mass distributions, as determined from Monte Carlo calculations.  The
correspondence between a parameterization of the phenomenological
prediction and the data is pictured in Figure~\ref{fig:mpipi}, where a
sideband subtraction was used to remove the background to the
$\psi(2S)$ candidate decays.  A recent precision measurement of the
$\psi(2S)\to J/\psi\,\pi^+\,\pi^-$ branching fraction reports a
similar $M(\pi^+\,\pi^-)$ distribution~\cite{armstrong:psi2s_br}.  A
conservative cut of $M(\pi^+\pi^-) > 0.35$~GeV/$c^2$ (refer to
Section~\ref{sect:pipi_select}) was applied to the dipion candidates
prior to the application of any vertex, vertex-plus-mass, or
momentum-pointing constrained fits.  The $M(\pi^+\pi^-)$ requirement
was found to be efficient and capable of rejecting a significant
number of background candidates.

\section{Simulation of the Detector Response}

Once Monte Carlo $B$ mesons were generated and decayed into their
daughter states, a full Monte Carlo simulation of the CDF detector
response was invoked.  The simulation produced raw data structures
that were in most cases identical to those read out from the DAQ
system using collider data (refer to Section~\ref{sect:daq}); this
enabled the subsequent processing of Monte Carlo events in a manner
almost identical to that used for the data.  The corrections for local
nonuniformities in the magnetic field, described in
Section~\ref{sect:mag}, however, were switched off at the detector
simulation and production (see Section~\ref{sect:data_reduct})
processing stages.

As outlined in Section~\ref{sect:data_sample}, the CDF detector
geometry changed between Runs 1A and 1B.  For the purposes of the
detector simulation in the present study, the Run 1B geometry
information was used to represent the CDF detector for the entire Run
1 period.  Within the Monte Carlo statistics used for this analysis,
this assumption appeared to be reasonable for the calculation of
relative geometric acceptances and kinematic selection efficiencies.

Greater care was necessary in the characterization of the beam
profiles near the primary vertex.  The simulated transverse beam
profile was tuned to match approximately that observed in the data.
The longitudinal beam profile, however, was more important to this
study due to its implications for the fiducial SVX acceptance.  The
$z$ location of the simulated primary vertex was forced to be
distributed according to the sum of two Gaussian probability density
functions with means and standard deviations of 2.3~cm and 2.1~cm, and
18.4~cm and 35.6~cm, respectively.  The relative normalization of
these two Gaussian distributions was approximately 2:3.

The Monte Carlo events were processed by the same production machinery
(Section~\ref{sect:data_reduct}) used on collider data.  This ensured
that the data and the Monte Carlo events were both subjected to any
biases, should they exist, inherent to the reconstruction algorithms.

\section{Simulation of the Level 1 and 2 Triggers}
\label{sect:trigsim_eff}

After the CDF detector response to the Monte Carlo decay fragments had
been simulated, the events were passed through a Level~1 and Level~2
dimuon trigger simulation module.  (Refer to
Sections~\ref{sect:trigger_l1} and \ref{sect:level2} for descriptions
of the Level~1 and Level~2 triggers, respectively.)  This module used
parameterizations of the measured muon trigger efficiencies to
determine the probability that a given candidate dimuon event
satisfied the trigger requirement.  The Monte Carlo events were then
selected using a Monte Carlo procedure based on this probability.

\begin{figure}[p]
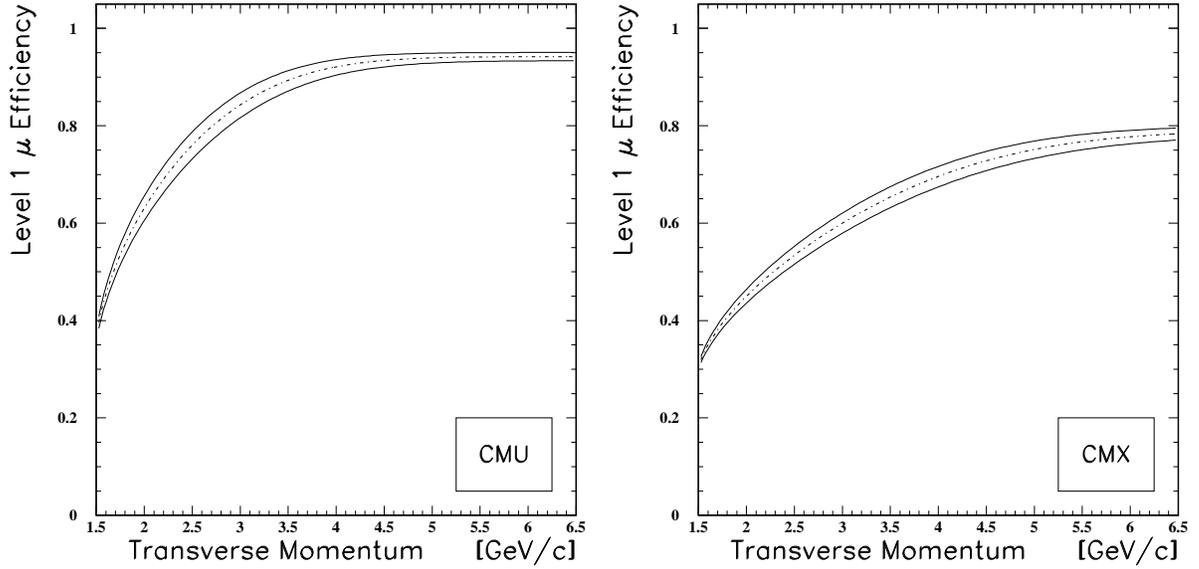

\begin{center}
\leavevmode
\hbox{%
\epsfxsize=3.0in
\epsffile{diagrams/l1prbu.epsi}
\hspace{0.05in}
\epsfxsize=3.0in
\epsffile{diagrams/l1prbx.epsi}}
\end{center}
\caption
[Level 1 low-$p_{\rm T}$ CMU and CMX trigger efficiency parameterizations.]
{The Level 1 low-$p_{\rm T}$ CMU and CMX trigger efficiency
parameterizations.  The dashed-dotted curves are the central values and
the solid curves represent shifts of the plateau efficiencies and
$p_{\rm T}$ values by one standard deviation.}
\label{fig:l1prbux}
\end{figure}

Figure~\ref{fig:l1prbux} illustrates the Level~1 CMU and CMX trigger
efficiency parameterizations as functions of muon transverse momentum.
These efficiencies were assumed to be functions of $p_{\rm T}$ alone and to
remain unchanged during the course of Run~1.  In the case of the CMX
efficiency, the veto scintillation counters (refer to
Section~\ref{sect:cmx}) degraded the plateau efficiency uniformly.

The Level 2 dimuon trigger efficiencies were found to depend not only
on $p_{\rm T}$, but also on charge, pseudorapidity ($\eta$), azimuth
($\varphi$), and time-integrated luminosity ($\int\!{\cal L}\,dt$).
Due to the geometry of the CTC, positive muon tracks were accepted by
the trigger more efficiently than negative tracks.  The $\eta$
dependence of the efficiency was observed to be parabolic due to the
increased hit efficiency of higher-$\eta$ tracks\footnote{Muon tracks
with large $|\eta|$ values deposited more charge on the CTC wires;
this increased the pulse height, which in turn enhanced the hit
efficiency.}.  The $\varphi$ dependence of the efficiency was observed
to be sinusoidal due to the offset of the beam from the geometrical
centre of the CTC\footnote{The CFT (Section~\ref{sect:level2}) did not
measure the impact parameter, thereby introducing a false curvature
term, which varied with $\varphi$.}.  Finally, the decline in the CTC
hit efficiencies as a function of $\int\!{\cal L}\,dt$ caused a
corresponding degradation in the Level~2 CFT efficiency.  This effect
also caused an increase in the curvature of the $\eta$-dependent
efficiency parabola.

\begin{figure}[p]
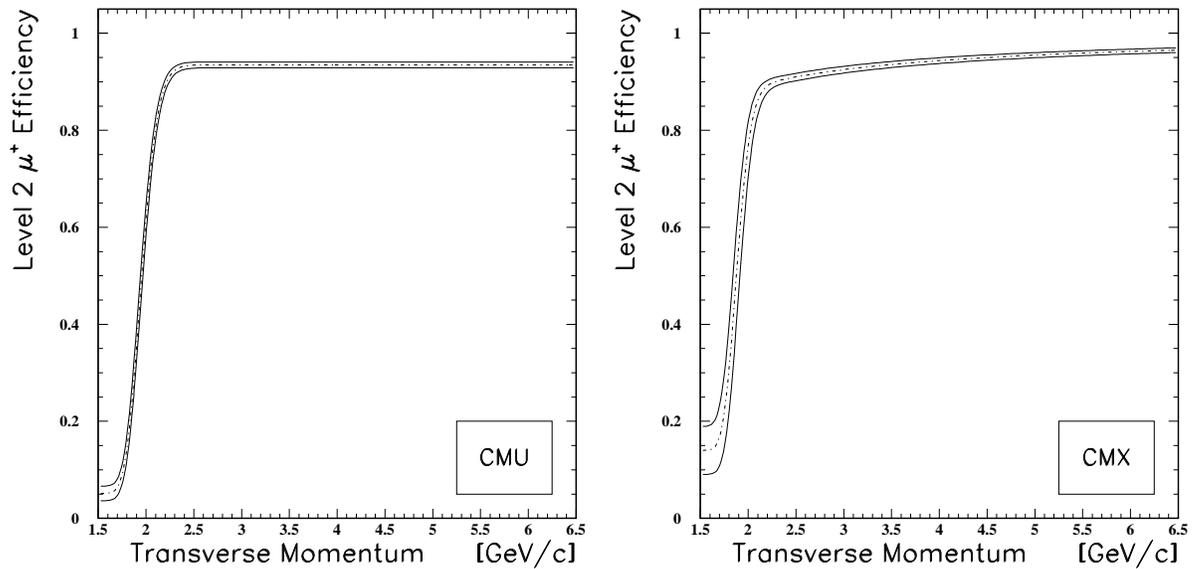

\begin{center}
\leavevmode
\hbox{%
\epsfxsize=3.0in
\epsffile{diagrams/l2prbu.epsi}
\hspace{0.05in}
\epsfxsize=3.0in
\epsffile{diagrams/l2prbx.epsi}}
\end{center}
\caption
[Level 2 low-$p_{\rm T}$ CMU and CMX trigger efficiency parameterizations.]
{The Level 2 low-$p_{\rm T}$ CMU and CMX trigger efficiency parameterizations
for positively charged muons in the 1B run range $65\,001 - 66\,000$
and the lowest-$p_{\rm T}$ CFT bin.  The dashed-dotted curves are the
central values and the solid curves represent shifts of the plateau
efficiencies and $p_{\rm T}$ values by one standard deviation.}
\label{fig:l2prbux}
\end{figure}

Figure~\ref{fig:l2prbux} depicts the Level~2 CMU and CMX trigger
efficiency parameterizations as functions of muon transverse momentum.
The curves shown are based on 8.89~pb$^{-1}$ of positively charged
muon candidates that fell in a particular 1B run range ($65\,001 -
66\,000$) and the lowest $p_{\rm T}$ bin ($\sim$2.2~GeV/$c$) of the
CFT.

Note that the Level~3 dimuon trigger efficiency, which was measured to
be $0.97 \pm 0.02$ and independent of $p_{\rm T}$, was not explicitly
simulated in the geometric and kinematic acceptance calculations.

\section{The Acceptance Calculations}
\label{sect:accept_calc}

In the final step of the default Monte Carlo procedure, the events
were processed by the same analysis-specific user algorithms used on
the collider data.  The reconstructed Monte Carlo events were used to
estimate the geometric and kinematic acceptances for the decays under
study.  The products of these quantities are listed in
Table~\ref{tab:abs_geom_eff}.  The Monte Carlo statistical
uncertainties on the absolute acceptance values listed in
Table~\ref{tab:abs_geom_eff} ranged from $1.2\%$ to $2.9\%$.

\begin{table}
\begin{center}
\begin{tabular}{|l|l|c|}        \hline
\multicolumn{1}{|c}{$B$ Mode} & \multicolumn{1}{|c|}{$c\bar{c}$ Mode} &
			Acceptance \\ \hline\hline
\rule{0mm}{5mm}
$B^+\to J/\psi\,K^+$ & $J/\psi\to\mu^+\,\mu^-$ & $0.01920\pm 0.00025$ \\
\rule{0mm}{5mm}
$B^+\to\psi(2S)\,K^+$ & $\psi(2S)\to\mu^+\,\mu^-$ & $0.02183\pm 0.00027$ \\
\rule[-3mm]{0mm}{8mm}
$B^+\to\psi(2S)\,K^+$ & $\psi(2S)\to J/\psi\,\pi^+\,\pi^-$ &
			$0.00631\pm 0.00015$ \\ \hline\hline
\rule{0mm}{5mm}
$B^0\to J/\psi\,K^*(892)^0$ & $J/\psi\to\mu^+\,\mu^-$ & $0.00777\pm 0.00016$ \\
\rule{0mm}{5mm}
$B^0\to\psi(2S)\,K^*(892)^0$ & $\psi(2S)\to\mu^+\,\mu^-$ &
			$0.00828\pm 0.00017$ \\
\rule[-3mm]{0mm}{8mm}
$B^0\to\psi(2S)\,K^*(892)^0$ & $\psi(2S)\to J/\psi\,\pi^+\,\pi^-$ &
			$0.00385\pm 0.00011$ \\ \hline
\end{tabular}
\end{center}
\caption
[Absolute products of the geometric and kinematic acceptances.]
{A summary of the absolute products of the geometric and kinematic
acceptances, calculated for each decay mode for $B$ mesons produced
with $k_{\rm T}(b) > 5.0$~GeV/$c$ and $-1.1 < y_b < 1.1$.  The grouping is on
the basis of common selection criteria, and the uncertainties given are
due to Monte Carlo statistics alone.}
\label{tab:abs_geom_eff}
\end{table}


Table~\ref{tab:rel_geom_eff} lists the appropriate relative geometric
and kinematic acceptances necessary for the branching-fraction ratio
calculations.  The syntax in Table~\ref{tab:rel_geom_eff} follows that
introduced in Table~\ref{tab:ratios_to_measure}, but with particular
entries for the two different $\psi(2S)$ modes.  The trends in
Table~\ref{tab:rel_geom_eff} are consistent with differences in
daughter-track multiplicity and $c\bar{c}$ meson mass.  When the
relative acceptance values were used in the branching-fraction
calculations, a systematic uncertainty of $3.0\%$ was applied to
account for differences between two different CDF detector simulations
and between the Run 1A and 1B detector
geometries~\cite{metzler_huffman:cdf3541}.  This systematic
uncertainty in the acceptance-measurement method precluded a need for
increased Monte Carlo statistics.

\begin{table}
\begin{center}
\begin{tabular}{|l|ccc|}        \hline
\multicolumn{1}{|c|}{Relative} &
			\multicolumn{1}{|c|}{$B^+\to J/\psi\,K^+$} &
			\multicolumn{1}{|c|}{$B^0\to J/\psi\,K^*(892)^0$} &
			\multicolumn{1}{|c|}{$B^+\to \psi(2S)\,K^+$} \\
\multicolumn{1}{|c|}{Acceptance}&
			\multicolumn{1}{|c|}{$J/\psi\to\mu^+\,\mu^-$} &
			\multicolumn{1}{|c|}{$J/\psi\to\mu^+\,\mu^-$} &
			\multicolumn{1}{|c|}{$\psi(2S)\to\mu^+\,\mu^-$}
				\\ \hline\hline
$B^0\to J/\psi\,K^*(892)^0$ &
	\raisebox{-0.10in}[0.0in][0.0in]{$0.405\pm 0.010$}& & \\
$J/\psi\to\mu^+\,\mu^-$     & & & \\ \cline{1-1}

$B^+\to \psi(2S)\,K^+$      &
	\raisebox{-0.10in}[0.0in][0.0in]{$1.137\pm 0.020$} &
	\raisebox{-0.10in}[0.0in][0.0in]{$2.810\pm 0.068$}& \\
$\psi(2S)\to\mu^+\,\mu^-$   &  & & \\ \cline{1-1}

$B^+\to \psi(2S)\,K^+$      & 
	\raisebox{-0.10in}[0.0in][0.0in]{$0.329\pm 0.009$}&
	\raisebox{-0.10in}[0.0in][0.0in]{$0.813\pm 0.025$}&
	\raisebox{-0.10in}[0.0in][0.0in]{$0.289\pm 0.008$}\\
$\psi(2S)\to J/\psi\,\pi^+\,\pi^-$ &  & & \\ \cline{1-1}

$B^0\to \psi(2S)\,K^*(892)^0$ &
	\raisebox{-0.10in}[0.0in][0.0in]{$0.431\pm 0.010$}&
	\raisebox{-0.10in}[0.0in][0.0in]{$1.066\pm 0.031$}&
	\raisebox{-0.10in}[0.0in][0.0in]{$0.379\pm 0.009$}\\
$\psi(2S)\to\mu^+\,\mu^-$   &  & & \\ \cline{1-1}

$B^0\to \psi(2S)\,K^*(892)^0$ &
	\raisebox{-0.10in}[0.0in][0.0in]{$0.201\pm 0.006$}&
	\raisebox{-0.10in}[0.0in][0.0in]{$0.496\pm 0.018$}&
	\raisebox{-0.10in}[0.0in][0.0in]{$0.177\pm 0.006$}\\
$\psi(2S)\to J/\psi\,\pi^+\,\pi^-$ &  & & \\ \hline
\end{tabular}
\end{center}
\caption
[Relative products of the geometric and kinematic acceptances.]
{A summary of the relative products of the geometric and kinematic
acceptances, calculated for each ratio of decay modes for $B$ mesons produced
with $k_{\rm T}(b) > 5.0$~GeV/$c$ and $-1.1 < y_b < 1.1$.  The indicated
uncertainties represent the Monte Carlo statistics alone.}
\label{tab:rel_geom_eff}
\end{table}

\subsection{Differential Production Cross Section of Generated $b$ Quarks}

A significant source of systematic uncertainty on the relative
acceptance calculations reported in Section~\ref{sect:accept_calc} was
that due to the assumed generated $b$-quark momentum spectrum.
Although the measured $B$-meson differential cross section has been
observed to have a shape in agreement with theoretical predictions,
the absolute rate was near the limits of that predicted by typical
variations in the theoretical parameters~\cite{abe:bxsec}.  An
estimate of the acceptance uncertainty due to the calculated $b$-quark
momentum spectrum was taken by varying the default Monte Carlo
generation parameters detailed in Section~\ref{sect:mc_gen} from their
nominal values.  The $b$-quark mass was varied from its central value
of $m_b = 4.75$~GeV/$c^2$ to 4.50~GeV/$c^2$ and 5.00~GeV/$c^2$, and
the renormalization scale was varied from a nominal value of $\mu =
\mu_0$ to $\mu_0/4$ and $2\mu_0$.  The effects of these variations on
the acceptance ratios were examined, and one half of each maximum
deviation was taken as the appropriate systematic uncertainty on the
relative acceptance.  Table~\ref{tab:pt_spectrum_syst} summarizes the
systematic uncertainties attributed to the generated differential
production cross section of the $b$ quarks.

\begin{table}
\begin{center}
\begin{tabular}{|l|ccc|}        \hline
\multicolumn{1}{|c|}{$b$-Quark Production} &
			\multicolumn{1}{|c|}{$B^+\to J/\psi\,K^+$} &
			\multicolumn{1}{|c|}{$B^0\to J/\psi\,K^*(892)^0$} &
			\multicolumn{1}{|c|}{$B^+\to \psi(2S)\,K^+$} \\
\multicolumn{1}{|c|}{Spectrum Systematics (\%)}&
			\multicolumn{1}{|c|}{$J/\psi\to\mu^+\,\mu^-$} &
			\multicolumn{1}{|c|}{$J/\psi\to\mu^+\,\mu^-$} &
			\multicolumn{1}{|c|}{$\psi(2S)\to\mu^+\,\mu^-$}
				\\ \hline\hline
$B^0\to J/\psi\,K^*(892)^0$ &
	\raisebox{-0.10in}[0.0in][0.0in]{$4.0$}& & \\
$J/\psi\to\mu^+\,\mu^-$     & & & \\ \cline{1-1}

$B^+\to \psi(2S)\,K^+$      &
	\raisebox{-0.10in}[0.0in][0.0in]{$2.4$} &
	\raisebox{-0.10in}[0.0in][0.0in]{$5.9$}& \\
$\psi(2S)\to\mu^+\,\mu^-$   &  & & \\ \cline{1-1}

$B^+\to \psi(2S)\,K^+$      & 
	\raisebox{-0.10in}[0.0in][0.0in]{$2.8$}&
	\raisebox{-0.10in}[0.0in][0.0in]{$5.0$}&
	\raisebox{-0.10in}[0.0in][0.0in]{$5.5$}\\
$\psi(2S)\to J/\psi\,\pi^+\,\pi^-$ &  & & \\ \cline{1-1}

$B^0\to \psi(2S)\,K^*(892)^0$ &
	\raisebox{-0.10in}[0.0in][0.0in]{$5.5$}&
	\raisebox{-0.10in}[0.0in][0.0in]{$1.9$}&
	\raisebox{-0.10in}[0.0in][0.0in]{$5.8$}\\
$\psi(2S)\to\mu^+\,\mu^-$   &  & & \\ \cline{1-1}

$B^0\to \psi(2S)\,K^*(892)^0$ &
	\raisebox{-0.10in}[0.0in][0.0in]{$6.7$}&
	\raisebox{-0.10in}[0.0in][0.0in]{$3.4$}&
	\raisebox{-0.10in}[0.0in][0.0in]{$9.6$}\\
$\psi(2S)\to J/\psi\,\pi^+\,\pi^-$ &  & & \\ \hline
\end{tabular}
\end{center}
\caption
[Systematic uncertainty on acceptance due to variations in the
differential production cross section of generated $b$ quarks.]
{A summary of the systematic uncertainties (expressed in \%) on the
relative geometric and kinematic acceptance due to variations in the
differential production cross section of the generated $b$ quarks.}
\label{tab:pt_spectrum_syst}
\end{table}

\subsection{Trigger Effects}
\label{sect:trig_effects}

Even though the numerators and denominators of the branching-fraction
ratios in Table~\ref{tab:ratios_to_measure} involve muon pairs that
originate from common data samples and the $p_{\rm T}$ spectra of the
$B^+$ and $B^0$ mesons are expected to be comparable, some care is
required due to the possibility of topological effects when
considering relative trigger acceptances in branching-fraction ratio
studies.  In approximately 75\%\ of the selected events, the two muon
candidates that were identified as charmonium daughters were also the
muon candidates identified by the dimuon trigger system.  In most of
the remaining events, an additional muon candidate in the event
satisfied the dimuon trigger requirements.  These `volunteers' were
included in the analysis in order to maximize the sensitivity of the
data sample.

In order to determine systematic uncertainties on the relative
acceptance values that accounted for candidates that were triggered in
ways not treated by the dimuon trigger simulation, the relative
topology dependence of the unsimulated triggers was investigated by
examining some representative yield ratios.  Three different
topological scenarios with trigger implications were identified as (a)
mass-difference effects in ratios involving the dimuon decays of both
$J/\psi$ and $\psi(2S)$ daughters, (b) polarization effects in ratios
involving both $K^*(892)^0$ and $K^+$ daughters, and (c) effects in
ratios involving both of the aforementioned phenomena.  Of the six
decay topologies in the analysis, the three with the most populous
statistics ($B^+\to J/\psi\,K^+$, $B^0\to J/\psi\,K^*(892)^0$, and
$B^+\to\psi(2S)\,K^+$ (dimuonic)) were used to form three yield ratios
representing the three topological scenarios described above.  The
relative yields for these ratios were examined for the full data
sample, the subsample that passed the dimuon trigger simulation, and
the subsample that failed the trigger simulation.  Within the
available statistics, the yield ratios remained consistent for all
three trigger samples.  Comparison of the yield ratios between the
full sample and the subsample that passed the dimuon trigger
simulation suggested a 6\% systematic uncertainty for ratios with a
dimuon $\psi(2S) / J/\psi$ mass difference, a 2\% uncertainty for
ratios with a $K^*(892)^0 / K^+$ polarization difference, and a 7\%
uncertainty for ratios with both differences.  The application of
these results to the acceptance ratios is summarized in
Table~\ref{tab:trig_syst}.

\begin{table}
\begin{center}
\begin{tabular}{|l|ccc|}        \hline
\multicolumn{1}{|c|}{Trigger} &
                        \multicolumn{1}{|c|}{$B^+\to J/\psi\,K^+$} &
                        \multicolumn{1}{|c|}{$B^0\to J/\psi\,K^*(892)^0$} &
                        \multicolumn{1}{|c|}{$B^+\to \psi(2S)\,K^+$} \\
\multicolumn{1}{|c|}{Systematics (\%)}&
                        \multicolumn{1}{|c|}{$J/\psi\to\mu^+\,\mu^-$} &
                        \multicolumn{1}{|c|}{$J/\psi\to\mu^+\,\mu^-$} &
                        \multicolumn{1}{|c|}{$\psi(2S)\to\mu^+\,\mu^-$}
                                \\ \hline\hline
$B^0\to J/\psi\,K^*(892)^0$ &
        \raisebox{-0.10in}[0.0in][0.0in]{$2.0$}& & \\
$J/\psi\to\mu^+\,\mu^-$     & & & \\ \cline{1-1}

$B^+\to \psi(2S)\,K^+$      &
        \raisebox{-0.10in}[0.0in][0.0in]{$6.0$} &
        \raisebox{-0.10in}[0.0in][0.0in]{$7.0$}& \\
$\psi(2S)\to\mu^+\,\mu^-$   &  & & \\ \cline{1-1}

$B^+\to \psi(2S)\,K^+$      & 
        \raisebox{-0.10in}[0.0in][0.0in]{$2.0$}&
        \raisebox{-0.10in}[0.0in][0.0in]{$2.0$}&
        \raisebox{-0.10in}[0.0in][0.0in]{$6.0$}\\
$\psi(2S)\to J/\psi\,\pi^+\,\pi^-$ &  & & \\ \cline{1-1}

$B^0\to \psi(2S)\,K^*(892)^0$ &
        \raisebox{-0.10in}[0.0in][0.0in]{$7.0$}&
        \raisebox{-0.10in}[0.0in][0.0in]{$6.0$}&
        \raisebox{-0.10in}[0.0in][0.0in]{$2.0$}\\
$\psi(2S)\to\mu^+\,\mu^-$   &  & & \\ \cline{1-1}

$B^0\to \psi(2S)\,K^*(892)^0$ &
        \raisebox{-0.10in}[0.0in][0.0in]{$2.0$}&
        \raisebox{-0.10in}[0.0in][0.0in]{$2.0$}&
        \raisebox{-0.10in}[0.0in][0.0in]{$7.0$}\\
$\psi(2S)\to J/\psi\,\pi^+\,\pi^-$ &  & & \\ \hline
\end{tabular}
\end{center}
\caption
[Systematic uncertainty on acceptance due to trigger requirement.]
{A summary of the systematic uncertainties (expressed in \%) on the
relative geometric and kinematic acceptance due to the effects of the
dimuon triggers on the relative acceptance ratios.}
\label{tab:trig_syst}
\end{table}

It is useful to investigate the sensitivity of the relative acceptance
measurements to uncertainties in the Level~1 and Level~2 CMU and CMX
trigger efficiency parameterizations.  Table~\ref{tab:trig_param_syst}
details the effects of the one-standard-deviation parameterization
shifts, pictured in Figures~\ref{fig:l1prbux} and \ref{fig:l2prbux},
on the relative geometric and kinematic acceptance values.  A
comparison of Tables~\ref{tab:rel_geom_eff} and
\ref{tab:trig_param_syst} indicates that, for each ratio of decay-mode
acceptances, the effect of varying the trigger efficiency
parameterizations is insignificant when compared with the accuracy to
which each relative geometric and kinematic acceptance ratio was
determined.  These variations were therefore not used in the
calculation of the systematic uncertainties listed in
Table~\ref{tab:trig_syst}.

\begin{table}
\begin{center}
\begin{tabular}{|l|ccc|}        \hline
\multicolumn{1}{|c|}{$+1\sigma$ Trigger Shift} &
			\multicolumn{1}{|c|}{$B^+\to J/\psi\,K^+$} &
			\multicolumn{1}{|c|}{$B^0\to J/\psi\,K^*(892)^0$} &
			\multicolumn{1}{|c|}{$B^+\to \psi(2S)\,K^+$} \\
\multicolumn{1}{|c|}{$-1\sigma$ Trigger Shift}&
			\multicolumn{1}{|c|}{$J/\psi\to\mu^+\,\mu^-$} &
			\multicolumn{1}{|c|}{$J/\psi\to\mu^+\,\mu^-$} &
			\multicolumn{1}{|c|}{$\psi(2S)\to\mu^+\,\mu^-$}
				\\ \hline\hline
$B^0\to J/\psi\,K^*(892)^0$ &
	$0.403$& & \\
$J/\psi\to\mu^+\,\mu^-$     & $0.409$ & & \\ \hline

$B^+\to \psi(2S)\,K^+$      &
	$1.139$ &
	$2.814$& \\
$\psi(2S)\to\mu^+\,\mu^-$   & $1.138$ & $2.782$ & \\ \hline

$B^+\to \psi(2S)\,K^+$      & 
	$0.326$&
	$0.808$&
	$0.287$\\
$\psi(2S)\to J/\psi\,\pi^+\,\pi^-$ &$0.330$ & $0.807$ & $0.290$ \\ \hline

$B^0\to \psi(2S)\,K^*(892)^0$ &
	$0.428$&
	$1.061$&
	$0.376$\\
$\psi(2S)\to\mu^+\,\mu^-$   & $0.435$ & $1.064$ & $0.382$ \\ \hline

$B^0\to \psi(2S)\,K^*(892)^0$ &
	$0.199$&
	$0.493$&
	$0.175$\\
$\psi(2S)\to J/\psi\,\pi^+\,\pi^-$ & $0.202$  & $0.495$&$0.178$ \\ \hline
\end{tabular}
\end{center}
\caption
[Summary of trigger parameterization shift effects on relative acceptance.]
{A summary of the effects of $+1\sigma$ (upper number) and $-1\sigma$
(lower number) trigger efficiency parameterization shifts on the
relative geometric and kinematic acceptance quotients.}
\label{tab:trig_param_syst}
\end{table}

\subsection{Helicity Distributions}

The longitudinal polarization fractions inherent to the $B$-meson
decays with vector-vector final states can have implications for the
relative acceptance measurements.  The systematic uncertainties
associated with the longitudinal polarization fractions were estimated
by varying the nominal measured world-average $\Gamma_L/\Gamma$ value
for the $B^0\to J/\psi\,K^*(892)^0$ mode (refer to
Reference~\cite{browder:review} and Section~\ref{sect:mc_decay}) by
its standard deviation, $\pm 0.07$.

The value of $\Gamma_L/\Gamma$ has not been measured for the decay
$B^0\to\psi(2S)\,K^*(892)^0$; however, it is expected to be similar to
that for the $B^0\to J/\psi\,K^*(892)^0$ mode.  Two standard
deviations of the world-average $\Gamma_L/\Gamma$ value for the
$B^0\to J/\psi\,K^*(892)^0$ measurements, $\pm 0.14$, were therefore
used as the $\Gamma_L/\Gamma$ variation range for the
$B^0\to\psi(2S)\,K^*(892)^0$ modes.  This variation range was applied
to each of the $\psi(2S)\,K^*(892)^0$ reconstructions, {\it i.e.},
those involving $\psi(2S)\to\mu^+\,\mu^-$ and $\psi(2S)\to
J/\psi\,\pi^+\,\pi^-$ decays.

The consequent systematic uncertainties are summarized in
Table~\ref{tab:helicity_syst}.  The `absolute' systematic
uncertainties are used in cases where the vector-vector acceptance is
considered relative to that for one of the vector-pseudoscalar modes.

\begin{table}
\begin{center}
\begin{tabular}{|l|c|c|}        \hline
\multicolumn{1}{|c|}{$\Gamma_L/\Gamma$} &
			\multicolumn{1}{|c|}{$B^0\to J/\psi\,K^*(892)^0$} &
			\multicolumn{1}{|c|}{Absolute} \\
\multicolumn{1}{|c|}{Systematics (\%)}&
			\multicolumn{1}{|c|}{$J/\psi\to\mu^+\,\mu^-$} &
			\multicolumn{1}{|c|}{Acceptance}
				\\ \hline\hline
$B^0\to J/\psi\,K^*(892)^0$ &
	& \raisebox{-0.10in}[0.0in][0.0in]{$2.0$} \\
$J/\psi\to\mu^+\,\mu^-$     & & \\ \cline{1-1}

$B^0\to \psi(2S)\,K^*(892)^0$ &
	\raisebox{-0.10in}[0.0in][0.0in]{$4.2$}&
	\raisebox{-0.10in}[0.0in][0.0in]{$3.3$}\\
$\psi(2S)\to\mu^+\,\mu^-$   & & \\ \cline{1-1}

$B^0\to \psi(2S)\,K^*(892)^0$ &
	\raisebox{-0.10in}[0.0in][0.0in]{$2.2$}&
	\raisebox{-0.10in}[0.0in][0.0in]{$2.6$}\\
$\psi(2S)\to J/\psi\,\pi^+\,\pi^-$ & & \\ \hline
\end{tabular}
\end{center}
\caption
[Systematic uncertainty on acceptance due to variations in the
longitudinal polarization fractions of those decays with vector-vector
final states.]
{A summary of the systematic uncertainties (expressed in \%) on the
relative and absolute geometric and kinematic acceptance due to
variations in the longitudinal polarization fractions
($\Gamma_L/\Gamma$) of those decays with vector-vector final states.
The column on the right lists the systematic uncertainties on the
relevant absolute geometric and kinematic acceptance values.}
\label{tab:helicity_syst}
\end{table}

%% file: effic.tex
\label{chapt:effic}

The topological and candidate-selection similarities amongst the six
decay channels in this study afforded the cancelation of several
common efficiencies and systematic uncertainties in the
branching-fraction ratios.  Only those efficiencies that were unique
to one of the decay modes in a given ratio or that differed between
the channels required careful study.  This chapter details the
determination of event-yield corrections and their systematic
uncertainties due to inefficiencies relevant to the detection and
reconstruction of the candidate $B$-meson decay modes.

\section{Constrained Fit Confidence Level Criteria}
\label{sect:cl_efficiencies}

The vertex and vertex-plus-mass constrained fit confidence level
criteria imposed on the charmonium and $B$-meson event selection were
described in Section~\ref{sect:charm_reconstruct}.  In the present
section, several different techniques are used to study the effects of
the $CL(\chi^2)$ criteria on the yield efficiencies.

\subsection{Dimuon $CL(\chi^2)$ Criteria}
\label{sect:dimuon_cl_efficiencies}

In a branching-fraction ratio expression such as the one given in
Equation~\ref{eqn:ratio_br_fract}, most of the dimuon selection
criteria, including the $CL(\chi^2)$ requirements, are expected to
divide out of the calculation.  It is nevertheless useful to
investigate the effects of the dimuon $CL(\chi^2)$ criteria to test
this assumption, examine the background rejection, and test the
efficiency measurement techniques.

\begin{figure}[p]
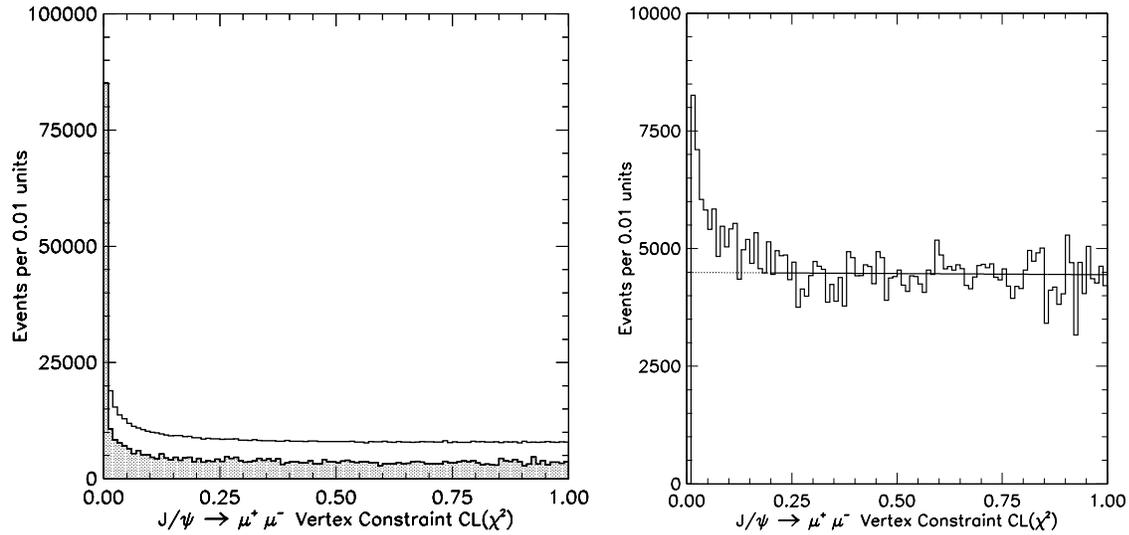

\begin{center}
\leavevmode
\hbox{%
\epsfysize=2.75in
\epsffile{diagrams/prob_jpsi_vtx_tot.epsi}
\hspace{0.05in}
\epsfysize=2.75in
\epsffile{diagrams/prob_jpsi_vtx_sig.epsi}}
\end{center}
\caption
[$CL(\chi^2)$ distribution for vertex-constrained $J/\psi$ dimuon fit.]
{The $CL(\chi^2)$ distribution of the vertex-constrained $J/\psi$
dimuon fit.  The unshaded histogram in the left plot contains the
combined signal and background events; the shaded histogram in the
left plot represents the like-sign dimuon ($\mu^\pm\,\mu^\pm$)
candidates, normalized to the background yield.  The right plot is the
difference of the two histograms in the left plot.}
\label{fig:prob_jpsi_vtx}
\end{figure}

\begin{figure}[p]
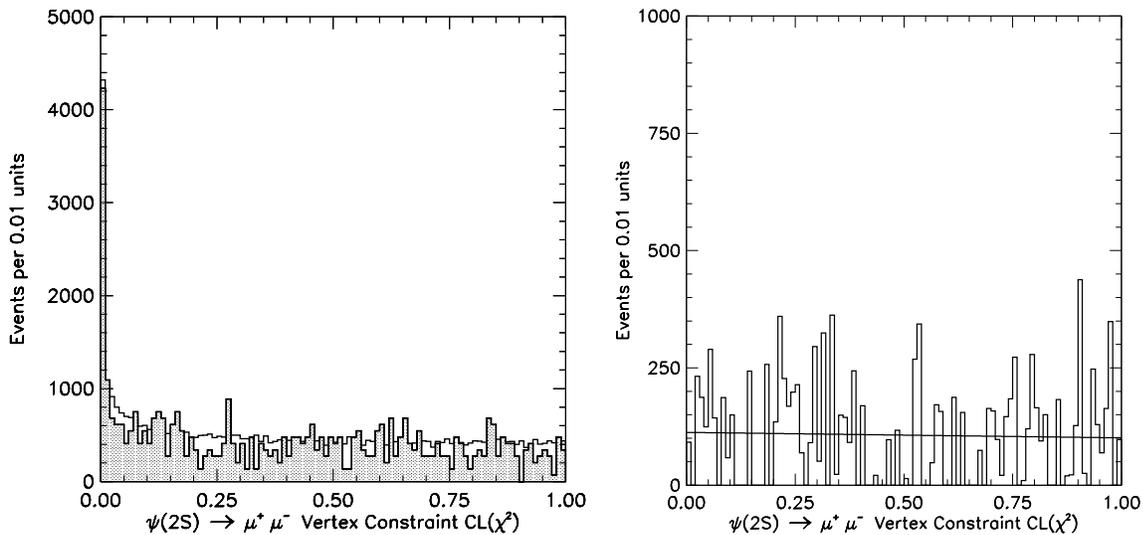

\begin{center}
\leavevmode
\hbox{%
\epsfysize=2.75in
\epsffile{diagrams/prob_mumu_vtx_tot.epsi}
\hspace{0.05in}
\epsfysize=2.75in
\epsffile{diagrams/prob_mumu_vtx_sig.epsi}}
\end{center}
\caption
[$CL(\chi^2)$ distribution for vertex-constrained $\psi(2S)$ dimuon fit.]
{The $CL(\chi^2)$ distribution of the vertex-constrained $\psi(2S)$
dimuon fit.  The unshaded histogram in the left plot contains the
combined signal and background events; the shaded histogram in the
left plot represents the like-sign dimuon ($\mu^\pm\,\mu^\pm$)
candidates, normalized to the background yield.  The right plot is the
difference of the two histograms in the left plot.}
\label{fig:prob_mumu_vtx}
\end{figure}

Figures~\ref{fig:prob_jpsi_vtx} and \ref{fig:prob_mumu_vtx} illustrate
the distribution of the $CL(\chi^2)$ variable for vertex constrained
fits of $J/\psi$ and $\psi(2S)$ dimuon candidates, respectively.  The
shaded histograms consist of like-sign ($\mu^\pm\,\mu^\pm$) dimuon
candidate combinations, normalized to the measured background yield.
Dimuon candidates that clearly did not originate from a common vertex
are expected to result in constrained fits with low $CL(\chi^2)$
values, as exemplified by the populous lower $CL(\chi^2)$ bins in the
left-hand plots of Figures~\ref{fig:prob_jpsi_vtx} and
\ref{fig:prob_mumu_vtx}.  The relative `flatness' of the $CL(\chi^2)$
distributions over most of the abscissa suggests that the covariance
matrices of the individual candidate track fits are realistic for most
vertexing situations.


Figures~\ref{fig:prob_jpsi_vtx_mss} and \ref{fig:prob_mumu_vtx_mss}
are analogous to Figures~\ref{fig:prob_jpsi_vtx} and
\ref{fig:prob_mumu_vtx}, with the difference that the former are due to fits
with simultaneous vertex and mass constraints.  Similar conclusions to
those for the vertex-constrained fits may be drawn for the
vertex-plus-mass constraint $CL(\chi^2)$ distributions.


\begin{figure}[p]
\begin{center}
\leavevmode
\hbox{%
\epsfysize=2.75in
\epsffile{diagrams/prob_jpsi_vtx_mss_tot.epsi}
\hspace{0.05in}
\epsfysize=2.75in
\epsffile{diagrams/prob_jpsi_vtx_mss_sig.epsi}}
\end{center}
\caption
[$CL(\chi^2)$ distribution for vertex-plus-mass-constrained $J/\psi$
dimuon fit.]
{The $CL(\chi^2)$ distribution of the vertex-plus-mass-constrained
$J/\psi$ dimuon fit.  The unshaded histogram in the left plot contains
the combined signal and background events; the shaded histogram in the
left plot represents the like-sign dimuon ($\mu^\pm\,\mu^\pm$)
candidates normalized to the background yield.  The right plot is the
difference of the two histograms in the left plot.}
\label{fig:prob_jpsi_vtx_mss}
\end{figure}

\begin{figure}[p]
\begin{center}
\leavevmode
\hbox{%
\epsfysize=2.75in
\epsffile{diagrams/prob_mumu_vtx_mss_tot.epsi}
\hspace{0.05in}
\epsfysize=2.75in
\epsffile{diagrams/prob_mumu_vtx_mss_sig.epsi}}
\end{center}
\caption
[$CL(\chi^2)$ distribution for vertex-plus-mass-constrained $\psi(2S)$
dimuon fit.]
{The $CL(\chi^2)$ distribution of the vertex-plus-mass-constrained
$\psi(2S)$ dimuon fit.  The unshaded histogram in the left plot
contains the combined signal and background events; the shaded
histogram in the left plot represents the like-sign dimuon
($\mu^\pm\,\mu^\pm$) candidates normalized to the background yield.
The right plot is the difference of the two histograms in the left
plot.}
\label{fig:prob_mumu_vtx_mss}
\end{figure}

The $CL(\chi^2) > 0.01$ criterion efficiency for vertex-constrained
$J/\psi$ candidate dimuon fits, $\epsilon^{\rm v}_{J/\psi \to
\mu\mu}$, may be calculated directly by examining the effect of the
cut on the $J/\psi$ candidate yield.  Figure~\ref{fig:jpsi_vtx}
depicts the $J/\psi$ invariant mass distribution before and after the
application of a $CL(\chi^2) > 0.01$ requirement.  Similarly,
Figure~\ref{fig:mumu_vtx} illustrates the analogous $\psi(2S)$
invariant mass distributions used in the calculation of $\epsilon^{\rm
v}_{\psi(2S) \to \mu\mu}$.

\begin{figure}[p]
\begin{center}
\leavevmode
\hbox{%
\epsfysize=2.45in
\epsffile{diagrams/jpsi_vtx_before.epsi}
\hspace{0.05in}
\epsfysize=2.45in
\epsffile{diagrams/jpsi_vtx_after.epsi}}
\end{center}
\caption
[$J/\psi$ dimuon invariant mass distribution and the effect of a
vertex-constraint criterion.]
{The $J/\psi$ dimuon candidate invariant mass distribution before (left)
and after (right) a vertex-constraint criterion of $CL(\chi^2) > 0.01$
was applied.  The fit is to a double Gaussian signal with linear amplitudes
and a linear background term.}
\label{fig:jpsi_vtx}
\end{figure}

\begin{figure}[p]
\begin{center}
\leavevmode
\hbox{%
\epsfysize=2.8in
\epsffile{diagrams/mumu_vtx_before.epsi}
\hspace{0.05in}
\epsfysize=2.8in
\epsffile{diagrams/mumu_vtx_after.epsi}}
\end{center}
\caption
[$\psi(2S)$ dimuon invariant mass distribution and the effect of a
vertex-constraint criterion.]
{The $\psi(2S)$ dimuon candidate invariant mass distribution before
(left) and after (right) a vertex-constraint criterion of $CL(\chi^2)
> 0.01$ was applied.  The fit is to a Gaussian signal and a linear
background term.  A requirement of $p_{\rm T} > 2.5$~GeV/$c$ was
imposed on each muon candidate to control the background.}
\label{fig:mumu_vtx}
\end{figure}

\begin{figure}[p]
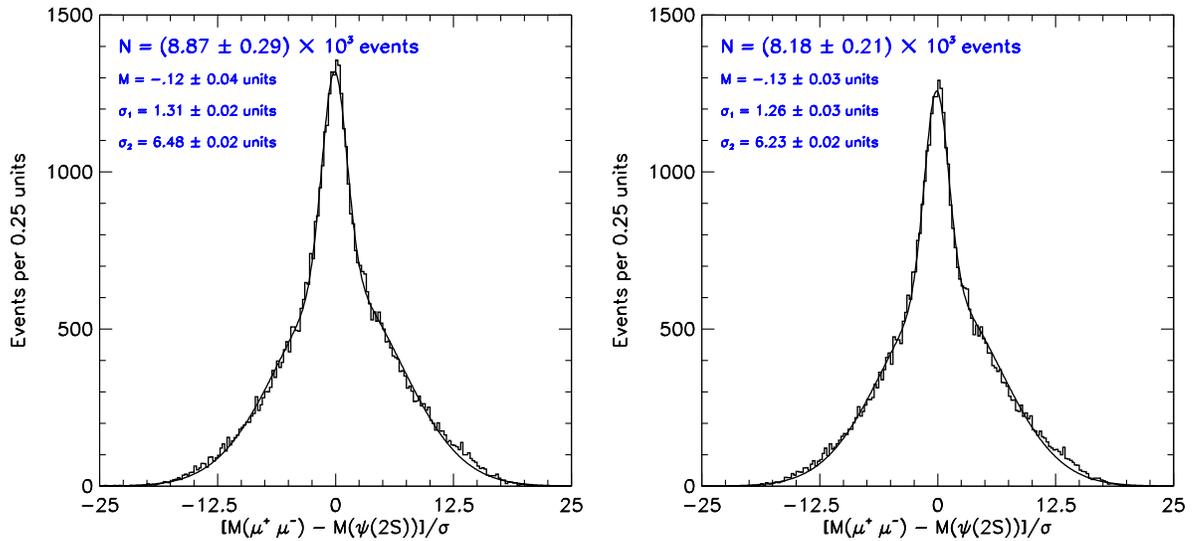

\begin{center}
\leavevmode
\hbox{%
\epsfxsize=3.0in
\epsffile{diagrams/mumu_vtx_nrml_before.epsi}
\hspace{0.05in}
\epsfxsize=3.0in
\epsffile{diagrams/mumu_vtx_nrml_after.epsi}}
\end{center}
\caption
[$\psi(2S)$ dimuon normalized invariant mass distribution and the effect of a
vertex-constraint criterion.]
{The $\psi(2S)$ dimuon normalized candidate invariant mass
distribution before (left) and after (right) a vertex-constraint
criterion of $CL(\chi^2) > 0.01$ was applied.  The fit is to a
Gaussian signal and a Gaussian background term.  A requirement of
$p_{\rm T} > 2.5$~GeV/$c$ was imposed on each muon candidate to
control the background.}
\label{fig:mumu_vtx_nrml}
\end{figure}

An alternate method to measure the charmonium event yields is
available by constructing normalized invariant mass distributions.
Figures~\ref{fig:jpsi_vtx_nrml} and \ref{fig:mumu_vtx_nrml} depict the
normalized invariant mass distributions, or `pulls', before and after
the $CL(\chi^2)$ criterion was imposed, for the $J/\psi$ and
$\psi(2S)$ dimuon candidate distributions, respectively.  Note that
the widths of the Gaussian signals exceed unity in both cases,
indicating some underestimation of the track helix uncertainties.  The
distributions in Figures~\ref{fig:jpsi_vtx_nrml} and
\ref{fig:mumu_vtx_nrml} may also be used to estimate $\epsilon^{\rm
v}_{J/\psi \to \mu\mu}$ and $\epsilon^{\rm v}_{\psi(2S) \to \mu\mu}$.

The extra inefficiency arising from the additional constraint to the
appropriate world average charmonium mass, $\epsilon^{\rm m}$, may be
estimated from the normalized invariant mass distributions.  By
relating the peak area obtained with a double-Gaussian fit in a
$\pm2.58$ standard deviation window about the mean of the normalized
invariant mass signal (with the $CL(\chi^2) > 0.01$ criterion on the
vertex-only constrained fit already applied) to the entire signal fit
area, the efficiency of an additional mass-constraint $CL(\chi^2) >
0.01$ requirement can be computed.  The right-hand plots in
Figures~\ref{fig:jpsi_vtx_nrml} and \ref{fig:mumu_vtx_nrml} were used
in this way to measure $\epsilon^{\rm m}_{J/\psi \to \mu\mu}$ and
$\epsilon^{\rm m}_{\psi(2S)
\to \mu\mu}$, respectively.

\begin{figure}[p]
\begin{center}
\leavevmode
\hbox{%
\epsfxsize=3.0in
\epsffile{diagrams/jpsi_vtx_nrml_before.epsi}
\hspace{0.05in}
\epsfxsize=3.0in
\epsffile{diagrams/jpsi_vtx_nrml_after.epsi}}
\end{center}
\caption
[$J/\psi$ dimuon normalized invariant mass distribution and the effect of a
vertex-constraint criterion.]
{The $J/\psi$ dimuon normalized candidate invariant mass distribution
before (left) and after (right) a vertex-constraint criterion of
$CL(\chi^2) > 0.01$ was applied.  The fit is to a double Gaussian
signal with linear amplitudes and a Gaussian background term.}
\label{fig:jpsi_vtx_nrml}
\end{figure}

Table~\ref{tab:dimuon_cl_eff} presents a summary of the various
$CL(\chi^2)$ criterion efficiencies measured in this section.  As
expected, the various measurements, using either $J/\psi$ or
$\psi(2S)$ dimuon reconstructions, are largely mutually consistent,
thereby supporting the assumption that dimuon $CL(\chi^2)$ selection
effects divide out of the branching-fraction ratio calculations.  The
average vertex and mass constraint requirements were measured to be
$\epsilon^{\rm v}_{\mu\mu} = 0.967 \pm 0.003$ and $\epsilon^{\rm
m}_{\mu\mu} = 0.963 \pm 0.002$, respectively.

\begin{table}
\begin{center}
\begin{tabular}{|c|c|c|}        \hline
\rule[-3mm]{0mm}{8mm}
Efficiency & Technique & Measurement \\ \hline \hline
\rule{0mm}{5mm}
$\epsilon^{\rm v}_{J/\psi \to \mu\mu}$ &
	mass & $0.966 \pm 0.003$ \\
\rule{0mm}{5mm}
$\epsilon^{\rm v}_{J/\psi \to \mu\mu}$ &
	normalized mass & $0.969 \pm 0.004$ \\
\rule{0mm}{5mm}
$\epsilon^{\rm v}_{\psi(2S) \to \mu\mu}$ &
	mass & $0.956 \pm 0.036$ \\
\rule[-3mm]{0mm}{8mm}
$\epsilon^{\rm v}_{\psi(2S) \to \mu\mu}$ &
	normalized mass & $0.922 \pm 0.039$ \\ \hline
\rule[-3mm]{0mm}{8mm}
$\epsilon^{\rm v}_{\mu\mu}$ &
	weighted mean & $0.967 \pm 0.003$ \\ \hline \hline
\rule{0mm}{5mm}
$\epsilon^{\rm m}_{J/\psi \to \mu\mu}$ &
	normalized mass & $0.963 \pm 0.002$ \\
\rule[-3mm]{0mm}{8mm}
$\epsilon^{\rm m}_{\psi(2S) \to \mu\mu}$ &
	normalized mass & $0.958 \pm 0.035$ \\ \hline
\rule[-3mm]{0mm}{8mm}
$\epsilon^{\rm m}_{\mu\mu}$ &
	weighted mean & $0.963 \pm 0.002$ \\ \hline
\end{tabular}
\end{center}
\caption
[Dimuon $CL(\chi^2)$ criterion efficiencies.]
{Dimuon vertex and vertex-plus-mass $CL(\chi^2) > 0.01$ criterion
efficiency summary.  The quantities $\epsilon^{\rm v}_{\mu\mu}$ and
$\epsilon^{\rm m}_{\mu\mu}$ were computed by performing a weighted
mean of the measured values.}
\label{tab:dimuon_cl_eff}
\end{table}

\subsection{Dipion $CL(\chi^2)$ Criteria}
\label{sect:dipion_cl_eff}

Unlike in the dimuon cases, charmonium reconstructions performed via
the decay $\psi(2S) \to J/\psi\,\pi^+\,\pi^-$ did not necessarily have
$CL(\chi^2)$ requirement efficiencies that divided out of the
branching-fraction ratios.  The vertex and vertex-plus-mass
$CL(\chi^2)$ cut efficiencies for this hadronic-cascade charmonium
mode were taken to be factorizable according to the expressions
$\epsilon^{\rm v}_{\psi(2S)\to J/\psi\,\pi\,\pi} =
\epsilon^{\rm v}_{\mu\mu} \cdot \epsilon^{\rm v}_{\pi\pi}$ and
$\epsilon^{\rm m}_{\psi(2S)\to J/\psi\,\pi\,\pi} =
\epsilon^{\rm m}_{\mu\mu} \cdot \epsilon^{\rm m}_{\pi\pi}$, respectively.
The fact that the $J/\psi$ dimuon candidate was first vertex-plus-mass
constrained separately, before the $\pi^+\,\pi^-$ candidates were
included in the fit, permitted this factorization of $CL(\chi^2)$ cut
efficiencies.  The principal advantage of factorizing the efficiency
in this way is that the dimuon efficiency factors, $\epsilon^{\rm
v}_{\mu\mu}$ and $\epsilon^{\rm m}_{\mu\mu}$, can still divide to
unity in the branching-fraction ratio expressions.

\begin{figure}
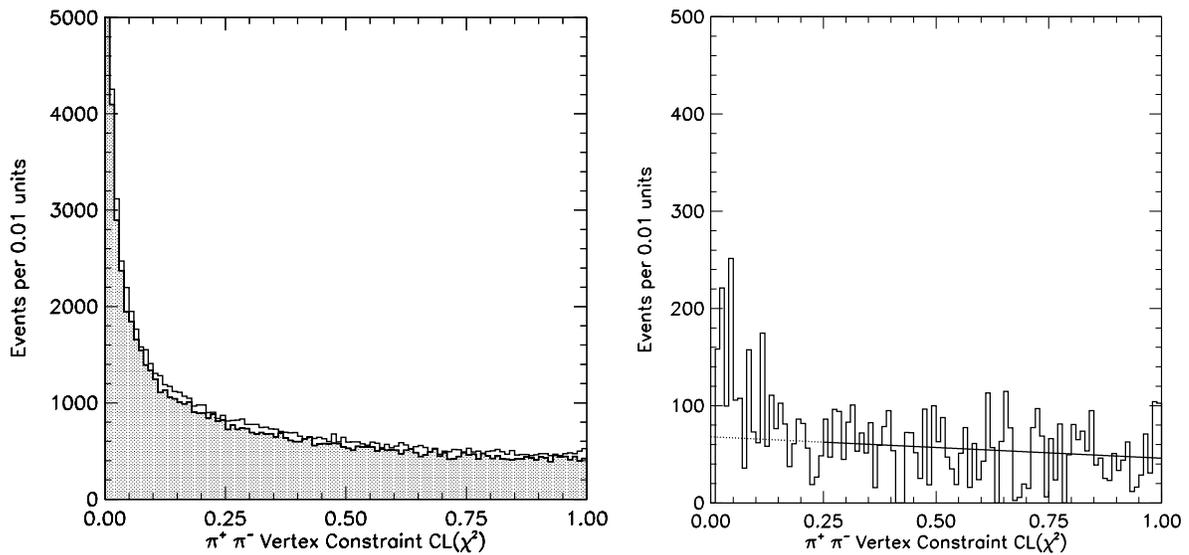

\begin{center}
\leavevmode
\hbox{%
\epsfysize=2.85in
\epsffile{diagrams/prob_pipi_vtx_tot.epsi}
\hspace{0.05in}
\epsfysize=2.85in
\epsffile{diagrams/prob_pipi_vtx_sig.epsi}}
\end{center}
\caption
[$CL(\chi^2)$ distribution for vertex-constrained $\pi^+\,\pi^-$ fit.]
{The $CL(\chi^2)$ distribution of the vertex-constrained $\pi^+\,\pi^-$
fit.  The unshaded histogram in the left plot contains the
combined signal and background events; the shaded histogram in the
left plot represents the like-sign dipion ($\pi^\pm\,\pi^\pm$)
candidates normalized to the background yield.  The right plot is the
difference of the two histograms in the left plot.}
\label{fig:prob_pipi_vtx}
\end{figure}

Figure~\ref{fig:prob_pipi_vtx} illustrates the distribution of the
$CL(\chi^2)$ variable for vertex-constrained fits of
$J/\psi\,\pi^+\,\pi^-$ candidates performed after vertex-plus-mass
constrained fits (and the corresponding $CL(\chi^2)$ criteria) were
applied to the $J/\psi$ dimuon candidates alone.  The shaded histogram
was created from like-sign-pion ($J/\psi\,\pi^\pm\,\pi^\pm$) candidate
combinations, normalized to the measured background yield.  The
features in Figure~\ref{fig:prob_pipi_vtx} are similar to those
observed in Figures~\ref{fig:prob_jpsi_vtx} and
\ref{fig:prob_mumu_vtx} in Section~\ref{sect:dimuon_cl_efficiencies}.

\begin{figure}
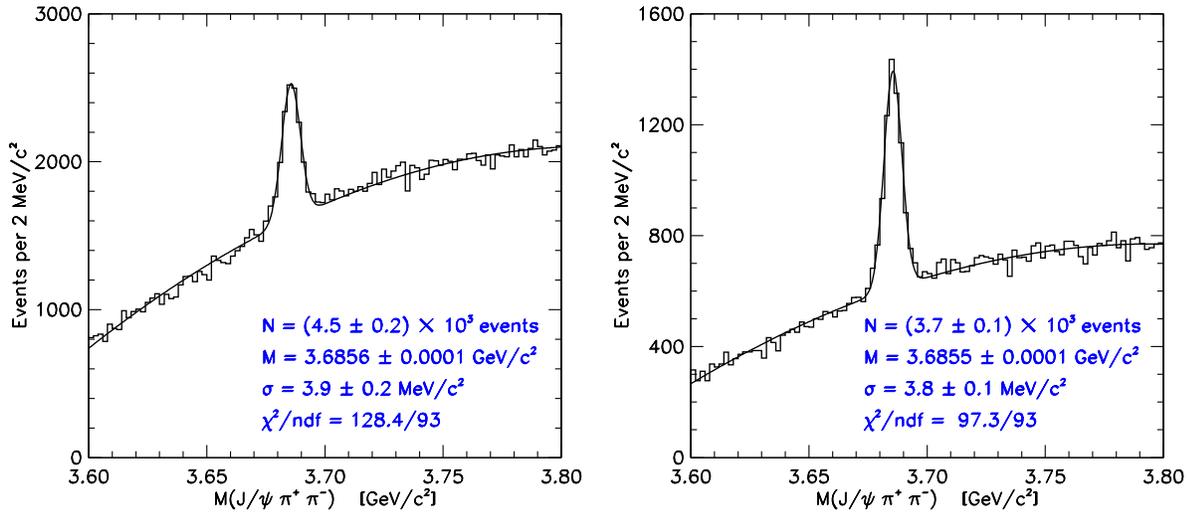

\begin{center}
\leavevmode
\hbox{%
\epsfxsize=3.0in
\epsffile{diagrams/pipi_vtx_before.epsi}
\hspace{0.05in}
\epsfxsize=3.0in
\epsffile{diagrams/pipi_vtx_after.epsi}}
\end{center}
\caption
[$J/\psi\,\pi^+\,\pi^-$ invariant mass distribution and the effect
of a $\pi^+\,\pi^-$ vertex-constraint criterion.]
{The $J/\psi\,\pi^+\,\pi^-$ candidate invariant mass distribution
before (left) and after (right) a vertex-constraint criterion of
$CL(\chi^2) > 0.01$ was applied.  The fit is to a Gaussian signal and
a cubic polynomial background term.}
\label{fig:pipi_vtx}
\end{figure}

Figures~\ref{fig:pipi_vtx} and \ref{fig:pipi_vtx_nrml} respectively
depict the $J/\psi\,\pi^+\,\pi^-$ invariant mass and normalized
invariant mass distributions before and after the additional
$CL(\chi^2) > 0.01$ criterion was applied to the $\pi^+\,\pi^-$
vertex-constrained fit quality.  The techniques used to measure the
$\epsilon^{\rm v}_{\pi\pi}$ and $\epsilon^{\rm m}_{\pi\pi}$
efficiencies were similar to those described in
Section~\ref{sect:dimuon_cl_efficiencies}.
Table~\ref{tab:dipion_cl_eff} summarizes the $\epsilon^{\rm
v}_{\pi\pi}$ and $\epsilon^{\rm m}_{\pi\pi}$ results.  The
uncertainties on the efficiencies were estimated by conservatively
combining in quadrature the yield uncertainties before and after the
$CL(\chi^2) > 0.01$ criterion was imposed.


\begin{figure}
\begin{center}
\leavevmode
\hbox{%
\epsfxsize=3.0in
\epsffile{diagrams/pipi_vtx_nrml_before.epsi}
\hspace{0.05in}
\epsfxsize=3.0in
\epsffile{diagrams/pipi_vtx_nrml_after.epsi}}
\end{center}
\caption
[$J/\psi\,\pi^+\,\pi^-$ normalized invariant mass distribution and the
effect of a $\pi^+\,\pi^-$ vertex-constraint criterion.]
{The $J/\psi\,\pi^+\,\pi^-$ normalized candidate invariant mass
distribution before (left) and after (right) a vertex-constraint
criterion of $CL(\chi^2) > 0.01$ was applied.  The fit is to a
Gaussian signal and a cubic polynomial background term.}
\label{fig:pipi_vtx_nrml}
\end{figure}

\begin{table}
\begin{center}
\begin{tabular}{|c|c|c|}        \hline
\rule[-3mm]{0mm}{8mm}
Efficiency & Technique & Measurement \\ \hline \hline
\rule{0mm}{5mm}
$\epsilon^{\rm v}_{\pi\pi}$ &
	mass & $0.816 \pm 0.039$ \\
\rule[-3mm]{0mm}{8mm}
$\epsilon^{\rm v}_{\pi\pi}$ &
	normalized mass & $0.858 \pm 0.045$ \\ \hline
\rule[-3mm]{0mm}{8mm}
$\epsilon^{\rm v}_{\pi\pi}$ &
	weighted mean & $0.834 \pm 0.039$ \\ \hline \hline
\rule[-3mm]{0mm}{8mm}
$\epsilon^{\rm m}_{\pi\pi}$ &
	normalized mass & $0.945 \pm 0.031$ \\ \hline
\end{tabular}
\end{center}
\caption
[Dipion $CL(\chi^2)$ criterion efficiencies.]
{Dipion vertex and vertex-plus-mass $CL(\chi^2) > 0.01$ criterion
efficiency summary.}
\label{tab:dipion_cl_eff}
\end{table}

\subsection{$B$-Meson $CL(\chi^2)$ Criteria}

The efficiency of the vertex and momentum-pointing $CL(\chi^2)>0.01$
criterion on the candidate $B$ decays was expected to be similar to
those calculated for the inclusive charmonium reconstructions.  In
practice, limited statistics made it difficult to examine the effects
of $CL(\chi^2)$ requirements on `global' vertex-constrained fits of
$B$-meson candidates.  To accommodate any possible noncancelation of
the global $B$ $CL(\chi^2)$ cut efficiencies for a given ratio of
branching fractions, a systematic uncertainty of $2.0\%$ was assigned
on each ratio.  This systematic uncertainty was estimated by taking
the standard deviation of the measured $\epsilon^{\rm v}$ and
$\epsilon^{\rm m}$ values in Table~\ref{tab:dimuon_cl_eff}.

\section{Tracking Efficiencies}
\label{sect:track_effic}

An advantage of measuring ratios of branching fractions was that the
tracking efficiencies largely divided out of the calculations.  Such
was the case for ratios involving the dimuonic charmonium modes and
like species of strange meson in both numerator and denominator.
Ratios of branching fractions involving the decay $\psi(2S)\to
J/\psi\,\pi^+\,\pi^-$ and unlike kaon species of the form $K^*(892)^0
/ K^+$, however, possessed residual dipion and single-pion tracking
efficiency factors, respectively.

Over the course of the entire Run 1 period, several known effects with
potential implications for tracking performance were noted: the
instantaneous luminosity varied by amounts in excess of an order of
magnitude (refer to Section~\ref{sect:data_sample}), the propensity
for multiple primary vertices in a given event increased, the Tevatron
beam optics were modified, the wire geometry in the CTC was altered by
gravity-induced sagging, and the electrostatic and gas-quality
characteristics of the CTC changed.  

Motivated not only by the requirements of the present
ratio-of-branching-fraction analysis, but also by the needs of other
CDF analyses that made use of low momentum tracking information, a
detailed study~\cite{warburton:trk_effic} of single- and double-track
CTC pattern recognition efficiencies was conducted.
Appendix~\ref{app:patt_rec_eff} summarizes the methodology and results
of the pattern recognition efficiency measurements.  The single-pion
pattern recognition efficiency was measured to be $0.928 \pm 0.020$
for $p_{\rm T} > 0.4$~GeV/$c$ and $r^{\rm exit}_{\rm CTC} > 110$~cm.
The double-pion efficiency was measured to be $0.881 \pm 0.043$, also
for $p_{\rm T} > 0.4$~GeV/$c$ and $r^{\rm exit}_{\rm CTC} > 110$~cm.

\section{Proper Decay Length Criteria}
\label{sect:ctau_cut_eff}

In this analysis, both the numerator and the denominator in each ratio
of branching fractions contained $B^+$ or $B^0$ meson candidates.  The
measured world-average proper decay lengths ($c\tau_B$) of $B^+$ and
$B^0$ mesons are ($492 \pm 18$)~$\mu$m and ($465 \pm
15$)~$\mu$m~\cite{pdg96}, respectively.  The fact that the accepted
lifetimes for the two $B$-meson flavours were mutually consistent lead
to the expectation that the $c\tau_B > 100$~$\mu$m criterion (refer to
Section~\ref{sect:bmeson_select}) efficiencies would be similar for
all the modes.  The world-average $B$ lifetimes, however, were short
enough that the efficiency of a $100$-$\mu$m proper decay length cut
was potentially sensitive to the resolution of the $c\tau_B$
measurement itself, which, in turn, was sensitive to the momentum and
vertex resolution of the final-state tracks.  Here the different decay
topologies could play a measurable r\^{o}le in the efficiency, owing
to the $c\tau_B$ resolution dependence on the multiplicity of
candidate tracks that made use of SVX (Section~\ref{sect:svx})
tracking information.

The sideband regions of the $B$-meson candidate invariant mass
distributions were expected to be populated primarily by prompt,
$i.e.$, low proper-decay-length, processes.  The regions $5.145 -
5.220$~GeV/$c^2$ and $5.340 - 5.415$~GeV/$c^2$, with the kinematic
selection criteria relaxed to enhance the statistics, were therefore
used to examine the $c\tau_B$ resolution as a function of the number
of daughter-candidate tracks with SVX hit information used in their
fits.  The results for the $B^+$ and $B^0$ decay modes are given in
Tables~\ref{tab:ctau_res_k} and \ref{tab:ctau_res_kst}, respectively.

\begin{table}
\begin{center}
\begin{tabular}{|c|c|c|c|}        \hline
Number of&$B^+\to J/\psi\,K^+$&$B^+\to\psi(2S)\,K^+$&$B^+\to\psi(2S)\,K^+$ \\
tracks with& $J/\psi\to\mu^+\,\mu^-$ &$\psi(2S)\to\mu^+\,\mu^-$ &
		$\psi(2S)\to J/\psi\,\pi^+\,\pi^-$ \\
SVX hits   & & & $J/\psi\to\mu^+\,\mu^-$ \\ \hline\hline
\rule{0mm}{5mm}
0 &$575\pm 6$&$783\pm 49$&$520\pm 26$ \\
\rule{0mm}{5mm}
1 &$171\pm 4$&$181\pm 1$ &$189\pm 35$ \\
\rule{0mm}{5mm}
2 &$94\pm 1$ &$99\pm 3$  &$243\pm 122$ \\
\rule{0mm}{5mm}
3 &$57\pm 1$ &$52\pm 1$  &$55\pm 9$ \\
\rule{0mm}{5mm}
4 &          &           &$78\pm 10$ \\
\rule[-3mm]{0mm}{8mm}
5 &          &           &$49\pm 2$ \\ \hline
\end{tabular}
\end{center}
\caption
[Proper decay length resolutions for $B^+$ candidates.]
{Sideband $c\tau_B$ resolutions, measured in units of $\mu$m, for each $B^+$
decay mode and SVX track multiplicity.}
\label{tab:ctau_res_k}
\end{table}

\begin{table}
\begin{center}
\begin{tabular}{|c|c|c|c|}        \hline
Number of&$B^0\to J/\psi\,K^*(892)^0$&$B^0\to\psi(2S)\,K^*(892)^0$ &
		$B^0\to\psi(2S)\,K^*(892)^0$ \\
tracks with & $J/\psi\to\mu^+\,\mu^-$ & $\psi(2S)\to\mu^+\,\mu^-$ &
		$\psi(2S)\to J/\psi\,\pi^+\,\pi^-$ \\
SVX hits    & $K^*(892)^0\to K^+\,\pi^-$ & $K^*(892)^0\to K^+\,\pi^-$ &
		$J/\psi\to\mu^+\,\mu^-$ \\
 & & & $K^*(892)^0\to K^+\,\pi^-$ \\ \hline\hline
\rule{0mm}{5mm}
0 &$591\pm 3$   &$762\pm 22$    &$584\pm 13$ \\
\rule{0mm}{5mm}
1 &$202\pm 3$   &$210\pm 7$     &$291\pm 52$ \\
\rule{0mm}{5mm}
2 &$117\pm 1$   &$102\pm 3$     &$168\pm 21$ \\
\rule{0mm}{5mm}
3 &$68\pm 1$    &$68\pm 1$      &$74\pm 8$ \\
\rule{0mm}{5mm}
4 &$49\pm 1$    &$56\pm 1$      &$61\pm 4$ \\
\rule{0mm}{5mm}
5 &             &               &$59\pm 2$ \\
\rule[-3mm]{0mm}{8mm}
6 &             &               &$48\pm 1$ \\ \hline
\end{tabular}
\end{center}
\caption
[Proper decay length resolutions for $B^0$ candidates.]
{Sideband $c\tau_B$ resolutions, measured in units of $\mu$m, for each $B^0$
decay mode and SVX track multiplicity.}
\label{tab:ctau_res_kst}
\end{table}

The relative frequencies of occurrence of the different SVX
multiplicity possibilities were estimated for each decay mode with a
Monte Carlo calculation (described in Chapter~\ref{chapt:geom_kin}),
for which the detector simulation had been tuned on data to determine
the SVX hit efficiencies and hit association characteristics.  The
measured $c\tau_B$ resolutions in Tables~\ref{tab:ctau_res_k} and
\ref{tab:ctau_res_kst} were then convolved with $c\tau_B$ distributions
that used the world average $B^+$ and $B^0$ proper
lifetimes~\cite{pdg96} in their decay constants, respectively.  A toy
Monte Carlo programme was employed to perform $10^5$ convolutions for
each decay, and the resolution contributions from the various SVX
track multiplicities were administered according to weighting
probabilities determined by the relative frequencies calculated from
the Monte Carlo samples.  The procedure was then repeated for the
world-average $c\tau_B$ values shifted high and low by their standard
deviations ($\sigma$)~\cite{pdg96}.  The results of these shifts,
along with the central values, are listed in
Table~\ref{tab:ctau_cut_eff}.  The systematic uncertainties on the
efficiency values correspond to the maximum efficiency deviations
observed in the $-1\sigma$ and $+1\sigma$ studies.  As expected, the
measured $c\tau_B$ efficiencies were similar, with all values near
0.75.  The small efficiency differences amongst the various channels
were consistent with differences in the daughter-track multiplicities.

\begin{table}
\begin{center}
\begin{tabular}{|l|l|c|c|c|c|}        \hline
\multicolumn{1}{|c|}{$B$ Mode} & \multicolumn{1}{c|}{$c\bar{c}$ Mode} &
	Central & $-1\sigma$ & $+1\sigma$ & Efficiency \\ \hline\hline
\rule{0mm}{5mm}
$B^+\to J/\psi\,K^+$ & $J/\psi\to\mu^+\,\mu^-$ & .753 & .741 &
	.764 & $.753\pm .012$ \\
\rule{0mm}{5mm}
$B^+\to\psi(2S)\,K^+$ & $\psi(2S)\to\mu^+\,\mu^-$ &
        .738 & .726 & .748 & $.738\pm .013$ \\
\rule[-3mm]{0mm}{8mm}
$B^+\to\psi(2S)\,K^+$ & $\psi(2S)\to J/\psi\,\pi^+\,\pi^-$ &
        .762 & .750 & .774 & $.762\pm .012$ \\ \hline\hline
\rule{0mm}{5mm}
$B^0\to J/\psi\,K^*(892)^0$ & $J/\psi\to\mu^+\,\mu^-$ & .755 & .742 &
	.766 & $.755\pm .013$ \\
\rule{0mm}{5mm}
$B^0\to\psi(2S)\,K^*(892)^0$ & $\psi(2S)\to\mu^+\,\mu^-$ &
        .749 & .736 & .760 & $.749\pm .013$ \\
\rule[-3mm]{0mm}{8mm}
$B^0\to\psi(2S)\,K^*(892)^0$ & $\psi(2S)\to J/\psi\,\pi^+\,\pi^-$ &
        .761 & .748 & .773 & $.761\pm .013$ \\ \hline
\end{tabular}
\end{center}
\caption
[Measured $c\tau_B$ cut efficiencies in the $B^+$ and $B^0$ decay modes.]
{Measured $c\tau_B$ cut efficiencies in the $B^+$ and $B^0$ decay
modes, showing the effects of variations of the world-average $B$ lifetimes
by their standard deviations.}
\label{tab:ctau_cut_eff}
\end{table}

\section{$B$-Candidate Isolation Criterion}

The $B$-meson candidate isolation criterion, $I_B = {\displaystyle
\frac{\sum\limits_{i \in\!\!\!/ B}^{R} \vec{p}_i \cdot
\hat{p}_B} {\left| \vec{p}_B \right|} < \frac{7}{13}}$,
was defined and described in Section~\ref{sect:bmeson_select}.  This
requirement rejected significant contributions of background to the
$B$-candidate signals, namely backgrounds arising from invariant mass
combinations of the dimuon candidates with non-$B$ tracks from the
underlying event (which is represented by $X$ in
Equation~\ref{eqn:b_bbar_hadroproduction}).  The large
$\eta$-$\varphi$ cone radius and the stipulation that the non-$B$
tracks that contribute to the summation in the isolation expression be
consistent with having originated from the same primary vertex as the
$B$ candidate suggested that the isolation criterion would be
relatively efficient, as exemplified in Figure~\ref{fig:isolation}.
Investigations of $B$-meson isolation requirements indicate that the
efficiencies of these criteria are largely independent of the type of
$B$ decay mode and the $p_{\rm T}$ of the $B$
candidate~\cite{mit:isolation}.

As a consistency test of the observation~\cite{mit:isolation} that the
$B$-candidate isolation criterion was independent of decay topology
(therefore obviating a need for individual efficiencies in the
branching-fraction ratio calculations), the effective efficiency of
the isolation requirement was examined for each of the six $B$-meson
reconstructions investigated in this study.
Table~\ref{tab:pbfrac_eff} lists the efficiencies, which were measured
by comparing the appropriate candidate yields after and before the
isolation criterion was imposed.  The combined efficiency of the
isolation requirement was found to be $0.928 \pm 0.054$ and the
efficiencies for the individual topologies were observed to be
mutually consistent.

\begin{table}
\begin{center}
\begin{tabular}{|l|l|c|}        \hline
\multicolumn{1}{|c|}{$B$ Mode} & \multicolumn{1}{c|}{$c\bar{c}$ Mode} &
	Efficiency \\ \hline\hline
\rule{0mm}{5mm}
$B^+\to J/\psi\,K^+$ & $J/\psi\to\mu^+\,\mu^-$ & $0.900\pm 0.063$ \\
\rule{0mm}{5mm}
$B^+\to\psi(2S)\,K^+$ & $\psi(2S)\to\mu^+\,\mu^-$ &
	$1.000 {+0.000 \atop -0.300}$ \\
\rule[-4mm]{0mm}{9mm}
$B^+\to\psi(2S)\,K^+$ & $\psi(2S)\to J/\psi\,\pi^+\,\pi^-$ &
	$1.000 {+0.000 \atop -0.294}$ \\ \hline\hline
\rule{0mm}{5mm}
$B^0\to J/\psi\,K^*(892)^0$ & $J/\psi\to\mu^+\,\mu^-$ &
	$0.937 {+0.063 \atop -0.092}$ \\
\rule{0mm}{5mm}
$B^0\to\psi(2S)\,K^*(892)^0$ & $\psi(2S)\to\mu^+\,\mu^-$ &
	n/a \\
\rule[-4mm]{0mm}{9mm}
$B^0\to\psi(2S)\,K^*(892)^0$ & $\psi(2S)\to J/\psi\,\pi^+\,\pi^-$ &
        $0.907 {+0.093 \atop -0.268}$ \\ \hline\hline
\multicolumn{2}{|c|}{Combined-Channel Measurement} & $0.928 \pm 0.054$ \\
	\hline
\end{tabular}
\end{center}
\caption
[Measured isolation criterion efficiencies in the $B^+$ and $B^0$ decay modes.]
{Measured isolation criterion efficiencies in the $B^+$ and $B^0$
decay modes.  The domination of background in the
$B^0\to\psi(2S)\,K^*(892)^0$, $\psi(2S)\to\mu^+\,\mu^-$ decay-chain
reconstruction prevented an individual measurement; however, the
invariant mass distributions for all six modes were used to
compute the combined efficiency in the bottom row.}
\label{tab:pbfrac_eff}
\end{table}


\section{Daughter Branching Fractions} 

Depending on the ratio of $B$-meson branching fractions being
calculated in this study, different combinations of daughter branching
fractions did not divide out of the calculation.  These quantities are
listed in Table~\ref{tab:daught_br_frac}.  The charmonium branching
fractions were taken to be the most recent accepted world-average
values~\cite{pdg96}; however, the ${\cal B}(J/\psi\to\mu^+\,\mu^-)$
average did not take into account the measurement reported in
Reference~\cite{bai:jpsi_br_frac}.  For the value of ${\cal
B}(\psi(2S)\to\mu^+\,\mu^-)$, the world average for ${\cal
B}(\psi(2S)\to e^+\, e^-)$, which has a significantly smaller
uncertainty, was used under the assumption of lepton universality.
The ${\cal B}(K^*(892)^0\to K^+\,\pi^-)$ value is based on isospin
symmetry.  The uncertainties on the world-average charmonium branching
fractions constitute the largest contributions to the total systematic
uncertainties in the present analysis.

\begin{table}
\begin{center}
\begin{tabular}{|l|c|}        \hline
\multicolumn{1}{|c|}{Decay Mode} &
		\multicolumn{1}{c|}{Branching Fraction} \\ \hline\hline
\rule{0mm}{5mm}
$J/\psi\to\mu^+\,\mu^-$   & $(6.01 \pm 0.19) \times 10^{-2}$ \\
\rule{0mm}{5mm}
$\psi(2S)\to\mu^+\,\mu^-$ & $(8.5  \pm 0.7)  \times 10^{-3}$ \\
\rule[-3mm]{0mm}{8mm}
$\psi(2S)\to J/\psi\,\pi^+\,\pi^-$ &
		$(3.07 \pm 0.19) \times 10^{-1}$ \\ \hline\hline
\rule[-3mm]{0mm}{8mm}
$K^*(892)^0\to K^+\,\pi^-$ & $2/3$ \\ \hline
\end{tabular}
\end{center}
\caption
[Branching fractions of the daughter meson decay modes.]
{Branching fractions of the daughter meson decay modes that were
reconstructed in the present analysis.  For the charmonia, the
world-average branching fractions were used, with the assumption of
lepton universality in the ${\cal B}(\psi(2S)\to\mu^+\,\mu^-)$
case~\cite{pdg96}.  The $K^*(892)^0$ branching fraction was based on
isospin symmetry.}
\label{tab:daught_br_frac}
\end{table}

%% file: br_fract.tex
\label{chapt:br_fract}
\section{$B$-Meson Candidate Event Yields}
\label{sect:yields}
\begin{figure}
\begin{center}
\leavevmode
\hbox{%
\epsfxsize=3.0in
\epsffile{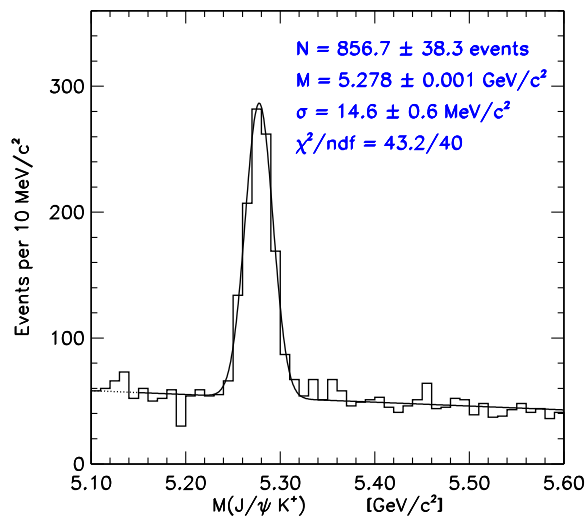}}
\end{center}
\caption
[The $J/\psi\,K^+$ invariant mass distribution.]
{The $J/\psi\,K^+$ invariant mass distribution.  The fit is to a Gaussian
signal lineshape and a linear background parameterization.}
\label{fig:bjpkg}
\end{figure}

\begin{figure}
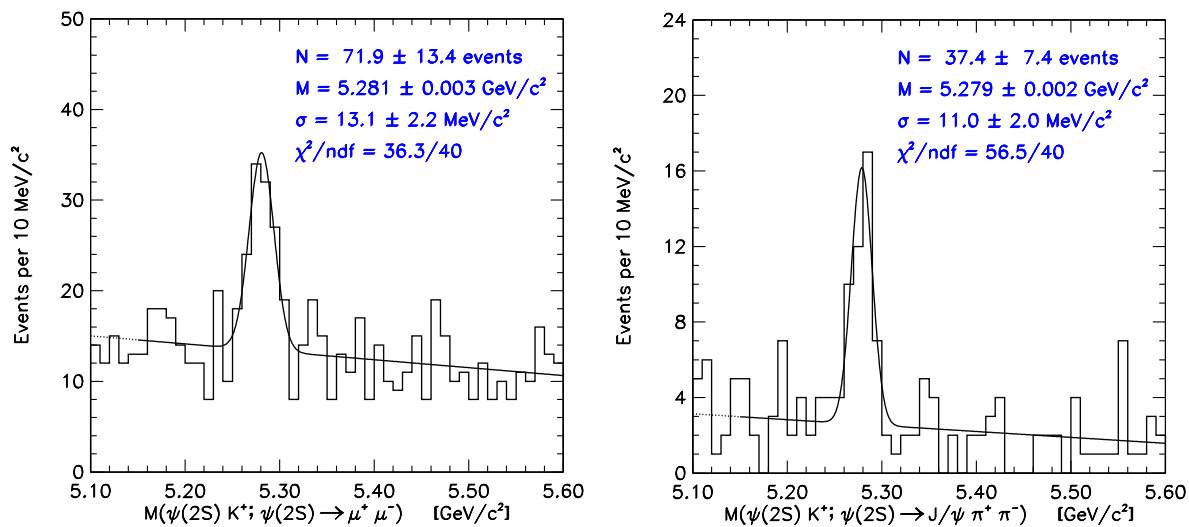

\begin{center}
\leavevmode
\hbox{%
\epsfxsize=3.0in
\epsffile{diagrams/buukg.epsi}
\hspace{0.05in}
\epsfxsize=3.0in
\epsffile{diagrams/bppkg.epsi}}
\end{center}
\caption
[The $\psi(2S)\,K^+$ invariant mass distributions using the
$\psi(2S)\to\mu^+\,\mu^-$ and $\psi(2S)\to J/\psi\,\pi^+\,\pi^-$ modes.]
{The $\psi(2S)\,K^+$ invariant mass distributions using the
$\psi(2S)\to\mu^+\,\mu^-$ (left plot) and $\psi(2S)\to
J/\psi\,\pi^+\,\pi^-$ (right plot) modes.  Each fit is to a Gaussian
signal lineshape and a linear background parameterization.}
\label{fig:buukg_bppkg}
\end{figure}

The $B$-meson candidate event yields in this study were computed by
fitting binned invariant mass distributions to signal and background
lineshapes using a maximum $\log$-likelihood technique.  In order to
avoid any contributions of partially reconstructed $B$ mesons to the
lower-sideband fit region, invariant masses $<$5.15~GeV/$c^2$ were
excluded from all the fits.  Figure~\ref{fig:bjpkg} features the
observed $B^+ \to J/\psi\,K^+$ ($J/\psi \to \mu^+\,\mu^-$) invariant
mass distribution.  Figure~\ref{fig:buukg_bppkg} illustrates the
analogous $B^+ \to \psi(2S)\,K^+$ invariant mass distributions, with
both the $\psi(2S) \to \mu^+\,\mu^-$ and $\psi(2S) \to
J/\psi\,\pi^+\,\pi^-$ charmonium daughter modes.

\begin{figure}
\begin{center}
\leavevmode
\hbox{%
\epsfxsize=3.0in
\epsffile{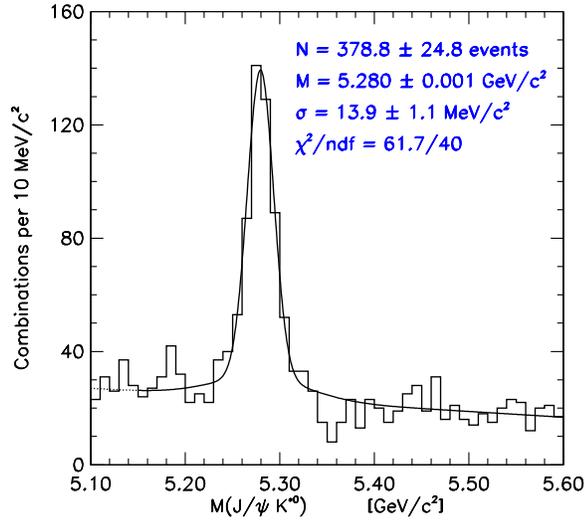}}
\end{center}
\caption
[The $J/\psi\,K^*(892)^0$ invariant mass distribution.]
{The $J/\psi\,K^*(892)^0$ invariant mass distribution.  The fit is to a
double Gaussian signal lineshape and a linear background parameterization.
The indicated fit parameters describe the Gaussian distribution of those
combinations with correct $K$-$\pi$ mass hypotheses.}
\label{fig:bjpkstg}
\end{figure}

The fits for the $B^0$ candidate event yields embodied the relative
double-Gaussian parameterizations described in
Table~\ref{tab:kst_rc_wc} in Section~\ref{sect:kst_rc_wc}.
Figure~\ref{fig:bjpkstg} shows the $J/\psi\,K^*(892)^0$ ($J/\psi
\to\mu^+\,\mu^-$) invariant mass distribution, whereas
Figure~\ref{fig:buukstg_bppkstg} presents both the $\psi(2S) \to
\mu^+\,\mu^-$ and $\psi(2S) \to J/\psi\,\pi^+\,\pi^-$ reconstructed
$\psi(2S)\,K^*(892)^0$ invariant mass distributions.


\begin{figure}
\begin{center}
\leavevmode
\hbox{%
\epsfxsize=3.0in
\epsffile{diagrams/buukstg.epsi}
\hspace{0.05in}
\epsfxsize=3.0in
\epsffile{diagrams/bppkstg.epsi}}
\end{center}
\caption
[The $\psi(2S)\,K^*(892)^0$ invariant mass distributions using the
$\psi(2S)\to\mu^+\,\mu^-$ and $\psi(2S)\to J/\psi\,\pi^+\,\pi^-$ modes.]
{The $\psi(2S)\,K^*(892)^0$ invariant mass distributions using the
$\psi(2S)\to\mu^+\,\mu^-$ (left plot) and $\psi(2S)\to
J/\psi\,\pi^+\,\pi^-$ (right plot) modes.  Each fit is to a double
Gaussian signal lineshape and a linear background parameterization.
The indicated fit parameters describe the Gaussian distribution of
those combinations with correct $K$-$\pi$ mass hypotheses.}
\label{fig:buukstg_bppkstg}
\end{figure}

The event yields and their statistical uncertainties are summarized in
Table~\ref{tab:yields}.  The relative yields of those final states
with $\psi(2S)$ candidates are consistent with expectations from the
acceptance and efficiency measurements.  An uncertainty of 4.0\% was
assigned on the branching-fraction ratios to account for differences
in the event yields due to different fitting intervals and different
right-mass-combination counting techniques in the $K^*(892)^0$ modes.
The fit interval was varied from $5.1 - 5.6$~GeV/$c^2$ to $5.2 -
5.6$~GeV/$c^2$ and an algorithm that selected $K^+\,\pi^-$
combinations nearest the world-average $K^*(892)^0$ pole
mass~\cite{pdg96} were used in lieu of the default techniques (fitting
over the interval $5.15 - 5.6$~GeV/$c^2$ and using the double-Gaussian
method) to estimate this systematic uncertainty.

\begin{table}
\begin{center}
\begin{tabular}{|l|l|c|}        \hline
\multicolumn{1}{|c|}{$B$ Mode} & \multicolumn{1}{c|}{$c\bar{c}$ Mode} &
	Event Yield \\ \hline\hline
\rule{0mm}{5mm}
$B^+\to J/\psi\,K^+$ & $J/\psi\to\mu^+\,\mu^-$ & $856.7\pm 38.3$ \\
\rule{0mm}{5mm}
$B^+\to\psi(2S)\,K^+$ & $\psi(2S)\to\mu^+\,\mu^-$ &
	$71.9\pm 13.4$ \\
\rule[-4mm]{0mm}{9mm}
$B^+\to\psi(2S)\,K^+$ & $\psi(2S)\to J/\psi\,\pi^+\,\pi^-$ &
	$37.4\pm 7.4$ \\ \hline\hline
\rule{0mm}{5mm}
$B^0\to J/\psi\,K^*(892)^0$ & $J/\psi\to\mu^+\,\mu^-$ &
	$378.8\pm 24.8$ \\
\rule{0mm}{5mm}
$B^0\to\psi(2S)\,K^*(892)^0$ & $\psi(2S)\to\mu^+\,\mu^-$ &
	$20.9\pm 7.3$ \\
\rule[-4mm]{0mm}{9mm}
$B^0\to\psi(2S)\,K^*(892)^0$ & $\psi(2S)\to J/\psi\,\pi^+\,\pi^-$ &
        $29.1\pm 7.5$ \\ \hline
\end{tabular}
\end{center}
\caption
[Summary of $B$-meson candidate event yields.]
{Summary of $B$-meson candidate event yields.  The indicated
uncertainties are statistical only.}
\label{tab:yields}
\end{table}

\section{Sample Branching-Fraction Ratio Calculation}
\label{sect:sample_calc}

The event yields, relative geometric and kinematic acceptances,
efficiencies, daughter-meson branching fractions, and systematic
uncertainties were combined into the calculations of the ratios of
branching fractions.  An example of the calculation of the ratio
${\displaystyle R^{\psi(2S)\,K^+}_{J/\psi\,K^*(892)^0}}$ in
Table~\ref{tab:ratios_to_measure} can be illustrated by expanding the
form of the expression in Equation~\ref{eqn:ratio_br_fract} to treat
both of the $\psi(2S)$ daughter reconstructions:
\begin{eqnarray}
\label{eqn:sample_calc_form}
\lefteqn{\rule{0mm}{7mm} \frac{{\cal B}(B^+\to\psi(2S)\,K^+)}
		{{\cal B}(B^0\to J/\psi\,K^*(892)^0)} =} \\
& & \rule{0mm}{7mm}
\frac{N(B^+\to\psi(2S)\,K^+,\psi(2S)\to\mu^+\,\mu^-)\ \  +\ \ 
      N(B^+\to\psi(2S)\,K^+,\psi(2S)\to J/\psi\,\pi^+\,\pi^-)}
     {N(B^0\to J/\psi\,K^*(892)^0,J/\psi\to\mu^+\,\mu^-)} \,\times \nonumber \\
& & \rule{0mm}{7mm}
\frac{\varepsilon^{\rm tot}(B^0\to J/\psi\,K^*(892)^0,J/\psi\to\mu^+\,\mu^-)}
 {\varepsilon^{\rm tot}(B^+\to\psi(2S)\,K^+,\psi(2S)\to\mu^+\,\mu^-)\ \ + \ \ 
  \varepsilon^{\rm tot}(B^+\to\psi(2S)\,K^+,\psi(2S)\to J/\psi\,\pi^+\,\pi^-)}
  \nonumber,
\end{eqnarray}
where the $N$ symbols represent the event yields and the
$\varepsilon^{\rm tot}$ quantities represent products of absolute
geometric and kinematic acceptances, efficiencies, and daughter-meson
branching fractions\footnote{The method of combining the results from
the two $\psi(2S)$ channels was motivated by the low efficiencies
involved in identifying candidate events, thereby requiring the use of
Poisson probability distributions.}.  Resolving the expression in
Equation~\ref{eqn:sample_calc_form} into the relevant measured factors
gives the form
\begin{eqnarray}
\label{eqn:sample_calc_full}
\lefteqn{\rule{0mm}{7mm} \frac{{\cal B}(B^+\to\psi(2S)\,K^+)}
		{{\cal B}(B^0\to J/\psi\,K^*(892)^0)} =} \\
& & \rule{0mm}{7mm}
\frac{N(B^+\to\psi(2S)\,K^+,\psi(2S)\to\mu^+\,\mu^-) +
      N(B^+\to\psi(2S)\,K^+,\psi(2S)\to J/\psi\,\pi^+\,\pi^-)}
     {N(B^0\to J/\psi\,K^*(892)^0,J/\psi\to\mu^+\,\mu^-)} \,\times \nonumber \\
& &
\rule[-2.5mm]{0mm}{7mm} \varepsilon^{\rm trk}_\pi \,\cdot\,
	\varepsilon^{c\tau}(B^0\to J/\psi\,K^*(892)^0) \,\cdot\,
	{\cal B}(K^*(892)^0\to K^+\,\pi^-) \,\times \nonumber \\
& &
\left[ \rule{0mm}{7mm}
\frac{{\cal A}(B^+\to\psi(2S)\,K^+,\psi(2S)\to\mu^+\,\mu^-)}
     {{\cal A}(B^0\to J/\psi\,K^*(892)^0)} \,\cdot\,
    \varepsilon^{c\tau}(B^+\to\psi(2S)\,K^+,\psi(2S)\to\mu^+\,\mu^-)
    \right. \nonumber \\
& & \rule{0mm}{7mm}
\times \frac{{\cal B}(\psi(2S)\to\mu^+\,\mu^-)}
           {{\cal B}(J/\psi\to\mu^+\,\mu^-)}
\ \ \ \ \ \ \ \ \ + \ \ \ \ \ \ 
\ \ \ \frac{{\cal A}(B^+\to\psi(2S)\,K^+,\psi(2S)\to J/\psi\,\pi^+\,\pi^-)}
       {{\cal A}(B^0\to J/\psi\,K^*(892)^0)}  \nonumber\\
& & \times \left. \rule{0mm}{7mm}
   \varepsilon^{c\tau}(B^+\to\psi(2S)\,K^+,\psi(2S)\to J/\psi\,\pi^+\,\pi^-)
\,\cdot\, \varepsilon^{\rm v}_{\pi\pi} \,\cdot\, \varepsilon^{\rm m}_{\pi\pi}
\,\cdot\, \varepsilon^{\rm trk}_{\pi\pi} \,\cdot\,
{\cal B}(\psi(2S)\to J/\psi\,\pi^+\,\pi^-)\right]^{-1}, \nonumber
\end{eqnarray}
where $\varepsilon^{\rm trk}_\pi$ represents the single-pion pattern
recognition efficiency (refer to Section~\ref{sect:track_effic} and
Appendix~\ref{app:patt_rec_eff}), $\varepsilon^{c\tau}$ represents the
proper-decay-length requirement efficiencies (refer to
Section~\ref{sect:ctau_cut_eff}), the quantities ${\cal A}$ represent
the geometric and kinematic acceptances, $\varepsilon^{\rm
v}_{\pi\pi}$ and $\varepsilon^{\rm m}_{\pi\pi}$ represent the
$CL(\chi^2)$ requirement efficiencies (refer to
Section~\ref{sect:dipion_cl_eff}), and $\varepsilon^{\rm
trk}_{\pi\pi}$ represents the double-pion pattern recognition
efficiency (refer to Section~\ref{sect:track_effic}).  The numerical
ingredients entering into the calculation of
Equation~\ref{eqn:sample_calc_full} are listed in
Table~\ref{tab:sample_calc_full}.

\begin{table}
\begin{center}
\begin{tabular}{|l|c|}        \hline
\multicolumn{1}{|c|}{Quantity} & Value \\ \hline\hline
\rule{0mm}{5mm}
$N(B^+\to\psi(2S)\,K^+,\psi(2S)\to\mu^+\,\mu^-)$ &
		$71.9\pm 13.4$ \\
\rule{0mm}{5mm}
$N(B^+\to\psi(2S)\,K^+,\psi(2S)\to J/\psi\,\pi^+\,\pi^-)$ &
		$37.4\pm 7.4$ \\
\rule[-3mm]{0mm}{8mm}
$N(B^0\to J/\psi\,K^*(892)^0,J/\psi\to\mu^+\,\mu^-)$ &
		$378.8\pm 24.8$ \\ \hline\hline
\rule{0mm}{9mm}
${\displaystyle \frac{{\cal A}(B^+\to\psi(2S)\,K^+,\psi(2S)\to\mu^+\,\mu^-)}
     {{\cal A}(B^0\to J/\psi\,K^*(892)^0)}}$ &
		$2.810\pm 0.068$ \\
\rule[-6mm]{0mm}{14mm}
${\displaystyle \frac{{\cal A}(B^+\to\psi(2S)\,K^+,\psi(2S)\to
J/\psi\,\pi^+\,\pi^-)} {{\cal A}(B^0\to J/\psi\,K^*(892)^0)}}$ &
		$0.813\pm 0.025$ \\ \hline\hline
\rule{0mm}{5mm}
$\varepsilon^{c\tau}(B^+\to\psi(2S)\,K^+,\psi(2S)\to\mu^+\,\mu^-)$ &
		$0.738\pm 0.013$ \\
\rule{0mm}{5mm}
$\varepsilon^{c\tau}(B^+\to\psi(2S)\,K^+,\psi(2S)\to J/\psi\,\pi^+\,\pi^-)$
	      & $0.762\pm 0.012$ \\
\rule[-3mm]{0mm}{8mm}
$\varepsilon^{c\tau}(B^0\to J/\psi\,K^*(892)^0)$ & $0.755\pm .013$
		\\ \hline\hline
\rule{0mm}{5mm}
$\varepsilon^{\rm trk}_\pi$ & $0.928\pm 0.020$ \\
\rule{0mm}{5mm}
$\varepsilon^{\rm trk}_{\pi\pi}$ & $0.881\pm 0.043$ \\ \hline\hline
\rule{0mm}{5mm}
$\varepsilon^{\rm v}_{\pi\pi}$ & $0.834\pm 0.039$ \\
\rule[-3mm]{0mm}{8mm}
$\varepsilon^{\rm m}_{\pi\pi}$ & $0.945\pm 0.031$ \\ \hline\hline
\rule{0mm}{5mm}
${\cal B}(\psi(2S)\to\mu^+\,\mu^-)$ & $(8.5  \pm 0.7)  \times 10^{-3}$ \\
\rule{0mm}{5mm}
${\cal B}(\psi(2S)\to J/\psi\,\pi^+\,\pi^-)$ &
		$(3.07 \pm 0.19) \times 10^{-1}$ \\
\rule{0mm}{5mm}
${\cal B}(J/\psi\to\mu^+\,\mu^-)$ & $(6.01 \pm 0.19) \times 10^{-2}$ \\
\rule[-3mm]{0mm}{8mm}
${\cal B}(K^*(892)^0\to K^+\,\pi^-)$ & $2/3$ \\ \hline\hline
\rule{0mm}{5mm}
$b$-Quark Production Spectrum Systematics & 5.9\%, 5.0\% \\
\rule{0mm}{5mm}
Trigger Systematics & 7.0\%, 2.0\% \\
\rule{0mm}{5mm}
Helicity Systematic & 2.0\% \\
\rule{0mm}{5mm}
Simulation Systematic & 3.0\% \\
\rule{0mm}{5mm}
Fitting Procedure Systematic & 4.0\% \\
\rule[-3mm]{0mm}{8mm}
Global $CL(\chi^2)$ Systematic & 2.0\% \\ \hline
\end{tabular}
\end{center}
\caption
[Numbers entering into sample calculation of
${\displaystyle R^{\psi(2S)\,K^+}_{J/\psi\,K^*(892)^0}}$.]
{Summary of event yields, geometric and kinematic acceptances,
efficiencies, daughter-meson branching fractions, and systematic
uncertainties used in the sample calculation of ${\displaystyle
R^{\psi(2S)\,K^+}_{J/\psi\,K^*(892)^0}}$ by way of
Equation~\ref{eqn:sample_calc_full}.}
\label{tab:sample_calc_full}
\end{table}

\section{Branching-Fraction Ratio Results}

The results for the ratios of the four $B$-meson final states are
presented in Table~\ref{tab:ratio_results}.  Where appropriate, the
$\psi(2S)\to\mu^+\,\mu^-$ and $\psi(2S)\to J/\psi\,\pi^+\,\pi^-$
contributions were consolidated to produce single measurements using
approaches similar to that of the sample calculation in
Section~\ref{sect:sample_calc}.  In ratios that contained unlike
species of $B$ meson in the numerator and denominator, the equality
$f_u = f_d$ was assumed (refer to Section~\ref{sect:overview}).  It is
interesting to note from Table~\ref{tab:ratio_results} that
${\displaystyle R^{\psi(2S)\,K^+}_{J/\psi\,K^+} \sim
R^{\psi(2S)\,K^*(892)^0}_{J/\psi\,K^*(892)^0}}$, which suggests that
the flavour of the light spectator quark (refer to
Figure~\ref{fig:bdecay_feyn}) has little effect on the decay width.

\begin{table}
\begin{center}
\begin{tabular}{|c|l|}        \hline
Branching-Fraction Ratio & \multicolumn{1}{|c|}{Measurement} \\ \hline\hline
\rule{0mm}{7mm}$R^{J/\psi\,K^*(892)^0}_{J/\psi\,K^+}$ &
  $1.76\pm 0.14 {\rm [stat]}\pm 0.15 {\rm [syst]}$\\
\rule{0mm}{7mm}$R^{\psi(2S)\,K^+}_{J/\psi\,K^+}$ &
  $0.558\pm 0.082 {\rm [stat]}\pm 0.056 {\rm [syst]}$\\
\rule{0mm}{7mm}$R^{\psi(2S)\,K^*(892)^0}_{J/\psi\,K^+}$ &
  $0.908\pm 0.194 {\rm [stat]}\pm 0.100 {\rm [syst]}$\\
\rule{0mm}{7mm}$R^{\psi(2S)\,K^+}_{J/\psi\,K^*(892)^0}$ &
  $0.317\pm 0.049 \rm{[stat]}\pm 0.036 \rm{[syst]}$\\
\rule{0mm}{7mm}$R^{\psi(2S)\,K^*(892)^0}_{J/\psi\,K^*(892)^0}$ &
  $0.515\pm 0.113 \rm{[stat]}\pm 0.052 \rm{[syst]}$ \\
\rule[-5mm]{0mm}{12mm}$R^{\psi(2S)\,K^*(892)^0}_{\psi(2S)\,K^+}$ &
  $1.62\pm 0.41 \rm{[stat]}\pm 0.19 \rm{[syst]}$\\ \hline
\end{tabular}
\end{center}
\caption
[Branching-fraction ratio results.]
{The branching-fraction ratios measured for the various $B$-meson
final states, corresponding to the elements of
Table~\ref{tab:ratios_to_measure}.  Note that the ratios containing
$\psi(2S)$ mesons are composed of contributions from two separate
$B$-meson reconstructions.  The first uncertainties are statistical
and the second are systematic.}
\label{tab:ratio_results}
\end{table}

\begin{figure}
\begin{center}
\leavevmode
\hbox{%
\epsfxsize=5.0in
\epsffile{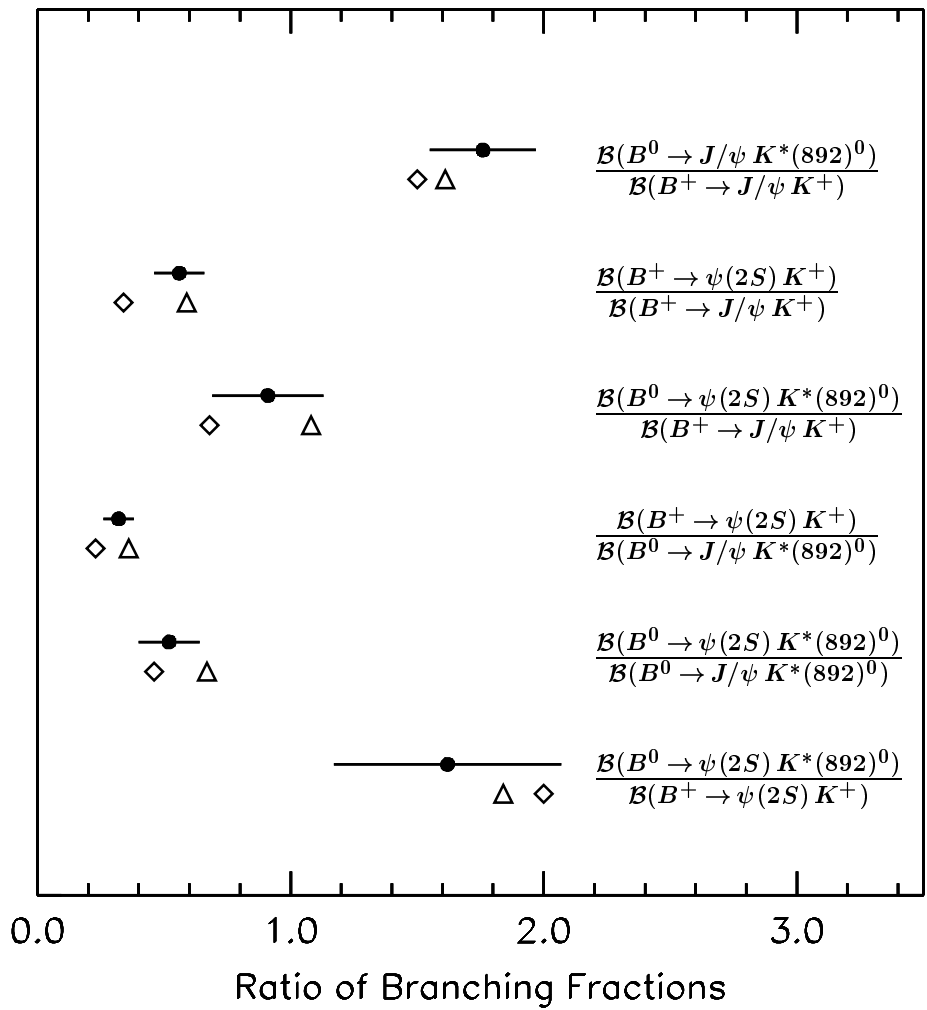}}
\end{center}
\caption
[Comparison of measured branching-fraction ratios with theoretical
predictions.]
{Comparison of the measured branching-fraction ratios (filled circles)
with theoretical predictions by Neubert {\it et
al.}~\cite{neubert:heavy_flavours} (triangles) and Deandrea {\it et
al.}~\cite{deandrea:predictions} (diamonds).  The CDF measurements
were taken from Table~\ref{tab:ratio_results}, with the statistical
and systematic uncertainties combined in quadrature.  In ratios
involving both $B^+$ and $B^0$ mesons, $f_u = f_d$ was assumed.}
\label{fig:ratio_predictions}
\end{figure}

Figure~\ref{fig:ratio_predictions} compares the measured
branching-fraction ratios with predictions derived using two different
factorization approaches.  As discussed in
Section~\ref{sect:factorization}, several factors divide out of the
theoretical expressions for the decay-rate ratios, including the CKM
elements and the parameter $a_2$ in Equation~\ref{eqn:amplitude}.  The
Neubert {\it et al.}~\cite{neubert:heavy_flavours} predictions used
the same form factors as BSW~\cite{bsw:semi,bsw:nonleptonic}, which
were calculated using a relativistic harmonic oscillator potential
model; however, whereas the BSW form factors were given a single-pole
dependence on $q^2$ in References~\cite{bsw:semi,bsw:nonleptonic}, the
Neubert {\it et al.}~\cite{neubert:heavy_flavours} predictions were
based on most of the form factors having a dipole behaviour.  As
mentioned in Section~\ref{sect:predictions}, the Deandrea {\it et
al.}~\cite{deandrea:predictions} predictions employed monopole form
factors that were derived from $D$-meson decay measurements.

Inspection of Figure~\ref{fig:ratio_predictions} indicates that
neither of the two phenomenological approaches is discounted by the
data.  The agreement between prediction and data, however, is favoured
somewhat in the Neubert {\it et al.}~\cite{neubert:heavy_flavours}
case, for which all but one of the predictions are consistent with the
measured ratios to within one standard deviation.  The Deandrea {\it et
al.}~\cite{deandrea:predictions} predictions are only consistent
within one standard deviation of the measured values for the last two
ratios in Figure~\ref{fig:ratio_predictions}; in the first four
entries, the Deandrea {\it et al.} predictions agree with the data
within two standard deviations.  Note that Deandrea {\it et al.}
predict a somewhat lower $B^+\to\psi(2S)\,K^+$ rate, which accounts
for much of the observed difference.

Although the consistency between data and prediction in
Figure~\ref{fig:ratio_predictions} does not constitute proof for the
validity of the factorization {\it Ansatz} in colour-suppressed
$B$-meson decays, the data do appear to favour the approach where the
majority of the form factors have a dipole dependence in $q^2$ and
where semileptonic $D$-meson measurements were not used to estimate
the heavy-to-light form factors.

\section{Derived Absolute Branching-Fraction Results}

The measured ratios of branching fractions listed in
Table~\ref{tab:ratio_results} were used to extract absolute branching
fractions for three of the four final states by employing the world
average value for ${\cal B}(B^+\to J/\psi\,K^+)$, which is the best
measured quantity\footnote{The structure of the matrix in
Table~\ref{tab:ratios_to_measure}, however, suggested that an
alternate approach could have been effected by performing weighted
averages of multiple ratio measurements and normalizing to
world-average values for the modes other than the one for which an
absolute value was to be derived.  This approach was taken in
References~\cite{abe:george_prd,sganos:thesis}.}.  The technique of
normalizing to a single world-average branching fraction was adopted
in the present study for several reasons: a weighted-mean approach
would have required a more complicated method of combining correlated
systematic uncertainties, the world-average values for those modes
other than $B^+\to J/\psi\,K^+$ had comparatively large uncertainties,
no previous measurements using the CDF data sample were used in the
determination of the world average value of ${\cal B}(B^+\to
J/\psi\,K^+)$, and the extracted absolute branching-fraction results
derived using a single world-average normalization could be updated
easily with subsequent improvements to the world-average value of
${\cal B}(B^+\to J/\psi\,K^+)$.

The world-average value~\cite{pdg96}
\begin{equation}
\label{eqn:bf_bjpk_pdg}
{\cal B}(B^+\to J/\psi\,K^+) = (1.01 \pm 0.14) \times 10^{-3}
\end{equation}
was therefore used with the first three entries in
Table~\ref{tab:ratio_results} to compute the derived absolute
branching fractions listed in Table~\ref{tab:abs_results}.  The
calculations of the `branching fraction' uncertainties for the two
modes involving $\psi(2S)$ daughters assumed that the world-average
charmonium branching-fraction uncertainties were independent of the
uncertainty in Equation~\ref{eqn:bf_bjpk_pdg}; this was the most
conservative assumption.

\begin{table}
\begin{center}
\begin{tabular}{|l|l|c|}        \hline
\multicolumn{1}{|c|}{Branching Fraction ($\cal{B}$)} &
		\multicolumn{1}{|c|}{CDF Measurement $[\times 10^{-3}]$} &
		PDG $[\times 10^{-3}]$ \\ \hline\hline
\rule{0mm}{6mm}
$B^0\to J/\psi\,K^*(892)^0$ &
	$1.78 \pm 0.14 {\rm [stat]} \pm 0.29 {\rm [syst]}$ & $1.49\pm 0.22$\\
\rule{0mm}{6mm}
$B^+\to\psi(2S)\,K^+$ &
	$0.56 \pm 0.08 {\rm [stat]} \pm 0.10 {\rm [syst]}$ & $0.69\pm 0.31$ \\
\rule[-4mm]{0mm}{10mm}
$B^0\to\psi(2S)\,K^*(892)^0$ &
	$0.92 \pm 0.20 {\rm [stat]} \pm 0.16 {\rm [syst]}$ & $1.4\pm 0.9$\\ \hline
\end{tabular}
\end{center}
\caption
[Derived absolute branching-fraction results.]
{The derived CDF absolute branching fractions calculated by relating
the entries in Table~\ref{tab:ratio_results} to the world-average
value in Equation~\ref{eqn:bf_bjpk_pdg}.  The first uncertainties are
statistical and the second are systematic (including the uncertainty
in Equation~\ref{eqn:bf_bjpk_pdg} and, if applicable, the
uncertainties in the world-average values of the charmonium branching
fractions~\cite{pdg96}).  The right column lists the previous
world-average values compiled by the Particle Data Group
(PDG)~\cite{pdg96}.}
\label{tab:abs_results}
\end{table}

\begin{figure}
\begin{center}
\leavevmode
\hbox{%
\epsfxsize=5.0in
\epsffile{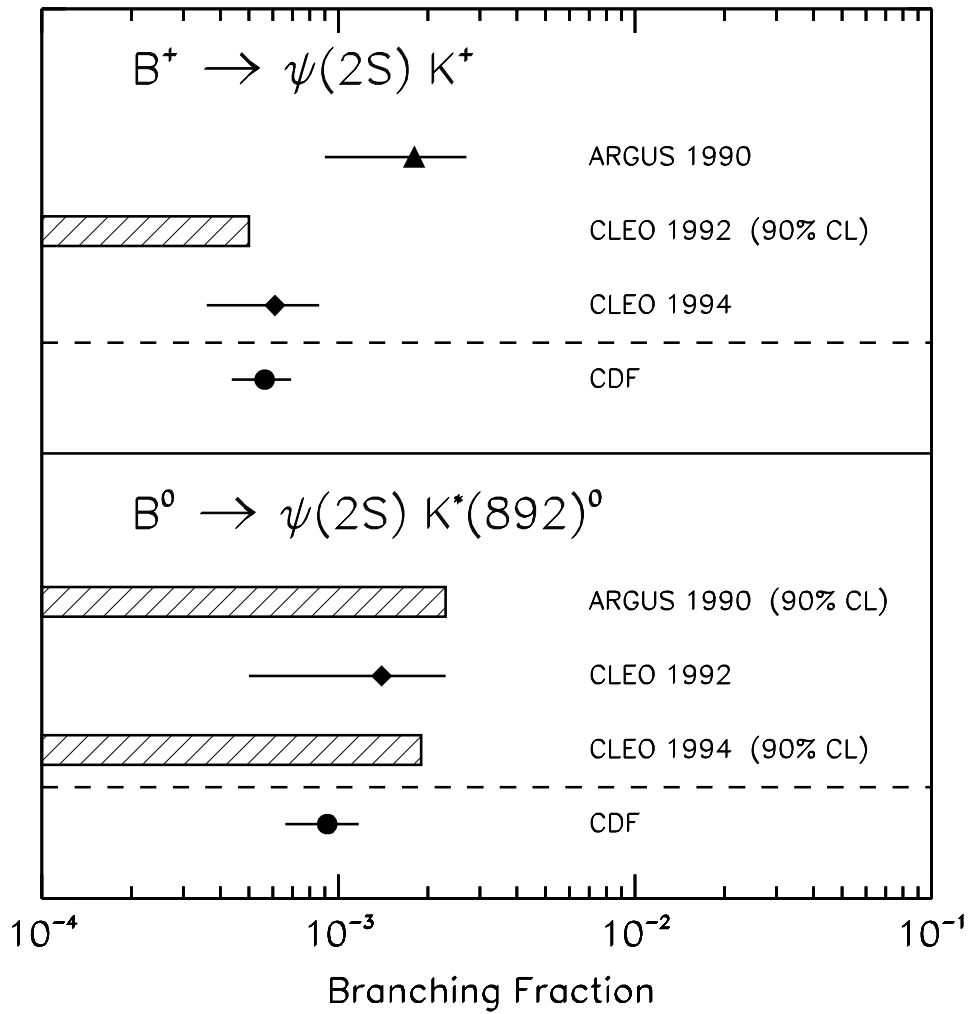}}
\end{center}
\caption
[Comparison with $e^+\,e^-$ absolute branching fractions for the $\psi(2S)$
modes.]
{A comparison of the derived CDF ${\cal B}(B^+\to\psi(2S)\,K^+)$ and
${\cal B}(B^0\to\psi(2S)\,K^*(892)^0)$ absolute branching fractions
with measurements and limits from the ARGUS~\cite{albrecht:argus_bf},
CLEO~\cite{bortoletto:cleo_bf}, and CLEO~II~\cite{alam:cleo2_bf}
experiments.  The hatched bars denote 90\% CL upper limits and the
error bars represent the statistical, systematic, and
branching-fraction uncertainties added in quadrature.}
\label{fig:br_exp_plot}
\end{figure}

The derived absolute branching fractions for the decays
$B^+\to\psi(2S)\,K^+$ and $B^0\to\psi(2S)\,K^*(892)^0$ are compared
with measurements and limits from $e^+\,e^-$ colliders in
Figure~\ref{fig:br_exp_plot}.  These results are the world's most
precise measurements of these branching fractions.  In the case of the
mode $B^+\to\psi(2S)\,K^+$, the result from the present study is
consistent within one standard deviation of the measurement reported
by the CLEO II collaboration\cite{alam:cleo2_bf}, but is only
marginally consistent with the measurement reported by the ARGUS
collaboration~\cite{albrecht:argus_bf}.  For the
$B^0\to\psi(2S)\,K^*(892)^0$ channel, the CDF result is seen to be
consistent with the previous measurement and limits\footnote{The
reversion of the CLEO result from a measurement to a limit (refer to
Figure~\ref{fig:br_exp_plot}) has been attributed to statistical
effects~\cite{kreinick:pc}.}.

\begin{figure}
\begin{center}
\leavevmode
\hbox{%
\epsfxsize=5.0in
\epsffile{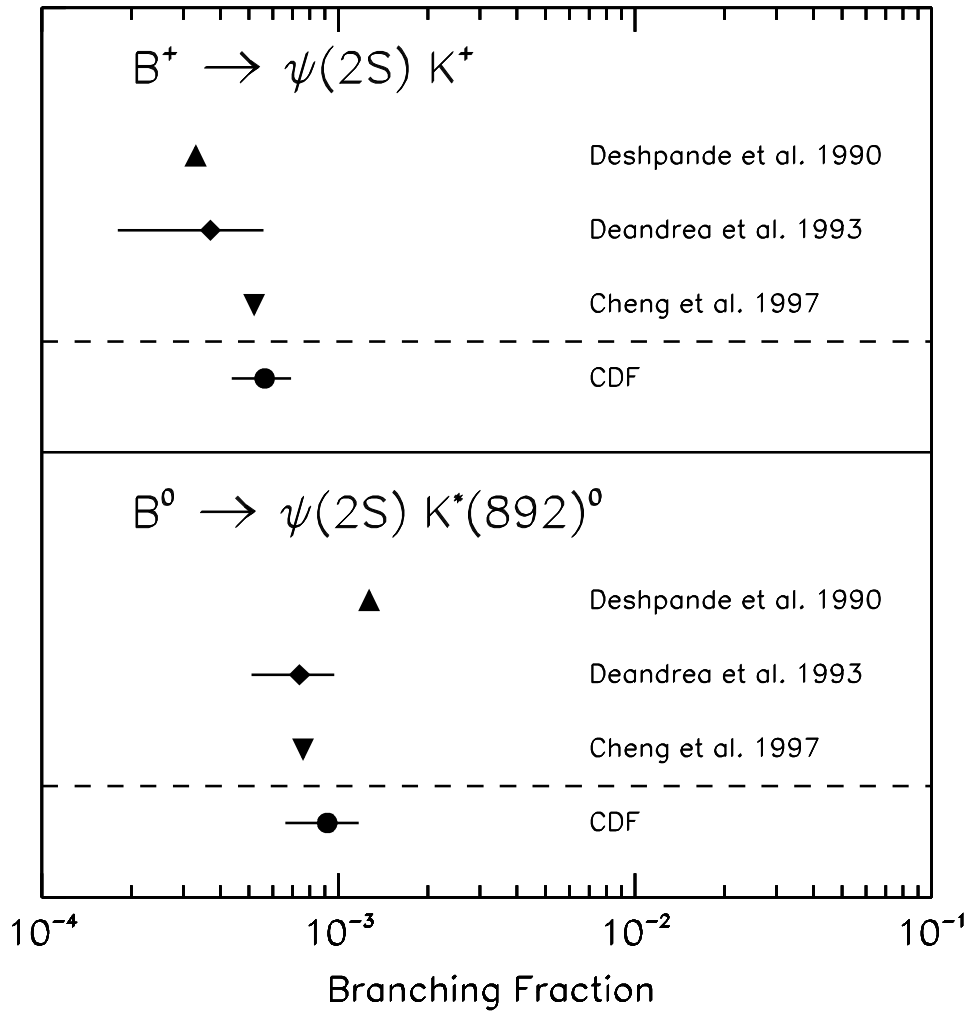}}
\end{center}
\caption
[Comparison with theoretical predictions for the $\psi(2S)$
final-state modes.]
{A comparison of the derived CDF ${\cal B}(B^+\to\psi(2S)\,K^+)$ and
${\cal B}(B^0\to\psi(2S)\,K^*(892)^0)$ absolute branching fractions
with theoretical predictions due to Deshpande and
Trampetic~\cite{deshpande:bpsik}, Deandrea {\it et
al.}~\cite{deandrea:predictions}, and Cheng {\it et
al.}~\cite{cheng:nonfactorization}.  The error bars in the Deandrea
{\it et al.} results represent only the uncertainties on the
experimental inputs to their form factor estimations.}
\label{fig:br_th_plot}
\end{figure}

The derived absolute $B\to\psi(2S)$ branching fractions may also be
compared against the predictions detailed in
Table~\ref{tab:predictions}, as is illustrated in
Figure~\ref{fig:br_th_plot}.  Whereas the less recent Deshpande and
Trampetic~\cite{deshpande:bpsik} predictions do not lie within one
standard deviation of the measured CDF values, the Deandrea {\it et
al.}~\cite{deandrea:predictions} and Cheng {\it et
al.}~\cite{cheng:nonfactorization} predictions are consistent with the
measurements.  The general agreement amongst the measurements and
predictions in Figure~\ref{fig:br_th_plot} is relatively good in spite
of the large model dependencies and uncertainties in the predicted
values.

%% file: concl.tex
\label{chapt:concl}


Exclusive nonleptonic decays of $B$ mesons into strange mesons and
$J/\psi$ or $\psi(2S)$ mesons were studied using the CDF detector to
observe a sample corresponding to $\int\!{\cal L}\,dt = (109 \pm
7)$~pb$^{-1}$ of 1.8-TeV proton-antiproton collisions produced by the
Fermilab Tevatron collider.  The investigated decays were

\begin{itemize}
\item[]
$B^+\to J/\psi\,K^+$
\item[]
$B^0\to J/\psi\,K^*(892)^0$
\item[]
$B^+\to\psi(2S)\,K^+$
\item[]
$B^0\to\psi(2S)\,K^*(892)^0$,
\end{itemize}

which were reconstructed via the daughter meson decay modes

\begin{itemize}
\item[]
$J/\psi\to\mu^+\,\mu^-$
\item[]
$\psi(2S)\to\mu^+\,\mu^-$
\item[]
$\psi(2S)\to J/\psi\,\pi^+\,\pi^-$
\item[]
$K^*(892)^0\to K^+\,\pi^-$.
\end{itemize}

With statistical significances of $\sim$5 and $\sim$3 standard
deviations, respectively, the observations of the decay modes
$B^+\to\psi(2S)\,K^+$ and $B^0\to\psi(2S)\,K^*(892)^0$ constituted the
first measurements of these processes in a hadron collider environment
and amounted to the world's largest recorded sample of these decays.

\section{Branching Fraction Measurements}

Several similarities between the decay modes were exploited to make
possible the precise measurement of the relative branching fractions
between the $B$-meson decay modes studied.  The calculated relative
geometric and kinematic acceptances, measured efficiency corrections,
observed yields, world-average daughter-meson branching fractions, and
systematic uncertainties were used to compute the ratios of branching
fractions.  For ratios that involved both $B^+$ and $B^0$ mesons,
equal production rates were assumed for both meson types.  The
measured ratios are
\begin{eqnarray}
\frac{{\cal B}(B^0\to J/\psi\,K^*(892)^0)}
		{{\cal B}(B^+\to J/\psi\,K^+)} &=&
		1.76\pm 0.14 [\rm{stat}]\pm 0.15 [\rm{syst}]\\[0.25in]
\frac{{\cal B}(B^+\to\psi(2S)\,K^+)}
		{{\cal B}(B^+\to J/\psi\,K^+)} &=&
		0.558\pm 0.082 [\rm{stat}]\pm 0.056 [\rm{syst}]\\[0.25in]
\frac{{\cal B}(B^0\to\psi(2S)\,K^*(892)^0)}
		{{\cal B}(B^+\to J/\psi\,K^+)} &=&
		0.908\pm 0.194 [\rm{stat}]\pm 0.100 [\rm{syst}]\\[0.25in]
\frac{{\cal B}(B^+\to\psi(2S)\,K^+)}
		{{\cal B}(B^0\to J/\psi\,K^*(892)^0)} &=&
		0.317\pm 0.049 [\rm{stat}]\pm 0.036 [\rm{syst}]\\[0.25in]
\frac{{\cal B}(B^0\to\psi(2S)\,K^*(892)^0)}
		{{\cal B}(B^0\to J/\psi\,K^*(892)^0)} &=&
		0.515\pm 0.113 [\rm{stat}]\pm 0.052 [\rm{syst}]\\[0.25in]
\frac{{\cal B}(B^0\to\psi(2S)\,K^*(892)^0)}
		{{\cal B}(B^+\to\psi(2S)\,K^+)} &=&
		1.62\pm 0.41 [\rm{stat}]\pm 0.19 [\rm{syst}],
\end{eqnarray}
where the first uncertainties are statistical and the second are systematic.

The branching-fraction ratio measurements were compared with two
different sets of theoretical predictions that each employed the
factorization {\it Ansatz}, but that used form factors derived in
different ways with different parameterizations in $q^2$.  All of the
predictions were consistent with the measured ratios to within two
standard deviations, thereby supporting, but not proving, the
applicability of the factorization {\it Ansatz} to colour-suppressed
$B\to\psi\,K$ decays.

Of the two sets of predictions, one was consistent with the
measurements to within one standard deviation for all but one of the
six branching-fraction ratios.  This set of predictions, due to
Neubert {\it et al.}~\cite{neubert:heavy_flavours}, assumed a dipole
behaviour for the majority of the form factors, which were calculated
from solutions to a relativistic harmonic oscillator potential
model~\cite{bsw:semi,bsw:nonleptonic}.  The somewhat less-favoured set
of predictions assumed single-pole form factors that were estimated
from measurements of semileptonic $D$-meson (heavy-to-light)
decays~\cite{deandrea:predictions}.

Relative branching-fraction measurements were combined with the
world-average branching fraction for the decay $B^+\to
J/\psi\,K^+$~\cite{pdg96}, $(1.01\pm 0.14)\times 10^{-3}$, to extract
the following absolute branching fractions:
\begin{itemize}
\item[]
${\cal B}(B^0\to J/\psi\,K^*(892)^0) =
	(1.78 \pm 0.14 [\rm{stat}] \pm 0.29 [\rm{syst}]) \times 10^{-3}$
\item[]
${\cal B}(B^+\to\psi(2S)\,K^+) =
	(0.56 \pm 0.08 [\rm{stat}] \pm 0.10 [\rm{syst}]) \times 10^{-3}$
\item[]
${\cal B}(B^0\to\psi(2S)\,K^*(892)^0) =
	(0.92 \pm 0.20 [\rm{stat}] \pm 0.16 [\rm{syst}]) \times 10^{-3}$,
\end{itemize}
where the first uncertainties are statistical and the second are
systematic.  The derived absolute branching fractions for the two
$B$-meson decays to $\psi(2S)$ final states were shown to be
consistent with both recent measurements and limits at $e^+\,e^-$
colliders and recent phenomenological predictions.

The branching-fraction measurements reported here are consistent with
the world-average values compiled by the Particle Data
Group~\cite{pdg96} (refer to Table~\ref{tab:abs_results}).  The
measured branching fractions that involve $\psi(2S)$ final states
constitute the world's most precise measurements of these quantities.

\section{Future Prospects}

Using the same data sample as that described in the present study,
measurements of several other similar exclusive $B$-meson decay
processes are feasible.  These processes include the analogous decays
$B^0\to J/\psi\,K^0$, $B^0\to\psi(2S)\,K^0$, $B^+\to
J/\psi\,K^*(892)^+$, and $B^+\to\psi(2S)\,K^*(892)^+$, which, when
combined with the results of the present work, could yield a
measurement of the ratio of fragmentation fractions, $f_u / f_d$.
Decays of the $B^0_s$ meson into $c\bar{c}$ final states, $B^0_s\to
J/\psi\,\phi(1020)$ and $B^0_s\to\psi(2S)\,\phi(1020)$, also have
similar topologies and reconstruction techniques.  Relative branching
fractions of some of these modes were reported for Run 1A
data~\cite{abe:julio_prl,abe:george_prd}.  Other recent
work~\cite{abe:bjpkst_helicity} also used the Run 1A data set to
measure the longitudinal polarization fractions in the decays $B^0\to
J/\psi\,K^*(892)^0$ and $B^0_s\to J/\psi\,\phi(1020)$, a study that
can be extended to use the full Run 1 sample and to measure
$\Gamma_L/\Gamma$ for the other vector-vector decays mentioned above.
In particular, improved measurements of the quantity ${\displaystyle
\frac{\Gamma_L}{\Gamma} (B^0\to J/\psi\,K^*(892)^0)}$ are needed for
comparison with measurements from $e^+\,e^-$
colliders~\cite{pdg96,jessop:jpsi_kst}.  Finally, the large samples of
$\psi(2S)\to J/\psi\,\pi^+\,\pi^-$ and $\psi(2S)\to\mu^+\,\mu^-$
decays in Run 1 (refer to Chapter~\ref{chapt:effic}) could be used to
measure the relative branching fraction between the two charmonium
modes, thereby reducing the total systematic uncertainty of the
$B$-meson results reported in this dissertation.

The Fermilab Tevatron is currently being upgraded with a `Main
Injector' to operate at a centre-of-mass energy of $\sqrt{s} =
2.0$~TeV with instantaneous luminosities of up to $2\times
10^{32}$~cm$^{-2}$s$^{-1}$.  Scheduled to begin in 1999, the next run
of the Tevatron (Run 2) is expected to accumulate a data sample with
time-integrated luminosity $\int\!{\cal L}\,dt = 2$~fb$^{-1}$, a
twenty-fold increase over that for the sample used in this study.
Commensurate with the improvements to the Tevatron, the CDF detector
is being rebuilt into the upgraded `CDF II' detector~\cite{blair:tdr}.
The tracking systems will have redesigned and integrated silicon and
drift-chamber detectors with improvements in the combined pattern
recognition, momentum resolution, and pseudorapidity coverage.  The
muon systems will be extended to cover nearly all angles in azimuth
and pseudorapidities in the range $|\eta| \leq 2.0$.  The data
acquisition system is expected to handle 132-ns bunch crossings, and
the trigger will have no dead time with the added improvements of
tracking information at Level 1, impact parameter discrimination at
Level 2, and 300-Hz operation at Level 3.  The CDF II and Tevatron
upgrades are expected to augment significantly both the scope and
reach of inquiry into the properties and interactions of the bottom
quark well into the next millennium.

%% file: l2_dimuon_trig.tex
\label{app:l2_dimuon_trig}

This appendix describes the Level 2 dimuon triggers that were used to
construct the data samples studied for this analysis.  Refer to
Section~\ref{sect:level2} for a description of the Level 2 system.
For each Level 2 trigger, the associated Level 1 prerequisite dimuon
trigger is given.  A description of the Run 1A and Run 1B running
periods is given in Section~\ref{sect:data_sample}.

\section{Run 1A Level 2 Dimuon Triggers}

\begin{enumerate}

\item {\sc two\_cmu\_one\_cft}:  Two CMU Level 2 muon clusters are
necessary, with one of the clusters required to match a CFT track.
The Level 1 prerequisite trigger is {\sc two\_cmu\_3pt3}.  This
trigger is dynamically prescaled.

\item {\sc cmx\_one\_cmu\_cft}:  One CMX Level 2 muon cluster and one
CMU Level 2 muon cluster are necessary, with the CMU cluster required
to match a CFT track.  The Level 1 prerequisite trigger is {\sc
two\_cmu\_cmx\_3pt3}.  This trigger is dynamically prescaled.

\item {\sc cmu\_or\_cmx\_one\_cmx\_cft}:  Either two CMX Level 2 muon
clusters or both a CMU cluster and a CMX cluster are necessary, with a
CMX cluster required to match a CFT track.  The Level 1 prerequisite
trigger is {\sc two\_cmu\_cmx\_3pt3}.  This trigger is dynamically
prescaled.

\end{enumerate}

\section{Run 1B Level 2 Dimuon Triggers}

\begin{enumerate}

\item {\sc two\_cmu\_two\_cft}: Two CMU Level 2 muon clusters are
necessary, with each cluster required to match a CFT track.  The Level
1 prerequisite trigger is {\sc two\_cmu\_3pt3}.  This trigger is not
prescaled, and the two muon clusters are required to be in
noncontiguous modules.  If one cluster is in the $+z$ region of the CDF
detector and the other cluster is in the $-z$ region, then the two
clusters are required to have different $\varphi$ values.

\item {\sc cmx\_cmu\_two\_cft}:  One CMU Level 2 muon cluster and one
CMX Level 2 muon cluster are necessary, with each cluster required to
match a CFT track.  The Level 1 prerequisite trigger is {\sc
two\_cmu\_cmx\_3pt3}.  This trigger is dynamically prescaled.

\item {\sc two\_cmx\_two\_cft}:  Two CMX Level 2 muon clusters are
necessary, with each cluster required to match a CFT track.  The Level
1 prerequisite trigger is {\sc two\_cmu\_cmx\_3pt3}.  This trigger is
dynamically prescaled, and, if both CMX muon clusters are in the same
hemisphere (in $z$) of the CDF detector, then they are required to be
separated by at least one wedge.

\item {\sc two\_cmu\_one\_cft\_6tow}: Two adjacent CMU Level 2
muon stubs, which together form a single cluster that spans six or
more calorimeter trigger towers, are necessary.  The single cluster is
required to match a CFT track, and the Level 1 prerequisite trigger is
{\sc two\_cmu\_3pt3}.  This trigger is dynamically prescaled and is
intended to offset losses due to the discontiguities imposed in the
{\sc two\_cmu\_two\_cft} trigger.

\end{enumerate}

%% file: trk_char.tex
\label{app:trk_char}

This appendix introduces the mathematical
formulation~\cite{marriner:cdf1996,gonzalez:thesis}, adapted for use
in the present study, that describes helical trajectories of charged
particles in the CDF detector.  Five parameters are used to define a
track's helix: the curvature, $c$; the azimuthal angle, $\varphi_0$,
at the point of closest approach to $(x,y) = (0,0)$; the cotangent of
the polar angle, $\cot\theta$; the impact parameter with respect to
$(x,y) = (0,0)$, $d_0$; and the local $z$ coordinate with respect to
$(x,y) = (0,0)$, $z_0$.  In Section~\ref{sect:helix_to_global}, the
global coordinates of a particle's trajectory are derived from the
track helical parameters; in Section~\ref{sect:p_to_helix}, a method
to derive the track helix using only point-of-origin and momentum
information is outlined.

\section{Global Coordinates from the Track Helix}
\label{sect:helix_to_global}

The curvature of a track is defined as
\begin{eqnarray}
\label{eqn:c}
c & \equiv & \frac{Q}{2\,r_0},
\end{eqnarray}
where $r_0$ is the radius of curvature and $Q$ denotes the sign of the
electric charge of the particle.  It is assumed here that $|Q| = 1$.

In order to determine the global coordinates of a track in the CDF
detector, the azimuthal deflection of the particle's momentum,
$\Delta\varphi$, is defined as
\begin{eqnarray}
\label{eqn:dphi}
\Delta\varphi & \equiv & \frac{s\,Q}{r_0},
\end{eqnarray}
where $s$ is the arc length of the helix, projected onto the transverse
$(x,y)$ plane.  By substitution of Equation~\ref{eqn:c} into
Equation~\ref{eqn:dphi}, the global azimuthal angle, $\varphi$, is
defined in radians as
\begin{eqnarray}
\varphi \equiv (\Delta\varphi + \varphi_0) \bmod 2\pi = (2\,c\,s + \varphi_0) \bmod 2\pi.
\end{eqnarray}
If the coordinates $(x_0,y_0)$ define the transverse location of the
centre of the track helix, then the global transverse coordinates of the
helix are
\begin{eqnarray}
\label{eqn:xy}
x & = & x_0 + r_0\,Q\,\sin\varphi \\
y & = & y_0 - r_0\,Q\,\cos\varphi. \nonumber
\end{eqnarray}
The same form as that in Equation~\ref{eqn:xy} may be used to
calculate the coordinates of the point in the track trajectory that is
nearest, in the transverse plane, to the geometric centre of the
detector, $(x^\prime,y^\prime)$:
\begin{eqnarray}
\label{eqn:xyprime}
x^\prime & = & x_0 + r_0\,Q\,\sin\varphi_0 \\
y^\prime & = & y_0 - r_0\,Q\,\cos\varphi_0. \nonumber
\end{eqnarray}
Since the magnitude of the impact parameter, $|d_0|$, is defined as
$\left|d_0\right| \equiv \left|\sqrt{x^2_0 + y^2_0} - r_0\right|$, and
its sign is defined as
\begin{eqnarray}
\frac{d_0}{|d_0|} & \equiv & \frac{Q\,\left(\sqrt{x_0^2 + y_0^2} - r_0\right)}{\left|\sqrt{x_0^2 + y_0^2} - r_0\right|},
\end{eqnarray}
the
coordinates $(x^\prime,y^\prime)$ may also be written as
\begin{eqnarray}
\label{eqn:d_def}
x^\prime & = & -d_0\,\sin\varphi_0 \\
y^\prime & = &  d_0\,\cos\varphi_0. \nonumber
\end{eqnarray}
Solving for $(x_0,y_0)$ in Equations~\ref{eqn:xyprime} and \ref{eqn:d_def}
yields
\begin{eqnarray}
\label{eqn:x0y0}
x_0 & = & -(r_0\,Q + d_0)\,\sin\varphi_0 \\
y_0 & = & (r_0\,Q + d_0)\,\cos\varphi_0. \nonumber
\end{eqnarray}

Substitution of Equation~\ref{eqn:x0y0} into Equation~\ref{eqn:xy},
and translation of the $z_0$ helix parameter into its analogous global
coordinate, furnishes the following expressions for a track's global
coordinates:
\begin{eqnarray}
\label{eqn:glob_coord}
x & = & r_0\,Q\,\sin\varphi - (r_0\,Q + d_0)\,\sin\varphi_0 \nonumber \\
y & = & -r_0\,Q\,\cos\varphi + (r_0\,Q + d_0)\,\cos\varphi_0 \\
z & = & z_0 + s\,\cot\theta. \nonumber
\end{eqnarray}

\section{Helix from Momentum and Point of Origin}
\label{sect:p_to_helix}

In studies that involved the embedding of Monte Carlo generated
particle tracks into actual data events, a transformation from
momentum, electric charge, and spatial-origin coordinates to local
helical parameter coordinates was necessary.  The transformations
used, as well as any assumptions made, are detailed in the following.

The curvature parameter was calculated with the formula
\begin{eqnarray}
c & = & \frac{(1.49898 \times 10^{-3})\,Q\,B}{p_{\rm T}},
\end{eqnarray}
where $B$ is the magnetic field in units of Tesla and $p_{\rm T}$ is the
transverse momentum, $p_{\rm T} \equiv |\vec{p}|\,\sin\theta$, in units of
GeV/$c$.

The azimuth parameter was determined with the expression
\begin{eqnarray}
\label{eqn:p0_def}
\varphi_0 & = & \left\{ \begin{array}{ll}
	\arccos\left(\frac{p_x}{p_{\rm T}}\right) & \mbox{if $p_y > 0$} \\
	2\pi - \arccos\left(\frac{p_x}{p_{\rm T}}\right) & \mbox{if $p_y \leq 0$}
		    \end{array} \right.,
\end{eqnarray}
where $p_x$ and $p_y$ are the $x$ and $y$ components of the momentum
vector, respectively, and an approximation\footnote{The stated
approximation is justified in track embedding studies where only a
subset of the Monte Carlo tracks generated for a given event is
considered and there is therefore no need to preserve the precise
kinematics of the parent particle.} that $\varphi_0 = \varphi$ is
assumed.

The cotangent of the polar angle was simply computed using
\begin{eqnarray}
\cot\theta & = & \frac{p_z}{p_{\rm T}},
\end{eqnarray}
where $p_z$ is the $z$ component of the momentum vector.

The calculation of the impact parameter, $d_0$, required care to avoid
a numerical instability.  The impact parameter may be defined
as
\begin{eqnarray}
\label{eqn:d0_defn}
d_0 & \equiv & Q\,(\xi - r_0),
\end{eqnarray}
where $\xi \equiv \sqrt{x_0^2 + y_0^2}$.  In practice,
Equation~\ref{eqn:d0_defn} is difficult to calculate since $|d_0| \ll
\xi$.  This was remedied by rewriting Equation~\ref{eqn:d0_defn} in the
form~\cite{gonzalez:thesis}
\begin{eqnarray}
\label{eqn:d0_new_defn}
d_0 = Q\,(\xi - r_0) = \frac{Q\,(\xi - r_0)\,(\xi + r_0)/r_0}{(\xi + r_0)/r_0}.
\end{eqnarray}
The coordinates $(x^{\prime\prime},y^{\prime\prime})$ were defined to
represent the displacement of the track's point of origin.
A substitution of an analogous form of Equation~\ref{eqn:xy} was made
to convert Equation~\ref{eqn:d0_new_defn} to the form
\begin{eqnarray}
d_0&=&\frac{Q\,\left[(x^{\prime\prime})^2+(y^{\prime\prime})^2\right] +
2r_0\,\left[y^{\prime\prime}\,\cos\varphi_0 -
x^{\prime\prime}\,\sin\varphi_0\right]}
{r_0 + \sqrt{\left(x^{\prime\prime}-r_0\,Q\,\sin\varphi_0\right)^2
+ \left(y^{\prime\prime}+r_0\,Q\,\cos\varphi_0\right)^2}}.
\end{eqnarray}
Further substitution using Equation~\ref{eqn:c} yielded the practical
form
\begin{eqnarray}
d_0&=&\frac{2c\,\left[(x^{\prime\prime})^2+(y^{\prime\prime})^2\right] +
2\left[y^{\prime\prime}\,\cos\varphi_0 -
x^{\prime\prime}\,\sin\varphi_0\right]}
{1 + 2c\,Q\,\sqrt{\left(x^{\prime\prime}-r_0\,Q\,\sin\varphi_0\right)^2
+ \left(y^{\prime\prime}+r_0\,Q\,\cos\varphi_0\right)^2}}.
\end{eqnarray}

Following the same justification as that used in the $\varphi_0 =
\varphi$ assumption in Equation~\ref{eqn:p0_def}, the value of $z_0$,
which according to Equation~\ref{eqn:glob_coord} is $z_0 =
z^{\prime\prime} - s\,\cot\theta$, was approximated as
\begin{eqnarray}
z_0 & = & z^{\prime\prime}.
\end{eqnarray}

%% file: patt_rec_eff.tex
\label{app:patt_rec_eff}

This appendix summarizes a study~\cite{warburton:trk_effic} of the
single- and double-track pion pattern recognition efficiencies in the
CDF central tracking chamber (CTC) over the course of the Run 1
data-taking period.

\section{Data Sample}

The sample used in this efficiency study consisted of $\sim$36\,000
raw dimuon events from 40 data-taking runs, 10 from Run 1A and 30 from
Run 1B.  The runs represented in this sample are detailed in
Table~\ref{tab:runs}.

\begin{table}
\begin{center}
\begin{tabular}{|c||c|c|c|c|}	\hline
Experiment & Run 1A & \multicolumn{3}{c|}{Run 1B} \\ \cline{1-1}
Run &  & Low & Medium & High \\
Range & 40\,100$-$47\,835 & 55\,242$-$60\,900 & 60\,901$-$67\,390 & 67\,391$-$71\,023 \\ \hline \hline
R & 40\,512 & 55\,571 & 61\,024 & 68\,040 \\
u & 41\,085 & 56\,002 & 61\,548 & 68\,206 \\
n & 42\,030 & 57\,006 & 63\,171 & 68\,517 \\
  & 43\,016 & 57\,513 & 63\,502 & 68\,758 \\
N & 44\,366 & 58\,093 & 64\,041 & 69\,036 \\
u & 45\,047 & 58\,564 & 65\,004 & 69\,550 \\
m & 46\,031 & 59\,042 & 65\,560 & 69\,809 \\
b & 46\,519 & 59\,517 & 66\,020 & 70\,019 \\
e & 47\,008 & 60\,004 & 66\,539 & 70\,559 \\
r & 47\,552 & 60\,597 & 66\,615 & 71\,013 \\ \hline \hline
$\int\!{\cal L}\,dt$ $[{\rm pb}^{-1}]$& 20.90 & 9.46 & 47.08 & 35.41 \\ \hline
\end{tabular}
\end{center}
\caption
[Run list for pattern recognition efficiency embedding study.]
{A list of runs representing four logical run-range divisions of the
Run 1 data-taking period.  Each of these four divisions contributed
$\sim$9\,000 events to the sample.  The bottom row of the table lists
the time-integrated luminosities represented by the various run-range bins.}
\label{tab:runs}
\end{table}

\begin{figure}[p]
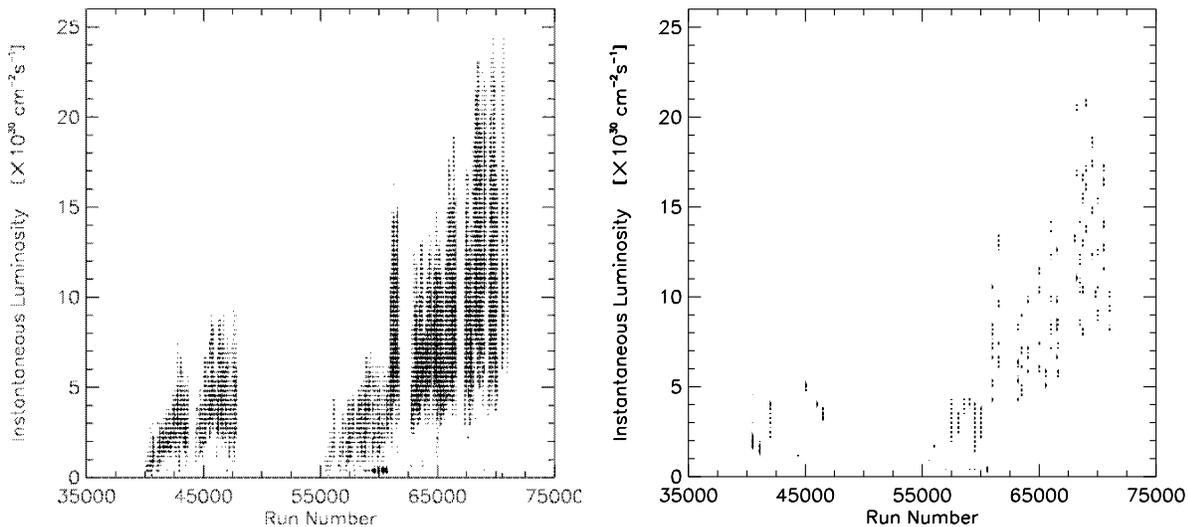

\begin{center}
\leavevmode
\hbox{%
\epsfxsize=3.0in
\epsffile{diagrams/lum_bjpk.epsi}
\hspace{0.05in}
\epsfxsize=3.0in
\epsffile{diagrams/lum_used.epsi}}
\end{center}
\caption
[Instantaneous luminosity profiles of the data and embedded samples.]
{Instantaneous luminosity vs.~run number profile for (left)
$\sim$530\,000 $B^+\to J/\psi\,K^+$ candidates and that for (right)
$\sim$36\,000 raw dimuon events used in the present efficiency study.}
\label{fig:lum_bjpk_used}
\end{figure}

The four loose run-range bins in Table~\ref{tab:runs} are delimited by
the changeover period between Runs 1A and 1B, a major reorientation of
the Tevatron beam optics, and a prolonged shutdown of the collider,
respectively.  The left side of Figure~\ref{fig:lum_bjpk_used}
illustrates the instantaneous luminosity profile of $\sim$530\,000
$B^+\to J/\psi\,K^+$ candidates, including background, and clearly
features the natural divisions between the four run bins.  Portrayed
on the right side of Figure~\ref{fig:lum_bjpk_used} is the profile of
the event sample used in the present efficiency study.

\section{Track Embedding Procedure}
\label{sect:embedding_procedure}

The crux of this study was the embedding of hits representing two
Monte-Carlo-generated pions into raw data events.  $B^+$ mesons were
generated in a manner similar to that described in
Section~\ref{sect:mc_gen}.  The CLEO {\sc qq}
programme~\cite{avery:cleoqq} was used, as in
Section~\ref{sect:mc_decay}, to decay the $B^+$ mesons in the chain
$B^+ \to \psi(2S)\,K^+$, $\psi(2S)\to J/\psi\,\pi^+\,\pi^-$, and
$J/\psi\to\mu^+\,\mu^-$.  The dipion invariant mass distribution was
generated according to the customized matrix element described in
Section~\ref{sect:mpipi}.  For events where the $\pi^+$ and $K^+$
particles each passed a $p_{\rm T} > 250$~MeV/$c$ transverse momentum
requirement, the 3-momentum and charge of the two pions in the decay
chain were extracted from the CLEO {\sc qq} output and stored for use
in the embedding stage.

The curvature ($c$), azimuth ($\varphi_0$), and polar angle
($\cot\theta$) helical parameters of the tracks to be embedded were
derived from the 3-momentum and charge information furnished in the
Monte Carlo track generation stage.  The impact ($d_0$) and
longitudinal displacement ($z_0$) helical parameters were calculated
with additional information on the primary vertex location of the
particular event into which the tracks were to be embedded.
Appendix~\ref{app:trk_char} describes the method by which the track
helices were determined.

Once all five helical parameters for each of the two pion tracks were
known, a set of routines~\cite{aseet:ctaddh} was used to embed hits
into the CTC and VTX data structures.  The values of the embedded
parameters were recorded for later comparisons, and the full default
production executable package (described in
Section~\ref{sect:data_reduct}) was used to reconstruct the tracks in
the events.

\begin{figure}[p]
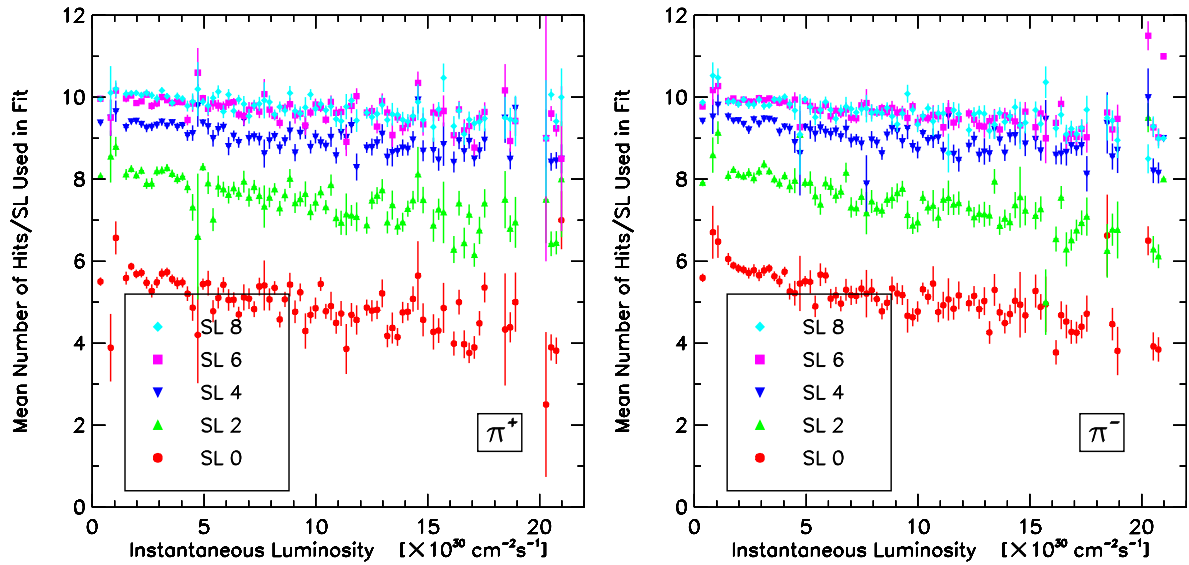

\begin{center}
\leavevmode
\hbox{%
\epsfxsize=3.0in
\epsffile{diagrams/hits_SL_axial_pos.epsi}
\hspace{0.05in}
\epsfxsize=3.0in
\epsffile{diagrams/hits_SL_axial_neg.epsi}}
\end{center}
\caption
[Profile of mean number of used hits per axial superlayer.]
{The mean number of hits per axial superlayer used in fits of $\pi^+$
(left) and $\pi^-$ (right) tracks embedded with uniform 100\% wire hit
efficiencies, as a function of instantaneous luminosity.}
\label{fig:hits_SL_axial}
\end{figure}

\begin{figure}[hp]
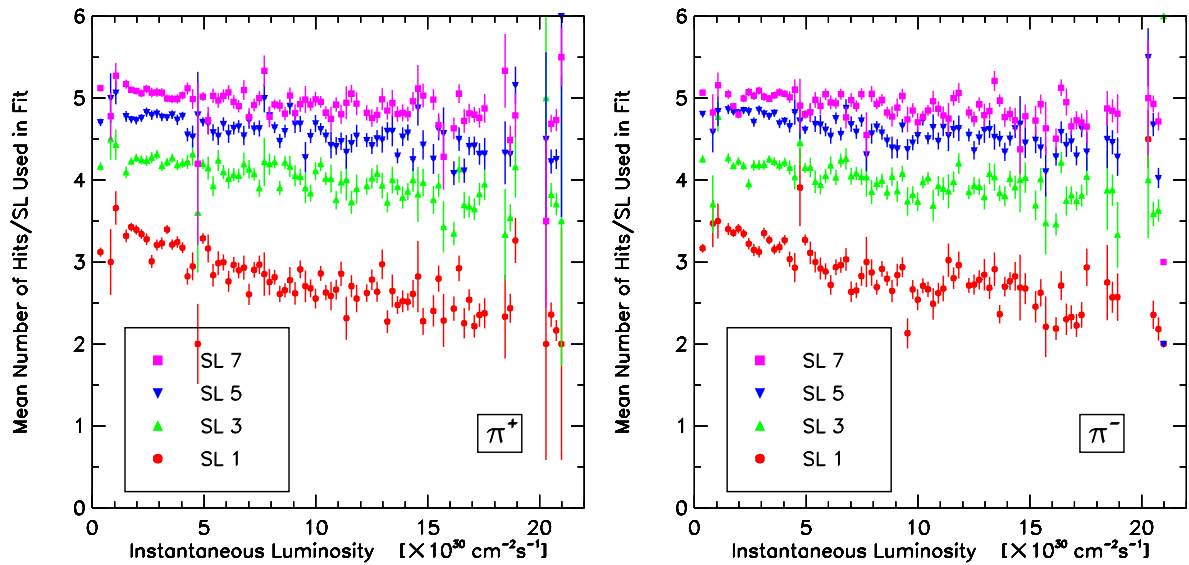

\begin{center}
\leavevmode
\hbox{%
\epsfxsize=3.0in
\epsffile{diagrams/hits_SL_stereo_pos.epsi}
\hspace{0.05in}
\epsfxsize=3.0in
\epsffile{diagrams/hits_SL_stereo_neg.epsi}}
\end{center}
\caption
[Profile of mean number of used hits per stereo superlayer.]
{The mean number of hits per stereo superlayer used in fits of $\pi^+$
(left) and $\pi^-$ (right) tracks embedded with uniform 100\% wire hit
efficiencies, as a function of instantaneous luminosity.}
\label{fig:hits_SL_stereo}
\end{figure}

In order to determine the wire hit efficiencies to use in the
embedding software, the mean numbers of hits used by the reconstructed
track fits in each CTC superlayer were examined, as functions of
instantaneous luminosity, for both embedded and data tracks.
Figures~\ref{fig:hits_SL_axial} and
\ref{fig:hits_SL_stereo} show the mean numbers of hits per
superlayer used in fits of tracks embedded with 100\% wire hit
efficiencies, as functions of instantaneous luminosity, for the axial
and stereo superlayers, respectively.

The mean numbers of hits per superlayer used in the embedded track
fits were tuned to match those observed in data by iteratively
adjusting the superlayer-by-superlayer wire hit efficiencies in the
embedding routines.  Table~\ref{tab:wire_eff} summarizes the reduced
wire efficiencies calculated using this technique.

\begin{table}
\begin{center}
\begin{tabular}{|c|c|c||c|}	\hline
Axial & Stereo & Wire & Wire \\
Superlayer & Superlayer & Numbers & Efficiency \\ \hline\hline
8 &   & $72-83$ & 1.00 \\
  & 7 & $66-71$ & 0.99 \\
6 &   & $54-65$ & 1.00 \\
  & 5 & $48-53$ & 0.96 \\
4 &   & $36-47$ & 0.97 \\
  & 3 & $30-35$ & 0.89 \\
2 &   & $18-29$ & 0.80 \\
  & 1 & $12-17$ & 0.82 \\
0 &   & \ \,$0-11$  & 0.74 \\ \hline
\end{tabular}
\end{center}
\caption
[Wire efficiencies used in the embedding process.]
{The wire efficiencies used to embed hits in the 9 CTC superlayers.
These were calculated by tuning the mean number of fitted hits per
superlayer to match in embedded and data samples.}
\label{tab:wire_eff}
\end{table}

It is evident from Figures~\ref{fig:hits_SL_axial} and
\ref{fig:hits_SL_stereo} that there was negligible sensitivity to the
charge of the embedded track.  Consequently, the wire efficiencies
featured in Table~\ref{tab:wire_eff} were applied to both positively
and negatively charged tracks.  Within the axial and stereo superlayer
categories in Table~\ref{tab:wire_eff}, the wire efficiencies were
observed to diminish monotonically from the outer to the inner
superlayers.  Due to the geometry of the CTC and the $p_{\rm T}$
distribution of charged tracks~\cite{abe:charged_particles}, increased
charged-track occupancies near the inner-superlayer radii accounted
for the bulk of the observed relative trends in the
superlayer-dependent wire efficiencies.

\section{Definition of a ``Found'' Track or Track Pair}
\label{sect:found}

The technique used to identify a ``found'' track attempted to account
for interparameter correlations in the reconstructed helices.
Specifically, for a given Monte Carlo track with embedded helical
parameter vector
\mbox{\boldmath $\alpha$}, where $\alpha_i \in
\left(\cot\theta, c, z_0, d_0, \varphi_0\right)$, each reconstructed track
in the event, $\widehat{\mbox{\boldmath $\alpha$}}$, that shared the
same sign of curvature with \mbox{\boldmath $\alpha$}, was used to
calculate a matching $\chi^2$ quantity via the expression~\cite{pdg96}
\begin{eqnarray}
\label{eqn:chisq}
\chi^2 & \equiv & \left(\mbox{\boldmath $\alpha$} - \widehat{\mbox{\boldmath $\alpha$}}\right)^{\rm T} \widehat{\mbox{\boldmath V}}^{-1} \left(\mbox{\boldmath $\alpha$} - \widehat{\mbox{\boldmath $\alpha$}}\right),
\end{eqnarray}
where the matrix $\widehat{\mbox{\boldmath V}}$ was the $5\times5$
covariance matrix of the reconstructed track helix.  For each embedded
track charge, the reconstructed track in the event that formed the
lowest matching $\chi^2$ value with the embedded track, $\chi^2_{\rm
min}$, was retained for consideration as a ``found'' track.

There was some speculation about the relative effects of the stereo
and axial contributions to inefficiencies in the CTC.  Tracks were
initially identified in the CTC by seeking segments in the axial
superlayers and forming 2-dimensional track objects in the
$r$-$\varphi$ view.  Stereo reconstruction then took place to form
3-dimensional tracks, using the information from the stereo
superlayers and from the VTX, which provided the initial information
on the $z_0$ helix parameter of the tracks.  It was therefore
predicted that, under increased instantaneous luminosity conditions
that precipitated higher primary vertex multiplicities in the VTX, the
reliability of the $z_0$ determination, and hence the stereo
reconstruction efficiency, would suffer.  Occupancy-related
luminosity-dependent efficiency losses in the inner CTC superlayers
were also expected to afflict the stereo more than the axial
reconstruction because of the lesser number of superlayers and wires
per superlayer in the stereo view (refer to Section~\ref{sect:ctc} for
a description of the CTC).

The foregoing considerations about axial and stereo contributions to
tracking performance motivated the construction of an axial-only
matching $\chi^2$ quantity.  Specifically, the vector
\mbox{\boldmath $\alpha$} in Equation~\ref{eqn:chisq} was modified such that
$\alpha_i \in \left(c, d_0, \varphi_0\right)$ and the matrix
$\widehat{\mbox{\boldmath V}}$ became the corresponding $3\times3$
track parameter covariance matrix.

\begin{figure}[p]
\begin{center}
\leavevmode
\hbox{%
\epsfysize=3.1in
\epsffile{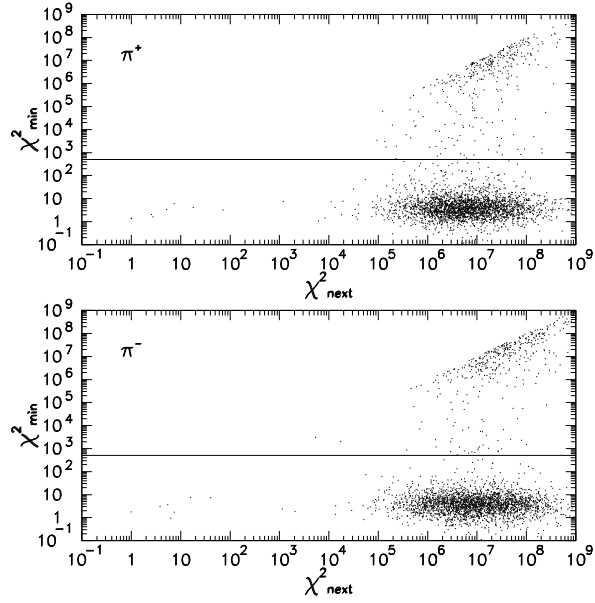}}
\end{center}
\caption
[Minimum matching $\chi^2$ vs.~next-to-minimum matching $\chi^2$.]
{The minimum 3-dimensional matching $\chi^2$, $\chi^2_{\rm min}$,
plotted against the 3-dimensional next-to-minimum matching $\chi^2$,
$\chi^2_{\rm next}$, for 3\,000 embedded $\pi^+$ mesons (top) and $\pi^-$
mesons (bottom).  The horizontal line designates the selection
criterion, $\chi^2_{\rm min} < 500$.}
\label{fig:chsq_nchsq}
\end{figure}

In the case of 3-dimensional track matching, a reconstructed track was
designated as ``found'' if its matching $\chi^2$ value satisfied the
criterion \[\chi^2_{\rm min} < 500.\] The analogous criterion in the
case of 2-dimensional axial-only matching was \[\chi^2_{\rm min} <
300.\] The efficacy of these cuts in discriminating between the
embedded track and other tracks in the event is exemplified in
Figure~\ref{fig:chsq_nchsq}, which plots the minimum matching
$\chi^2$, $\chi^2_{\rm min}$, versus the next-to-minimum matching
$\chi^2$, $\chi^2_{\rm next}$, for the 3-dimensional case.
Figure~\ref{fig:chsq_nchsq} furnishes two useful observations: first,
the matching $\chi^2$ selection criteria, indicated by the horizontal
lines, were relatively efficient at determining the distinction
between the found track and any other track in the event; and second,
the majority of events where an embedded track was not found at all
had $\chi^2_{\rm min} \sim \chi^2_{\rm next}$, as expected.

\begin{figure}[p]
\begin{center}
\leavevmode
\hbox{%
\epsfysize=3.1in
\epsffile{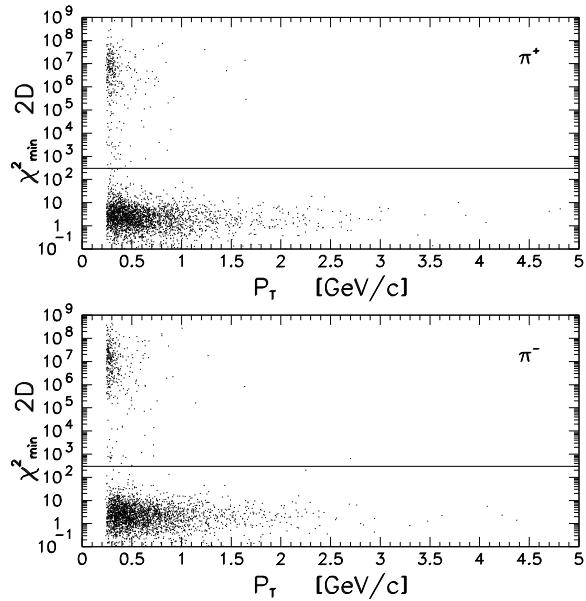}}
\end{center}
\caption
[Minimum matching $\chi^2$ vs.~the embedded transverse momentum.]
{The minimum 2-dimensional matching $\chi^2$, $\chi^2_{\rm min}$,
plotted against the embedded track transverse momentum,
$p_{\rm T}$, for 3\,000 embedded $\pi^+$ mesons (top) and $\pi^-$
mesons (bottom).  The horizontal line designates the selection
criterion, $\chi^2_{\rm min} < 300$.}
\label{fig:ch3q_pt}
\end{figure}

The $p_{\rm T}$ dependence of tracks that passed and failed the
``found''-track identification criteria is pictured in
Figure~\ref{fig:ch3q_pt}, this time for 2-dimensional matching.  These
plots indicate that any inefficiencies in the $\chi^2_{\rm min}$
matching criteria were largely confined to embedded tracks with
extremely low $p_{\rm T}$.  These figures also suggest that the majority of
embedded tracks that failed to be ``found'' had low $p_{\rm T}$.


The embedding of both $\pi^+$ and $\pi^-$ mesons into the same event
enabled the determination of a ``found'' track pair.  A track pair was
considered to be ``found'' if, in the case of 3-dimensional track
matching, \[\max\left(\chi^2_{\rm min}[\pi^+],\chi^2_{\rm
min}[\pi^-]\right) < 500\] and, in the case of 2-dimensional track
matching, \[\max\left(\chi^2_{\rm min}[\pi^+],\chi^2_{\rm
min}[\pi^-]\right) < 300.\]

\section{Pattern Recognition Efficiency}

\subsection{Definition}

The single- and double-track CTC pattern recognition efficiencies may
be expressed as a function of one of several observables.
Equation~\ref{eqn:eff_defn} defines the bin-by-bin pattern recognition
efficiency for a given bin, $\Delta\xi$, in an observable $\xi$:
\begin{equation}
\label{eqn:eff_defn}
\varepsilon\left(\Delta\xi\right) \equiv
\left.\frac{\int\limits_{\Delta\xi} N_{\rm found}(\xi)\,d\xi}
{\int\limits_{\Delta\xi}\left[N_{\rm found}(\xi) +
                              N_{\overline{\rm found}}(\xi)\right]\,d\xi}
\right|_{{p_{\rm T}\,>\,p_{\rm T}^{\rm min}\ \ \ \ \ \ \ \ \atop
r^{\rm exit}_{\rm CTC}\,>\,\left(r^{\rm exit}_{\rm CTC}\right)^{\rm min}}
             \atop    }.
\end{equation}
The $N_{\rm found}(\xi)$ symbol represents the number of tracks (or
track pairs) at a given value of $\xi$ that are identified as
``found'' according to the criteria discussed in
Section~\ref{sect:found}.  The $N_{\overline{\rm found}}(\xi)$ symbol
represents the corresponding ``found''-track (or track pair) failures,
and the quantity $r^{\rm exit}_{\rm CTC}$ is the CTC exit radius, or
the radius at which tracks intersect one of the two endplate planes of
the CTC.  Note that the $p_{\rm T}$ or $r^{\rm exit}_{\rm CTC}$
requirements in Equation~\ref{eqn:eff_defn} are not applied when $\xi
\equiv p_{\rm T}$ or $\xi \equiv r^{\rm exit}_{\rm CTC}$,
respectively.  Also, when paired track efficiencies are determined as
a function of $p_{\rm T}$, $\xi
\equiv \min\left(p_{\rm T}[\pi^+], p_{\rm T}[\pi^-]\right)$ in
Equation~\ref{eqn:eff_defn}.


An aggregate pattern recognition efficiency may be calculated in terms
of $\xi \equiv p_{\rm T}$, for $p_{\rm T} > p_{\rm T}^{\rm min}$, as
\begin{equation}
\label{eqn:tot_eff_defn}
\varepsilon\left(p_{\rm T}\,>\,p_{\rm T}^{\rm min}\right) \equiv
\left.\frac{\sum\limits_{p_{\rm T}\,>\,p_{\rm T}^{\rm min}} N_{\rm found}(\Delta p_{\rm T})}
{\sum\limits_{p_{\rm T}\,>\,p_{\rm T}^{\rm min}}\left[N_{\rm found}(\Delta p_{\rm T}) +
                              N_{\overline{\rm found}}(\Delta p_{\rm T})\right]}
\right|_{r^{\rm exit}_{\rm CTC}\,>\,
		\left(r^{\rm exit}_{\rm CTC}\right)^{\rm min}},
\end{equation}
with binomial statistics used to compute the statistical
uncertainty on $\varepsilon$.

\subsection{Efficiency Dependence on Kinematic Observables}
\label{sect:kinematic}

Employing the definition described in Equation~\ref{eqn:eff_defn}, the
pattern recognition efficiency was studied for tracks in several
different kinematic observables.  Unless noted otherwise, $p_{\rm T}^{\rm
min} \equiv 0.4$~GeV/$c$ in Equation~\ref{eqn:eff_defn}.

\begin{figure}
\begin{center}
\leavevmode
\hbox{%
\epsfysize=3.3in
\epsffile{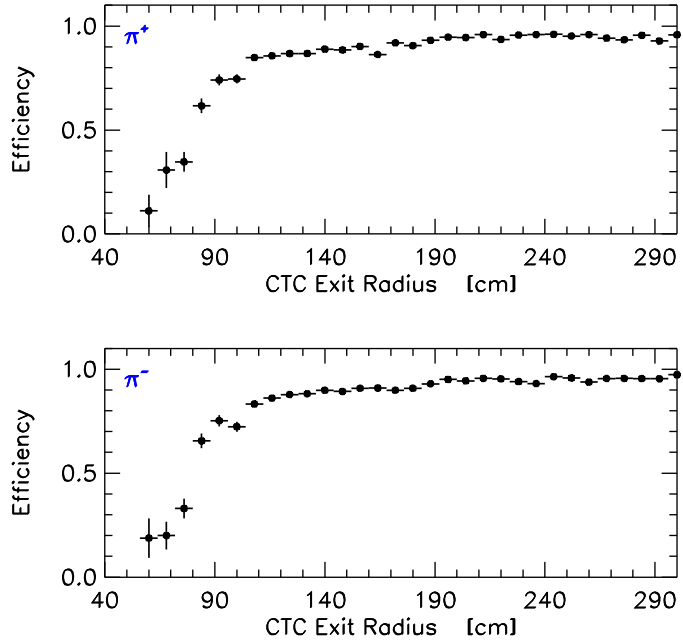}}
\end{center}
\caption
[Pattern recognition efficiency dependence on CTC exit radius.]
{The efficiency dependence on the CTC exit radius for $\pi^+$ (top)
and $\pi^-$ (bottom) tracks.}
\label{fig:eff_rctcx}
\end{figure}

Figure~\ref{fig:eff_rctcx} depicts the efficiency dependence on
$r^{\rm exit}_{\rm CTC}$ for both $\pi^+$ and $\pi^-$ tracks with no
$\left(r^{\rm exit}_{\rm CTC}\right)^{\rm min}$ criterion.  The shape
of the turn-on motivated the use of a $\left(r^{\rm exit}_{\rm
CTC}\right)^{\rm min}$ criterion less than the often-used 132-cm one,
which is conservative and corresponds to the outer edge of superlayer
8.  Figure~\ref{fig:eff_rctcx} suggests that a conservative criterion
of $\left(r^{\rm exit}_{\rm CTC}\right)^{\rm min} = 110$~cm was
appropriate for the provision of reliable tracking efficiency
information for muon tracks, without any significant compromise in the
CMX acceptance.  The remaining figures in this appendix (except the
left side of Figure~\ref{fig:stereo_kinematic}) have been prepared
using $\left(r^{\rm exit}_{\rm CTC}\right)^{\rm min} = 110$~cm, which
corresponds to a radius just outside the outer edge of superlayer 6.

\begin{figure}
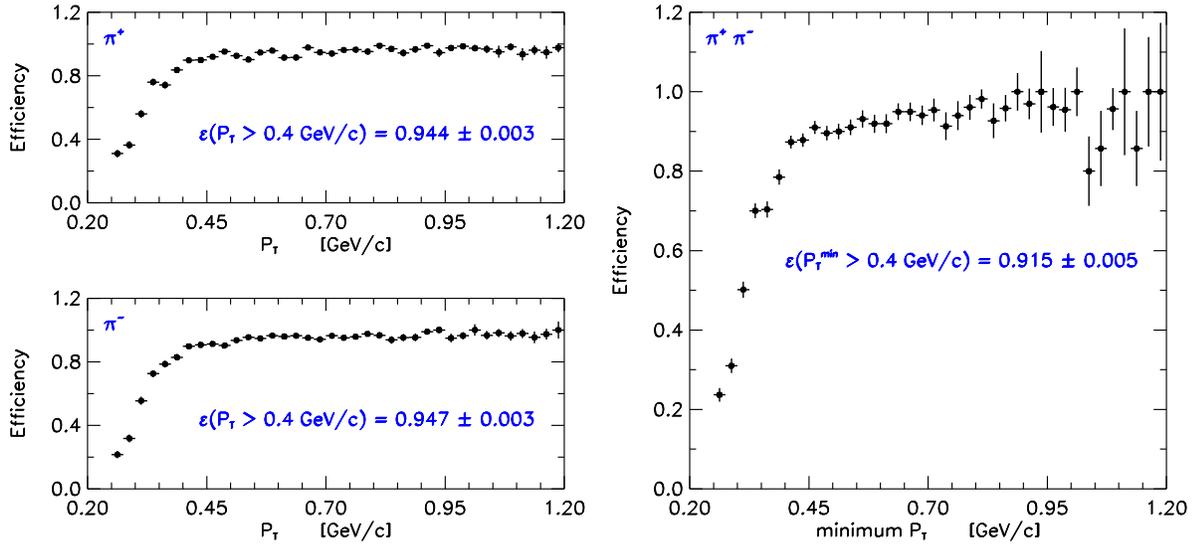

\begin{center}
\leavevmode
\hbox{%
\epsfxsize=3.0in
\epsffile{diagrams/eff_pt_1b_1_low_p_n.epsi}
\hspace{0.05in}
\epsfxsize=3.0in
\epsffile{diagrams/eff_pt_1b_1_low_pn.epsi}}
\end{center}
\caption
[Pattern recognition efficiency dependence on transverse momentum.]
{An example of the efficiency dependence on transverse momentum for
single tracks (left) and track pairs (right), in this case for the Run
1B ``Low'' run-range sample.  The indicated measured aggregate efficiencies
represent this run range alone; the uncertainties are statistical.}
\label{fig:eff_pt_1b_1_low}
\end{figure}

Figure~\ref{fig:eff_pt_1b_1_low} presents an example of the
$p_{\rm T}$-dependent pattern recognition efficiencies for single $\pi^+$
and $\pi^-$ tracks and $\pi^+\,\pi^-$ track pairs.  The data points
were calculated using Equation~\ref{eqn:eff_defn} with $p_{\rm T}^{\rm min}
= 0$ and $\left(r^{\rm exit}_{\rm CTC}\right)^{\rm min} = 110$~cm.
The indicated aggregate efficiencies were calculated using
Equation~\ref{eqn:tot_eff_defn}, but with $p_{\rm T}^{\rm min} =
0.4$~GeV/$c$; the uncertainties are statistical only.  Note that the
product of the two single-track efficiencies, $0.894 \pm 0.004$,
differs from the two-track efficiency by $\sim$3.3 standard
deviations, indicating that the pattern recognition efficiencies for two
tracks in a single event are correlated.  This observation is
discussed further in Section~\ref{sect:occupancy}.

\subsection{Efficiency Dependence on Occupancy-Related Observables}
\label{sect:occupancy}

\begin{figure}[p]
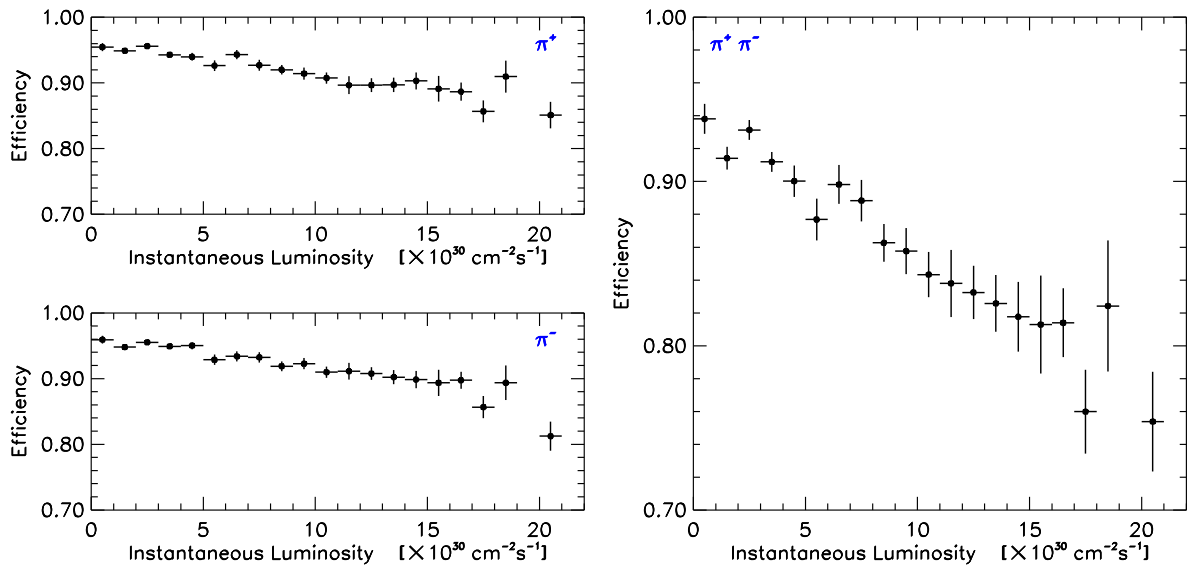

\begin{center}
\leavevmode
\hbox{%
\epsfxsize=3.0in
\epsffile{diagrams/eff_lum_p_n.epsi}
\hspace{0.05in}
\epsfxsize=3.0in
\epsffile{diagrams/eff_lum_pn.epsi}}
\end{center}
\caption
[Pattern recognition efficiency dependence on instantaneous luminosity.]
{The efficiency dependence on instantaneous luminosity for single
tracks (left) and track pairs (right).}
\label{fig:eff_lum}
\end{figure}

The effects of CTC occupancy on the pattern recognition efficiency
were traditionally contemplated in terms of instantaneous
luminosity dependence.  Figure~\ref{fig:eff_lum} shows plots of the
single- and two-track efficiencies as functions of instantaneous
luminosity.

\begin{figure}[p]
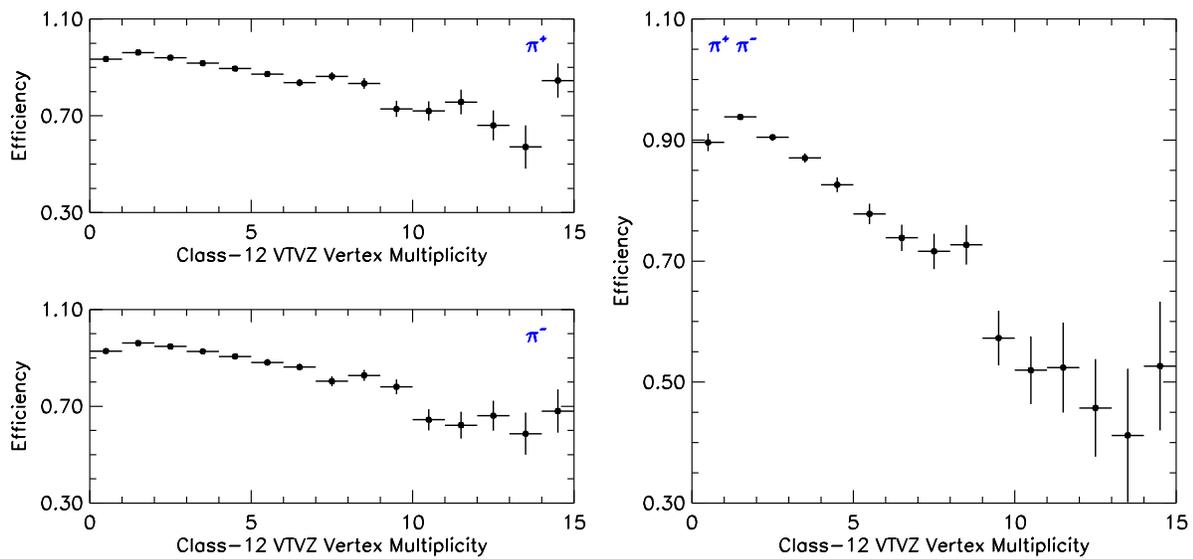

\begin{center}
\leavevmode
\hbox{%
\epsfxsize=3.0in
\epsffile{diagrams/eff_nvtz12_p_n.epsi}
\hspace{0.05in}
\epsfxsize=3.0in
\epsffile{diagrams/eff_nvtz12_pn.epsi}}
\end{center}
\caption
[Pattern recognition efficiency dependence on primary vertex multiplicity.]
{The efficiency dependence on multiplicity of Class-12 (high-quality)
vertices for single tracks (left) and track pairs (right).}
\label{fig:eff_nvtz12}
\end{figure}

For a given event, the instantaneous luminosity was only indirectly
related to the true track occupancy in the CTC fiducial volume.  A
more direct measure of occupancy is provided by the multiplicity of
high-quality primary vertices in an event.  In
Figure~\ref{fig:eff_nvtz12}, the single- and two-track efficiencies as
functions of the number of Class-12 vertices\footnote{The highest
quality $z$ vertex identifiable using the VTX was deemed a
`Class-12' vertex.  Class-12 vertices were defined as such when the
number of associated hits in the VTX exceeded 180.} are illustrated.
Note that the efficiency dependence on primary vertex multiplicity is
greater than that on instantaneous luminosity by more than a factor of
two.

\begin{figure}
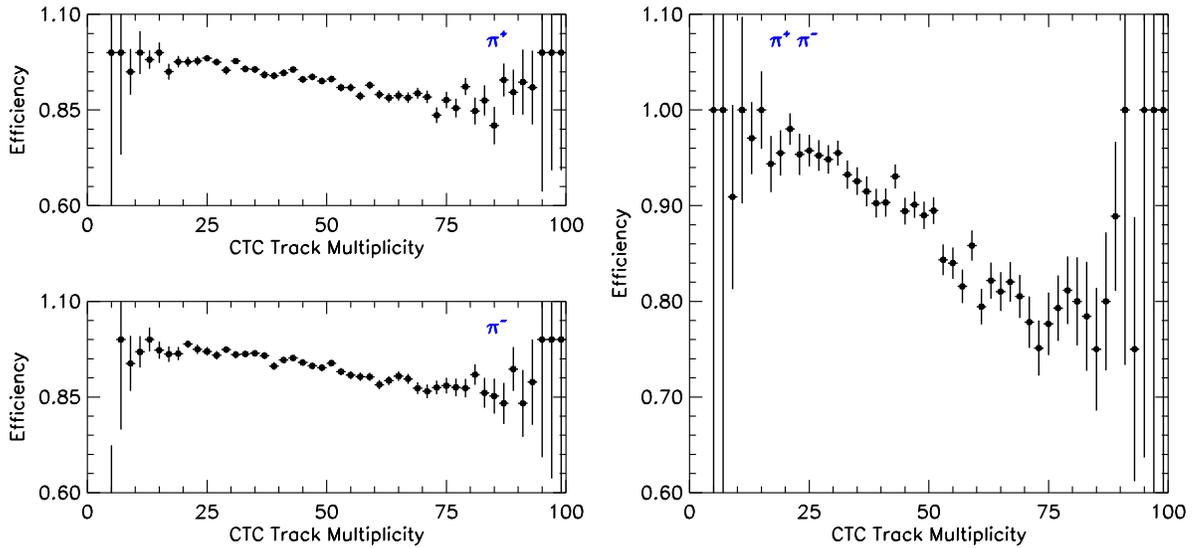

\begin{center}
\leavevmode
\hbox{%
\epsfxsize=3.0in
\epsffile{diagrams/eff_ntrks_p_n.epsi}
\hspace{0.05in}
\epsfxsize=3.0in
\epsffile{diagrams/eff_ntrks_pn.epsi}}
\end{center}
\caption
[Pattern recognition efficiency dependence on CTC-track multiplicity.]
{The efficiency dependence on multiplicity of CTC tracks for single
tracks (left) and track pairs (right).}
\label{fig:eff_ntrks}
\end{figure}

It is useful to examine the efficiency dependence on variables that
are considered to be even more correlated than primary vertex
multiplicity with the true occupancy of the CTC.  An example of one of
these is the total number of fitted CTC tracks in each event.  In
spite of the fact that the CTC track fits themselves used primary
$z$-vertex information, the CTC track multiplicity was expected to be
more correlated with the CTC occupancy because the track multiplicity
was measured in the CTC itself and, for example, would be largely
independent of those primary vertices that represented tracks at high
absolute pseudorapidities, outside the fiducial acceptance of the CTC.
Figure~\ref{fig:eff_ntrks} portrays the single- and double-track
efficiencies as functions of the number of CTC tracks.  Although the
distributions are limited by poor statistics at low and high
multiplicities, the efficiencies exhibit a significant dependence on
CTC track multiplicity.

\begin{figure}
\begin{center}
\leavevmode
\hbox{%
\epsfxsize=6.0in
\epsffile{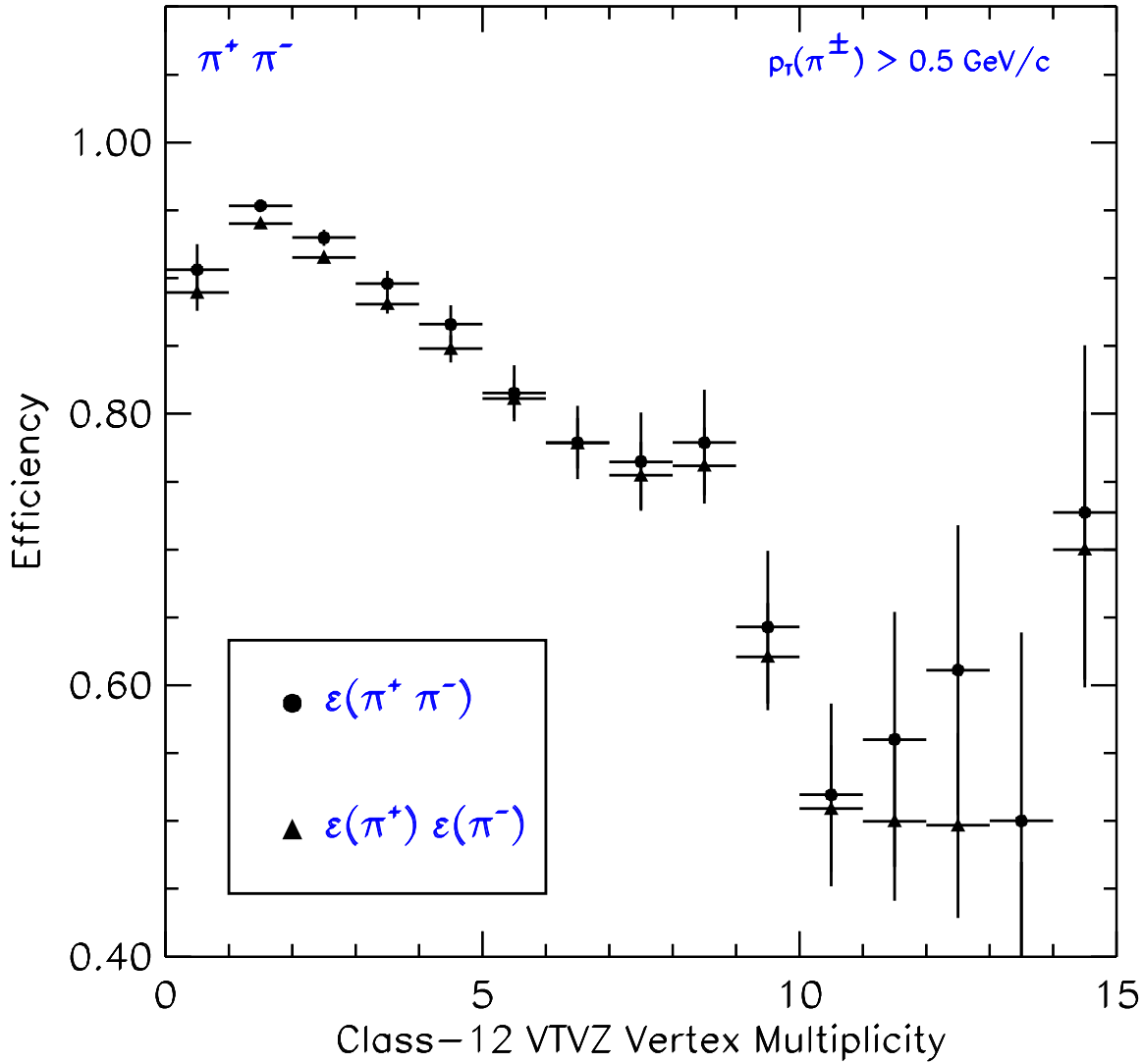}}
\end{center}
\caption
[Two-track efficiency correlations.]
{A comparison of the two-track efficiency and the product of
single-track efficiencies, as a function of primary vertex
multiplicity.  A tighter $p_{\rm T}(\pi^\pm) > 0.5$~GeV/$c$ criterion was
imposed to reduce the tracking differences between positively and
negatively charged tracks, but the exit radius cut was kept at $r^{\rm
exit}_{\rm CTC} > 110$~cm.  The indicated uncertainties are
statistical only.}
\label{fig:eff_nvtz12_pn_corr}
\end{figure}

The correlation between the pattern recognition efficiencies of two
tracks in a given event was discussed briefly in
Section~\ref{sect:kinematic}.  Figure~\ref{fig:eff_nvtz12_pn_corr}
compares the efficiency of a track pair, $\varepsilon(\pi^+\,\pi^-)$,
with the efficiency product of two individual tracks,
$\varepsilon(\pi^+)\,\varepsilon(\pi^-)$, as a function of Class-12
primary vertex multiplicity.  For the purposes of this figure, a more
restrictive $p_{\rm T}(\pi^\pm) > 0.5$~GeV/$c$ criterion was imposed to
reduce the tracking differences between positively and negatively
charged tracks; the exit radius cut, however, was kept at $r^{\rm
exit}_{\rm CTC} > 110$~cm.  The central values of the
$\varepsilon(\pi^+\,\pi^-)$ measurements were consistently greater
than those for the $\varepsilon(\pi^+)\,\varepsilon(\pi^-)$
measurements, but the two quantities were only statistically
inconsistent (by at least one standard deviation) in events with only
one or two Class-12 primary vertices.  In spite of the fact that the
events with only one or two Class-12 vertices constituted a major
fraction of the total number of events in the sample, the degree of
correlation between two tracks in a single event, while interesting,
was unobservable when systematic uncertainties were considered.

\subsection{Time-Dependent Effects}

Section~\ref{sect:track_effic} mentioned several time-dependent
factors that may have been detrimental to the CTC tracking
efficiencies in Run 1.  Time-dependent effects can be organized into
two broad categories: effects due to occupancy and those due to aging
and other sources.  Although the time dependence of the latter
category is typically studied in terms of run number, time-integrated
luminosity is a more meaningful quantity for this purpose.

It is apparent from the results presented in
Section~\ref{sect:occupancy} that the CTC pattern recognition
efficiencies depended strongly on the occupancy conditions inside the
CTC.  The results from Section~\ref{sect:occupancy} also suggest that
variables similar to primary-vertex and CTC-track multiplicity be used
to constrain occupancy levels in studies of run-dependent aging
effects in the CTC.

As outlined in Section~\ref{sect:embedding_procedure}, hits were
embedded into the CTC with wire efficiencies that were tuned from the
data.  The embedding techniques used in this study therefore did
account for some time-dependent performance effects (in addition to
occupancy), to the extent that the wire efficiencies used in the
embedding were derived from data quantities (the numbers of used hits
per superlayer) that were expected to be sensitive to time-dependent
performance effects in the CTC.  It is important to note, however,
that the present study cannot be expected to resolve {\it a
posteriori} any time-dependencies that affected the embedded hits
directly, since the wire efficiencies were calculated using
run-averaged effects and were administered uniformly for all the
events in the sample.  It should be emphasized that, with the
exception of the embedded hits, this technique can potentially resolve
all other time-dependent effects in each of the $\sim$36\,000 events.

\begin{figure}
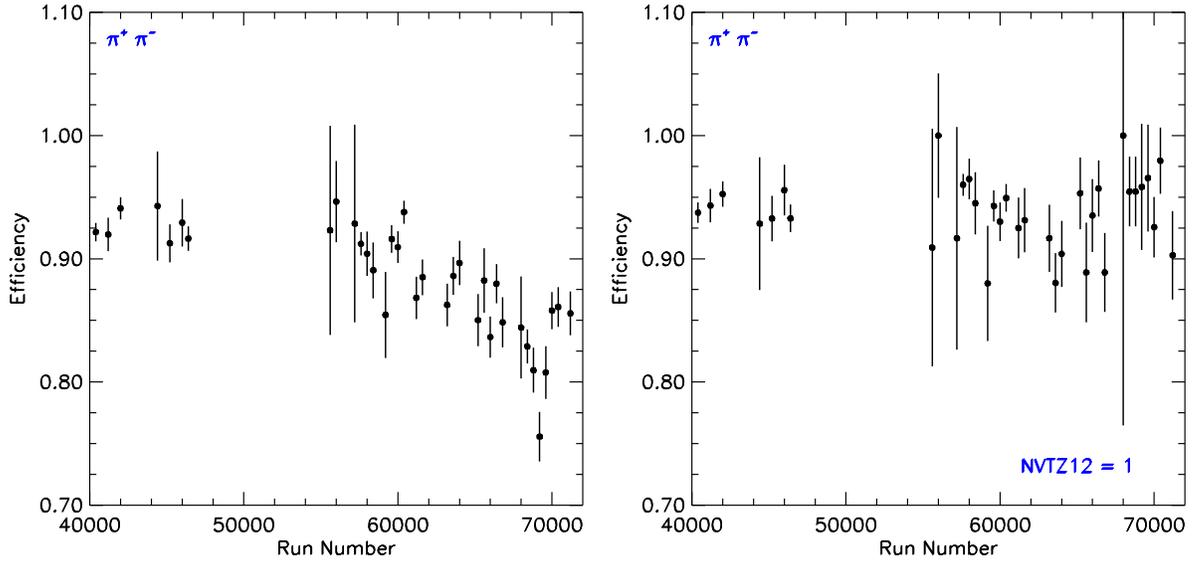

\begin{center}
\leavevmode
\hbox{%
\epsfxsize=3.0in
\epsffile{diagrams/eff_nrun.epsi}
\hspace{0.05in}
\epsfxsize=3.0in
\epsffile{diagrams/eff_nrun_1vtz12.epsi}}
\end{center}
\caption
[Dipion pattern recognition efficiency dependence on run number.]
{The dipion efficiency dependence on run number for all Class-12
primary vertex multiplicities (left) and for one and only one Class-12
primary vertex (right).}
\label{fig:eff_nrun}
\end{figure}

Figure~\ref{fig:eff_nrun} illustrates the dependence of the pattern
recognition efficiency on run number.  For the case where all primary
vertex multiplicities are permitted, a negative slope in the
distribution is discernible.  When the efficiency is examined in
events with only one Class-12 vertex, the slope is observed to vanish
within the available statistics.  The trends in
Figure~\ref{fig:eff_nrun} are consistent with the expectations of this
embedding technique and the conclusion that the most pernicious effect
on the pattern recognition is one of occupancy.

\subsection{Stereo Efficiency}
\label{sect:stereo}

\begin{figure}
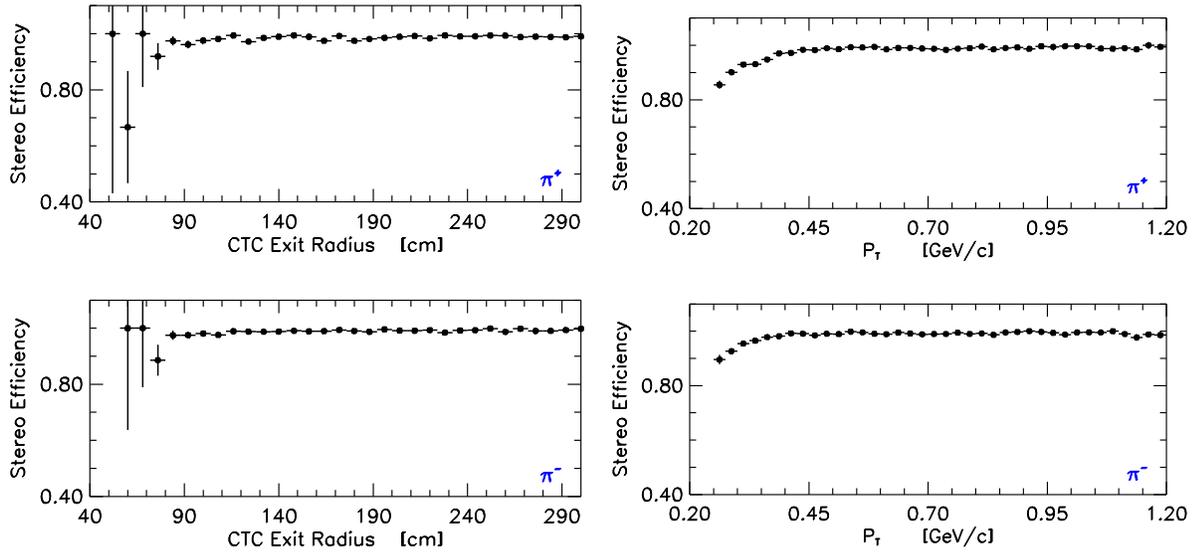

\begin{center}
\leavevmode
\hbox{%
\epsfxsize=3.0in
\epsffile{diagrams/stereo_eff_rctcx.epsi}
\hspace{0.05in}
\epsfxsize=3.0in
\epsffile{diagrams/stereo_eff_pt.epsi}}
\end{center}
\caption
[Kinematic dependencies of the stereo pattern recognition efficiency.]
{The stereo pattern recognition efficiency for $\pi^+$ (top) and
$\pi^-$ (bottom) tracks as a function of two kinematic observables:
(left) the CTC exit radius, $r^{\rm exit}_{\rm CTC}$, and (right) the
transverse momentum, $p_{\rm T}$.}
\label{fig:stereo_kinematic}
\end{figure}

\begin{figure}
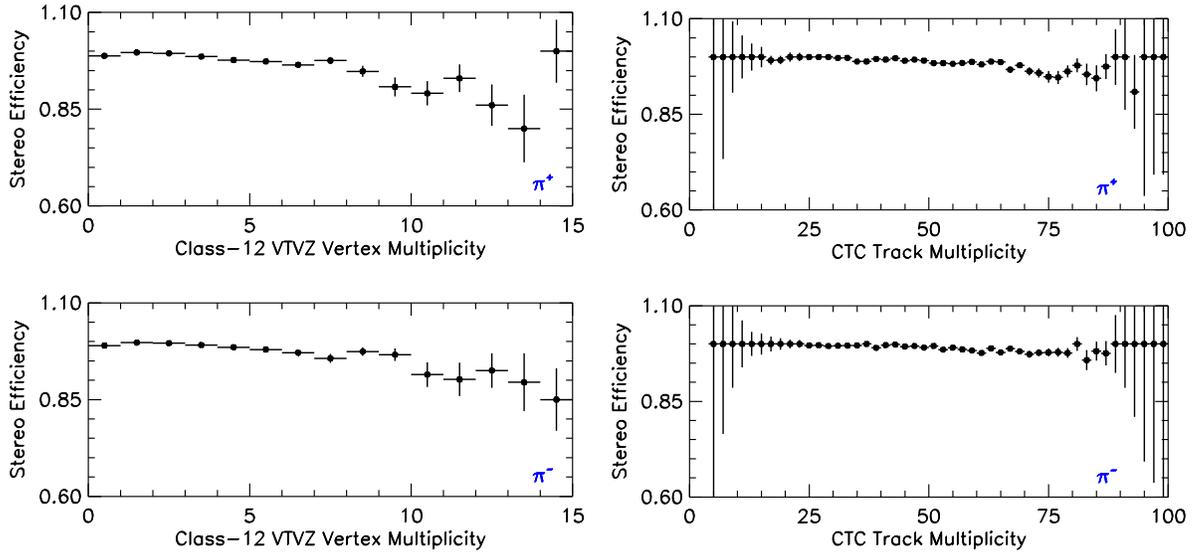

\begin{center}
\leavevmode
\hbox{%
\epsfxsize=3.0in
\epsffile{diagrams/stereo_eff_nvtz12.epsi}
\hspace{0.05in}
\epsfxsize=3.0in
\epsffile{diagrams/stereo_eff_ntrks.epsi}}
\end{center}
\caption
[Occupancy dependencies of the stereo pattern recognition efficiency.]
{The stereo pattern recognition efficiency for $\pi^+$ (top) and
$\pi^-$ (bottom) tracks as a function of two occupancy-related
observables: (left) the Class-12 {\sc vtvz} primary vertex
multiplicity and (right) the CTC track multiplicity.}
\label{fig:stereo_occupancy}
\end{figure}

Pursuant to the discussion in Section~\ref{sect:found} about relative
stereo and axial contributions to the tracking inefficiency, a {\it
stereo} pattern recognition efficiency was defined, also using
Equation~\ref{eqn:eff_defn}, but with the denominator populated by the
subsample of events that satisfied the 2-dimensional axial-only
matching criteria.  In this expression of the stereo efficiency,
$N_{\rm found}$ represented the number of events that, in addition to
meeting the axial-only criterion, satisfied the 3-dimensional matching
requirement.  Figure~\ref{fig:stereo_kinematic} shows plots of the
stereo pattern recognition efficiency as a function of two kinematic
observables: the CTC exit radius, $r^{\rm exit}_{\rm CTC}$, and the
transverse momentum, $p_{\rm T}$.  The stereo efficiency in
Figure~\ref{fig:stereo_kinematic} was observed to remain constant near
100\% in $r^{\rm exit}_{\rm CTC}$, but to diminish appreciably in
$p_{\rm T}$ for $p_{\rm T} < 0.4$~GeV/$c$.  The dependence of the stereo pattern
recognition efficiency on occupancy-related observables is illustrated
in Figure~\ref{fig:stereo_occupancy}, which features plots of the
stereo efficiency as a function of the primary-vertex and CTC-track
multiplicities.  The stereo pattern recognition efficiency appears to
drop by $\sim$10\% with increasing primary-vertex multiplicity and by
$\sim$5\% with increasing CTC-track multiplicity.  No significant
differences in stereo efficiencies between $\pi^+$ and $\pi^-$ tracks
are apparent in Figures~\ref{fig:stereo_kinematic} and
\ref{fig:stereo_occupancy}.

\section{Results}
\label{sect:results}

Single- and double-track pattern recognition efficiencies were
calculated for each of the four run-number bins using
Equation~\ref{eqn:tot_eff_defn} with $p_{\rm T}^{\rm min} = 0.4$~GeV/$c$.
Tables~\ref{tab:eff_3d_110} and
\ref{tab:eff_2d_110} list the results of these calculations for
$r^{\rm exit}_{\rm CTC} > 110$~cm with 3-dimensional and 2-dimensional
matching criteria, respectively.  Similarly,
Tables~\ref{tab:eff_3d_132} and \ref{tab:eff_2d_132} list the results
of the calculations for $r^{\rm exit}_{\rm CTC} > 132$~cm with
3-dimensional and 2-dimensional matching criteria, respectively.
Tables~\ref{tab:eff_tot_run1_110} and \ref{tab:eff_tot_run1_132}
summarize the aggregate Run 1 single- and double-track pattern
recognition efficiencies for $r^{\rm exit}_{\rm CTC} > 110$~cm and
$r^{\rm exit}_{\rm CTC} > 132$~cm, respectively.  The totals were
computed by calculating mean values weighted in time-integrated
luminosity (refer to Table~\ref{tab:runs}) and statistical
uncertainty.  For each aggregate efficiency, the mean statistical
uncertainty was added in quadrature with the systematic uncertainty,
which was taken to be equal to the maximum difference between the
tracking efficiencies calculated with full (100\%) wire hit
efficiencies and those calculated with reduced wire hit efficiencies.

\begin{table}[p]
\begin{center}
\begin{tabular}{|c|l||c|c|c|}	\hline
\multicolumn{2}{|c||}{3-D Efficiency} & $\varepsilon(\pi^+)$ &
	$\varepsilon(\pi^-)$ &
	$\varepsilon(\pi^+\,\pi^-)$ \\ \hline\hline
\multicolumn{2}{|c||}{Run 1A} & $.952\pm.003\pm.017$ &
			        $.956\pm.003\pm.018$ &
				$.924\pm.004\pm.029$ \\ \cline{1-2}
	& Low		      & $.944\pm.003\pm.016$ &
				$.947\pm.003\pm.014$ &
				$.915\pm.005\pm.023$ \\
Run 1B&   Medium	      & $.921\pm.004\pm.023$ &
				$.923\pm.003\pm.023$ &
				$.869\pm.006\pm.036$ \\
	& High		      & $.899\pm.004\pm.030$ &
				$.900\pm.004\pm.026$ &
				$.827\pm.006\pm.043$ \\ \hline
\end{tabular}
\end{center}
\caption
[3-D pattern recognition efficiencies for tracks with
$r^{\rm exit}_{\rm CTC} > 110$~cm.]
{The single- and double-track 3-dimensional pattern recognition
efficiencies measured in the four run-number bins for tracks with $p_{\rm T}
> 0.4$~GeV/$c$ and $r^{\rm exit}_{\rm CTC} > 110$~cm.  The first
uncertainties shown are statistical and the second are systematic.}
\label{tab:eff_3d_110}
\end{table}

\begin{table}[p]
\begin{center}
\begin{tabular}{|c|l||c|c|c|}	\hline
\multicolumn{2}{|c||}{2-D Efficiency} & $\varepsilon(\pi^+)$ &
	$\varepsilon(\pi^-)$ &
	$\varepsilon(\pi^+\,\pi^-)$ \\ \hline\hline
\multicolumn{2}{|c||}{Run 1A} & $.957\pm.003\pm.017$ &
			        $.960\pm.003\pm.017$ &
				$.931\pm.004\pm.030$ \\ \cline{1-2}
	& Low		      & $.949\pm.003\pm.016$ &
				$.952\pm.003\pm.013$ &
				$.922\pm.005\pm.023$ \\
Run 1B&   Medium	      & $.933\pm.003\pm.021$ &
				$.931\pm.003\pm.020$ &
				$.883\pm.005\pm.035$ \\
	& High		      & $.918\pm.004\pm.025$ &
				$.916\pm.004\pm.019$ &
				$.858\pm.006\pm.035$ \\ \hline
\end{tabular}
\end{center}
\caption
[2-D pattern recognition efficiencies for tracks with
$r^{\rm exit}_{\rm CTC} > 110$~cm.]
{The single- and double-track 2-dimensional pattern recognition
efficiencies measured in the four run-number bins for tracks with $p_{\rm T}
> 0.4$~GeV/$c$ and $r^{\rm exit}_{\rm CTC} > 110$~cm.  The first
uncertainties shown are statistical and the second are systematic.}
\label{tab:eff_2d_110}
\end{table}

\begin{table}[p]
\begin{center}
\begin{tabular}{|c|c||c|c|c|} \hline
Aggregate Efficiency & Matching & $\varepsilon(\pi^+)$ &
	$\varepsilon(\pi^-)$ & $\varepsilon(\pi^+\,\pi^-)$ \\ \hline\hline
Run 1B & 3-D & $.917\pm.030$ & $.920\pm.026$ & $.860\pm.043$ \\
only   & 2-D & $.931\pm.025$ & $.930\pm.020$ & $.880\pm.035$ \\ \hline
Runs   & 3-D & $.927\pm.030$ & $.928\pm.026$ & $.881\pm.043$ \\
1A + 1B& 2-D & $.937\pm.025$ & $.936\pm.020$ & $.895\pm.035$ \\ \hline
\end{tabular}
\end{center}
\caption
[Aggregate Run 1 efficiencies for tracks with
$r^{\rm exit}_{\rm CTC} > 110$~cm.]
{The total Run 1B and 1A + 1B single- and double-track pattern
recognition efficiencies, each calculated by combining the relevant
run-number bins into a single mean, weighted by the appropriate
statistical uncertainties and time-integrated luminosities.  Values
are given for both the 3-dimensional and 2-dimensional matching
techniques for tracks with $p_{\rm T} > 0.4$~GeV/$c$ and $r^{\rm exit}_{\rm
CTC} > 110$~cm, and the uncertainties are quadratic sums of the
statistical and systematic components.}
\label{tab:eff_tot_run1_110}
\end{table}

\begin{table}[p]
\begin{center}
\begin{tabular}{|c|l||c|c|c|}	\hline
\multicolumn{2}{|c||}{3-D Efficiency} & $\varepsilon(\pi^+)$ &
	$\varepsilon(\pi^-)$ &
	$\varepsilon(\pi^+\,\pi^-)$ \\ \hline\hline
\multicolumn{2}{|c||}{Run 1A} & $.959\pm.003\pm.015$ &
			        $.961\pm.003\pm.016$ &
				$.937\pm.004\pm.028$ \\ \cline{1-2}
	& Low		      & $.951\pm.003\pm.013$ &
				$.952\pm.003\pm.014$ &
				$.929\pm.005\pm.024$ \\
Run 1B&   Medium	      & $.929\pm.004\pm.021$ &
				$.931\pm.003\pm.021$ &
				$.887\pm.006\pm.032$ \\
	& High		      & $.906\pm.004\pm.029$ &
				$.907\pm.004\pm.023$ &
				$.849\pm.007\pm.040$ \\ \hline
\end{tabular}
\end{center}
\caption
[3-D pattern recognition efficiencies for tracks with
$r^{\rm exit}_{\rm CTC} > 132$~cm.]
{The single- and double-track 3-dimensional pattern recognition
efficiencies measured in the four run-number bins for tracks with $p_{\rm T}
> 0.4$~GeV/$c$ and $r^{\rm exit}_{\rm CTC} > 132$~cm.  The first
uncertainties shown are statistical and the second are systematic.}
\label{tab:eff_3d_132}
\end{table}

\begin{table}[p]
\begin{center}
\begin{tabular}{|c|l||c|c|c|}	\hline
\multicolumn{2}{|c||}{2-D Efficiency} & $\varepsilon(\pi^+)$ &
	$\varepsilon(\pi^-)$ &
	$\varepsilon(\pi^+\,\pi^-)$ \\ \hline\hline
\multicolumn{2}{|c||}{Run 1A} & $.963\pm.003\pm.015$ &
			        $.964\pm.003\pm.016$ &
				$.944\pm.004\pm.027$ \\ \cline{1-2}
	& Low		      & $.955\pm.003\pm.014$ &
				$.957\pm.003\pm.012$ &
				$.937\pm.004\pm.021$ \\
Run 1B&   Medium	      & $.942\pm.003\pm.017$ &
				$.938\pm.003\pm.019$ &
				$.900\pm.005\pm.032$ \\
	& High		      & $.926\pm.004\pm.022$ &
				$.923\pm.004\pm.016$ &
				$.881\pm.006\pm.031$ \\ \hline
\end{tabular}
\end{center}
\caption
[2-D pattern recognition efficiencies for tracks with
$r^{\rm exit}_{\rm CTC} > 132$~cm.]
{The single- and double-track 2-dimensional pattern recognition
efficiencies measured in the four run-number bins for tracks with $p_{\rm T}
> 0.4$~GeV/$c$ and $r^{\rm exit}_{\rm CTC} > 132$~cm.  The first
uncertainties shown are statistical and the second are systematic.}
\label{tab:eff_2d_132}
\end{table}

\begin{table}[p]
\begin{center}
\begin{tabular}{|c|c||c|c|c|} \hline
Aggregate Efficiency & Matching & $\varepsilon(\pi^+)$ &
	$\varepsilon(\pi^-)$ & $\varepsilon(\pi^+\,\pi^-)$ \\ \hline\hline
Run 1B & 3-D & $.925\pm.029$ & $.927\pm.023$ & $.882\pm.040$ \\
only   & 2-D & $.939\pm.022$ & $.936\pm.019$ & $.901\pm.032$ \\ \hline
Runs   & 3-D & $.934\pm.029$ & $.935\pm.023$ & $.901\pm.040$ \\
1A + 1B& 2-D & $.944\pm.022$ & $.942\pm.019$ & $.913\pm.032$ \\ \hline
\end{tabular}
\end{center}
\caption
[Aggregate Run 1 efficiencies for tracks with
$r^{\rm exit}_{\rm CTC} > 132$~cm.]
{The total Run 1B and 1A + 1B single- and double-track pattern
recognition efficiencies, each calculated by combining the relevant
run-number bins into a single mean, weighted by the appropriate
statistical uncertainties and time-integrated luminosities.  Values
are given for both the 3-dimensional and 2-dimensional matching
techniques for tracks with $p_{\rm T} > 0.4$~GeV/$c$ and $r^{\rm exit}_{\rm
CTC} > 132$~cm, and the uncertainties are quadratic sums of the
statistical and systematic components.}
\label{tab:eff_tot_run1_132}
\end{table}

\section{Conclusions}

This study of Run 1A and 1B single- and double-track pattern
recognition efficiencies yielded several qualitative and
quantitative conclusions:
\begin{enumerate}

\item To model accurately the CTC performance in Run 1, it was
necessary to employ reduced wire hit efficiencies in the track
embedding procedure.

\item Throughout this study, $\pi^+$ and $\pi^-$ tracks were
treated separately.  Within the statistical uncertainties alone, no
significant differences between efficiencies for positively and
negatively charged tracks were observed.

\item A CTC track exit radius criterion of $r^{\rm exit}_{\rm CTC} >
110$~cm ensured that the track was in a region of well-understood
pattern recognition efficiency without undue compromise to the CTC and
CMX fiducial acceptance (refer to Section~\ref{sect:kinematic}).

\item The deleterious effects of CTC occupancy on the pattern
recognition constituted the principal source of tracking inefficiency
in Run 1.

\item The pattern recognition efficiencies were studied for
single and double tracks as a function of three occupancy-related
observables, which were, in order of increasing correlation with the
true CTC occupancy, the instantaneous luminosity, the multiplicity of
high-quality primary vertices, and the multiplicity of reconstructed
tracks in the CTC (refer to Section~\ref{sect:occupancy}).  It was
recommended that investigations of non-occupancy time-dependent
effects be performed with constraints using variables that were as
closely correlated with the CTC occupancy as possible.

\item The pattern recognition efficiencies of two tracks within a single
event were correlated, a conclusion that followed from the observation
that these efficiencies were driven by the event-by-event
track-multiplicity environment conditions inside the CTC.  In the
context of primary-vertex multiplicity, the intertrack correlations
were only statistically significant at low primary-vertex
multiplicities.  Any double-track correlations became insignificant
when systematic uncertainties were included in the efficiency
measurements.

\item CTC tracking performance in the stereo view alone appeared to
contribute little to the observed pattern recognition inefficiencies.
The stereo efficiencies possessed a weak dependence on occupancy and a
strong dependence on $p_{\rm T}$ in the region $p_{\rm T} < 0.4$~GeV/$c$, which
lay outside the practical kinematic range of this investigation
(refer to Section~\ref{sect:stereo}).

\item Single- and double-track efficiencies were calculated in
2 dimensions (axial-only) and 3 dimensions (axial and stereo) in four
run-range bins spanning the Run 1A and 1B data-taking periods.
Measurements were presented for tracks with $p_{\rm T} > 0.4$~GeV/$c$ and
either $r^{\rm exit}_{\rm CTC} > 110$~cm or $r^{\rm exit}_{\rm CTC} >
132$~cm.  Aggregate Run 1 single- and double-track 3-dimensional and
2-dimensional pattern recognition efficiency measurements were also
computed for both the $r^{\rm exit}_{\rm CTC} > 110$-cm and
$r^{\rm exit}_{\rm CTC} > 132$-cm selection criteria.  Refer to
Tables~\ref{tab:eff_3d_110}, \ref{tab:eff_2d_110},
\ref{tab:eff_tot_run1_110}, \ref{tab:eff_3d_132}, \ref{tab:eff_2d_132},
and \ref{tab:eff_tot_run1_132} in Section~\ref{sect:results}.

\item The total 3-dimensional Run 1A + 1B single-track efficiency for
$p_{\rm T} > 0.4$~GeV/$c$ and $r^{\rm exit}_{\rm CTC} > 110$~cm was measured
to be $0.928 \pm 0.020$.

\item The total 3-dimensional Run 1A + 1B double-track efficiency for
$p_{\rm T} > 0.4$~GeV/$c$ and $r^{\rm exit}_{\rm CTC} > 110$~cm was measured
to be $0.881 \pm 0.043$.

\end{enumerate}

%% file: cdf_runi_authors.tex
\font\eightit=cmti8
\def\r#1{\ignorespaces $^{#1}$}
\hfilneg
\begin{sloppypar}
\noindent
F.~Abe,\r {17} H.~Akimoto,\r {39}
A.~Akopian,\r {31} M.~G.~Albrow,\r 7 A.~Amadon,\r 5 S.~R.~Amendolia,\r {27} 
D.~Amidei,\r {20} J.~Antos,\r {33} S.~Aota,\r {37}
G.~Apollinari,\r {31}
% ...
\end{sloppypar}
\vskip .026in
\begin{center}
(CDF Collaboration)
\end{center}

\vskip .026in
\begin{center}
\r 1  {\eightit Argonne National Laboratory, Argonne, Illinois 60439} \\
\r 2  {\eightit Istituto Nazionale di Fisica Nucleare, University of Bologna,
I-40127 Bologna, Italy} \\
\r 3  {\eightit Brandeis University, Waltham, Massachusetts 02254} \\
\r 4  {\eightit University of California at Los Angeles, Los 
Angeles, California  90024} \\  
\r 5  {\eightit University of Chicago, Chicago, Illinois 60637} \\
\r 6  {\eightit Duke University, Durham, North Carolina  27708} \\
\r 7  {\eightit Fermi National Accelerator Laboratory, Batavia, Illinois 
60510} \\
\r 8  {\eightit University of Florida, Gainesville, FL  32611} \\
\r 9  {\eightit Laboratori Nazionali di Frascati, Istituto Nazionale di Fisica
               Nucleare, I-00044 Frascati, Italy} \\
\r {10} {\eightit University of Geneva, CH-1211 Geneva 4, Switzerland} \\
\r {11} {\eightit Harvard University, Cambridge, Massachusetts 02138} \\
\r {12} {\eightit Hiroshima University, Higashi-Hiroshima 724, Japan} \\
\r {13} {\eightit University of Illinois, Urbana, Illinois 61801} \\
\r {14} {\eightit Institute of Particle Physics, McGill University, Montreal 
H3A 2T8, and University of Toronto,\\ Toronto M5S 1A7, Canada} \\
\r {15} {\eightit The Johns Hopkins University, Baltimore, Maryland 21218} \\
\r {16} {\eightit Institut f\"{u}r Experimentelle Kernphysik, 
Universit\"{a}t Karlsruhe, 76128 Karlsruhe, Germany} \\
\r {17} {\eightit National Laboratory for High Energy Physics (KEK), Tsukuba, 
Ibaraki 305, Japan} \\
\r {18} {\eightit Ernest Orlando Lawrence Berkeley National Laboratory, 
Berkeley, California 94720} \\
\r {19} {\eightit Massachusetts Institute of Technology, Cambridge,
Massachusetts  02139} \\   
\r {20} {\eightit University of Michigan, Ann Arbor, Michigan 48109} \\
\r {21} {\eightit Michigan State University, East Lansing, Michigan  48824} \\
\r {22} {\eightit University of New Mexico, Albuquerque, New Mexico 87131} \\
\r {23} {\eightit The Ohio State University, Columbus, OH 43210} \\
\r {24} {\eightit Osaka City University, Osaka 588, Japan} \\
\r {25} {\eightit Universita di Padova, Istituto Nazionale di Fisica 
          Nucleare, Sezione di Padova, I-36132 Padova, Italy} \\
\r {26} {\eightit University of Pennsylvania, Philadelphia, 
        Pennsylvania 19104} \\   
\r {27} {\eightit Istituto Nazionale di Fisica Nucleare, University and Scuola
               Normale Superiore of Pisa, I-56100 Pisa, Italy} \\
\r {28} {\eightit University of Pittsburgh, Pittsburgh, Pennsylvania 15260} \\
\r {29} {\eightit Purdue University, West Lafayette, Indiana 47907} \\
\r {30} {\eightit University of Rochester, Rochester, New York 14627} \\
\r {31} {\eightit Rockefeller University, New York, New York 10021} \\
\r {32} {\eightit Rutgers University, Piscataway, New Jersey 08855} \\
\r {33} {\eightit Academia Sinica, Taipei, Taiwan 11530, Republic of China} \\
\r {34} {\eightit Texas A\&M University, College Station, Texas 77843} \\
\r {35} {\eightit Texas Tech University, Lubbock, Texas 79409} \\
\r {36} {\eightit Istituto Nazionale di Fisica Nucleare, University of Trieste/
Udine, Italy} \\
\r {37} {\eightit University of Tsukuba, Tsukuba, Ibaraki 315, Japan} \\
\r {38} {\eightit Tufts University, Medford, Massachusetts 02155} \\
\r {39} {\eightit Waseda University, Tokyo 169, Japan} \\
\r {40} {\eightit University of Wisconsin, Madison, Wisconsin 53706} \\
\r {41} {\eightit Yale University, New Haven, Connecticut 06520} \\
\end{center}

%% file: arbeit.bbl
\begin{thebibliography}{100}
\bibitem{sganos:pc}
Some of the analysis codes used in the reconstruction of meson candidates in
  this study were adapted from routines prepared by G.~S.~Sganos.
\end{thebibliography}
